\documentclass[a4paper, 11pt]{book}
\usepackage{amsmath,amsfonts,amsthm,amssymb,stmaryrd}
\usepackage{graphicx,color}
 \usepackage{bbold}
\usepackage{mathrsfs}
\usepackage{setspace}
\usepackage{tikz}
\usepackage{comment}
\usepackage{appendix}
\usepackage[british,german]{babel}
\usepackage[active]{srcltx}
\usepackage{subfig}
\def\a{\alpha}
\def\b{\beta}
\def\g{\gamma}

\def\d{\delta}
\def\D{\Delta}
\def\e{\epsilon}

\def\z{\zeta}
\def\h{\eta}

\def\l{\lambda}
\def\L{\Lambda}
\def\m{\mu}
\def\n{\nu}
\def\x{\xi}

\def\s{\sigma}
\def\S{\Sigma}
\def\t{\tau}

\def\O{\Omega}

\def\cH{{\cal H}}

\def\cJ{{\cal J}}

\def\cL{{\cal L}}
\def\cM{{\cal M}}

\def\cP{{\cal P}}
\def\cQ{{\cal Q}}

\def\cS{{\cal S}}

\def\cV{{\cal V}}
\def\cW{{\cal W}}

\def\be{\begin{equation}}
\def\ee{\end{equation}}
\def\bali{\begin{align}}
\def\eali{\end{align}}
\def\bea{\begin{eqnarray}}
\def\eea{\end{eqnarray}}
\def\ba{\begin{array}}
\def\ea{\end{array}}
\def\ben{\begin{enumerate}}
\def\een{\end{enumerate}}
\def\bi{\begin{itemize}}
\def\ei{\end{itemize}}

\def\eal{\end{align}}

\def\bra{\langle }
\def\ket{ \rangle}

\def\12{\frac{1}{2}}

\def\ra{\rightarrow}

\def\Lra{\Longrightarrow}

\newcommand{\CFT}{{\small{CFT }}}
\newcommand{\WZW}{{\small{WZW }}}

\newcommand{\SCA}{{\small{SCA }}}
\newcommand{\MM}[1]{\text{MM$_{#1}$}}

\newcommand{\N}{\mathbb{N}}
\newcommand{\Z}{\mathbb{Z}}

\newcommand{\R}{\mathbb{R}}
\newcommand{\C}{\mathbb{C}}

\newcommand{\id}{\mathbb{1}}

\newcommand{\Hilb}{\mathcal{H}}
\newcommand{\T}{\mathbb{T}}

\def\ibra{\langle \!\langle}
\def\iket{ \rangle\!\rangle}
\def\sgn{\mathop{\rm sgn}}
\def\Tr{\text{Tr}}

\newcommand{\de}{\partial}
\newcommand{\debar}{\bar{\partial}}

\newcommand{\order}[1]{\mathcal{O}\!\left( #1 \right)}
\newcommand{\integer}[1]{\lfloor #1\rfloor}
\newcommand{\fract}[1]{\left\{ #1\right\}}
\newcommand{\wzw}[3]{\widehat{#1}(#2)_{#3}}
\newcommand{\kmalg}[1]{\widehat{ #1}}
\newcommand{\torus}[2]{\text{{\scriptsize{$#1$}}}\,\,\! \underset{#2}
{
\begin{array}{|c|}
\hline \hspace{7pt} \\[3pt] 
\hline 
\end{array}
}
}

\def\={\overset{!}{=}}

\begin{document}
\thispagestyle{empty}
\pagenumbering{roman}

{\large

\begin{center}\parindent0pt



\vspace*{0cm}

%
{\LARGE \bf Limit theories and continuous orbifolds}\medskip\\[30pt]


{\large D i s s e r t a t i o n}\\[25 pt]
zur Erlangung des akademischen Grades\\[10pt]
d o c t o r\quad r e r u m\quad n a t u r a l i u m\\[10pt]
(Dr. rer. nat.)\\[10pt]
im Fach Physik\\[10pt]
eingereicht an der\\[20pt]
Mathematisch-Naturwissenschaftlichen Fakult\"at I\\[10pt]
der Humboldt-Universit\"at zu Berlin\\[10pt]
von\\[20pt]
Cosimo Restuccia\\[1 pt]
geb.\ am 02.02.1984 in Florenz (Italien)\\[20pt]
Pr\"asident der Humboldt-Universit\"at zu Berlin:\\[1pt]
Prof.~Dr.~Jan-Hendrik Olbertz\\[20pt]
Dekan der Mathematisch-Naturwissenschaftlichen Fakult\"at I:\\[1pt]
Prof.~Stefan Hecht,~Ph.D.\\[40pt]
\begin{flushleft}
\begin{tabular}{lll}  
Gutachter/innen: 
\quad&1.\quad Prof.~Dr.~Hermann Nicolai& 
\\[10pt]
 &2.\quad  Prof.~Dr.~Volker Schomerus& 
 \\[10pt]
 &3.\quad  Prof.~Dr.~Matthias Staudacher& 
\\[30pt]
 \multicolumn{3}{l}{
Tag der m\"undlichen Pr\"ufung:\quad 5.~Juli 2013
}
 \end{tabular}
\end{flushleft}

\end{center}

}
\thispagestyle{empty}

\newpage
\thispagestyle{empty}
\begin{center}
\phantom{physik macht spass}
\end{center}
Typeset in \LaTeX
\vskip 23cm
\newpage

\selectlanguage{british}

\thispagestyle{empty}

\begin{center}
\section*{Abstract}
\end{center}



\vspace{1cm}

Relativistic quantum field theories are in general defined by a collection of effective actions, describing the dynamics of quantum fields at different energy scales.
The consequent natural idea of a space of theories is still nowadays a rather imprecise notion, since a detailed knowledge or a classification of quantum field theories is out of reach.
In two space-time dimensions the situation is in a better shape: in this context conformal field theories are under control in many instances, and we know sequences of rational theories emerging as end-points of renormalisation group flows.

The present thesis explores the behaviour of sequences of rational two-dimensional conformal field theories when the central charge approaches its supremum.
In particular, after a review of various notions useful to study limit theories, like the concept of averaged fields and the construction of continuous orbifolds, we analyse in detail the limit of sequences of $N=2$ supersymmetric conformal field theories that are connected by renormalisation group flows.

Remarkably, as we show explicitly for $N=2$ minimal models in the $c\to 3$ regime, the limit is not unique, since with extended Virasoro symmetry one can choose different scalings for the labels of the spectrum in the limit.
We construct explicitly two conformal field theories emerging as the large level limit of minimal models, and we identify both of them: one is the $N=2$ superconformal field theory of two uncompactified free real bosons and two free real fermions, the other is its continuous orbifold by $U(1)$.
We compare spectrum, torus partition function, correlators and boundary conditions.
The neatest interpretation of this result is given by studying the realisation of $N=2$ minimal models as gauged Wess Zumino Witten models: taking the two different limits amounts to zooming into two different regions of the target-space geometry.

We furthermore conjecture that one of the limit theories emerging from the limit of $su(n+1)$ Kazama-Suzuki models ($n=1$ being the minimal models case) is a $U(n)$ continuous orbifold of a free theory.
We motivate this idea by comparing the boundary spectra.

At the end we speculate about the possible extensions and generalisations of the idea presented in this thesis.

\vspace{2cm}
\subsubsection*{Keywords:}
%

%
Two-dimensional conformal field theories, string theory, supersymmetric minimal models, continuous orbifolds

\newpage
\thispagestyle{empty}
\begin{center}
\phantom{physik macht spass}
\end{center}

\selectlanguage{german}
\newpage
\thispagestyle{empty}

\begin{center}
\section*{Zusammenfassung}
\end{center}
\vspace{1cm}

\vspace{0.8cm}

Relativistische Quantenfeldtheorien sind im Allgemeinen durch eine Vielzahl effektiver Wirkungen definiert. 
Diese beschreiben die Dynamik der Quantenfelder bei gewissen Energieskalen.
Die daraus folgende Idee von einem Raum von Theorien ist heutzutage immer noch eine eher unpr\"azise Bezeichnung, da eine genaue Kenntnis oder Klassifizierung von Quantenfeldtheorien sich au\ss er Reichweite befindet.

In zwei Dimensionen ist die Situation vorteilhafter: hier sind konforme Feldtheorien in vielen F\"allen besser verstanden, und wir kennen Folgen von rationalen Theorien, die durch Renormierungsgruppenfl\"ussen verbunden sind.

Diese Dissertation untersucht das Verhalten von Folgen rationaler zweidimensionaler konformer Feldtheorien wenn sich die zentrale Ladung ihrem Supremum n\"ahert.

Nach einer Diskussion von verschiedenen, sich f\"ur das Studium von Grenzwerttheorien 
n\"utzlich erweisenden Aspekten wie das Konzept gemittelter Felder und der Konstruktion von kontinuierlichen Orbifolds, analysieren wir im Detail den Grenzwert einer Folge von $N=2$ supersymmetrischen konformen Feldtheorien.
Wir zeigen explizit f\"ur $ N=2$ minimale Modelle im Limes $c \to 3$, dass bemerkenswerterweise der Grenzwert nicht eindeutig ist, da mit erweiterter Virasoro-Symmetrie verschiedene Skalierungen f\"ur die Quantenzahlen des Spektrums im Grenzwert gew\"ahlt werden k\"onnen.
Wir konstruieren explizit zwei konforme Feldtheorien, die aus dem Grenzwert f\"ur gro\ss e Level der minimalen Modelle hervorgehen, und identifizieren beide: Eine ist die $N=2$ superkonforme Feldtheorie von zwei nichtkompaktifizierten freien reellen Bosonen und zwei freien reellen Fermionen, die andere ist dessen kontinuierlicher $U(1)$-Orbifold.
Wir vergleichen Spektrum, Torus-Zustandsummen, Korrelatoren und Randbedingungen.
Die eing\"angigste Interpretation dieses Ergebnisses wird die Realisierung von $N=2$ Minimalen Modellen als geeichte Wess-Zumino-Witten-Modelle gegeben: das Auftauchen zweier verschiedener Grenzwerte kann als das Hineinzoomen in zwei verschiedene Regionen der Geometrie im Zielraum interpretiert werden.

Desweiteren vermuten wir, dass eine der Grenzwerttheorien, die aus dem Grenzwert von $su(n+1)$ Kazama-Suzuki Modellen ($n=1$ ist der Fall f\"ur Minimale Modelle) ein kontinuierlicher $U(n)$-Orbifold einer freien Theorie ist.
Wir motivieren diese Idee durch den Vergleich der Randspektren.

Am Ende spekulieren wir \"uber die M\"oglichkeit des Auftretens von kontinuierlichen Orbifolds in unterschiedlichen F\"allen.

\vspace{1.4cm}
\subsubsection*{Schlagw\"orter:}
%

%
Zweidimensionale Konforme Feldtheorien, Stringtheorie, Supersymmetrische Minimale Modelle, Kontinuierliche Orbifolds

\newpage
\thispagestyle{empty}
\begin{center}
\phantom{physik macht spass}
\end{center}
\selectlanguage{british}

\tableofcontents
\newpage
%
\pagenumbering{arabic}
\chapter*{Introduction}\markboth{INTRODUCTION}{INTRODUCTION}
\addcontentsline{toc}{chapter}{Introduction}

The discovery of a Higgs-like boson at CERN last year has produced a renewed excitement in the world of subatomic physics. 
Once again the Standard Model of particle physics has shown its merits in forecasting the existence of new fundamental particles.
The conceptual and mathematical scheme on which the Standard Model is based is the framework of relativistic quantum field theories (QFTs).

Since the early days of their (roughly) 70 years long history, QFTs have been plagued by infinities, ubiquitously appearing as outcome of computations of observables.
The modern approach to deal conceptually with this problem, pioneered by Wilson~\cite{Wilson:1974mb}, consists in looking at QFTs as families of effective actions. 
At a fixed energy (or length) scale, the field modes of a QFT that are more energetic than the scale are unlikely to be excited, but crucially modify the form of the action, and therefore also the dynamics of the modes with energies comparable with the scale.

This point of view suggests the picture of a space of actions, and the speculative idea that a given theory (a description of a physical system) manifests itself as a trajectory in this space, by going from higher to lower energy scales.

Unfortunately, this concept is nowadays still rather heuristic, since we do not have a proper classification of QFTs, or a general definition of a metric and of a topology for a space of quantum theories~\cite{Douglas:2010ic}.

In this introduction we describe the point of view of our work in this ideal framework.
This thesis is devoted to explore the limits of sequences of $N=2$ supersymmetric two dimensional conformal field theories.
We start by briefly illustrating the importance of renormalisation group flows, specifically in two dimensions, where they can be explored more efficiently.
In this context we stress the importance of sequences of rational two-dimensional conformal field theories, and of their limits, also for string theory, where naturally the notion of $N=2$ supersymmetric conformal field theories appears.

\section*{Limits of two-dimensional conformal field theories}

\subparagraph{RG-flows}
Given a generic QFT, we can imagine to concentrate the attention only on excitations with wavelength much shorter than all the scales present, or, using a common jargon, on the ultraviolet behaviour of the theory: if there exists a consistent QFT at very high energy (and this is not granted in general) the effective action must be scale invariant.
In contrast with that, we can imagine to excite the system with waves much longer than any typical scale present in the action, namely to look at the infrared behaviour of the theory: In this other regime the system is too small to feel our perturbation, and the effective action is again scale invariant.
If an ultraviolet QFT exists, we can therefore imagine the system to be described as an interpolation between two scale invariant QFTs.

The flow in the space of actions for a QFT from a short wave length regime, to a longer one is called renormalisation group (RG) flow.
RG-flows are not necessarily continuous, neither bounded.
It is in general out of reach to concretely follow the flow in the space of actions, namely to explicitly write down the effective action corresponding to a precise point, since the degrees of freedom tend to reorganise themselves in a non-trivial fashion along the flow, and their dynamical properties also tend to change drastically.

The only analytic way we know to investigate the flow is to start from a well-known QFT, and to perturb it: the flow can be explored in the vicinity of the starting point, as a perturbative expansion.

\subparagraph{Conformal field theories}

Quantum scale invariant theories are prominent points of the trajectories in the space of actions: they represent fixed points of the RG-flows.

Scale transformations represent a subset of a larger group, called conformal group, which in addition to scale transformations includes translations, rotations and the so-called special conformal transformations: the important class of scale invariant theories, based on the conformal group are called~\textit{conformal field theories}.
These QFTs are in general easier to handle than others, since the symmetry algebra they possess is very large.
This feature opens the possibility to study RG-flows in a perturbative way starting from an ultraviolet conformal field theory (this technology is called conformal perturbation theory): as in ordinary perturbation theory one treats interactions as small correction over the known and solved free theory, in this case one perturbs the (supposedly known and solved) conformal field theory by small modifications in the action that bring the theory away from the conformal point. 
This is a way to provide a concrete perturbative description, in terms of a conformal action modified by some relevant operators, of theories which are perturbatively inaccessible starting from a free theory.

This whole program is still out of reach for $(3+1)$-dimensional theories, since we still cannot (completely) solve any non-free four-dimensional QFT.
Nevertheless in recent years there has been some progress towards a general better understanding of flows in four dimensions (for example the very interesting work of Luty, Polchinski and Rattazzi~\cite{Luty:2012ww}), and tremendous steps have been taken towards a complete understanding of the $N=4$ superconformal Yang-Mills theory (for an extensive recent review see~\cite{Beisert:2010jr}), bringing hope that the programme of conformal perturbation theory in four dimensions could progress much further during the next decades.
But this is not our concern in this thesis.

\subparagraph{Two dimensions}

In two dimensions the situation is in a much better shape: our focus in this thesis is on two-dimensional conformal field theories\footnote{Extensive accounts for this very vast and intriguing subject are for instance~\cite{Ginsparg:1988ui,DiFrancesco:1997nk,Blumenhagen:2009zz}.} (which will be called CFTs from here on), which are arguably the best understood interacting QFTs that one can study.
In two dimensions the conformal group becomes infinite dimensional, making conformal theories so constrained, to yield, in some lucky instances, complete solvability.
Furthermore, conformal invariance in two dimensions is equivalent to scale invariance~\cite{Zamolodchikov:1986gt,Polchinski:1987dy}.
This means that RG-flows in two dimensions naturally interpolate between CFTs.

Since the first attempts to formulate CFTs the rich symmetry structure has been exploited to constrain spectra and correlators~\cite{Belavin:1984vu}: the basic infinite dimensional symmetry characterising a CFT is the one of the modes of the energy momentum tensor, the so-called Virasoro algebra, characterised by a central element called the central charge~$c$.
It turns out that imposing unitarity the central charge cannot be negative, and can take only a countable set of values if $0<c<1$.
Moreover, the number of unitary irreducible representations is finite for any value of $c$ in this range; correlators of fields associated with these representations can be determined explicitly.
The theories realised in this way are the celebrated unitary Virasoro minimal models, which have been object of deep investigation and are nowadays completely under control: we are able to compute every correlator explicitly (with the due patience, as explained in detail in~\cite{DiFrancesco:1997nk}).
Virasoro minimal models come in a discrete sequence, of increasing central charges from $c=1/2$ to $c$ infinitesimally close to 1.

\subparagraph{RG-flows between minimal models}
Our interest in these models is driven by two profound results by Alexander Zamolodchikov.

The first one is his famous $c$-theorem~\cite{Zamolodchikov:1986gt}, which states that following every RG-flow in two dimensions, there is an always decreasing quantity defined in terms of the energy momentum tensor of the theory, which equates the central charge $c$ when the flow hits conformal points.
As a consequence, ultraviolet CFTs have larger central charge than any CFT obtained as an RG-flow starting from them\footnote{As an aside we mention here that very recently Komargodski and Schwimmer have proposed a proof of the so-called $a$-theorem~\cite{Komargodski:2011vj,Komargodski:2011xv}, a four dimensional analogue of the $c$-theorem of Zamolodchikov. This important result opens new avenues for the study of RG-flows in four-dimensional theories.}.

The second one is the result coming from the detailed first order conformal perturbation theory in the vicinity of minimal models~\cite{Zamolodchikov:1987ti}: for large enough central charge (still smaller than~1), choosing a suitable relevant perturbation among the representations of the minimal models themselves, these CFTs are all connected by RG-flows.
This is an explicit treatable exploration of the space of theories; of a very restrict class of theories of course (perturbed unitary Virasoro minimal models), but still it is an encouraging result in trying to speculate further.

It is worth mentioning here that, very recently, new avenues for the study if RG-flows between CFTs have been opened, with the explicit construction of Gaiotto~\cite{Gaiotto:2012np} of a conformal interface\footnote{For a general review on the fascinating topic of conformal interfaces see~\cite{Bachas:2008jd}.} (whose existence had been long conjectured~\cite{Brunner:2007qu}), which seems to correctly interpolate between minimal models separated by RG-flows (such constructions seem to be possible also in three dimensions~\cite{Billo:2013jda}).
The ideas used by Gaiotto involve the interesting boundary CFT construction of generalised permutation branes~\cite{Fredenhagen:2005an,Fredenhagen:2006qw,Fredenhagen:2009hx}.

\subparagraph{Limits of sequences}
Now that we have a sequence in the space of theories, a natural question to ask is whether there exists a notion of convergence.
The answer is positive in this case: the sequence of Virasoro minimal models converges to a specific $c=1$ interacting CFT if one scales appropriately the representation labels, as firstly described by Runkel and Watts~\cite{Runkel:2001ng}.
The choice of the scaling function is crucial, as shown by Roggenkamp and Wendland in~\cite{Roggenkamp:2003qp}, where a different $c=1$ limit theory could be obtained by scaling differently the representations: this apparent non-uniqueness of the limit can be understood by looking at the large radius limit of the free boson on a circle (as explained in detail in chapter~\ref{ch:limit-theories}): we can choose to keep all the modes finite while sending the radius to infinity, or to scale the Kaluza-Klein modes with the radius. 
The first choice is the Roggenkamp-Wendland analogue, while the second one corresponds to the limit of Runkel and Watts.

Schomerus has furthermore shown~\cite{Schomerus:2003vv} that the theory of Runkel and Watts is a common limit of minimal models and Liouville theory\footnote{The same considerations hold for $N=1$ supersymmetric minimal models~\cite{Fredenhagen:2007tk}, and recently were extended for $W_N$ minimal models as well: in this last example the structures of $A_{N-1}$ Toda theory in the $c=N-1$ case were matched with the limit of $W_N$ minimal models~\cite{Fredenhagen:2010zh}.}.
 
Runkel-Watts (RW) theory  has appeared to be a mysterious but consistent $c=1$ CFT, escaping any classification scheme for more than a decade. 
Its relationship with Roggenkamp-Wendland theory has not been clear for long time as well.
Until very recently, when Gaberdiel and Suchanek~\cite{Gaberdiel:2011aa} robustly argued that RW theory admits a natural interpretation in terms of the twisted sector of a continuous orbifold of a free theory, whence Roggenkamp-Wendland limit is given by the untwisted sector.
As extensively explained in chapter~\ref{ch:orbifolds}, an orbifold of a parent CFT is a CFT, whose fields take values on a space obtained by modding out the action of a group on the image of the parent fields.

To appreciate this proposal, and other interesting points of view, we need now a detour into string theory.

\section*{The stringy picture}
\subparagraph{CFTs in string theory}
String theory is a physical framework based on the idea of substituting the elementary particles of relativistic QFTs, with quantum relativistic one-dimensional laces of energy (strings), fluctuating in space-time\footnote{\label{intro:footnote:reference-string-theory}For extensive accounts on string theory, the reader can consult any of the very good textbooks on the market. We used mostly~\cite{GreenSchwarzWittenBookI:1987,GreenSchwarzWittenBookII:1987},~\cite{PolchinskiBookI:1998,PolchinskiBookII:1998} and the very recent~\cite{Blumenhagen:2013fgp}.}.

The different quantised vibrational modes of the strings represent different particles, whose interactions are constrained by the way in which strings can join and split.
This simple idea has great unifying power, which comes from the different nature of open and closed string amplitudes: the former realise interactions of gauge fields, the latter interactions of (among others) gravitational waves.

As the motion of a particle in a $d$-dimensional space-time can be classically described by an embedding of a (time) interval into a $d$-dimensional space, the motion of a string in space-time is classically given by an embedding of a two-dimensional surface (the world-sheet) into a $d$-dimensional space (the target space).
Each of the coordinates is a function defined on a two-dimensional space, namely a classical field living in two dimensions: in the formulation of Polyakov,
the world-sheet theory is (after bringing the metric in the so-called conformal gauge) classically conformal.
The string scattering matrix is obtained as the path integral over the coordinates (which are fields defined in two dimensions) supplemented by the integration over all metrics one can define on a two-dimensional Riemann surface, which can be classified by their topology and their moduli spaces.
The amplitudes are therefore a sum of CFT correlators over Riemann surfaces of increasing genera.
CFT fields have values representing the string's coordinates embedded in the target space.

Also the boundaries of the two-dimensional world-sheet have their string-theoretical interpretation: the boundary conditions on the fields of the CFT can be formulated by specifying a submanifold in the target space (a D-brane), that encodes the possible boundary values of the fields.
Open strings have their endpoints attached to D-branes, and their fluctuations are confined on the world-volume (the multidimensional analogue of the world-sheet) of the brane.

To summarise in one slogan, CFTs are the building blocks of string theory: the type of strings, the geometry of the space-time in which they propagate (closed string modes contain gravitons), and the presence of gauge fields on D-branes (open string modes contain gauge fields) are encoded in the CFTs and in their boundary theory. 
From the CFT perspective, D-branes are constructed from boundary conditions, by imposing gluing conditions on the chiral holomorphic and anti-holomorphic currents on the boundary (for a brief introduction to this topic see appendix~\ref{app:vanilla}. 
For the more demanding reader we refer to the lecture notes~\cite{Schomerus:2002dc}).
The concrete computation of the perturbative open string spectrum is based on world-sheet duality, which relates one-loop open string amplitudes to tree-level closed string exchange between special coherent states that couple to the bulk, called boundary states.

The (continuous) orbifold CFT mentioned before describes a string moving in a background obtained by identifying the points described by the parent CFT, under the action of the orbifold (continuous) group.

\subparagraph{SUSY}
The necessity coming from string theory of modelling fermionic fields in the target space (and the interest alone in defining new CFTs as well), leads to the definition of CFTs of fermionic fields.
The theory constructed defining on the same world-sheet a free bosonic CFT with a free fermionic one possesses a remarkable new feature: supersymmetry, the symmetry under transformations that map the boson into the fermion and vice versa.
Supersymmetric string theories based on supersymmetric CFTs are more appealing than bosonic string theories for many reasons, which we do not discuss here.
We only want to mention that closed string backgrounds generated by theories based only on the Virasoro algebra are doomed to decay: the simplest mode of vibration of both closed and open strings is a tachyon, a particle of negative squared mass, which signals an instability of the ground states~\cite{Sen:1998sm}.
Five perturbative superstring theories do not have tachyons in the spectrum, making stable the background they move in.
In this thesis we consider superstrings based on $N=2$ extended supersymmetric CFTs, which are the basic ingredient for type II (and type 0) strings, and whose knowledge helps in the construction of heterotic strings as well.

We cannot resist mentioning here that (super)string theory is the most prominent candidate for a unifying theory of quantum gravity, since it is to date the only conceptual framework we know to handle (super)gravity and gauge theories with the same language.
String theory gives a profound, consistent and unified description of the universe at all scales, once the discovery of the role of D-branes~\cite{Polchinski:1995mt} has brought strongly coupled and non-perturbative aspects into the game as well (we refer to the textbooks cited in footnote~\ref{intro:footnote:reference-string-theory} of this introduction for all the details).

Now that we have given an interpretation of the fields of a CFT as coordinates of a world-sheet in some (stable) target space, we go back to our main story.
We can appreciate another nice feature of RG-flows in two dimensions interpolating between conformal fixed points: they can be seen as dynamical processes from a geometry to another, and/or from a D-brane to another, in the target space (see~\cite{Schomerus:2004ds} for a nice informal account of these concepts).
From this we come to the limits of $N=2$ CFTs, which will be our central concern in this thesis.

The $N=2$ extension of Virasoro algebra, as an operator algebra, contains one $N=2$ supermultiplet of generators: the energy momentum tensor and its two supersymmetric partners called supercurrents, plus a $U(1)$ current, which plays the role in two dimensions of the R-symmetry generator of extended supersymmetric theories (among the plethora of textbooks on the subject, we have especially enjoyed the approach of~\cite{Hori:2003ic}).
Restricting to the holomorphic sector, representations are labeled by the conformal dimension (the eigenvalue of the zero mode of the holomorphic energy momentum tensor), and by the $U(1)$ charge (the eigenvalue of the zero mode of the holomorphic $U(1)$ current), as explained in chapter~\ref{ch:minmod}.

\subparagraph{Kazama-Suzuki models}
The first concern is to recognise a sequence of known (rational) CFTs, which enjoy $N=2$ supersymmetry: a class of theories with these characteristics have been constructed by Kazama and Suzuki in~\cite{Kazama:1988uz}.
Kazama-Suzuki models are realised as coset CFTs, and come parameterised by two positive integers, the rank of the numerator algebra, and its level.
In particular the so-called coset Grassmannian models belong to the class of Kazama-Suzuki models, are based on an $\frac{\wzw{su}{n+1}{}}{\wzw{u}{n}{}}$ coset construction (for details see section~\ref{ch3:app:sec:KS-models}), and are the best known among this class.

In this thesis we concentrate on limit theories emerging from the large level behaviour of coset Grassmannian models.

\section*{$N=2$ Supersymmetric limit CFT and continuous orbifolds}

We will present in the following the results obtained by taking the limit of the sequence of the most prominent and known representative of Kazama-Suzuki models: the $N=2$ unitary superconformal minimal models. 
They are based on the coset construction $\frac{\wzw{su}{2}{}}{\wzw{u}{1}{}}$, but they can be also defined as ``minimal'' unitary CFTs based on $N=2$ superconformal algebra, in the same spirit as Virasoro minimal models (the details are in chapter~\ref{ch:minmod}).

$N=2$ minimal models are relevant for string theory for phenomenological reasons in type II string compactifications: they serve as building blocks for Gepner models~\cite{Gepner:1987qi}, which are CFT backgrounds that model type II string propagation on certain Calabi-Yau varieties (the simplest compact target spaces on which type II strings can move~\cite{Greene:1996cy}).

From a world-sheet perspective, they are rational, and they are the best known interacting $N=2$ supersymmetric CFTs. 
They come in a family parametrised by a positive integer $k$, the level of the $\wzw{su}{2}{}$ in the numerator of the coset, and the central charge approaches $c=3$ for large $k$. 
Different bulk models are connected by RG-flows, with countable positive central charges ranging from $c=\12$ to $c=3$.
Unlike Virasoro minimal models, the Zamolodchikov distance~\cite{Kutasov:1988xb} between them is large, hence the anomalous dimensions are not accessible through conformal perturbation theory.

In the large level regime the bulk and the boundary theory admit a target space interpretation~\cite{Maldacena:2001ky} (as in general for any coset~\cite{Fredenhagen:2001kw}) by looking at the fields as coordinates of a (super)string, moving on a specific geometry, obtained by modding out the adjoint action of the denominator group on the numerator from the group geometry of the numerator itself (we explain these ideas in section~\ref{ch3:sec:geometry}).

Equipped with these concepts, we can study the limit of the sequence.

\subparagraph{$c=3$ limits of $N=2$ minimal models}

It is non-trivial that the theory obtained as the inverse limiting point of an infinite number of flows can be defined at all, and whether it is unique. 
It is necessary to be careful in introducing averaged fields (in the spirit of~\cite{Runkel:2001ng}, and as explained in detail in chapter~\ref{ch:limit-theories}) as candidates for being primaries in the limit theory: before taking the limit the presence of the $U(1)$ charge in the $N=2$ superconformal algebra allows different consistent rescaling of the primary fields for the limit theory.
We have discovered through many detailed conformal field theory computations that there are indeed two different limit theories emerging by scaling differently the $U(1)$ charge: one is a free theory of two uncompactified bosons and their fermionic superpartners (chapter~\ref{ch:free-limit}), the other is new and has some similarity with the RW $c=1$ theory of~\cite{Runkel:2001ng} (details in chapter~\ref{ch:new-theory}): the spectrum of primary fields is continuous and unbounded, with continuous $U(1)$ charge in the interval $[-1,1]$; for every chosen charge we have a discrete series of primaries whose weight scales in half-integer multiples of the charge. 
There is therefore some residual discreteness, but the theory is essentially non-rational.

\subparagraph{Geometry of the limit}
This outcome is best understood once we make use of the geometric picture (as explained in chapter~\ref{ch:geometry}): using the coset description one can visualise $N=2$ minimal models as the $N=2$ supersymmetric sigma-model of a string moving on an $SU(2)$ group manifold with a gauged $U(1)$ subgroup. 
The target has the geometry of a disc of non-trivial metric, with coordinate singularity on the circular boundary. 
Taking the two different $k\to\infty$ limits mentioned before corresponds to zooming into two different regions of the disc, one in the neighbourhood of the centre, and the other close to the boundary. 
Concentrating on the centre of the disc gives rise to a free theory, since the metric becomes flat in the limit, and the wave functions of the sigma model approach free waves; the singular boundary instead can be mapped by T-duality to the conical singularity of a $\mathbb Z_{k+2}$ orbifold of the minimal model itself (as explained in section~\ref{ch3:sec:geometry}). 

Sending $k\to\infty$ suggests the picture of a continuous orbifold of the complex plane by $U(1)$, plus two fermions on it. 
In chapter~\ref{ch:our-contorbi} we give strong evidence that this is indeed the case: we match data of the non-rational limit theory of chapter~\ref{ch:new-theory} with the ones obtained directly from the continuous orbifold description.

\subparagraph{Continuous orbifolds}
The example described in chapter~\ref{ch:our-contorbi}, together with the one given in~\cite{Gaberdiel:2011aa}, lends support to the assumption that continuous orbifold theories are indeed well defined CFTs, and it is suggestive of the conjecture that a relevant class of limit theories might be generically described by continuous orbifolds. 

These models have one or more continuous twist parameters, and as a consequence they naturally describe non-rational theories, since the twisted sectors come in infinite families; as a by-product the untwisted sectors contribute with zero measure to the partition function.
The one presented in chapter~\ref{ch:our-contorbi} is the simplest and most natural example of continuous orbifold theory constructed until now; given the formal and physical relevance of these models, it is appealing to look further for generalisations and applications.

The last chapter of this thesis is devoted to the preliminary study of the second simplest $N=2$ large level CFT that one can construct: the non-minimal Kazama-Suzuki coset $\frac{\wzw{su}{3}{}}{\wzw{u}{2}{}}$.
The elements that we have collected there point towards the conjecture that the whole class of $\frac{\wzw{su}{n+1}{}}{\wzw{u}{n}{}}$ supersymmetric Grassmannian coset tend in the large level limit to a continuous orbifold theory, specifically to the supersymmetric $\C^n/U(n)$.

\section*{Outline}
This thesis is organised in two parts.

\paragraph{Introductory part}
The first one consists of chapters~\ref{ch:limit-theories},~\ref{ch:orbifolds} and~\ref{ch:minmod}: it is a review of known material useful for a proper definition of the problems we consider in the second part.

In chapter~\ref{ch:limit-theories} we explain what we mean specifically when we talk about limit CFTs: in section~\ref{ch1:sec:generalities} we give an overview on the problems and the methods used.
We give then an explicit account of two examples: in section~\ref{ch1:sec:free-boson-limit} the limit of the free boson on a circle, and in section~\ref{ch1:sec:RW} a brief review of Runkel-Watts theory.

In chapter~\ref{ch:orbifolds} we review the technology used for constructing discrete and continuous orbifold~CFTs in general terms in sections~\ref{ch2:sec:discrete-orbifold-generalities} and~\ref{ch2:sec:contorbi}, and giving two simple examples involving free bosons; we recall in subsection~\ref{ch2:sec:RW-as-contorbi} the construction of the continuous orbifold of~\cite{Gaberdiel:2011aa} as the limit of Virasoro minimal models.

In chapter~\ref{ch:minmod} we give an extensive account of $N=2$ minimal models: we write down the $N=2$ superconformal algebra starting from free fields in subsection~\ref{ch3:sec:SCA:subs:free-fields}, we look into representations and present three different ways of studying the spectrum of these models (the ``minimal'' construction in~\ref{ch3:sec:unitarity-spectrum-minmod}, the parafermionic construction in~\ref{ch3:sec:MM-parafermions}, and the general coset description for Grassmannian coset in section~\ref{ch3:app:sec:KS-models}.
We give expressions for bulk correlators in section~\ref{ch3:sec:correlators}, for superconformal boundary conditions in section~\ref{ch3:sec:BC}, and we give details about the geometric interpretation in section~\ref{ch3:sec:geometry}.

\paragraph{Research part}
The second part of the thesis, consisting of chapters~\ref{ch:geometry}-\ref{ch:limit-KS} is based on our original contributions.

In chapter~\ref{ch:geometry} we give a heuristic justification of the results presented in the following: we analyse the limits of the geometry of $N=2$ minimal models, in general terms in section~\ref{ch4:sec:intro}, concentrating more specifically on the bulk geometry in section~\ref{ch4:sec:bulk-geometry}, and on D-branes in section~\ref{ch4:sec:limit-geometry-BC}.

In chapter~\ref{ch:free-limit} we present a detailed analysis of the free limit theory emerging in the large level regime of $N=2$ minimal models.

In chapter~\ref{ch:new-theory} we give account of many details of the new $c=3$ limit theory emerging in the limit by appropriately rescaling the labels of the representations of minimal models.

In chapter~\ref{ch:our-contorbi} we construct the $\C/U(1)$ continuous orbifold CFT, and compare the results with the limit of~\ref{ch:new-theory}: the concrete construction of the orbifold theory is given in section~\ref{ch7:sec:orbifold}, section~\ref{ch7:sec:partition-fct} is devoted to the analysis of the bulk partition function, and boundary conditions are analysed in section~\ref{ch7:sec:BC}.

In chapter~\ref{ch:limit-KS} we give some evidence that limits of more complicated Grassmannian cosets admit a continuous orbifold interpretation as well.
We state the correspondence in section~\ref{ch8:sec:correspondence}, we give arguments for the matching of boundary conditions in section~\ref{ch8:sec:BC-match}, we study in detail the limit of boundary conditions of the limit of the theory $\frac{\wzw{su}{3}{}}{\wzw{u}{2}{}}$ in section~\ref{ch8:sec:BC-su(3)}, and we compare with boundary conditions of the supersymmetric continuous orbifold $\C^2/U(2)$ in section~\ref{ch8:sec:BC-contorbi}.

We give in three appendices additional useful material: in appendix~\ref{app:characters} we collect information about characters of the representations that are relevant for our work.
In appendix~\ref{app:wigner} we explain the technology used to compute the limit of three-point functions, which consists in studying the asymptotic behaviour of Wigner 3j-symbols in various regimes.
In appendix~\ref{app:vanilla} we give a very brief account of boundary CFT, to make this work more accessible to non-practitioners.

\vspace{0.5cm}
 \begin{center}
  - - - - - - - - -
\end{center}

\vspace{0.5cm}
  
The second part of this thesis is mostly based on the papers~\cite{Fredenhagen:2012rb,Fredenhagen:2012bw}.
 
 \begin{itemize}
\item Chapter~\ref{ch:new-theory} and appendix~\ref{app:wigner} are based on \cite{Fredenhagen:2012bw}, written in pleasant collaboration with Stefan Fredenhagen and Rui Sun.
\item Chapters~\ref{ch:geometry},\ref{ch:free-limit},\ref{ch:our-contorbi} and appendix~\ref{app:characters} are based on~\cite{Fredenhagen:2012rb}, written together with Stefan Fredenhagen.
\item Chapter~\ref{ch:limit-KS}  collects new unpublished material.
   \end{itemize}


 \chapter{Limit theories}\label{ch:limit-theories}
  
In this chapter we give a more technical account of how to define and study limit theories.
We are interested in limits of sequences of two-dimensional rational conformal field theories. 
One outcome of such an analysis is a deeper understanding of non-rational models, since the limit theories obtained in this way are non-rational, and seem to be consistently defined.
Another motivation comes from the interest in the space of QFTs as outlined in the introduction to this thesis.

The first hint that such limit theories might be well-defined and non-trivial came from the work of Graham, Runkel and Watts~\cite{Graham:2001tg}: they observed that taking the $c\to 1$ limit of  Virasoro minimal models the fundamental conformal boundary conditions could be organised in an infinite discrete set, labeled by positive integers.
Since no~$c=1$ theory was known to have such a boundary spectrum, Runkel and Watts proposed in~\cite{Runkel:2001ng} that a new non-trivial interacting non-rational CFT exists as the~$c\to 1$ limit of Virasoro minimal models. 
In the mentioned reference they explicitly computed several bulk CFT data, and gave arguments of consistency.
Later, an observation by Schomerus~\cite{Schomerus:2003vv} and subsequent work of Schomerus and Fredenhagen~\cite{Fredenhagen:2004cj} gave evidence that the~$c\to 1$ limit of Liouville theory approaches the same theory formulated by Runkel and Watts\footnote{Another approach in taking limits has to be mentioned here, namely the one proposed by Roggenkamp and Wendland in~\cite{Roggenkamp:2003qp}. 
We will not discuss it in the following.
Within this approach it has been possible to obtain a different well-defined theory as the $c\to 1$ limit of minimal models.}.

This set of results strongly suggests that the procedure of generating non-rational theories starting from sequences of rational ones gives non-trivial answers: limits of these sequences may give rise to well-defined (possibly new) non-rational theories.

These ideas have been applied to~$N=1$ minimal models~\cite{Fredenhagen:2007tk} (and to~$N=1$ Liouville theory) as well as to~$W_N$ minimal models~\cite{Fredenhagen:2010zh}.
We have presented evidence in~\cite{Fredenhagen:2012rb,Fredenhagen:2012bw} that the same approach can be used to study the~$N=2$ superconformal case as well, and we collect our results in later chapters.

This chapter is organised as follows: in section~\ref{ch1:sec:generalities} we give an account of the technology used to define this kind of limit theories; in section~\ref{ch1:sec:free-boson-limit} we present a very easy example of limit theory, the decompactification limit of the free boson; in section~\ref{ch1:sec:RW} we briefly review the limit of minimal models as given by Runkel and Watts in~\cite{Runkel:2001ng}.
We have added a sub-appendix, section~\ref{ch1:sec:free-boson}, in order to fix notations for the well-known example of the free compactified boson.

   \section{Generalities}\label{ch1:sec:generalities}

In this section we become more precise about the limit theories we are interested in. 
In two dimensions several classes of known CFTs come in sequences, parameterised by one or more natural or real numbers. 
They represent important elements of the space of two-dimensional QFTs, having a large amount of symmetry, being easier to handle compared to others, and representing non-trivial end-points of renormalisation group (RG) flows of theories that are not scale-invariant.
    
Defining limits of sequences of CFTs is interesting also as we aim to look at what kind of topological or metric structures could possibly be defined in the space of theories~\cite{Douglas:2010ic}: for this reason, limits of sequences of CFTs that we know in detail are among the best species to explore if we want to better understand spaces of QFTs.
 
The natural candidates are sequences of (quasi)-rational CFTs, in particular sequences of free theories or of minimal models, for which an extensive knowledge of the defining structures exists. 
The basic idea of this programme is to ``take the limit towards the boundary'' of the parameter space defining the sequence, and to see what happens to the data of the QFT. 
\subsection{CFT data and consistency conditions}
 
The limit of the CFT structures may or may not define a sensible QFT. 
Firstly we need to collect a minimal set of data sufficient to define a CFT (in general with conformal boundaries). 
Secondly we have to be sure that these data are compatible with each other, namely the theory has to satisfy a set of consistency conditions.
\smallskip
\smallskip

\noindent
The following table summarises the sufficient data one has to collect to properly define a rational CFT; the extension to non-rational theories is subtle, as discussed e.g. in~\cite{Runkel:2001ng}, but it is a common belief that to collect these data is the best starting point.
The last lines tell us which consistency tests should be in principle performed.
\begin{center}
\begin{tabular}{c||c}
\bf{BULK DATA} & \bf{BOUNDARY DATA}\\
\hline
bulk spectrum & open string spectrum\\
torus partition function & annulus partition function\\
three-point function bulk fields & three-point function boundary fields\\
\hline  
\end{tabular}
\begin{center}
bulk-boundary correlators
\end{center}
\begin{tabular}{c}
\bf{CONSISTENCY CHECKS} \\
\hline
crossing symmetry  bulk four-point function \\ 
Cardy-Lewellen constraints
\end{tabular}  
\end{center}

\paragraph{Bulk}
The first thing we have to know to build a QFT is what kind of quantised excitations we have, namely the spectrum.
In CFTs the excitations are organised in terms of their behaviour under two-dimensional conformal transformations.
Conformal symmetry is very large, and in two dimensions is even infinite dimensional: any holomorphic or anti-holomorphic function on the complex plane generates a conformal transformation of the plane (plus a point at complex infinity) to itself.
This fact has the consequence that it is possible to define on the complex plane fields with definite behaviour under holomorphic and anti-holomorphic transformations independently.
The algebra generating two-dimensional conformal transformations is the Virasoro algebra, which comes therefore in two copies (holomorphic and anti-holomorphic). 
In a CFT the spectrum is given by its representation spaces (many extensions are possible, like the one analysed in chapter~\ref{ch:minmod}: if the group of transformations is extended, the spectrum is given by representations of the extended algebra): each highest-weight representation of the algebra can be mapped to a corresponding primary field (state-operator correspondence), namely a field with a definite tensor-like behaviour under conformal maps.

Correlators on the plane of purely holomorphic fields with purely anti-holomorphic ones completely factorise into holomorphic and anti-holomorphic factors.
We do not have restrictions in introducing non-chiral fields.
This is apparently non-physical, as argued in~\cite{DiFrancesco:1997nk}: the conformally invariant point is not isolated in the space of parameters defining the theory, and the spectrum should change continuously as we leave the critical point. 
We expect then to inherit, also at criticality, constraints coming from the theory away from the conformal point.
A way to reduce the number of possible left-right excitations is to impose conformal invariance for the same collection of excitations defined on a torus (impose modular invariance).
Technically this amounts to choosing the torus partition function, which is a spectrum generating function: modular invariance limits the possible choices of holomorphic-anti-holomorphic couplings, hence constraints the spectrum.

The same spectrum can obviously be shared by different theories: the correlators of (primary) fields discriminate among them.
The three-point function is central in CFT; conformal invariance constrains it to depend only on the so-called \textit{structure constants} (the coefficients which multiply the dependence on the coordinates of the three-point function, which is fixed by conformal invariance), that in turn give us the coefficients for the operator product expansion (OPE) for the primary fields. 
Out of the three-point function one can construct the other correlators: the two-point function can be inferred from the three-point correlator letting one of the field approach the identity on the complex plane.
The~$n$-point functions can be as well constructed by repeatedly evaluating the full OPEs among the various fields in the correlator.

Spectrum, modular invariant torus partition function and the three-point correlator are enough, in principle, to write down any bulk correlator for the theory (namely to ``solve'' it).
Nevertheless, starting from the data, we have to be sure that the four-point function satisfies crossing symmetry, which is a consequence of the associativity of the operator algebra.
Four-point functions depend on the structure constants of the theory (which can be read off the three-point functions and are specific for any CFT) and on the so-called conformal blocks (purely representation theoretical data).
Unfortunately conformal blocks are difficult to compute (we only know recursive equations for them), hence the test of crossing symmetry is often hard to perform.

\paragraph{Boundary}
Similar considerations apply to conformal field theories defined on two-dimensional Riemann surfaces with boundary (BCFTs).
The open string spectrum consists of modules of the (possibly extended) Virasoro algebra, since a conformal boundary must, by definition, preserve one copy of the conformal algebra (at least).
We have to be specific, and choose the amount of (chiral) symmetry that our BCFT preserves, namely the gluing conditions for the chiral currents on the boundary.

If we know the bulk theory, this is enough to construct Ishibashi states, sort of coherent representations of the chiral algebra, to whom bulk excitations can couple.
From combinations of Ishibashi states, and from the one-point functions of bulk fields in presence of a boundary, it is possible to write down a modular annulus partition function~\cite{Cardy:1989ir,Recknagel:1998sb}, which play an analogous role to the torus partition function of the bulk spectrum: modularity constraint the spectrum of open string excitations among the representations of the symmetry preserved by the boundary condition (some details on this construction are collected in appendix~\ref{app:vanilla}).

To each state in the boundary spectrum it is possible to associate a boundary field, in a boundary state-operator correspondence.
The boundary theory is then fixed by the three-point functions of boundary fields, in the same spirit as in the bulk case.

We also need bulk-boundary correlators to conclude the set of ingredients.

Four-point functions between boundary fields must satisfy a set of consistency equations, called Cardy-Lewellen sewing constraints~\cite{Cardy:1991tv,Lewellen:1992tb}, which are the analogues of crossing symmetry relations for the four-point bulk correlators.
We do not enter the discussion of this wide topic here.
Excellent references are~\cite{Behrend:1999bn,Schomerus:2002dc}.
 
 \subsection{Limits of sequences}
Starting from a sequence of (quasi)-rational CFTs we expect to lose some of the features of the set of theories we have started from: in particular (quasi)-rational theories are characterised by the discreteness of their structures (e.g. spectrum of primary fields and fusion rules). 
In all the known examples the limiting procedure spoils this property.

This fact has a mixed flavour: on one side it makes it difficult to check the consistency of the collected data (above all the test of crossing symmetry is complicated to perform); on the other it gives us a recipe to construct non-rational CFTs, which is known to be a very hard task, and interesting for many reasons.
In particular CFTs with continuous spectra are necessarily non-rational, and hint at non-compactness of the target space of the associated sigma-model; notably AdS backgrounds e.g. are described by non-rational theories.

\paragraph{Sequences, RG flows and central charges}
A sequence of rational CFTs is likely to originate from a sequence of RG-flows in the space of two-dimensional theories.
    
 Starting from a given CFT, if we perturb it with an exactly marginal deformation, then every point of trajectory in the space of theories represents a CFT; the sequence is continuous and spans the moduli space of the CFT~\cite{Kadanoff:1978pv,Dijkgraaf:1987vp,Ginsparg:1987eb,Ginsparg:1988ui}; in this case the central charge does not change along the sequence (like in the free boson example of section~\ref{ch1:sec:free-boson-limit} in this chapter) and the limit theory describes some boundary of the moduli space in the direction of the deformation. 
 
 If instead the perturbation is driven by a relevant operator, only the infrared end-point - when non-trivial - is a CFT; it might be free or not, surely its central charge is lower than the ultraviolet one, thanks to Zamolodchikov's c-theorem~\cite{Zamolodchikov:1986gt}. 
 As explained in the introduction, limits of minimal models are based on discrete families which are inverse sequences of RG-flows between nearby fixed points: in this case the central charge increases monotonically along the sequence, and the limit theory has as central charge the upper bound of the sequence of central charges.
 
\paragraph{Spectrum and torus partition function}
The spectrum of ground states of a CFT is given in terms of highest-weight representations of its chiral algebra. 

In the examples of this thesis the structure of the chiral algebra remains fixed along the sequence up to the central elements\footnote{This is in contrast with the limit theories analysed in the context of higher-spin holographic dualities~\cite{Gaberdiel:2012uj}; in that case to each CFT of the sequence corresponds a different $W$-algebra.}. 
The range of the labels of the representations get in general modified, since it depends on central elements.
Moreover, also conformal weights and possibly other labels characterising the spectrum are functions of the central elements: the spectrum gets modified along the sequence.
 
Rational theories are characterised by discrete sets of representations; we expect to lose this property approaching the limit of the sequence, hence to find ground states coming in continuous families.  
Furthermore we encounter the issue that the allowed weights of the representations are not bounded in general in the limit, leaving us the choice of rescaling the labels in different ways: inside the spectrum of the limit of a sequence we might be able to find different sectors recognisable as spectra of independent CFTs (this is the case with $N=2$ minimal models, where the limit theory organises itself into two different theories, as explained in chapter~\ref{ch:geometry} and in~\cite{Fredenhagen:2012bw}).
The theories are discriminated by the way one scales the labels of the representations, or, analogously, by which parts of the continuous spectrum one concentrates on.

From this point of view it is useful to check the limit of the torus partition function, since it may even happen that the different CFTs emerging in the limit get decoupled (although this does not seem to be the case of $N=2$ minimal models, as explained later in chapter~\ref{ch:free-limit} and~\ref{ch:new-theory} and in~\cite{Fredenhagen:2012bw}) 

\paragraph{Fields and correlators: concept of averaged or ``smeared'' field}
   
Every CFT of the sequence possesses a different Hilbert space, generated by the ladder operators of the chiral algebra, on top of a set of highest-weight states. 
In a (quasi)-rational CFT to any ground state is associated a primary field in virtue of the state-operator correspondence; this property is in general not true anymore in the non-rational case, since we might encounter the situation in which the Hilbert space of the CFT does not contain a proper vacuum, i.e. a ground state of conformal weight zero (Liouville theory being a notable example, see~\cite{Teschner:2001rv}).
Furthermore, once the labels of the primary fields become continuous, the normalised two-point function should approach a $\delta$-distribution in the continuous labels $x,y$
  \begin{equation}
  \bra\Phi_x(z_1,\bar z_1)\Phi_y(z_2,\bar z_2)\ket=\frac{\delta(x-y)}{|z_1-z_2|^{4h_x}}\ .
  \end{equation}
   It is not trivial to define such fields as limits of fields of rational theories.
  Runkel and Watts have proposed the construction of averaged fields~\cite{Runkel:2001ng}, which we will now explain.
  
  Choose a representative in the sequence of CFTs. 
  The primary fields (a finite number in a rational theory) are characterised by the value of their conformal weight.
  By going towards the limit of the sequence, the number of primary fields grows, and the values taken by their conformal weights get closer to each other: the limit weight will be some average over the weights of the primaries, once the number of fields grows in any fixed neighbourhood.
  We choose a smooth averaging function $f(h)$, and we define the averaged field as
  \begin{equation}\label{ch1:def-averaged-fields}
    \Phi_{\![f]}^{(k)} = \sum_{i} f (h_{i}) \,\phi^{(k)}_{i}\ ,
  \end{equation}
   where $\phi^{(k)}_{i}$ are primary fields with conformal weight $h_{i}$ in the $k^{\text{th}}$ CFT of the sequence.
   The fields of the limit theory are then given by the $k\to\infty$ limit of the averaged fields.
   As $k\to\infty$ the sum in~\eqref{ch1:def-averaged-fields} becomes an integral, since the number of primaries grows, and the averaging function can develop discontinuities.
   
 The choice of the averaging function is crucial in determining the spectrum: normally we choose it such that it takes non-zero value only in a small neighbourhood of the label that will play the role of the label characterising the fields of the limit theory. 
 For any small enough neighbourhood of that label, a large enough $k$ exists, which makes the neighbourhood of labels non-empty. 
 Therefore the rationale is to first populate a fixed neighbourhood by letting $k$ become large, and then send the neighbourhood to zero measure.

  
  
  The above procedure is ambiguous when quantum numbers other than the conformal weight are available. 
  In this case we also have the freedom to rescale them while taking the limit. 
  Different ways of treating these additional quantum labels can produce different non-trivial limit theories, as in the case of $N=2$ minimal models.
  
  The $n$-point functions for limit fields are obtained as limits of correlators of averaged fields. 
  To obtain finite and sensible correlators in the limit, we are free to appropriately rescale fields and correlators, and this can be done by studying the limit of the two- and three-point functions of averaged fields.
  
  The above considerations hold as well if one is interested in conformal boundary conditions. 
  The bulk one-point function is defined in the limit theory as any other correlator, by means of the averaging procedure; one has of course to be careful in performing the modular transformation to go from the boundary state overlap to the annulus amplitude (some details in appendix~\ref{app:vanilla}), since this involves integrals over characters. 
  We discuss these issues in detail when we analyse D-branes in the limit of $N=2$ minimal models.
  
\section{Large radius limit of one free boson on a circle}\label{ch1:sec:free-boson-limit}
     
In this section we describe the simplest limit CFT of the kind explored in this thesis, namely the decompactification limit of one free boson on a circle of radius $R$.
These free bosonic CFTs (briefly reviewed in section~\ref{ch1:sec:free-boson}) come naturally in a continuous infinite family parameterised by the radius of the circle. 
We can concretely construct the sequence by deforming the action in~\eqref{ch1:action-free-boson} with a very simple marginal deformation
\begin{equation}
 S_{R\sqrt{1+\lambda}}=S_R+\l \int d^2z\ \de\phi(z,\bar z)\debar\phi(z,\bar z)\ .
\end{equation} 
  As the notation suggests, we end up with a free boson at radius $R\sqrt{1+\l}$, always at $c=1$.
  
  We will now show how to use the techniques outlined in section~\ref{ch1:sec:generalities} to study the decompactification limit, $R\to\infty$, of the free boson on a circle: we get as limit theory the CFT of one free boson on the real line. 
     \paragraph{Spectrum} 
  As explained before in general terms, in the case at hand taking the $R\to\infty$ limit amounts to choosing how to scale the labels characterising the spectrum of the theory at finite radius; in other words, we must identify which sectors have to be kept at finite conformal dimensions after the limiting procedure.
  
  This is easily illustrated in this example by looking at the spectrum of ground states in equations~\eqref{ch1:weights-prim-free-bos}: the conformal weights of the primaries diverge as $R\to\infty$ unless we set identically to zero the winding modes $n$. 
  Furthermore, if the momenta~$m$ do not scale at least as fast as~$R$, all the primary fields are brought down to zero weight, and the theory becomes trivial and infinitely degenerate. 
  A condition we might thus impose is to set $n=0$ and $m=P R$, with $P$ some real label, chosen such that $m$ remains integer for large $R$. 
  In this way the spectrum of the primaries in the limit reads
 \begin{equation}
 h_P=\bar h _P=\frac12 P^2\ ,\qquad P\in\R\ ,
 \end{equation}
  as expected for a free boson on the real line. 
  \paragraph{Partition function}
  The limit of the partition function is also easy to compute: the contribution of the modes with $n\neq 0$ is exponentially suppressed for large $R$, as expected from the analysis of the spectrum; the sum over $m$ becomes at leading order in $R$ a gaussian integral over~$P$:
   \begin{equation}
     \begin{split}
     \lim\limits_{R\to\infty}Z_R=\ & \lim\limits_{R\to\infty}\frac{1}{|\eta(\tau)|^2}\sum_{n,m\in\Z}q^{\frac12(\frac mR+\frac12 Rn)^2}\bar q ^{\frac12(\frac mR-\frac12 Rn)^2}\\
     =\ &\lim\limits_{R\to\infty}\frac{1}{|\eta(\tau)|^2}\left(\sum_{m\in\Z}(q\bar q)^{\frac12(\frac m R)^2}+\order{(q\bar q)^{R^2}}\right)\\
     =\ &\lim\limits_{R\to\infty}\frac{R}{\sqrt 2} \frac{1}{|\eta(\tau)|^2}\frac{1}{\sqrt {\tau_2}}+\order{\frac{1}{R^2}}\ ,
     \end{split}
     \end{equation}
     where $\tau_2$ is the imaginary part of $\tau$.
     At leading order and up to a normalisation we recover the partition function of a free boson on the real line:
     \begin{equation}\label{ch1:free-boson-line-partition-fct}
     \lim\limits_{R\to\infty}\frac{Z_R}{R/\sqrt 2}=\frac{1}{\sqrt {\tau_2}}\frac{1}{|\eta(\tau)|^2}=Z_{\text{line}}\ .
     \end{equation}
  The infinite normalisation is due to the fact that in the momentum range~$P,P+\Delta P$ there are infinitely many states (a number that scales with $R\Delta P$, as explained in the following).
  Dividing by~$R/\sqrt 2$ in the left hand side accounts to avoiding the infinite overcounting.
  \paragraph{Averaged fields}
  Consider the set of labels
  \begin{equation}
  N(P,\epsilon,R)=\left\{m\ \middle|\  P-\frac{\epsilon}{2}<\frac{m}{R}<P+\frac{\epsilon}{2}\right\}\ 
  \end{equation}
  with $P$ real and $\epsilon>0$ small. This set describes the points belonging to a small neighbourhood of the limit label $P$, in the large $R$ regime. Since to every point of this set corresponds a primary field, the idea is to define the fields of the limit theory as an average in the neighbourhood of the field which would correspond to the limit label.
  In formulas:
  \begin{equation}
   \Phi^{\epsilon,R}_{P}=\frac{1}{|N(P,\epsilon,R)|}\sum_{m\in N(P,\epsilon,R)}\phi_{m,0}\ ,
   \end{equation}
where we have normalised by the cardinality of the set $N$, which scales for large $R$ as $|N(P,\epsilon,R)|=\epsilon R +\order{1}$. In this way we avoid infinite degeneracies, and we are left with only one representative for each weight. The primary fields of the limit theory are then formally defined in all the instances as the limit for small~$\epsilon$ and large~$R$ of $\Phi^{\e,R}_{P}$. 
  \paragraph{Correlators}
  The correlator of $r$ fields in the limit theory is defined as\footnote{The order of the limits is relevant here: if we took $\e\to0$ before $R\to\infty$ we would find zero, since the $R\to\infty$ limit is the one that populates the neighbourhoods.}
 \begin{equation}
 \bra \Phi_{P_1}(z_1,\bar z_1)\dots \Phi_{P_r}(z_r,\bar z_r)\ket = \lim\limits_{\epsilon\to 0}\lim\limits_{R\to \infty}\beta(R)^2\alpha(R)^r\bra \Phi^{\epsilon,R}_{P_1}(z_1,\bar z_1)\dots \Phi^{\epsilon,R}_{P_r}(z_r,\bar z_r)\ket\ .
 \end{equation}
 $\b(R)$ represents the normalisation of the vacuum and $\a(R)$ the individual normalisations of the fields (all equal in this example) while taking the limit; they are chosen in such a way that the correlators stay finite.
  \paragraph{Two-point function}   
 The two-point function for the primary fields of the CFT at radius~$R$ with zero winding is given in standard normalisations by
  \begin{equation}
   \bra \phi_{m_1,0}(z_1,\bar z_2)\phi_{m_2,0}(z_2,\bar z_2)\ket = \frac{\delta_{m_1,-m_2}}{|z_1-z_2|^{2m^2_1/R^2}}\ .
  \end{equation} 
  Therefore the two-point function of the limit theory reads at leading order
  \begin{equation}
  \begin{split}
   \bra \Phi_{P_1}(z_1,\bar z_1)\Phi_{P_2}(z_2,\bar z_2)\ket=&\ \lim\limits_{\epsilon\to0}\lim\limits_{R\to\infty}\frac{\b(R)^2\a(R)^2}{\e^2 R^2}\sum_{\substack{m_1\in N(P_1,\epsilon,R)\\m_2\in N(P_2,\epsilon,R)}}\frac{\delta_{m_1,-m_2}}{|z_1-z_2|^{2m^2_1/R^2}}\\
  =&\ \lim\limits_{\epsilon\to0}\lim\limits_{R\to\infty}\frac{\b(R)^2\a(R)^2}{\e^2 R^2}\frac{1}{|z_1-z_2|^{2P_1^2}}|N(P_1,\epsilon,R)\cap N(-P_2,\epsilon,R)|\ ,
  \end{split}
  \end{equation} 
  where in the second line we have saturated the Kronecker delta with the sum over $m_2$ and taken into account that $h_{m_1,0}\simeq \frac12 P_1^2$ if $m_1$ belongs to $N(P_1,\epsilon,R)$.
  
  The cardinality of the overlap can be easily estimated using the Heaviside theta function
  \begin{equation}
   |N(P_1,\epsilon,R)\cap N(-P_2,\epsilon,R)|=R(\epsilon - |P_1+P_2|)\theta(\epsilon - |P_1+P_2|)+\order{1}\ ,
  \end{equation} 
  In the small $\epsilon$ limit we find a representation of the $\d$-distribution 
  \begin{equation}
  \lim_{\epsilon\to 0} \frac{\epsilon-|P_1+P_2|}{\epsilon^2}\theta(\epsilon-|P_1+P_2|)\equiv\lim\limits_{\e\to 0}\d_{\e}(P_1+P_2)=\delta(P_1+P_2)\ ,
  \end{equation} 
  and we arrive to
  \begin{equation}
   \bra \Phi_{P_1}(z_1,\bar z_1)\Phi_{P_2}(z_2,\bar z_2)\ket = \lim\limits_{\e\to 0}\lim_{R\to\infty}\frac{\b^2(R)\a^2(R)}{R}\ \frac{\delta_{\e}(P_1+P_2)}{|z_1-z_2|^{4h_{P_1}}}\ .
  \end{equation} 
  In order to have finite correlators, and to be in conformity with standard normalisations, we set
  \begin{equation}\label{ch1:cond-on-alpha-beta-free-boson}
   \a(R)\b(R)\stackrel{!}{=}\sqrt{R}\ .
  \end{equation}
  In this way we recover the two-point function of a free boson on the real line by properly normalising our fields and the vacuum:
  \begin{equation}
  \bra \Phi_{P_1}(z_1,\bar z_1)\Phi_{P_2}(z_2,\bar z_2)\ket = \frac{\delta(P_1+P_2)}{|z_1-z_2|^{4h_{P_1}}}\ .
  \end{equation}
  \paragraph{Three-point function}
  The analysis of the three-point function is the same in spirit: the three-point function for a free boson on a circle with no winding is given by
  \begin{equation}
  \bra \phi_{m_1,0}(z_1,\bar z_1)\phi_{m_2,0}(z_2,\bar z_2)\phi_{m_3,0}(z_3,\bar z_3)\ket= \frac{\delta_{m_1+m_2,-m_3}}{|z_{12}|^{2(-h_3+h_2+h_1)}|z_{23}|^{2(h_3+h_2-h_1)}|z_{13}|^{2(h_3-h_2+h_1)}}\ .
  \end{equation}
  The definition of the three-point function of the limit theory is therefore (omitting the obvious $z$-dependences on the left hand side)
  \begin{equation}
  \bra \Phi_{P_1}\Phi_{P_2}\Phi_{P_3}\ket = \lim\limits_{\e\to 0}\lim\limits_{R \to\infty}\frac{\a (R)}{\e ^3 R^2}\sum_{\{m_i\in N(P_i,\e ,R)\}}\frac{\delta_{m_1+m_2,-m_3}}{|z_{12}|^{2(-h_3+h_2+h_1)}|z_{23}|^{2(h_3+h_2-h_1)}|z_{13}|^{2(h_3-h_2+h_1)}}\ ,
  \end{equation}
  where we have already inserted the result of equation~\eqref{ch1:cond-on-alpha-beta-free-boson}.
  As before we saturate the sum over~$m_3$ and put on-shell the conformal weights in the denominator on the right hand side. 
  We are left with the evaluation of the cardinality of the overlap $N_{123}$ between the three sets $N_i$, this time with $m_1+m_2+m_3=0$, again due to momentum conservation in the original free theory: 
  \begin{equation}
  N_{123}= \{(m_{1},m_{2},m_{3}) \in N (P_{1},\epsilon ,R) \times N (P_{2},\epsilon ,R) \times N (P_{3},\epsilon ,R):m_{1}+m_{2}+m_{3}=0
  \} \ .
  \end{equation}
\begin{figure}[h]
    \begin{center}
    \includegraphics{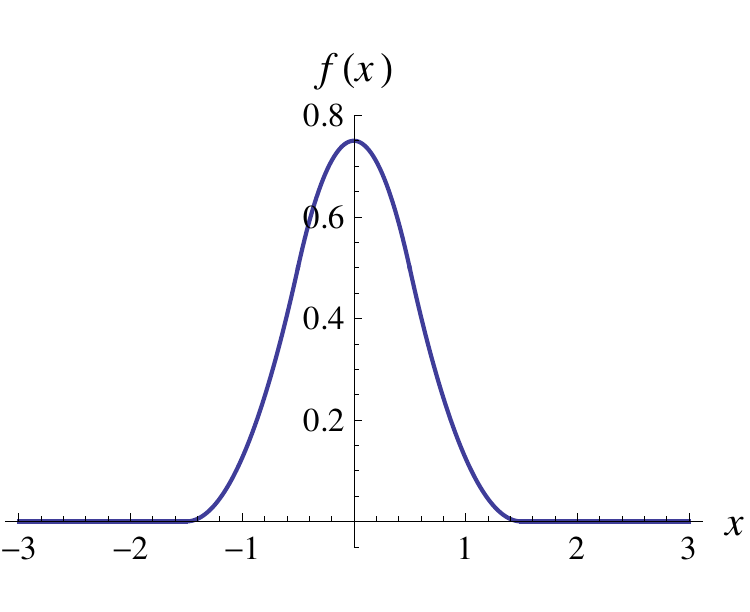}  
    \end{center}
    \caption{\label{fig:fPlot}An illustration of the function $f$ defined in~\eqref{def_f}.}
    \end{figure}
 The cardinality can be estimated as
  \begin{equation}
  |N_{123}| = R^{2}\epsilon^{2} f ({\textstyle \frac{1}{\epsilon}\sum_{i}P_{i}}) +\mathcal{O} (R) \ ,
  \end{equation}
  where the function $f$ is defined as
  \begin{equation}
  f (x) = \left\{\begin{array}{crr@{\,x\,}l}
  0 & \quad \text{for} & & < -\frac{3}{2} \\[1mm]
  \frac{1}{2} (x+\frac{3}{2})^{2}& \quad \text{for} & -\frac{3}{2}< & < -\frac{1}{2} \\[1mm]
  \frac{3}{4}-x^{2}& \quad \text{for} & -\frac{1}{2}< & < \frac{1}{2} \\[1mm]
  \frac{1}{2} (x-\frac{3}{2})^{2}& \quad \text{for} & \frac{1}{2}< & < \frac{3}{2} \\[1mm]
   0 & \quad \text{for} & \frac{3}{2}< & \ .
  \end{array} \right. 
  \label{def_f}
  \end{equation}
  The function $f$, displayed in figure~\ref{fig:fPlot}, has the property
  \begin{equation}
  \int dx\, f (x) = 1 \ .
  \end{equation}

  When we finally take the limit, we observe that the function $f$ leads to a $\d$-distribution for the sum of the momenta
  \begin{equation}
  \frac{1}{\epsilon}f ({\textstyle \frac{1}{\epsilon }\sum_{i}P_{i}}) \to \delta
  ({\textstyle\sum_{i}P_{i}}) \ .
  \end{equation}
   We can then write the result as:
  \begin{equation}
  \bra \Phi_{P_1}\Phi_{P_2}\Phi_{P_3}\ket  = \lim\limits_{\e\to 0}\lim\limits_{R\to \infty} \alpha(R) \frac{\delta_{\e}(P_1+P_2+P_3)}{|z_{12}|^{2(-h_3+h_2+h_1)}|z_{23}|^{2(h_3+h_2-h_1)}|z_{13}|^{2(h_3-h_2+h_1)}}\ .
  \end{equation}
  We are now in position to further specialise the coefficients as
  \begin{equation}
   \a(R)\stackrel{!}{=}1\ ,\qquad \b(R)\stackrel{!}{=}\sqrt{R}\ .
  \end{equation} 
  This fixes the definition of the correlators of the limit theory to be
  \begin{equation}
  \bra \Phi_{P_1}(z_1,\bar z_1)\dots \Phi_{P_r}(z_r,\bar z_r)\ket =\lim\limits_{\e\to 0} \lim\limits_{R\to \infty}R\bra \Phi^{\epsilon,R}_{P_1}(z_1,\bar z_1)\dots \Phi^{\epsilon,R}_{P_r}(z_r,\bar z_r)\ket\ .
  \end{equation}

  \smallskip
  
  In conclusion, with the method of averaged fields, we have been able to show that the free boson on the real line is indeed, as naively expected, the limit of the sequence of increasing radii of the free boson on a circle. 
  We have collected enough elements to reconstruct any bulk CFT correlator, in principle on any Riemann surface. 
  We leave to the reader the analysis of boundary conditions. Everything works similarly as well.
    
      \section{Large level limit of unitary Virasoro minimal models}\label{ch1:sec:RW}
      
  The second warm-up example that we want to discuss is the case of unitary Virasoro minimal models~\cite{Belavin:1984vu,DiFrancesco:1997nk}.
  \subsection{Virasoro minimal models: a minimal overview} 
   The very definition of any conformal field theory starts by the tracelessness of the conserved energy momentum tensor $T_{\m\n}$. 
   In two dimensions this fact has the consequence that $T_{\m\n}$ has only two non-vanishing components, one holomorphic ($T(z)$) and one anti-holomorphic ($\overline T(\bar z)$).
   After quantisation, the modes of the energy momentum tensor close two copies of the Virasoro algebra.
   If we restrict the CFT to these minimal assumptions and impose unitarity of the representations, after fixing a modular partition function we are already in the condition to say many things about the CFT.
   
   The recipe goes as follows: we determine the set of modules of the Virasoro algebra at fixed central charge, and impose unitarity by the requirement that every physical state has strictly positive norm.
   We find that only a particular discrete set of central charges is allowed
  \begin{equation}
  c_k=1-\frac{6}{(k+2)(k+3)}\ ,\qquad k\in\Z_{>0}\ ,
  \end{equation}
    and that for every choice of $k$, the corresponding model \MM{k} has only a finite number of primary fields, with spectrum
    \begin{equation}\label{ch1:spectrum-MM}
   h_{r,s}=\frac{\left[r(k+3)-s(k+2)\right]^2-1}{4(k+2)(k+3)}\ ,\quad h_{r,s}\equiv h_{k+2-r,k+3-s}\ ,\quad r=1\dots k+1\ , s=1\dots k+2\ .
    \end{equation}
    We define the quantities $d_{ab}=a-b\frac{k+2}{k+3}\equiv a-bt$, so that the spectrum can be rewritten as
    \begin{equation}\label{ch1:spectrum-mm-d}
    h^{(k)}_{r,s}=\frac{1}{4t}\left(d_{rs}^2-d_{11}^2\right)\ .
    \end{equation}
    We choose the partition function to be the diagonal one:
   \begin{equation}\label{ch1:minmod-part-fct}
   Z=\sum_{r,s}\chi_{r,s}(q)\bar{\chi}_{r,s}(\bar q)\ ,
   \end{equation} 
  with the characters of minimal model representations given in appendix~\ref{app:characters}, equation~\eqref{app:characters:vir-minmod-char}
    \begin{equation}\label{ch1:minmod-char}
    \chi^{(k)}_{r,s}(q)=\frac{q^{\frac{1-c}{24}}}{\eta(\tau)}\sum_{m\in \Z}\left(q^{h_{r+2m(k+2),s}}-q^{h_{r+2m(k+2),-s}}\right)\ .
    \end{equation}
    Explicit expressions for the correlators have been computed in~\cite{Dotsenko:1984nm,Dotsenko:1984ad,Dotsenko:1985hi}, and we omit their complicated expression here.
    \subsection{Virasoro minimal models: the Runkel-Watts $\MM{\infty}$ limit}
 We now look at the $k\to\infty$ limit of unitary Virasoro minimal models. The theory we obtain is non-trivial, non-free and non-rational. Its existence was proposed in~\cite{Runkel:2001ng}, and its interpretation in terms of orbifolds was given in~\cite{Gaberdiel:2011aa}.
There has been some debate on the uniqueness of this limit, especially after the proposal of~\cite{Roggenkamp:2003qp}.
The controversy has now been resolved, as we will explain in subsection~\ref{ch2:sec:RW-as-contorbi}, thanks to the continuous orbifold interpretation of~\cite{Gaberdiel:2011aa}.

\paragraph{Central charge and spectrum}
 The central charge approaches $c=1$ in the $k\to\infty$ limit. To study the limit of the conformal weights of the primaries we use the expression~\eqref{ch1:spectrum-mm-d}:
  \begin{equation}\label{ch1:RW-spectrum-lim}
  h^{(k)}_{r,s}=\frac{(r-s)^2}{4}+\frac{r^2-s^2}{4(k+3)}+\order{\frac{1}{k^2}}\ .
  \end{equation}
  If we keep both $r$ and $s$ finite while taking $k\to\infty$, the spectrum results to be discrete and infinitely degenerate. This is a consistent CFT analysed in details in~\cite{Roggenkamp:2003qp}.
  
  We describe here another limit theory, namely the one presented in~\cite{Runkel:2001ng}\footnote{The interpretation of the interplay between these two different limits has now been given in~\cite{Gaberdiel:2011aa}, as we will review in section~\ref{ch2:sec:RW-as-contorbi} in detail.}.
  We have the freedom to rescale the labels of the minimal models, in such a way that the spectrum becomes continuous, and the degeneracies are removed. 
  Let us sketch how: define the additional variables $x_{rs}\equiv 2\sqrt{h^{(k)}_{r,s}}$, expand them for large $k$, and identify the integer and fractional part:
  \begin{equation}\label{ch1:large-k-dec-RW}
  x_{rs}= \underbrace{r-s}_{ \integer{x_{rs}}}+\underbrace{\frac12 \frac{r+s}{k+3}+\order{\frac{1}{k^2}}}_{\fract{x_{rs}}}\ .
  \end{equation}
  Since the labels $(r,s)$ are positive integers, and span a limited range given in equation~\eqref{ch1:spectrum-MM}, we can constrain the fractional part in the range
  \begin{equation}
  \frac{1}{k+3}\left(1+\frac{\integer{x_{rs}}}{2}\right)\leq\fract{x_{rs}}\leq 1-\frac{1}{k+3}\left(2+\frac{\integer{x_{rs}}}{2}\right)\ .
  \end{equation}
  We observe that for all the possible choices of $k$, the fractional part of $x_{rs}$ is different from zero. 
  Furthermore we see from equation~\eqref{ch1:large-k-dec-RW} that at leading order the variables $\fract{x_{rs}}=\frac12 \frac{r+s}{k+3}$ are uniformly distributed, and the distance between two consecutive points is infinitesimal in~$k$.
  If we set $r-s=\integer x$ and $s=\frac12\integer x+(k+3)\fract x$, then $r+s=2\integer x+2\fract x(k+3)$, and as $k$ becomes large we obtain
  \begin{equation}
   h=\frac{x^2}{4}\ ,
  \end{equation} 
  with $x\in\R-\Z$, since $0<\fract x<1$. The conclusion is that the spectrum of the limit theory is dense in $\R$, but the integer values are missing.
  \paragraph{Averaged fields and correlators}
  As outlined in the previous paragraph, $r-s=\integer{x}$ labels the level of the representation, and the average should be taken summing over the allowed non-integer values of $x_{rs}$. In formulas (see~\cite{Runkel:2001ng} for details):
  \begin{equation}
  N(x,\e, k)=\left\{(\integer{x}+n,n)\ \middle|\  \frac12\integer{x}+\left(\fract{x}-\frac{\e}{2}\right)(k+3)<n< \frac12\integer{x}+\left(\fract{x}+\frac{\e}{2}\right)(k+3)\right\}\ .
  \end{equation}
  With this definition we can promptly normalise the averaged fields as follows
  \begin{equation}
  \Phi^{\epsilon,k}_{x}=\frac{1}{|N(x,\epsilon,k)|}\sum_{(r,s)\in N(x,\epsilon,R)}\phi_{r,s}\ ,
  \end{equation}
   and the correlators are therefore defined as in the example of section~\ref{ch1:sec:free-boson}
   \begin{equation}
     \bra \Phi_{x_1}(z_1,\bar z_1)\dots \Phi_{x_r}(z_r,\bar z_r)\ket = \lim\limits_{\epsilon\to 0}\lim\limits_{k\to \infty}\beta(R)^2\alpha(R)^r\bra \Phi^{\epsilon,k}_{x_1}(z_1,\bar z_1)\dots \Phi^{\epsilon,k}_{x_r}(z_r,\bar z_r)\ket\ .
     \end{equation}
   For the details of this limit (technically very involved) look at section 4 of~\cite{Runkel:2001ng}.     
     \paragraph{Partition Function}
    We take now the $k\to\infty$ limit of the Virasoro characters of equation~\eqref{ch1:minmod-char}, for $r-s$ fixed and $r+s$ scaling with $k+3$.
    The result is given in section~\ref{app:characters:sec:char-limits}:
    \begin{equation}
        \lim\limits_{k\to \infty}\chi^{(k)}_{r,s}(q)=\frac {q^{\frac{x^2}{4}}}{\h (\t)}=\vartheta_x(\t) \ .
   \end{equation}


Out of the limit of the characters we can reconstruct the limit of the partition function just taking the limit of the expression~\eqref{ch1:minmod-part-fct}.
\smallskip
   
In conclusion we have been able to show with the method of averaged fields that Virasoro minimal models tend for $k\to\infty$ to a non-rational theory, whose bulk data can be collected in the way we have sketched. 
In section~\ref{ch2:sec:RW-as-contorbi} we will give arguments along the lines of~\cite{Gaberdiel:2011aa} on the identification of this theory with a continuous orbifold.
\newpage
\begin{subappendices}  
\section{Free boson on a circle of radius $R$}\label{ch1:sec:free-boson}
    In this section, in order to fix the notations, we give some details of the well-known theory of a free boson with values on a circumference of radius $R$.
     The CFT is encoded in the free action
 \begin{equation}\label{ch1:action-free-boson}
    S_R= \int_{\Sigma} d^2 z\  \de \phi(z,\bar z) \debar \phi(z, \bar z)\ ,
     \end{equation}
     defined on the two-dimensional Riemann surface~$\Sigma$; the theory describes a closed string moving on a circumference of radius $R$, if we interpret the field $\phi:\Sigma\rightarrow S^1$ as the coordinate on the circle, and with the identification
        \begin{equation}\label{ch1:monod-free-boson}
        \phi(e^{2\pi i}z,e^{-2\pi i}\bar z) = \phi(z,\bar z)+2\pi R n\ ,\qquad n\in \Z\ .
       \end{equation}
     The theory is free, quasi-rational\footnote{The theory in not strictly rational, since for generic radius $R$ the number of primary fields is not bounded. This happens only if $R=\sqrt{2 k}$ with $k$ positive integer in our conventions. Nevertheless for any value of~$R$ only a finite number of fields appears in the fusion of two primaries, which is the definition of a quasi-rational CFT.}, and can be exactly solved.
     
     Expanding the chiral currents $j$ and $\bar j$ in terms of their normal modes
     \begin{equation}
    j(z)= i\de \phi(z,\bar z)=\sum_{l\in\Z}j_l z^{-l-1}\qquad  \bar{j}(\bar z)= i\debar \phi(z,\bar z)=\sum_{l\in\Z}\bar{j}_l \bar z^{-l-1}
     \end{equation}
     and integrating, we can explicitly solve the equations of motion:
     \begin{equation}\label{ch1:sol-free-bos}
     \phi(z,\bar z)=\phi_0-i\left(j_0\log{z}+\bar j_0\log \bar z\right)+i\sum_{l\neq 0}\frac 1l\left(j_lz^{-l}+\bar j _l \bar z ^{-l}\right)\ .
     \end{equation}
  By imposing the identification~\eqref{ch1:monod-free-boson} on the solution~\eqref{ch1:sol-free-bos} it follows that $j_0-\bar j _0=R n$; the ground states are thus also characterised by the integer $n$, namely the number of times that the closed string winds around the compact circle. 
  Upon canonical quantisation we find for the generic ground state
 \begin{equation}
   j_0|p,n\ket=p|p,n\ket \ ,\qquad \bar  j_0|p,n\ket=(p-Rn)|p,n\ket\  .
  \end{equation}
     Virasoro generators are defined as the square of the momenta; they close the Virasoro algebra at central charge $c=1$, independently from $R$. 
     The zero modes act on the ground states as follows:
     \begin{equation}
     \begin{split}
     L_0|p,n\ket=\ &\left(\frac12 j_0j_0+\sum_{k\geq 1}j_{-k}j_k\right)|p,n\ket=\frac12 p^2|p,n\ket \\
     \bar L_0|p,n\ket=\ &\left(\frac12 \bar j_0 \bar j_0+\sum_{k\geq 1}\bar j_{-k}\bar j_k\right)|p,n\ket=\frac12\left(p-Rn\right)^2|p,n\ket\ .
     \end{split}
     \end{equation}
     As we see the on-shell level matching condition is spoiled by the compact background. 
     As a consequence the partition function
     \begin{equation}\label{ch1:free-boson-partition-fct}
     Z_{R}=\frac{1}{\left|\eta(\tau)\right|^2}\sum_{p,n}q^{\frac12 p^2}\bar q^{\frac12(p-Rn)^2}
     \end{equation}
     is invariant under modular T-transformations only if the condition $p\equiv p_{m,n}=\frac mR+\frac{Rn}{2}$ with $m,n\in\Z$ is satisfied. 
     The spectrum is thus discrete, and given by
     \begin{equation}\label{ch1:weights-prim-free-bos}
     h_{m,n}=\frac12 \left( \frac mR+\frac{Rn}{2}\right)^2\ ,\qquad  \bar h_{m,n}=\frac12 \left( \frac mR-\frac{Rn}{2}\right)^2\ ,\qquad\text{with} \ m,n\in\Z\ ;
     \end{equation}
     the integers $n,m$ are called winding and Kaluza-Klein modes respectively. 
     The primary states are thus characterised by the two integers $m,n$ and are denoted~$|m,n\ket$.
     Correspondingly the primaries fields are indicated with~$\phi_{m,n}(z,\bar z)$.
     
It is known that this theory enjoys T-duality (see~\cite{Giveon:1994fu} for an extensive account), namely the property that the quantum theory at radius $R$ is completely equivalent to the theory at radius~$\frac2R$ (in our conventions), once we exchange the winding and the Kaluza-Klein modes.
     
If the radius is~$R=\sqrt{2k}$, then for left-chiral primary fields we have
\begin{equation}
\overline{L}_0|m,n\ket=\bar h_{m,n|m,n\ket}=\frac 14\left(\frac{m}{\sqrt k}-\sqrt k n\right)^2|m,n\ket \= 0\ .
\end{equation}
It follows that left-chiral primary fields are characterised by only one integer, since $m\= kn$.
This property allows us~(see e.g.\cite{Blumenhagen:2009zz}) to rewrite the modular invariant partition function~\eqref{ch1:free-boson-partition-fct} in terms of a finite sum of Ka\v c-Peterson $\Theta$-functions~(definition in appendix~\ref{app:characters})
\begin{equation}
Z_{\kmalg{u}(1)_{2k}}=\frac{1}{|\eta(\t)|^2}\sum_{m=-k+1}^{k}\left|\Theta_{m,k}(q)\right|^2\ .
\end{equation}        
The theory becomes therefore rational, and the representations are labeled by the $U(1)$-charge~$m$; its chiral algebra is conventionally called~$\kmalg{u}(1)_{2k}$.
The representations are identified modulo $2k$.
The case of $k=1$, self dual radius~$R=\sqrt 2$, possesses furthermore the enhanced symmetry given by the exchange of winding numbers with Kaluza-Klein ones; its chiral algebra becomes~$\kmalg{su}(2)_1$.
\end{subappendices}
\chapter{Discrete and continuous orbifolds in CFT}\label{ch:orbifolds}
 In this chapter we introduce some notions for orbifolds in CFT and in boundary CFT (BCFT) that will be useful in what follows.
 The basic idea is to allow the fields of the CFT to take values on spaces (in general not everywhere differentiable) obtained by modding out the action of a group from a smooth manifold.
Theories on orbifolds are very interesting for a variety of reasons, and have been used during the years for many different scopes, ranging from defining realistic string compactifications (seminal papers are e.g.~\cite{Casas:1988hb,Font:1988mm}) to making manifest difficult mathematical results (just to give an example, the string theoretical explanation of the McKay correspondence, see~\cite{Reid:1997zy} for an informal account).
The seminal idea dates back to the early days of the first superstring revolution with the works~\cite{Dixon:1985jw,Dixon:1986jc}; the authors realised that strings can propagate smoothly on discrete orbifold backgrounds, hence allowing the exploration of isolated conical singularities.
The original notion was generalised soon afterwards, with the detailed recipe to construct CFTs with values on orbifolds~\cite{Dixon:1986qv,Dijkgraaf:1989hb}, allowing for asymmetric group action on left and right movers~\cite{Narain:1986qm} and allowing different phases for each group element acting on the fields~\cite{Vafa:1986wx}.
Until very recently all analyses have been carried out only for (in general) non-abelian discrete groups; continuous group have been considered\footnote{To our surprise we could find the idea of continuous orbifolds in the literature in only two places: in Polchinski's first volume~\cite{PolchinskiBookI:1998} in section 8.5 (where a different concept is meant though), and in~\cite{Israel:2004ir} in subsection 2.3.} first in~\cite{Gaberdiel:2011aa}.

We do not have any pretension of completeness in this chapter. 
We limit ourselves here to the material needed in subsequent chapters.
This chapter is organised as follows: in section~\ref{ch2:sec:discrete-orbifold-generalities} we briefly review the notion of discrete orbifolds in CFTs (in general with boundaries) and we discuss a very simple example; in section~\ref{ch2:sec:contorbi} we introduce continuous orbifolds and we discuss two examples related to the limit theories of chapter~\ref{ch:limit-theories}: the T-dual of the decompactification limit of the free boson on a circle obtained through shrinking the radius to a point, and the construction of reference~\cite{Gaberdiel:2011aa} of Runkel-Watts theory (reviewed in section~\ref{ch1:sec:RW}) in terms of a non-abelian continuous orbifold of a free boson.

\section{Discrete orbifolds in CFT and in BCFT}\label{ch2:sec:discrete-orbifold-generalities}
  
Suppose we are given a CFT, whose classical fields take value on a smooth manifold~$\mathcal{M}$, and suppose we define a discrete (in general non-abelian) group~$G$ acting on~$\mathcal M$. 
We can consider the orbifold~$\mathcal M /G$, obtained by identifying the points of~$\mathcal{M}$ which lie in the same discrete orbit of~$G$. 
If the action of the group on the manifold is free, the resulting geometry is again a smooth manifold. If instead the manifold possesses fixed points under the action of~$G$, then the space develops curvature singularities localised on the fixed points.
The resulting CFT is well-defined also in the latter case, and this fact makes CFTs with values on orbifolds interesting objects to study, since as sigma-models they ``see singularities in a smooth way''.
\subsection{Bulk theory}
The points in the target space of the sigma-model are identified under the action of~$G$, hence the classical bulk fields undergo an identification of the form
\begin{equation}\label{ch2:orbifold-action-geom-general}
\phi(e^{2\pi i}z,e^{-2\pi i}\bar z)=U(g)\cdot\phi (z, \bar z)\ ,
\end{equation} 
where~$\phi$ is a coordinate on~$\cM$, $g\in G$, and $U(g)$ denotes its action on the target manifold. In other words, the closed string described by the parent~CFT closes up to a $g\in G$ action.
From the two-dimensional QFT perspective, not all the states in the theory can survive, and new sectors may appear. 
The bulk spectrum of the orbifold~CFT/$G$ is given by \textit{untwisted} and~\textit{twisted} sectors: the former are modules of the parent~CFT left invariant by the action of the group~$G$, while the latter are absent in the parent~CFT, and describe those closed string excitations which close only up to a $g$-twist.
  
\paragraph{Bulk: untwisted sector} 
The first step is to keep from the set of ground states of the parent theory only those states that are left invariant by the action of the group~$G$. 
In practice this is done by defining the action of the group~$G$ on the space of states of the~CFT, compatible with the constraint~\eqref{ch2:orbifold-action-geom-general}, and then to project out the states that are not invariant. 
The states that survive the projection constitute the~untwisted sector of the orbifold~CFT.
Clearly, we need to be sure that the chiral algebra admits~$G$ as a symmetry, namely that there exist non-empty submodules invariant under the projection; if this is not the case, the resulting theory is trivial. 
Let us suppose that the orbifold is not trivial, and show how to calculate the partition function.
 
The first thing we want to compute, is the partition function for the untwisted sectors. 
The projector onto invariant subsectors is given by (notations are summarised in section~\ref{ch2:sec-app:notations})
\begin{equation}\label{ch2:projector-discrete}
P:\mathcal{H}_{0}\mapsto\mathcal{H}_{\mathbb{1}}\ ,\qquad P=\frac{1}{|G|}\sum_{g\in G}g\ ,
\end{equation}
where $g$ denotes for simplicity the action of the group element on the Hilbert space and $|G|$ the order of the discrete group. 
The untwisted contribution to the partition function is obtained then by just inserting the projection inside the trace defining the sum over the states of the parent theory: it is given by
\begin{equation}
Z_{\text{untw}}=\frac{1}{|G|}\sum_{g\in G}\Tr_{\Hilb_0}\ g\,q^{L_0-\frac{c}{24}}\bar q^{\bar{L}_0-\frac{\bar c}{24}}=\frac{1}{|G|}\sum_{g\in G} \torus{g}{\id} \ .
\end{equation}
  
\paragraph{Bulk: twisted sectors} 
The second step consists in supplementing the resulting~CFT of those excitations that describe configurations of the closed string, closing only up to a $g\neq\mathbb 1$ action: we have to add the so-called~\textit{twisted sectors} to the orbifold~CFT.
Usually we do not know the explicit solutions for the fields, hence we cannot directly recognise the spectrum of the twisted sectors from equation~\eqref{ch2:orbifold-action-geom-general}.
Moreover, the sigma-model description can be extremely complicated, while the defining~CFT structures may be known and easy to handle.
The technology used is therefore an indirect one: we impose modular invariance on the partition function by adding the right combination of characters. 
These new characters represent the twisted sectors.

Let us show how this is done: the partition function
\begin{equation}\label{ch2:generic-partition-fct-orbifold}
Z=\frac{1}{|G|}\sum_{g,h\in G}\Tr_{\Hilb_h}\ g\,q^{L_0-\frac{c}{24}}\bar q^{\bar L_0-\frac{\bar c}{24}}=\frac{1}{|G|}\sum_{g,h\in G}\torus{g}{h}
\end{equation}
is automatically modular invariant, as one can easily prove using the modular transformation properties of the the blocks defined in section~\ref{ch2:sec-app:notations}.
For abelian groups, equation~\eqref{ch2:generic-partition-fct-orbifold} is also a useful starting point for explicit calculations.
For non-abelian groups several comments become necessary: the partition function written in equation~\eqref{ch2:generic-partition-fct-orbifold} is in this case ambiguous under modular T-transformations, since in general $[g,h]\neq 0$, and in equation~\eqref{ch2:S-T-modular-transf-box} we get in principle different results if we act with the twist on the left or on the right with respect to the projection inside the trace. 
It is furthermore redundant because
\begin{equation}\label{ch2:invariance-torus-gh-twisted}
\torus{g}{h}=\torus{g}{ghg^{-1}}\ .
\end{equation}
The last statement can be explained as follows~\cite{Dixon:1985jw,Dixon:1986jc}: consider a field~$\phi(z)$ belonging to the $h$-twisted sector $\mathcal{H}_{h}$. We know by definition $\phi(e^{2\pi i}z,e^{-2\pi i}\bar{z})= h\cdot\phi(z,\bar z)$; now act with $g$, as we do if we project onto the $G$-invariant subspace: we get $g\cdot \phi(e^{2\pi i}z,e^{-2\pi i}\bar{z})=gh\cdot\phi(z,\bar z)=(ghg^{-1})g\cdot\phi(z,\bar z)$; this means that $g\cdot\phi$ belongs to the space $\mathcal{H}_{ghg^{-1}}$, which is different from~$\mathcal{H}_h$ if $g$ and $h$ do not commute: under the action of the group, sectors in a given conjugacy class mix among each other.

Equation~\eqref{ch2:invariance-torus-gh-twisted} defines a map $\phi_g$ from the $h$-twisted Hilbert space to the $ghg^{-1}$-twisted one, which leaves invariant the full block.
Applying the map $\phi_{g'}$ to the $g$-twined $h$-twisted block we get
\begin{equation}\label{ch2:combined-cyclic-conj-invariance}
\torus{g}{h}=\torus{g'gg'^{-1}}{g'hg'^{-1}}\ .
\end{equation}
From equation~\eqref{ch2:invariance-torus-gh-twisted} we infer that in~\eqref{ch2:generic-partition-fct-orbifold} for every choice of~$h$ we must restrict the sum over~$g$ to those for which $ghg^{-1} = h$. 
Furthermore, equation~\eqref{ch2:combined-cyclic-conj-invariance} tells us that the contribution for the $h$-twisted sector depends only on the conjugacy class of~$h$.

We get then
\begin{align}
Z=\, \sum_{h\in G}\frac{1}{|G|}\sum_{g\in G}\torus{g}{h}=\sum_{[h]}\frac{|[h]|}{|G|}\sum_{g\in N_{[h]}}\torus{g}{[h]}\ .
\end{align}
We have used the following definitions: the conjugacy class of $h$
\begin{equation}\label{ch2:def-conjclass-discrete}
[h]\equiv\left\{h'\in G\ \middle | \ h'=ghg^{-1}\ ,\ \text{for some}\  g\in G\right\}\ ,
\end{equation}
and the centraliser of $h$
\begin{equation}\label{ch2:def-centraliser-discrete}
N_h\equiv\left\{g\in G\ \middle | \ hg=gh\ \right\}\ ,
\end{equation}
which is invariant for all the elements~$h$ which belong to the same conjugacy class.
Finally by the known result
\begin{equation}\label{ch2:thm-decomp-group}
|G|=|N_{[h]}|\times|[h]|\ ,\ \forall h\in G\ ,
\end{equation}
we get the formula
\begin{equation}\label{ch2:general-partition-orbifold}
Z=\sum_{[h]}\frac{1}{|N_{[h]}|}\sum_{g\in N_{[h]}}\torus{g}{[h]}\ .
\end{equation}
\smallskip
To compute the partition function in a non-abelian discrete orbifold~CFT one makes use of equation~\eqref{ch2:general-partition-orbifold}: it tells us that twisted sectors are labeled by conjugacy classes, and in the non-abelian case the projection for the twisted sectors has to be performed onto those subsectors which are invariant under the action of the elements commuting with the twist. 
Automatically we remove the ambiguities under modular transformations, since the centraliser is invariant under conjugation, so that the twists on the two cycles of the torus always commute (see remark under equation~\eqref{ch2:gen-modular-transf-box}).

\subsection{Boundary theory}
The analysis of the boundary theory is similar in spirit to the one for the bulk theory (for details the reader can consult~\cite{Douglas:1996sw,Johnson:1996py,Billo:2000yb,Bertolini:2001gq}): we search for superpositions of conformal boundaries in the parent theory that are invariant under the action of the orbifold group~$G$. 
This is done by studying the action of~$G$ on the boundary states. 
The orbifold group acts on open strings not only at the level of the chiral algebra as in the bulk case, but also changing the parent boundary labels (one can easily understand this making use of geometric intuition: since a brane effectively describes an embedding into a target manifold, the action of the orbifold group on the target manifold, changing the bulk geometry, maps the brane into another - not necessarily coincident - submanifold). 

Let us start with the example of point-like branes in a background with fixed points under the action of the orbifold group.
Geometrically  we can imagine to pick a point-like D-brane of the parent theory sitting at a generic (non-fixed) point, and to look at its image under the action of the whole orbifold group.
If we superpose all the images, the configuration we obtain is surely invariant by construction under the orbifold action.
We realise in this way a so-called \textit{bulk brane} of the orbifold theory.

If we pick instead a D-brane of the parent theory sitting at a fixed point under the action of the orbifold group, the action of the group is trivial on the brane, and the brane is by itself invariant.
We realise in this way the so-called \textit{fractional branes} of the orbifold theory: fractional branes correspond to boundary conditions already invariant under the orbifold action in the parent theory.

Let us be more specific: the action of the orbifold group $G$ on the boundary labels of the parent BCFT has in general non-trivial orbits: by picking up the whole orbit of a boundary condition of the parent theory we end up with a superposition of boundary states. 
The states of the boundary spectra associated to these superpositions are by construction invariant under the orbifold projection, and describe bulk branes.
Bulk branes are superpositions of branes coming from the parent~CFT. As a consequence they couple only to bulk representations of the original theory. 
The invariance under the orbifold action is given from one side by their ``space-time'' configuration, since their world-volumes span submanifolds given by $G$-mirrored geometries;  from the other by the orbifold projection in the open string channel. 
We see that they can only couple to the untwisted sectors in the closed string channel.

If the action of the group admits a stabiliser for certain submanifolds, then the irreducible representations of the stabiliser constitute the boundary labels of the fractional branes.
Fractional branes are localised at the orbifold fixed points where the twisted sectors live (in the~$\a'\to 0$ approximation): their moduli space is trivial in the directions of the orbifold.

In the following we will be mostly interested in fractional branes, since our main concern in this thesis is the discussion of continuous orbifolds, where untwisted states are outnumbered by the twisted ones (see section~\ref{ch2:sec:contorbi}).

\paragraph{Fractional branes}  


The quantity to analyse is the open string annulus amplitude appropriately weighted with the finite characters of the irreducible representations of the orbifold group.
The open string states circulating the loop are the states of the open string spectrum of the parent theory left invariant under the orbifold projection.
The self energy of an open string stretched between two fractional branes labeled by the group representations $R$ and $S$ reads therefore (see for example~\cite{Billo:2000yb}):
\begin{equation}\label{ch2:self-energy-open-string-discrete}
Z_{RS}(\tilde \t)=\frac{1}{|G|}\sum_{g\in G}\chi^*_R(g)\chi_S(g)\Tr_{\mathcal{H}_0^{\text{open}}}\,g\,\tilde{q}^{L_0-\frac{c}{24}} \ ,
\end{equation}
where $q^{2\pi i\tilde \t}$ is the open string modular nome, $\mathcal H^{\text{open}}$ is the open string Hilbert space of the parent~CFT, and $\chi_R(g)$ denotes the group character of the irreducible representation~$R$.
Using the Clebsch-Gordan decomposition of the representation $R^*\otimes S$ into $Q^*$ representations
\begin{equation}
\chi_R^*(g) \, \chi_S(g) = \sum_Q N_{SQ}^{R} \, \chi_Q^*(g) \ ,
\end{equation}
we can write 
\begin{equation}
Z_{RS} (\tilde t) =  \sum_Q N_{SQ}^{R} \frac{1}{|G|} \sum_{g\in G} 
\Tr_{\mathcal{H}_0^{\text{open}}}\left( g \, \tilde q^{L_0-\frac{c}{24}} \right) \, \chi_Q^*(g) \ .
\end{equation}
We can now grade the open string spectrum ${\mathcal H^{\text{open}}}$ with respect to the action of $G$, $\mathcal{H}^{\text{open}} = \bigoplus_{S} S \otimes \Hilb^{\text{open}}_{(S)}$, so that
\begin{equation}
\Tr_{\Hilb^{\text{open}}}\left( g \, \tilde q^{L_0-\frac{c}{24}} \right) 
= \sum_{S} \chi_S(g)\, \Tr_{\Hilb^{\text{open}}_{(S)}}\left( \tilde q^{L_0-\frac{c}{24}} \right) \ ,
\end{equation}
and using the orthogonality of group characters
\begin{equation}
 \frac{1}{|G|} \sum_{g\in G}  \chi^*_Q(g)\,  \chi_S(g) = \delta_{QS}
\end{equation}
the open string spectrum between fractional branes labeled by $R$ and $S$ consists then of those sectors $\Hilb^{(Q)}$ in $\Hilb_{\text{open}}$ that belong to the representation $Q$ of the orbifold group
\begin{equation}
Z_{RS} (\tilde \t) =  \sum_Q N_{SQ}^{R}  \Tr_{\Hilb^{\text{open}}_{(Q)}}\left( \tilde q^{L_0-\frac{c}{24}} \right)  \ .
\end{equation} 

To summarise, in the orbifold CFT/$G$ we find two species of D-branes, fractional and bulk.
The former are labeled by irreducible representations of the stabiliser subgroup of the orbifold group $G$, the latter are obtained as $G$-invariant superpositions of parent boundary conditions.

\subsection{Example: compactified boson as an orbifold}
As a very simple example of the orbifold construction, we illustrate how to get back the spectrum of a free boson compactified on a circle of radius~$R$ described in section~\ref{ch1:sec:free-boson} as a discrete orbifold of the free boson on the real line.
The identification of the field $\phi(e^{2\pi i}z,e^{-2\pi i}\bar z) = \phi(z,\bar z)+2\pi R n$ can be seen as the action on the fields of our theory of the discrete infinite abelian group of translations
\be
G=\left\{\mathbb{1},T_{2\pi R},T_{-2\pi R},T_{4\pi R},T_{-4\pi R},\dots,T_{2n\pi R},T_{-2n\pi R}, \dots\right\}\ ,
\ee 
where the action is defined as 
\be
T_{2n\pi R}\cdot \phi(z,\bar z)=\phi(z,\bar z)+2\pi R n\ .
\ee
We write~$\Hilb_{\R}$ to indicate the Hilbert space of the uncompactified boson, whose states~$|p;\{n_i\}\ket$ are parameterised by the real momentum $p$ and by the set of excited oscillator modes $\{n_i\}$. 
The trace over the ground states becomes therefore an integral over the real numbers with a suitable measure; the spectrum of the parent theory is $L_0|p;\{n_i\}\ket=\frac12 p^2|p;\{n_i\}\ket$, and the action of the translations is just $T_{2n\pi R}|p;\{n_i\}\ket=e^{2\pi i R n p}|p;\{n_i\}\ket$.  
The untwisted partition function is then
\begin{align}
Z_{\text{untw}}=\sum_{n\in\Z}\torus{T_{2\pi R n}}{\mathbb{1}}=&\ \sum_{n\in \Z}\Tr_{\Hilb_{0}}\,T_{2\pi nR}\ q^{L_0-\frac{1}{24}}\bar q^{\bar L_0-\frac{1}{24}}\\
=&\ \sum_{n\in\Z}\sum_{\{n_i\}}\int_{\R} dp\,e^{2\pi i R n p}\bra p;\{n_i\} | q^{L_0-\frac{1}{24}}\bar q^{\bar L_0-\frac{1}{24}} |p;\{n_i\}\ket\nonumber\\
=&\ \frac{1}{\left|\h(\t)\right|^2}\sum_{n\in\Z}
\int_{\R} dp\, e^{2\pi i R n p}e^{\pi i (\t-\bar\t)p^2}
= \frac{1}{\left|\h(\t)\right|^2}\frac{1}{\sqrt{2\t_2}}\sum_{n\in\Z}e^{-\frac{n^2\pi R^2}{2\t_2}}\label{ch2:free-bos-as-orbi-untw-1}\\
=\ &\frac{1}{\left|\h(\t)\right|^2}\frac{1}{R}\sum_{m\in\Z}(q\bar q)^{\frac12\frac{m^2}{R^2}}\ ,\label{ch2:free-bos-as-orbi-untw-2}
\end{align}
where going from line~\eqref{ch2:free-bos-as-orbi-untw-1} to the result~\eqref{ch2:free-bos-as-orbi-untw-2} we have used the Poisson formula
\begin{equation}\label{ch2:poisson-formula}
\sum_{n\in\Z}e^{-\pi a n^2+b n}=\frac{1}{\sqrt{a}}\sum_{m\in\Z}e^{-\frac{\pi}{a}(m+\frac{b}{2\pi i})^2}\ .
\end{equation}
We see that the Kaluza-Klein modes of equation~\eqref{ch1:weights-prim-free-bos} correspond to the untwisted sector in our example.

We perform now a modular S-transformation
\begin{equation}
\torus {T_{2\pi R n}}{\mathbb{1}}\quad\overset{S}{\longmapsto}\quad\torus{\mathbb{1}}{T_{2\pi R n}}=\frac{1}{|\h(\t)|^2}\frac{1}{\sqrt{2\t_2}}e^{-\frac{n^2\pi R^2}{2\t_2}|\t|^2}\  ,
\end{equation}
where we used the expression in~\eqref{ch2:free-bos-as-orbi-untw-1} leaving the sum over~$n$.
Now applying $l$ times the modular T-transformation we find
\begin{equation}
\torus {\id}{T_{2\pi R n}}\quad\overset{T^l}{\longmapsto}\quad\torus{T_{2\pi R l n}}{T_{2\pi R n}}=\frac{1}{|\h(\t)|^2}\frac{1}{\sqrt{2\t_2}}e^{-\frac{n^2\pi R^2}{2\t_2}|\t+l|^2}\  ,
\end{equation}
from which we can define an integer variable~$p=ln$, so that we get the expression
\begin{equation}
\torus{T_{2\pi R p}}{T_{2\pi R n}}=\frac{1}{|\h(\t)|^2}\frac{1}{\sqrt{2\t_2}}e^{-\frac{\pi R^2}{2\t_2}|n\t+p|^2}\  .
\end{equation}
In order to obtain the full partition function we have to sum over the integers~$n,p$.
We start with the sum over~$p$, and we perform again a Poisson resummation (this time $a=\frac{R^2}{2\t_2}, b=\frac{\pi R^2 n\t_1}{\t_2}$ in equation~\eqref{ch2:poisson-formula}):
\begin{equation}
Z=\sum_{n,p\in\Z}\torus{T_{2\pi R p}}{T_{2\pi R n}}=\frac{1}{|\h(\t)|^2}\frac{1}{R}\sum_{n,m\in\Z}q^{\frac 12\left(\frac{m}{R}+\frac{nR}{2}\right)^2}\bar q^{\frac 12\left(\frac{m}{R}-\frac{nR}{2}\right)^2}\ .
\end{equation}
We recover the free boson compactified on a circle of radius~$R$.
The winding modes appear as the twisted sectors.

 \section{Continuous orbifolds in CFT and in BCFT}\label{ch2:sec:contorbi}
In this section we introduce the due material to treat continuous orbifolds in detail. 
The basic idea is to quotient out of a parent~CFT a continuous compact Lie group~$G$, instead of a discrete one~\cite{Gaberdiel:2011aa}; we keep only invariant states under the identification by~$G$, and supplement the theory with suitable twisted sectors: we get the theory~$\text{CFT}/G$.
This construction can be at first sight mistaken with a coset construction; there is an important difference though: in coset theories we gauge a symmetry which is local with respect to the world-sheet coordinates. 
In the case at hand we are gauging a global symmetry, and supplementing the necessary twisted sectors. 
The~CFTs we produce by this procedure are naturally non-rational, since the twisted sectors come in continuous families. 
As a consequence, the untwisted sector is outnumbered by the twisted ones: bulk branes disappear, since they only couple to untwisted sectors.

To describe continuous orbifolds we have to generalise to continuous groups the operations that we have outlined for discrete groups. 
The sums over group elements have to be translated into integrals over submanifolds of the group manifold~$G$. We choose the Haar measure~$d\mu (g)$ to integrate. 
Let us start from the untwisted sector: the projection operator of equation~\eqref{ch2:projector-discrete} becomes
\begin{equation}
P=\frac{1}{|G|}\int_{G} d\mu(g)\, g\ ,
\end{equation}
where $|G|$ is the volume of~$G$ measured with the Haar measure $d\m (g)$. The untwisted partition function becomes then in this case
\begin{equation}
Z_{\text{untw}}=\frac{1}{|G|}\int_G d\m (g)\, \Tr_{\Hilb_{0}}\left(g\, q^{L_0-\frac{c}{24}}\bar{q}^{\bar {L}_0-\frac{\bar c}{24}}\right)\ .
\end{equation}
Every group element~$h$ is conjugated to some element in~$\T_{h}/\mathcal{W}$~\cite{Broecker_Dieck}, where~$\T_h$ is the Cartan torus passing by~$h$, i.e. the exponentiation of the $r$-dimensional Cartan subalgebra; this is one of the maximal tori of~$G$, which are all conjugated. $\mathcal W$ represents the Weyl group of~$G$, i.e. the set of inner automorphisms of the maximal tori.
Using the cyclicity of the trace we can thus simplify
\begin{equation}\label{ch2:untwisted-sector-contorbi}
Z_{\text{untw}}=\frac{1}{|G|}\int_{\T/\mathcal{W}} d\hat{\m} (h)\, \Tr_{\Hilb_{0}}\left(h\, q^{L_0-\frac{c}{24}}\bar{q}^{\bar {L}_0-\frac{\bar c}{24}}\right)\ ,
\end{equation}
where we have decomposed the measure in the same spirit as in equation~\eqref{ch2:thm-decomp-group}
\begin{equation}\label{ch2:Haar-meas-decomp}
d\hat{\mu}(h)=\text{Vol}\left([h]\right)\times d\mu(h)\ .
\end{equation}
The twisted sectors can also be defined accordingly, starting from equation~\eqref{ch2:general-partition-orbifold}. 
Since twisted sectors are labeled by the different conjugacy classes, they are now labeled by elements of~$\T_{h}/\mathcal{W}$. For a generic $h$, the centraliser~$N_h$ is now the Cartan torus passing through~$h$.
We get to the following general formula
\begin{equation}
Z=\int_{\T/\mathcal W}d\hat{\mu}(h)\,\frac{1}{|\T|}\int_{\T}d\m (t)\,\Tr_{\Hilb_h}\left(t\,q^{L_0-\frac{c}{24}}\bar{q}^{\bar {L}_0-\frac{\bar c}{24}}\right)\ .
\end{equation}
This is the starting point in constructing continuous orbifolds.

The boundary spectrum of a continuous orbifold is characterised by the fact that fractional branes are the only ones that survive the orbifold projection.
Equation~\eqref{ch2:self-energy-open-string-discrete} describing the spectrum for open strings stretched between two fractional branes, simply becomes in the continuous case
\begin{equation}\label{ch2:self-energy-open-string-continuous}
Z_{RS}(\tilde \t)=\frac{1}{|G|}\int _{g\in G}d\mu(g)\,\chi^*_R(g)\chi_S(g)\Tr_{\mathcal{H}_0^{\text{open}}}\,g\,\tilde{q}^{L_0-\frac{c}{24}} \ .
\end{equation}

\subsection{Example: continuous orbifold description of the limit of the free boson}\label{ch2:subs:boson-on-a-point}
In this section we want to analyse the simple example of the continuous orbifold $S^1/U(1)$, which could be named as ``boson on a point": the action of the orbifold group is geometrically equivalent to shrinking the radius of the free boson to zero.
The theory we obtain is the T-dual of the decompactification limit of the free boson analysed in section~\ref{ch1:sec:free-boson-limit}.

The parent theory is the free boson compactified on a circle (see section~\ref{ch1:sec:free-boson}). 
We recall here the eigenvalues of the zero modes and the torus partition function for reference:
\begin{equation}
j_0|m,n\ket=\left(\frac{m}{R}+\frac{n R}{2}\right)|m,n\ket\ ,\quad \bar j_0|m,n\ket=\left(\frac{m}{R}-\frac{n R}{2}\right)|m,n\ket\ ,
\end{equation}
and
\begin{equation}\label{ch2:partition-fct-free-boson-on-R}
\torus{\mathbb{1}}{\mathbb{1}}=\Tr_{\mathcal{H}_0}q^{L_0-\frac{1}{24}}\bar{q}^{\bar{L}_0-\frac{1}{24}}=\frac{1}{|\eta(\t)|^2}\sum_{m,n\in\Z}q^{\frac12(\frac m R+\frac{nR}{2})^2}\bar{q}^{\frac12(\frac m R-\frac{nR}{2})^2}\ .
\end{equation}
\paragraph{Continuous orbifold as a limit}
The problem at hand, dual to the limit of section~\ref{ch1:sec:free-boson-limit}, has a natural interpretation as a limit theory.
Consider the orbifold obtained by modding out translations along the circle of an angle~$2\pi k/K$, namely defined by the identification
\be
\phi(e^{2\pi i}z,e^{-2\pi i}\bar z)=\phi(z, \bar z)+2\pi R\frac{k}{K}\ ,\qquad k=0,1,\dots K-1\ .
\ee
Since the bosonic coordinate is~$2\pi R$-periodic the orbifold group\footnote{In the literature, in contrast with our terminology, it is common to find the expression~``$\Z_K$ orbifold'' to denote the orbifold theory obtained by modding out a discrete phase. $\Z_K$ acts in this other case as $\phi(e^{2\pi i}z,e^{-2\pi i}\bar z)=e^{\frac{2\pi i k}{K}}\phi(z, \bar z)$, with $k=0,1,\dots K-1$.} is~$\Z_K$.
It is easy to show, and to visualise geometrically, that modding out the action of the group on the circle of radius~$R$ accounts to reduce the radius of the circle to~$R/K$.
In particular one can show that the partition function of the free boson on a circle of radius~$R/K$ is equal to the partition function of the~$\Z_K$ orbifold of the free boson at radius~$R$.
Very heuristically, we can imagine to send~$K\to\infty$, in such a way that the group of discrete angular translations~$\Z_K$ approaches the group of continuous angular translations~$U(1)$.
We see then, in the spirit of the limits of sequences of chapter~\ref{ch:limit-theories}, that~$S^1/U(1)$ can be understood as the limit of the discrete infinite sequence of CFTs $\left\{S^1/\Z_2,S^1/\Z_3,\dots,S^1/\Z_K,\dots\right\}$.

\paragraph{Group action}
Let us go back to the actual construction of~$S^1/U(1)$.
We act with the orbifold group as a translation of~$a\in\R$ along the circle
\begin{equation}
 U(a)\cdot X(z,\bar z) = X(z,\bar z)+a\ .
\end{equation}
This is realised on the world-sheet through the operator $P_a=e^{i a \pi_0}$ with  $\pi_0=\frac12(j_0+\bar j_0)$, which is the generator of translations. 
The zero-modes of the boson are the only ones affected by the orbifold transformation, namely
\begin{equation}\label{ch2:action-group-boson-point}
P_a|m,n\ket=e^{ia\frac{m}{R}}|m,n\ket\ .
\end{equation}
\paragraph{Untwisted partition function}
The contribution of the twined untwisted partition function is
\begin{equation}\label{ch2:twined-untw-block-boson-point}
\begin{split}
\torus{P_a}{\mathbb{1}}=&\ \Tr_{\mathcal{H}_0}P_aq^{L_0-\frac{1}{24}}\bar q^{\bar{L}_0-\frac{1}{24}}=\sum_{m,n\in\Z}e^{ia\frac m R}\cdot\left[\frac{q^{\frac12(\frac m R+\frac{nR}{2})^2}}{\eta(\t)} \ \frac{\bar{q}^{\frac12(\frac m R-\frac{nR}{2})^2}}{\eta(\bar{\t})}\right]\\
 =&\sum_{m,n\in\Z}e^{ia\frac m R}\cdot \left[\vartheta_{\frac{\sqrt2 m}{ R}+\frac{nR}{\sqrt2}}(\t)\  \vartheta_{\frac{\sqrt2 m}{ R}-\frac{nR}{\sqrt2}}(\bar \t)\right]\ , 
\end{split}
\end{equation}
where function $\vartheta_p(\t)$ (also defined in appendix~\ref{app:characters}) reads
\begin{equation}
\vartheta_{p}(\t)=\frac{q^{\frac{p^2}{4}}}{\eta(q)}\ .
\end{equation}
\paragraph{Twisted sectors}
We apply a modular S-transformation ($\tau\mapsto-\frac{1}{\tau}$) to get the twisted untwined characters,
\begin{align}
\torus {P_a}{\mathbb{1}}\quad\overset{S}{\longmapsto}\quad\torus{\mathbb{1}}{P_a}\  ,
\end{align}
which is an easy task using (see again appendix~\ref{app:characters})
\begin{equation}\label{theta-S}
\vartheta_p(\tilde\t)=\frac{1}{\sqrt 2}\int_{-\infty}^{+\infty}ds\  e ^{\pi i p s}\vartheta_s(\t)\ .
\end{equation}
The result is
\begin{align}
\torus{\mathbb{1}}{P_a}=\Tr_{\mathcal{H}_a}q^{L_0-\frac{1}{24}}\bar q^{\bar{L}_0-\frac{1}{24}}=\sum_{r,p\in\Z}\vartheta_{\frac{Rr}{\sqrt 2}-\frac{a}{2\pi\sqrt2}+\frac{p\sqrt2}{R}}(\t)\ \vartheta_{\frac{Rr}{\sqrt 2}-\frac{a}{2\pi\sqrt2}-\frac{p\sqrt2}{R}}(\bar \t)\ ,
\end{align}
where we have used
\begin{equation}
\sum_{m\in\Z}e^{2\pi i mr}=\sum_{p\in\Z}\delta(r-p)\ ,
\end{equation}
and saturated the $\delta$s with the integrals coming from equation~\eqref{theta-S}.

If we now apply a modular T-transformation 
\begin{align}
\torus {\mathbb{1}}{P_a}\quad\overset{T}{\longmapsto}\quad\torus{P_a}{P_a}\  ,
\end{align}
since
\begin{equation}
\vartheta_{p}(\tau+1)=e^{\frac{i\pi}{2}(p^2-\frac16)}\vartheta_p(\tau)\ ,\qquad\vartheta_{p}(\bar{\tau}+1)=e^{-\frac{i\pi}{2}(p^2-\frac16)}\vartheta_p(\bar{\tau})\ ,
\end{equation}
we have
\begin{equation}
\begin{split}
\torus{P_a}{P_a}=&\sum_{r,p\in\Z}e^{2\pi i p(r-\frac{a}{2\pi R})}\vartheta_{\frac{Rr}{\sqrt 2}-\frac{a}{2\pi\sqrt2}+\frac{p\sqrt2}{R}}(\t)\ \vartheta_{\frac{Rr}{\sqrt 2}-\frac{a}{2\pi\sqrt2}-\frac{p\sqrt2}{R}}(\bar \t)\\
=&\sum_{r,p,\in\Z}e^{-i\frac{ap}{R}}\vartheta_{\frac{Rr}{\sqrt 2}-\frac{a}{2\pi\sqrt2}+\frac{p\sqrt2}{R}}(\t)\ \vartheta_{\frac{Rr}{\sqrt 2}-\frac{a}{2\pi\sqrt2}-\frac{p\sqrt2}{R}}(\bar \t)\ ,
\end{split}
\end{equation}
from which it is easy to guess the general block:
\begin{equation}
\torus{P_a'}{P_a}=\sum_{r,p,\in\Z}e^{-i\frac{a'p}{R}}\vartheta_{\frac{Rr}{\sqrt 2}-\frac{a}{2\pi\sqrt2}+\frac{p\sqrt2}{R}}(\t)\ \vartheta_{\frac{Rr}{\sqrt 2}-\frac{a}{2\pi\sqrt2}-\frac{p\sqrt2}{R}}(\bar \t)\ .
\end{equation}
The $a$-twisted contribution reads then:
\begin{align}\label{point-twisted}
Z_{a\text{-tw}}=\int_{0}^{2\pi R}\frac{da'}{2\pi R}\ \torus{P_a'}{P_a}=\sum_{r\in\Z}\vartheta_{\frac{Rr}{\sqrt 2}-\frac{a}{2\pi\sqrt2}}(\t)\vartheta_{\frac{Rr}{\sqrt 2}-\frac{a}{2\pi\sqrt2}}(\bar \t)\ ,
\end{align}
and the total one
\begin{align}
Z=&\int_{0}^{2\pi R}\frac{da}{2\pi R}\sum_{r\in\Z}\vartheta_{\frac{Rr}{\sqrt 2}-\frac{a}{2\pi\sqrt2}}(\t)\vartheta_{\frac{Rr}{\sqrt 2}-\frac{a}{2\pi\sqrt2}}(\bar \t)\nonumber\\
=&\int_{0}^{1}dx\ \vartheta_{\frac{Rx}{\sqrt 2}}(\t)\vartheta_{\frac{Rx}{\sqrt 2}}(\bar \t)+\int_{1}^{2}dx\ \vartheta_{\frac{Rx}{\sqrt 2}}(\t)\vartheta_{\frac{Rx}{\sqrt 2}}(\bar \t)+\dots\\
=&\int_{-\infty}^{\infty}dx\ \vartheta_{\frac{Rx}{\sqrt 2}}(\t)\vartheta_{\frac{Rx}{\sqrt 2}}(\bar \t)
=\frac{\sqrt 2}{R}\int_{-\infty}^{\infty}dy\ \vartheta_{y}(\t)\vartheta_{y}(\bar \t)\ .
\end{align}

\paragraph{Comments}
The last expression is the partition sum for a free uncompactified boson. 
By comparing with the partition function of the limit of the compactified boson of equation~\eqref{ch1:free-boson-line-partition-fct}, we see that sending $R\to \frac{2}{R}$ we obtain exactly the same result. 
The interpretation is clear: this is the T-dual of a free uncompactified boson. 
By shrinking the circle we make the Kaluza-Klein modes heavier and heavier, but we push the winding modes closer and closer to each other. 
In the limit~$R\to 0$ the winding modes become continuous, and the Kaluza-Klein excitations disappear. 
\subsection{Runkel-Watts theory as a continuous orbifold}\label{ch2:sec:RW-as-contorbi}
In reference~\cite{Gaberdiel:2011aa} it has been proposed that the theory obtained by Runkel and Watts~(\cite{Runkel:2001ng} and outlined in section~\ref{ch1:sec:RW}) as the large level limit of Virasoro minimal models can be obtained as a continuous orbifold of the free theory\footnote{In the mentioned paper~\cite{Gaberdiel:2011aa} it is furthermore conjectured that the $k\to\infty$ limits of~$W_N$ minimal models can be described as a continuous orbifolds of $\wzw{su}{N}{1}$.} $\wzw{su}{2}{1}$. The concrete proposal can be schematically summarised in the following formal limit
\begin{equation}
\left.\lim_{k\to\infty}\frac{\wzw{su}{2}{k}\oplus \wzw{su}{2}{1}}{\wzw{su}{2}{k+1}}\right|_{\text{coset}}\sim\ \left. \frac{\wzw{su}{2}{1}}{SO(3)}\right|_{\text{orbifold}}\ ,
\end{equation}
where the twisted sector corresponds to the theory of Runkel and Watts of~\cite{Runkel:2001ng}, and the untwisted sector describes the limit theory proposed by Roggenkamp and Wendland in~\cite{Roggenkamp:2003qp}.
Gaberdiel and Suchanek in~\cite{Gaberdiel:2011aa} perform various explicit tests of this proposal (comparison of the full spectrum, fusion rules, fractional and bulk branes), and the match is perfect. In this subsection we give a flavour of their computations, concentrating on the spectrum of primary fields.
\paragraph{Untwisted sector} The construction of the untwisted sector is performed explicitly by using the general equation~\eqref{ch2:untwisted-sector-contorbi}. One has to choose practical parameters for the orbifold group, identify the geometric quantities present in the integral, and then perform the integration over the group. We leave the details of this computation to a careful reading of the very explicit appendix~A of the aforementioned paper~\cite{Gaberdiel:2011aa}. 

The authors choose a standard~$SU(2)$ parameterisation, with a suitable angular identification to realise $SO(3)=SU(2)/\Z_2$. They construct explicitly the Cartan torus, starting from an equator of $SU(2)$ parameterised by the angle~$\psi\in[0,2\pi]$ in their notations; in the~$SO(3)$ case it becomes a half circumference (due to the~$\Z_2$ angular identification); the Weyl group~$\Z_2$ can be explicitly realised on the Cartan torus, so that the angle receives again a~$\Z_2$ identification, $\psi\in[0,\frac{\pi}{2} ]$. The Haar measure can now be written with the decomposition of equation~\eqref{ch2:Haar-meas-decomp} taken into account.

The parent theory~$\wzw{su}{2}{1}$ is the theory of a free boson compactified on a circle with self-dual radius~$R=\sqrt 2$; the partition functions is readily obtained by setting $R=\sqrt 2$ in equation~\eqref{ch2:partition-fct-free-boson-on-R}.
The action of the twining~$h$ in the formula~\eqref{ch2:untwisted-sector-contorbi} represents a translation over the Cartan torus, in this case a quarter of the original $SU(2)$~equator with the ends identified; its action on the primaries is thus very similar to the one described in subsection~\ref{ch2:subs:boson-on-a-point}, in equation~\eqref{ch2:action-group-boson-point}.

It is easy in this way to write the explicit expression for the untwisted sector of the orbifold~$\frac{\wzw{su}{2}{1}}{SO(3)}$:
\begin{equation}
Z_{\text{untw}}=\underbrace{\frac{1}{\pi^2}}_\frac{1}{\left|G\right|}\int_{0}^{\frac{\pi}{2}} \ \underbrace{d\psi}_{d\mu(h)}\ \overbrace{4\pi \sin^2\psi}^{\text{Vol([h])}}\ \frac{1}{|\eta(\t)|^2}\sum_{m,n\in\Z}\overbrace{e^{2im\psi}}^{h}\ q^{\frac{(m+n)^2}{4}}\bar q ^{\frac{(m-n)^2}{4}}=\sum_{r=1}^{\infty}\left|\hat\chi_r(q)\right|^2\ ,
\end{equation}
with $\hat\chi$ the characters of~$c=1$ Virasoro irreducible modules defined in appendix~\ref{app:characters}.

The computation of the twisted sectors is very similar to the one performed for the inverse decompactification limit of the free boson in subsection~\ref{ch2:subs:boson-on-a-point}, equation~\eqref{ch2:twined-untw-block-boson-point}: one starts with the twined block
\begin{equation}
\torus{h_{\a}}{\mathbb{1}}=\Tr_{\mathcal{H}_0}h_{\a}\,q^{L_0-\frac{1}{24}}\bar q^{\bar{L}_0-\frac{1}{24}}=\sum_{m,n\in\Z}e^{2\pi i m \a}\cdot \left[\vartheta_{m+n}(\t)\  \vartheta_{m-n}(\bar \t)\right]\ ,
\end{equation}
where we denoted the representatives of~$\mathcal{\T}/{\mathcal{W}}$ with~$\a=\frac{\psi}{\pi}\in[0,\frac12]$. Then one performs a modular S-transformation following the same strategy as before
\begin{equation}
\torus{\mathbb{1}}{h_{\a}}=\sum_{l,\bar l\in\Z}\sum_{l-\bar l\in 2 \Z}\vartheta_{-\a+l}(\t)\vartheta_{-\a+\bar l}(\bar \t)\ .
\end{equation}
Again by means of reiterated modular T-transformations, and by guessing the general twisted and twined block, the result is analogous to~\eqref{point-twisted}, and reads
\begin{equation}\label{ch2:gab-such-twisted-partition-fct}
Z_{\a \text {-tw}} =\sum_{l\in\Z}\vartheta_{-\a+l}(\t)\vartheta_{-\a+ l}(\bar \t)\ .
\end{equation}
If we integrate over all the twisted sectors we get
\be
Z_{\text{tw}}=\int_{-\infty}^{'\infty}dx\,\vartheta_x(\t)\vartheta_x(\bar \t)\ ,
\ee
where the apex indicates that we excluded the untwisted sectors from the integral, since~$\a$ in equation~\eqref{ch2:gab-such-twisted-partition-fct} is never integer. They are of course of measure zero, but strictly speaking cannot be absorbed in the twisted ones (this is relevant for boundary conditions for instance). 
We reproduce in this construction the spectrum of a free theory, with integer weights missing: this is the spectrum conjectured by Runkel and Watts as the limit of Virasoro minimal models. 
As already mentioned before, further analyses (again in reference~\cite{Gaberdiel:2011aa}), show that the agreement goes beyond the partition function.

The conclusion we can draw is that the twisted sectors of the continuous orbifold~$\frac{\wzw{su}{2}{1}}{SO(3)}$ describe the~$k\to\infty$ limit of Virasoro minimal models. 
This result is unexpected and relevant: RW theory finds finally its place in an interesting extension of the classification of rational~$c=1$ theories firstly given by Ginsparg many years ago~\cite{Ginsparg:1987eb}.
Moreover, it sheds light on its relation with the~$c=1$ limit theory of~\cite{Roggenkamp:2003qp}, and provides RW theory with a natural identity operator (now fully defined as the lowest weight untwisted sector of the orbifold).
Mostly interesting for us is that the correspondence between a continuous orbifold construction and the limit of Virasoro minimal models makes conceivable that the whole new class of continuous orbifold~CFTs might be indeed well-defined. 
Especially, in~$N=2$ supersymmetric realisations, orbifolds have interesting applications as non-trivial backgrounds for string propagation, and constitute important building blocks to geometrically engineer gauge theories on D-branes wrapped around these spaces. 
Continuous orbifolds might represent an interesting extension of these ideas.
\newpage
\begin{subappendices}
\section{Notations for orbifolds}\label{ch2:sec-app:notations}

We use the following notations for the spaces appearing in the construction of orbifold models: we denote with~$\Hilb_0$ the Hilbert space of the bulk of the parent theory. 
We denote with~$\Hilb_g$ the $g$-twisted sector, and consequently with~$\Hilb_{\id}$ the untwisted sector.
The open string spaces have ``open'' as an apex:~$\Hilb^{\text{open}}$.

We use the following box-notation for the partition functions of the different sectors, (see e.g.~\cite{Ginsparg:1987eb,Ginsparg:1988ui}) to keep easily track of modular transformations. With the symbol
\begin{align}\label{def-box}
\torus{g}{h}=\int_{T_{g,h}} \mathcal{D}\phi \ e^{-S[\phi]}=\Tr_{\mathcal{H}_{h}}\ g\,q^{L_0-\frac{c}{24}}\bar q^{\bar{L}_0-\frac{\bar c}{24}}
\end{align}
we mean: in the Lagrangian language (first equality) the euclidean path integral of the CFT defined by $S$ on the torus, with fields satisfying boundary conditions twisted by the action of $g,h\in G$ along the two cycles of the torus respectively; in the Hamiltonian language (second equality) the trace over the $h$-twisted Hilbert space~$\mathcal{H}_h$, with the insertion of a twining operator~$g$, which acts in general non-trivially on the twisted modules.
We write
\be
Z_{\text{untw}}=\frac{1}{|G|}\sum_{g\in G}\torus{g}{\id}\ ,\qquad 
Z_{h\text{-tw}}=\frac{1}{|G|}\sum_{g\in G}\torus{g}{h}\ .
\ee
Under a generic modular transformation $\tau\mapsto\frac{a\tau+b}{c\tau+d},\ ad-bc=1,$ the box of equation~\eqref{def-box} changes as follows in our conventions:
\begin{align}\label{ch2:gen-modular-transf-box}
 \torus{g}{h}\ \longmapsto \ \torus{g^ah^b}{g^ch^d}\ ,
 \end{align}
 given that $[g,h]=0$, to avoid ambiguities for T-modular transformations.
For instance, we have
\begin{align}\label{ch2:S-T-modular-transf-box}
\torus{g}{h}\overset{S}{\longmapsto}\torus{h}{g}\ ,\qquad \torus{g}{h}\overset{T}{\longmapsto}\ \torus{g h}{h}\ .
\end{align}
\end{subappendices}
\chapter{$N=2$ minimal models}\label{ch:minmod}

In this chapter we present a review of $N=2$ minimal models, whose large level limit will be our case study in this thesis.
These models represent an important example of $N=2$ CFTs, since they are interacting, rich and well-known models.

They realise $N=2$ supersymmetry on the world-sheet which is necessary (in the sigma-model string theoretical interpretation) to realise four-dimensional supersymmetry in target space~\cite{Banks:1987cy}.
In particular, since superstring theories are consistent only in ten space-time dimensions, in order to make contact with phenomenology we have to curl up six spatial dimensions in a compact space.
The simplest six-dimensional compact space that we can build is a six-torus, whose world-sheet description is given by six real bosons (plus fermions) compactified on six commuting circles.
The problem with this compactification, is that we realise too much supersymmetry for the theory to be comparable with experiments: we end up with $N=4$ or even $N=8$ in four dimensions in the closed string sector (depending on which superstrings we consider, heterotic or type I in the first case, type II in the second).
Hence, if the four non-compact dimensions can be realised as free fields in the simplest examples, we still have to consider some interacting world-sheet theory to get rid of the six remaining compact ones.
Minimal models are the easiest building blocks for compact backgrounds which produce $N=2$ supersymmetry in type II in the non-compact four-dimensional space-time.
In particular Gepner models~\cite{Gepner:1987vz,Gepner:1987qi} are realised as orbifolds of tensor products of minimal models' representations, and they are related to complete intersection Calabi-Yau manifolds, which are spaces suitable for string compactification, stable (no tachyons in the spectrum) and preserving a decent amount of supersymmetry.

Moreover minimal models realise the infrared end-point of $N=2$ two-dimensional non-conformal Landau-Ginzburg models~\cite{Witten:1993jg}, making them especially suitable to explore the space of two dimensional theories.

These CFTs can be constructed starting from the $N=2$ superconformal algebra in the same fashion as one construct Virasoro minimal models, and this procedure is reviewed in section~\ref{ch3:sec:unitarity-spectrum-minmod}, but they are better understood as coset models, as explained in section~\ref{ch3:sec:MM-parafermions}.
The coset description allows for a sigma-model geometric interpretation as strings moving on an $SU(2)$ group geometry (modded-out by the adjoint action of a $U(1)$), and we review this construction in section~\ref{ch3:sec:geometry}.
We give account of correlators and conformal boundary conditions in sections~\ref{ch3:sec:correlators} and~\ref{ch3:sec:BC} respectively.

\section{$N=2$ superconformal algebra}
Among the possible extensions of Virasoro algebra,~$N=2$ superconformal algebra~(SCA)~\cite{Ademollo:1975an} plays a prominent role since it serves (in different ways) as the fundamental symmetry for constructing all the five different perturbative string theories: world-sheet $N=2$ superconformal field theories realise four-dimensional supersymmetry in target space~\cite{Banks:1987cy}.
For this reason there has been an intense progress during the last decades to study its representation theory in detail, and the CFTs based on these representations.
\subsection{Free fields}\label{ch3:sec:SCA:subs:free-fields}
The simplest world-sheet realisation of~$N=2$ super-conformal algebra (at $c=3$) is the theory of two free uncompactified real bosons ($\phi,\phi^*$) and two free real fermions~($\psi^{\pm}$)~\cite{Blumenhagen:2009zz, Gaberdiel:2004nv, KlemmPhD}; the fields can be expanded in normal modes
 \begin{equation}\label{ch3:free-modes}
\begin{array}{ll}
\vspace{0.2cm}
\de\phi =-i\sum\limits_{m\in\Z}\alpha_m z^{-m-1}\quad&\quad\de\phi^* =-i\sum\limits_{m\in\Z}\alpha^*_m z^{-m-1}\\ 
\psi =\sum\limits_{r\in\Z+\eta}\psi_r z^{-r-\frac12}\quad&\quad\psi^* =\sum\limits_{r\in\Z+\eta}\psi^*_r z^{-r-\frac12}
\end{array}
\end{equation}
where $\eta=0,\frac12$ in the Ramond~(R), Neveu-Schwarz~(NS) sector respectively. 
The anti-holomorphic case is analogous. Upon canonical quantisation the modes respect the algebra of one free complex boson and one free Neveu-Schwarz complex fermion:
  \begin{equation}
  \begin{array}{l}
\left[\alpha_m, \alpha^*_n\right] = m\delta_{m,-n} \\
\left[\alpha_m,\alpha_n\right] = [\alpha^*_m, \alpha^*_n] = 0\ ,
  \end{array}
 \qquad
  \begin{array}{l}
  \{\psi_r,\psi^*_s\} = \delta_{r,-s}\,, \\
  \{\psi_r,\psi^*_s\}=\{\psi^*_r,\psi^*_s\}=0
  \end{array}
\end{equation}
The world-sheet equipped with these fields is automatically supplemented by a natural extension of the conformal algebra. Let us show how this is realised for holomorphic fields: the energy momentum tensor~$T(z)=\sum\limits_{n\in \Z}z^{-n-2}L_n$ can be explicitly written down in terms of the fields and modes of~\eqref{ch3:free-modes}
\begin{equation}
T=-\de\phi\de\phi^*-\frac12(\psi^*\de\psi+\psi\de\psi^*)\ ,\qquad  L_n = \sum_{m} :\alpha_{n-m}\alpha^*_{m}: +
    \frac{1}{2}\sum_{s}(2s-n):\psi^{*}_{n-s}\psi_s:\ .
\end{equation}
We include then also two fermionic superpartners of conformal dimension~$h=\frac32$:
\begin{align}
 \begin{array}{l}
G^+=i\sqrt 2\psi\de\phi^*\\
G^-=i\sqrt 2\psi^*\de\phi
 \end{array}
 \quad
\begin{array}{l}      
 G^+_r = \sqrt{2}\sum_m :\alpha^*_m\psi_{r-m}: \\
 G^-_r = \sqrt{2}\sum_m :\alpha_m\psi^*_{r-m}: 
  \end{array} \ .
\end{align} 
The modes of the energy momentum tensor satisfy the Virasoro algebra; now, by studying the commutation relations for the supercurrents one finds that the full algebra closes only if we add a further bosonic $U(1)$~current, which can be realised as
\be
J=-\psi^*\psi\qquad J_n=-\sum_s:\psi^*_{-s}\psi_{s+n}:\ ;
\ee
it represents the~$N=2$ $R$-symmetry generator in two dimensions.
\subsection{$N=2$ superconformal algebra}
Studying the OPE of the free field realisation of the operators~$T(z),G^{\pm}(z) , J(z)$, or analogously the commutators of their modes~$L_m,G^{\pm}_r , J_m$, we obtain the~$N=2$ superconformal algebra~\cite{Ademollo:1975an} at~$c=3$, which can be simply extended for~$c\in\R$ as follows:
\begin{equation}\label{ch3:super-conformal-algebra}
\begin{split}
  \left[L_m, L_n \right] &=(m-n)L_{m+n} + \tfrac{c}{12}(m^3 -
  m)\delta_{m,-n}  \\ 
  \left[L_m, J_n \right] &=-nJ_{m+n}\\
  \left[L_m, G_r^\pm \right] &=\left( \tfrac{1}{2}m-r \right)
  G_{m+r}^\pm  \\
  \left[J_m, J_n \right] &= \tfrac{c}{3} m\delta_{m,-n}\\
  \left[J_m, G_r^\pm \right] &=\pm G_{r+m}^\pm  \\
  \left\{ G^+_r, G^-_s\right\} &= 2 L_{r + s} + (r -s) J_{r + s} +
  \tfrac{c}{3}(r^2 - \tfrac{1}{4})\delta_{r, -s} \\
  \left\{G^+_r, G^+_s \right\} &=\left\{G^-_r, G^-_s\right\}=0\ ,
  \end{split}
\end{equation}
where $m,n$ are integers and $r,s$ are half-integers (integers) in the NS (R) sector.
The Cartan subalgebra is spanned by the generators~$L_0,J_0$ and by the central charge~$c$, and they can be simultaneously diagonalised with an appropriate choice of basis in the Hilbert space. 
The representations at fixed $c$ are thus characterised by two natural quantum numbers: the conformal weight~$h$, eigenvalue of~$L_0$, and the~$U(1)$ charge~$Q$, eigenvalue of~$J_0$.
\paragraph{Spectral flow}
The~$N=2$~SCA admits an interesting continuous automorphism~$\a_{\h}$, called spectral flow~\cite{Schwimmer:1986mf}. It acts on the generators as follows:
\be\label{ch3:spectral-flow-def}
\begin{split}
\a_{\h}(L_n)=&\ L^{\h}_n=L_n-\h J_n+\frac{c}{6}\eta^2\delta_{n,0}\\
\a_{\h}(G^{\pm}_r)=&\ G^{\h\pm}_r=G^{\pm}_{r\mp\h}\\
\a_{\h}(J_n)=&\ =J^{\h}_n=J_n-\frac{c}{3}\eta\delta_{n,0}
\end{split}\ ,
\ee
and~$\h\in\R$ in general.
Under the transformations given in~\eqref{ch3:spectral-flow-def} the~$N=2$~SCA of~\eqref{ch3:super-conformal-algebra} is left invariant. The action of~$L_n$ and~$J_n$ on the states gets instead modified. 
In particular if~$\h\in\Z+\frac12$ the transformation maps the NS to the R sector, and vice versa. If~$\h\in\Z$ the transformation acts separately on the two sectors:~$\a_{\Z}$ is an automorphism of both the~NS and~R subalgebras (spectral flows can be explicitly realised in all $N=2$ models. For the cases of interest in this thesis, namely Kazama-Suzuki models, the explicit realisation is given in section~\ref{ch3:app:sec:KS-models}).
\paragraph{Mirror automorphism}
The algebra is automorphic under the following transformation
\begin{equation}\label{ch3:mirror-automorphism}
\O_M\ :\  J_n\longmapsto -J_n\ ,\qquad G_{r}^{\pm}\longmapsto G_r^{\mp}\ ,
\end{equation}
which is called mirror automorphism.
The consequences of its presence in string theory has been a huge source of mathematical surprises (for a glimpse~\cite{Hori:2003ic}).
\paragraph{Chiral primaries}
In the following section we will analyse unitary representations for this algebra.
Among these representations, the class of the so-called (anti-)chiral primaries $(|\phi_a\ket)\ |\phi_c\ket$ is special for some reasons that we briefly mention\footnote{\label{ch3:sec:MM-as-parafermions:par:chiral-rings:footnote:four-rings}We restrict to the holomorphic sector here. All the statements in this paragraph can be carried over to the anti-holomorphic sector by simply replacing $G^{\pm}$ with $\overline{G}^{\pm}$. In general the (anti)-chiral primary states belong to four classes, namely $(c,c),(c,a),(a,c),(a,a)$, where the first (second) entry in parentheses denotes the chiral behaviour under (anti-)holomorphic generators.} 
(the details and proofs of our statements can be found in~\cite{Lerche:1989uy}).
(Anti-)chiral primaries are states belonging to the Hilbert space of an $N=2$ theory which satisfy the following condition
\begin{equation}
\begin{array}{rl}
\text{chiral primaries:} &\quad G^+_{n-\12}|\phi_c\ket=G^-_{n+\12}|\phi_c\ket=0\\
\text{anti-chiral primaries:} &\quad G^-_{n-\12}|\phi_a\ket=G^+_{n+\12}|\phi_a\ket=0
\end{array}\ .
\end{equation}
These states satisfy the properties:
\begin{itemize}
\item They saturate a general bound for ground states $|\psi\ket$, $h_{\psi}\geq \frac{|Q|}{2}$.
Then $h_{\phi}=\pm\frac{Q}{2}$, with the plus (minus) sign for (anti-)chiral primaries.
\item There is always a finite number of them.
\item The OPEs among them close on a non-singular ring \footnote{Corresponding to the classes of footnote~\ref{ch3:sec:MM-as-parafermions:par:chiral-rings:footnote:four-rings} of this paragraph the number of rings is four, pairwise conjugate by charge conjugation.}.
\item  There is a one-to-one correspondence between chiral primary states and Ramond ground states.
The map is given by a spectral flow of $\eta=\12$.
The same happens for anti-chiral primaries with flow of the opposite sign.
\end{itemize}

\section{Spectrum of~$N=2$ minimal models}\label{ch3:sec:unitarity-spectrum-minmod}

In this section we concentrate on the unitary irreducible highest-weight representations of the~$N=2$~SCA given in~\eqref{ch3:super-conformal-algebra}, with $c<3$. 
This is the range in which one constructs unitary~$N=2$ minimal models, once chosen a suitable modular invariant partition function. 
The approach we follow in this section puts the accent on the ``minimality'' of assumptions that we need to construct the minimal models, but it is not completely satisfying for many reasons.
We will fill the gaps in the next sections.
\subsection{Ground states and unitarity}\label{ch3:sec:unitarity-spectrum-minmod:subs:unitarity-Kac}
Given a value of the central charge~$c<3$, the set of highest-weight representations or ground states~$|h,Q\ket$ (labeled by the highest weights~$h,Q$ of the Cartan subalgebra~$\{L_0,J_0\}$) have corresponding Verma modules attached. These are generated from the highest-weight vectors by the action of the negative modes of the algebra.

The first thing to define is an NS~vacuum, $|0,0\ket$, invariant under the~SCA; 
this means $L_n|0,0\ket=J_n|0,0\ket=G^{\pm}_{n-\frac12}|0,0\ket\equiv0$ if $n\geq0$.
To this state correspond two different vacua in the R sector,~$|\frac{c}{24},\pm\frac12-1+\frac c3\ket$, which are mapped into each other with the corresponding modules by charge conjugation. 
They can be obtained by $\h=\pm\frac12$ units of spectral flow from the NS~vacuum.
Acting then with the negative modes of the generators of the~$N=2$~SCA we obtain the full module associated to the NS and R vacua, respectively~$\Hilb^{(c)}_{0,0}$ and~$\Hilb^{(c)}_{\frac{c}{24},\pm\frac 12-1+\frac c3}$.
In the same fashion it is possible to define Verma modules~$\Hilb^{(c)}_{h,Q}$ associated to highest-weight states with different eigenvalues of~$h$ and~$Q$,  and thus construct the full Hilbert space. 
A generic state~$|\chi\ket$ belonging to the NS~module~$\Hilb^{(c)}_{h,Q}$ 
can be therefore written as follows
\be\label{ch3:generic-state-N=2}
|\chi_{i,j,k,l}\ket=L(i,I)J(m,M)G^+(k,K)G^-(l,L)|h,Q\ket
\ee
where we have defined
\begin{align*}
L(i,I)\equiv&\ (L_{-1})^{i_1}(L_{-2})^{i_2}\dots(L_{-I})^{i_{I}}\\
J(m,M)\equiv&\ (J_{-1})^{m_1}(J_{-2})^{m_2}\dots(L_{-M})^{m_{M}}\\
G^{+}(k,K)\equiv&\ (G^+_{-1/2})^{k_1}(G^+_{-3/2})^{k_2}\dots(G^+_{-1/2-K+1})^{k_{K}}\\
G^{-}(l,L)\equiv&\ (G^-_{-1/2})^{l_1}(G^-_{-3/2})^{l_2}\dots(G^-_{-1/2-L+1})^{l_{L}}\ ,
\end{align*}
with $i_i,m_i\in\Z_{\geq0}$ and $k_i,l_i\in\{0,1\}$. Its weight and charge read
\begin{equation}\label{ch3:weight-charge-gen-state-N=2}
h_{\chi}=h+\sum_{a=1}^Na i_a+\sum_{a=1}^Ma m_a+\sum_{a=1}^{K}(a-\tfrac{1}{2}) k_a+\sum_{a=1}^{L}(a-\tfrac{1}{2}) l_a\ ,\qquad Q_{\chi}=Q+\sum_{a=1}^{K} k_a-\sum_{a=1}^{L} l_a\ .
\end{equation}
The states~$\chi$ defined in equation~\eqref{ch3:generic-state-N=2} serve as a (redundant) basis in the Hilbert space of our theories. 
We define also the \textit{level}~$n_{\chi}$ and the \textit{relative charge}~$p_{\chi}$ of the state~$|\chi\ket\in\Hilb^{(c)}_{h,Q}$ with respect to the quantum numbers~$h$ and~$Q$, as
\begin{equation}
n_{\chi}\equiv h_{\chi}-h\ ,\qquad p_{\chi}\equiv Q_{\chi}-Q\ .
\end{equation}
%
%
\paragraph{Unitarity and Ka\v c determinant} Unitarity imposes restrictions on the allowed states, since the modulus squared of a generic vector of our state space must be normalisable to one. 
If some of the states in the theory have negative norm, then the theory is non-unitary and non-unitarisable. 
A necessary condition for unitarity is that every vector in the space of states has non-negative squared norm. 

The problem can be formulated as follows: given~$\psi$ and~$\psi'$ arbitrary vectors of the space of states, they can be decomposed in the basis~$\{\chi\}$ defined in equation~\eqref{ch3:generic-state-N=2} as~$|\psi\ket=\sum\limits_{a}\psi_{a}|\chi_a\ket$, with $a$ multi-index for all the possible choices of~$n,m,k,l$. 
The scalar product of~$\psi$ with $\psi'$ becomes in the~$\chi$ basis
\begin{equation}\label{ch3:def-shapovalov-matrix}
\bra\psi|\psi'\ket=\sum_{a,b}\psi_a\psi'_b\bra \chi_a|\chi_b\ket=\vec{\psi}^T\cdot \Xi(c,h,Q)\cdot \vec{\psi}'\ .
\end{equation}
The matrix~$\Xi(c,h,Q)$ is called the Shapovalov form for the~$N=2$ SCA.
If the states~$\psi,\psi'$ do not belong to the same level and do not have the same relative charge, then the scalar product is surely vanishing, since in this case, after commuting and anticommuting among each other all the operators defining the two states, we are left only with modes with the same sign. In other words, the Shapovalov form is block diagonal, with blocks labeled by~$n,p$, and we indicate each block with $\Xi_{n,p}(c,h,Q)$. 
The quantity $\det \Xi_{n,p}(c,h,Q)$ is called Ka\v c determinant.

Now, if for some values of $(c,h,Q)$, $\det \Xi_{n,p}(c,h,Q) < 0$, then we cannot define a unitary CFT on the modules constructed on such ground states.
If $\det \Xi_{n,p}(c,h,Q) > 0$, we cannot say much, since two negative norm vector may result in a positive scalar product.

\paragraph{Unitarity and null vectors} 
The interesting locus for the construction of unitary theories is the one characterised by $\det \Xi_{n,p}(c,h,Q) = 0$.
In this case there must be vectors in the module associated to a given level which are orthogonal to any other vector in the same module, and we cannot say much about negative norm states in a given level.
Nevertheless, more detailed analyses reveal that these loci often lie at the boundary between unitary modules and non-unitarisable ones.
If this is the case, we can construct a unitary theory by just picking up a sector in this region, and removing by hand all the null vectors present. 
It might seem a formidable task, but in our instance this is doable (although fairly elaborate) for the reason that null vectors come in submodules of the superconformal algebra.
By eliminating these submodules we remove all the non-positive norm states, and we end up with a Hilbert space that can host a unitary CFT\footnote{To be completely correct, this analysis is not enough to really prove unitarity. To this aim the coset description reviewed in section~\ref{ch3:sec:MM-parafermions} is better suitable.}.

Let us give more details about this idea: a null vector is in general not annihilated by all the positive modes, and the action of a superconformal generator does not spoil the zero-norm property (see~\cite{DiFrancesco:1997nk,FuchsSchweigert} for further details); we see then that every null-vector is either annihilated by all the positive modes of the superconformal generators - and in this case is called~\textit{singular} - or is in a submodule generated by a singular vector.
A situation may appear as well (and this is not the case for unitary minimal models) in which a null vector is not a descendant of any singular vectors, but can be mapped to a singular vector by a lowering operator. 
These states are called \textit{subsingular} vectors~\cite{GatoRivera:1996ma,Semikhatov:1997gv,Feigin:1998sw}; they become singular after having divided out the singular vectors from the original modules. 
This structure can be obviously generalised to vectors which become subsingular after the removal of singular vectors, and are called~\textit{subsubsingular}, and so on.

The representations with singular (and subsingular) vectors are therefore reducible, since there are subspaces left invariant by the action of the full algebra, but non-decomposable, since the full module cannot be expressed as a direct sum of irreducible subspaces (in other words the singular submodule appears inside the original module). 
Therefore, in order to get a proper irreducible unitary representation, we have to subtract all the singular submodules from the representation.

\paragraph{Discrete series of $N=2$ minimal models}
Formulas for the Ka\v c determinant of~$N=2$ SCA have been firstly given in~\cite{Boucher:1986bh} and proven later in~\cite{Kato:1986td}. 
We quote here the result in the NS sector:
\begin{equation}
\det \Xi^{\text{NS}}_{n,p}=\prod_{\substack{1\leq rs\leq 2n\\s\text{ even}}}(f^{\text{NS}}_{r,s})^{P_{\text{NS}}(n-rs/2,p)}\prod_{k\in\Z+\frac12}(g^{\text{NS}}_{k})^{\tilde{P}_{\text{NS}}(n-|k|,p-\sgn k;k)}\ ,
\end{equation}
where $P_{\text{NS}}$ counts the states at level~$n$ and relative charge~$p$
\begin{equation}
\sum_{n,p}P_{\text{NS}}(n,p)q^nz^p=\prod\limits_{m=0}^{\infty}\frac{(1+q^{m+\frac12}z)(1+q^{m+\frac12}z^{-1})}{(1-q^{m+1})^2}=\frac{\vartheta_3(q,z)}{\eta^3(q)}\ ,
\end{equation}
and $\tilde{P}_{\text{NS}}$ counts the states built on charged fermionic highest-weight vectors
\begin{equation}
\sum_{n,p}\tilde{P}_{\text{NS}}(n,p,k)q^nz^p=(1+q^{|k|}z^{\sgn k})^{-1}\sum_{n,p}P_{\text{NS}}(n,p)q^nz^p\ .
\end{equation}
The functions $f^{\text{NS}}$ and $g^{\text{NS}}$ read
\begin{equation}
\begin{split}
f^{\text{NS}}_{r,s}(c,h,Q)=&\ 2\left(\frac c3 -1\right)h-Q^2-\frac14\left(\frac c3-1\right)^2+\frac14\left[\left(\frac c3-1\right)r+s\right]^2\\
g^{\text{NS}}_k(c,h,Q)=&\ 2h-2kQ+\left(\frac c3-1\right)\left(k^2-\frac14\right)\ .
\end{split}
\end{equation}
The reference~\cite{Boucher:1986bh} contains formulas for the R sector as well.
\paragraph{Bulk spectrum for $N=2$ minimal models}
A detailed analysis of the vanishing loci of the Ka\v c determinant has as an outcome that unitary representations of $N=2$~SCA exist for~$c<3$ only for a discrete set of values of the central charge, namely
\begin{equation}\label{ch3:minmod-central-charge}
c=3\left(1-\frac{2}{M}\right)\ ,\quad\ M\in\Z_{>2}\ .
\end{equation}
Allowed values of conformal weights and~$U(1)$ charges are given in the formulas~\cite{DiVecchia:1985iy,Waterson:1986ru,Boucher:1986bh,Nam:1985qe}
\begin{equation}\label{ch3:old-fashioned-minmod-spectrum}
\begin{array}{rlll}
h^{\text{NS}}_{j,k}= &\frac{1}{M}\left(jk-\tfrac{1}{4}\right)\ ,& Q^{\text{NS}}_{j,k}= \frac{j-k}{M}\ ,& j,k\in\Z+\12\ ,\ 0<j,k,j+k\leq M-1\ , \\
h^{\text{R}^{\pm}}_{j,k}= &\tfrac{c}{24}+\frac{jk}{M}\ , & Q^{\text{R}^{\pm}}_{j,k}= \left(\frac{j-k}{M}-\frac 12\right)\ , & j,k\in\Z\ ,\ 0\leq j-1,k,j+k\leq M-1\ .
\end{array}
\end{equation}
The expressions~\eqref{ch3:old-fashioned-minmod-spectrum} describe the spectrum of $N=2$ minimal models.
With this ``minimal'' approach it is very difficult nevertheless to prove that CFTs built using these representations are unitary~\cite{Eholzer:1996zi}, and it is also complicated to study other structures that will be useful in the following.
We do not write down explicitly the singular vectors and the characters in these notations (the interested reader can look at~\cite{Ravanini:1987yg}), since we prefer to give a more modern treatment of minimal models in sections~\ref{ch3:sec:MM-parafermions} and~\ref{ch3:app:sec:KS-models}.


\section{$N=2$ minimal models as supersymmetric parafermions}\label{ch3:sec:MM-parafermions}

In this section we get into the details of the modern way of looking at~$N=2$ minimal models: it has been indeed soon realised that the structures emerging from the analysis just outlined can be as well realised in the theory of a supersymmetric parafermion~\cite{Zamolodchikov:1986gh}.

\paragraph{Arguments for the  correspondence}

The first observation concerns the central charge: the formula~\eqref{ch3:minmod-central-charge} describes the central charge of~$\wzw{su}{2}{k}$ Wess-Zumino-Witten (WZW) models (if one identifies $k=M-2$), which suggests that~$N=2$ minimal models can be as well realised (in analogy with Virasoro minimal models) as some~$su(2)$ coset model, engineered in such a way that the central charge remains the one of~$\wzw{su}{2}{k}$, and which enjoys supersymmetry. 

The second observation concerns the spectrum: in~\cite{Ravanini:1987yg} it is shown that the expressions for the characters of these representations correspond to the following decomposition (using modern notations that will be explained momentarily)
\begin{equation}
\chi_{(l,m,s)}(q,z)=\sum_{j=1}^{k+2}C^{(k)}_{l,m-4j-s}(q)\Theta_{-2m+(4j+s)(k+2),2k(k+2)}(\t,-\tfrac{\nu}{2(k+2)})\ ,
\end{equation}
 which are recognisable as the branching functions for the super parafermion coset.
 
The third observation comes from a direct comparison of the correlators: they can be computed directly by means of the Coulomb gas formalism for minimal models (as done in~\cite{Yu:1986xc} and reviewed in~\cite{Mussardo:1988av}, extending the techniques explained in detail in~\cite[chapter 9]{DiFrancesco:1997nk}).
Correlators can be calculated as well in the parafermionic description, since they can be written explicitly in terms of~$\wzw{su}{2}{k}$ primaries~\cite{Zamolodchikov:1986gh}, whose correlators on the sphere are known~\cite{Zamolodchikov:1986bd}: the results agree~\cite{Mussardo:1988av}, giving us confidence that the two descriptions are equivalent.

\subsection{Spectrum of ground states}

The heuristic idea is to start from an~$\wzw{su}{2}{k}$~WZW model, to remove a free boson at one radius by dividing out the $U(1)$ from $SU(2)$ (and realising in this way a parafermionic coset model~$\frac{\wzw{su}{2}{k}}{\wzw{u}{1}{}}$) and to put a free boson back at a different radius.
The case at hand is a special instance of the so-called Grassmannian Kazama-Suzuki cosets~\cite{Kazama:1988uz,Kazama:1989qp}, whose construction is reviewed in section~\ref{ch3:app:sec:KS-models}. 
In this subsection we take a somewhat more ``physical'' approach. 
This will be helpful in the definition of bulk correlators.

The discrete series of $N=2$~minimal models~\MM{k} can be realised by the coset
\begin{equation}
\text{MM}_{k}=\frac{\wzw{su}{2}{k}\oplus\wzw{u}{1}{4}}{\wzw{u}{1}{2(k+2)}}\ .
\end{equation}
Coset models are based on the Sugawara construction of~CFTs with affine chiral symmetry algebra (in section~\ref{ch3:app:sec:KS-models} we try to be more precise, for details and references~\cite{DiFrancesco:1997nk}). 
In our case the Hilbert space is obtained by decomposing the space of states $\Hilb_{\kmalg{su}(2)_k}^{l}\otimes \Hilb_{\kmalg{u}(1)_4}^s$ in terms of a direct sum of representations of an embedded~$\kmalg{u}(1)_{2(k+2)}$ affine algebra. We restrict the attention to the holomorphic sector.
We choose the embedding for the Sugawara currents of the denominator chiral algebra in the numerator as 
\begin{equation}\label{ch3:emb-currents-paraf}
J^{2(k+2)}(z)=2J^3(z)+J^s(z)\ ,
\end{equation}
where $J^{2(k+2)}, J^s$ are the chiral $U(1)$ currents for the denominator and numerator respectively, $J^3$ is the chiral current associated with the Cartan generator of~$SU(2)$. All these currents can be bosonised as follows (prefactors are conventional):
\begin{equation}
J^{2(k+2)}(z)=i\sqrt{2(k+2)}\ \de g(z)\ ,\qquad J^s(z)=2i\ \de\sigma(z)\ ,\qquad J^3(z)=i\sqrt{\frac{k}{2}}\ \de X(z)\ ,
\end{equation}
where~$g,\s , X$ are scalar holomorphic fields.
Since we are aimed to realise an~$N=2$ superconformal algebra, we need a conserved $U(1)$ current surviving the coset decomposition. 
We choose it as a (at this point) generic combination of the two $U(1)$ currents already present in the numerator,
\begin{equation}\label{ch3:N=2-J-current-def}
J^{N=2}=\a J^3+\b J^s\ ,\qquad J^{N=2}(z)=i\sqrt{\frac{k}{k+2}}\de H(z)\ ,
\end{equation}
where the normalisation in the bosonisation is again conventional.
From these definitions, using the embedding of equation~\eqref{ch3:emb-currents-paraf}, and the choice~\eqref{ch3:N=2-J-current-def} we find
\begin{equation}
\begin{split}
\s(z)\,=&\,g(z)\sqrt{\frac{k+2}{2}}\frac{\a}{\a-2\b}\,+\,H(z)\sqrt{\frac{k}{k+2}}\frac{1}{2\b-\a}\ ,\\ X(z)\,=&\,g(z)\sqrt{\frac{k+2}{k}}\frac{2\b}{2\b-\a}\,+\,H(z)\sqrt{\frac{2}{k+2}}\frac{1}{\a-2\b}\ .
\end{split}
\end{equation}
We can then perform the decomposition of the primaries of the numerator in terms of the~$U(1)$ field in the denominator
\begin{equation}
\Phi^{SU(2)}_{l,n}(z)\  e^{i\frac s2 \s(z)}=\Psi_{l,n,s}(z)\  \exp\left[ig(z)\ \sqrt{\frac{2+k}{8k}}\left(\a \sqrt k s-\frac{8n\b}{\sqrt k}\right)\frac{1}{\a-2\b}\right]\ ,
\end{equation}
where~$\Phi_{l,n}^{SU(2)}$ is the~$\kmalg{su}(2)_k$ primary field corresponding to the highest-weight representation labeled by the integer highest weight~$l$, and $n$ is the weight of the state, i.e. the~$\hat J^3_0$~eigenvalue.
On the right hand side we have extracted the~$J^3$ contribution (following the Wakimoto construction, reviewed e.g. in~\cite[chapter 15]{DiFrancesco:1997nk}) $e^{i\sqrt{\frac{2}{k}}n X(z)}$, and fused it with the~$J^s$ bosonised field.
The field~$\Psi(z)$ represents the primary of our theory, and now we can explicitly extract its charge with respect to the current~$J^{N=2}$ defined in equation~\eqref{ch3:N=2-J-current-def}
\begin{equation}
\Psi_{l,n,s}(z)=\tilde{\Phi}_{l,n,s}(z)\exp \left[i H(z) \left(\frac12 \frac{4\frac{n}{\sqrt k}-s\sqrt k}{\sqrt{2+k}(\a-2\b)}\right)\right]\ .
\end{equation}
It remains to fix the constants~$\a$ and~$\b$: they are chosen in such a way that the theory just described closes the~$N=2$ SCA. Following~\cite{Zamolodchikov:1986gh} we define
\begin{equation}
G^+(z)=\sqrt{\frac{2}{k+2}}\Psi_{l,n,s}(z)e^{2i \sqrt{\frac{k+2}{k}}H(z)}\ ,\qquad G^-=\sqrt{\frac{2}{k+2}}\Psi^{\dagger}_{l,n,s}(z)e^{-2i \sqrt{\frac{k+2}{k}}H(z)}\ ,
\end{equation}
and using the OPE of the scalars we can ensure the closure of the algebra, provided that
\begin{equation}
\a=-\frac{2}{k+2}\ ,\qquad \b=\frac{k}{2(k+2)}\ .
\end{equation}
The resulting expression for the primaries of the parafermionic theory~$\Psi(z)$ is thus given by
\begin{equation}\label{ch3:primary-decomp-minmod}
\Phi^{SU(2)}_{l,n} e^{i\frac s 2 \s (z)}=\underbrace{\tilde\Phi(z)\exp\left[i\left(\frac s 2-\frac{s+2n}{k+2}\right)\sqrt{\frac{k+2}{k}} H(z)\right]}_{\Psi_{l,n,s}(z)}\ \exp\left[ i \frac{2n+s}{\sqrt{2(k+2)}}g(z)\right] \ .
\end{equation}
Let us now extract information about the spectrum we are interested in: from the representation theory of the $\wzw{u}{1}{2k}$ WZW model it is known that irreducible representations are labeled by an integer~$r$ defined modulo~$2k$, such that $r=-k+1,\dots ,k$. In our case we conclude that $s=-1,0,1,2$. 
Moreover, from the theory of~$\wzw{su}{2}{k}$ models, we know that  unitary highest-weight representations are labeled by the half integer spin~$l/2$ and by~$n/2$, the half integer eigenvalue of  the current~$\hat J_0^3$; their allowed range is $0\leq l\leq k\in\Z_{>0}$, $-l\leq n\leq l$, and $\frac{l+n}{2}$ is integer.
We conclude that we can define an integer label~$m=n+s\mod{2k+4}$; in terms of~$m$ we infer that $l+m+s=l+n+2s$ must be even, since~$s$ is integer.
Conformal dimensions and~$U(1)$-charges of the primaries can be easily inferred from equation~\eqref{ch3:primary-decomp-minmod}.

\paragraph{Summary of the results}
The discrete series of $N=2$ supersymmetric minimal models is parameterised by an integer $k$; they have central charge $c=\frac{3k}{k+2}$.
The primary states are labeled by three integers~$(l,m,s)$ with $l+m+s$ even, and their associated primary fields will be denoted~$\phi_{l,m,s}$.
The allowed range for the labels is
\be\label{ch3:allowed-range-minmod}
0\leq l\leq k\ ,\qquad m\equiv m+2k+4\ ,\qquad s\equiv s+4\ .
\ee
The identifications can be summarised in
\begin{equation}\label{ch3:identifications-minmod}
(l,m,s)\equiv (k-l,m+k+2,s+2)\ ,
\end{equation}
and primary states belonging to~$\Hilb_{l,m,s}$ have conformal weight and central charge (in the holomorphic sector) given by
\begin{align}\label{ch3:spectrum-N=2-min-mod}
\begin{array}{lll}
\vspace{0.3cm}
& h\in h_{l,m,s}+\N\ ,\qquad & h_{l,m,s}=\frac{l(l+2)-m^2}{4(k+2)}-\frac{s^2}{8}\\
& Q\in Q_{m,s}+2\Z\ , \qquad & Q_{m,s}=-\frac{m}{k+2}+\frac s2\ .
\end{array}
\end{align}
If one furthermore asks
\be 
|m-s|\leq l\ ,
\ee
the integral shifts in equation~\eqref{ch3:spectrum-N=2-min-mod} are zero and the states belong to the so-called~\textit{standard range}. 
We consider only models with diagonal spectrum~(see section~\ref{app:characters:sec:GSO}), for which~$\bar h=h,\  \bar Q=Q$. 
The index~$s$ takes values $s=-1,0,1,2$, and works as a book-keeping device to discriminate those states which differ from the primary field by the action of an even versus an odd number of supercurrents $G^{\pm}$.
In the NS sector, we take the values of $s=0,2$, the former referring to states which differ from the highest-weight state by the action of an even number of supercurrents (therefore including the original ground state) and the latter referring to states obtained by the action of an odd number of supercurrents.
In the R sector, we define $s$ to be $\pm1$, with these values playing an analogous role. 
The direct sum~$\Hilb_{l,m,0}\oplus\Hilb_{l,m,2}$ is the~NS representation of the full superconformal subalgebra, and the same for the~R sector with~$\Hilb_{l,m,1}\oplus\Hilb_{l,m,-1}$.

\subsection{Characters, modular S-matrix and partition function}\label{ch3:sec:paraf:subs:characters-partition-fct}

\paragraph{Characters and singular vectors}
As discussed in section~\ref{ch3:sec:unitarity-spectrum-minmod} the representations we deal with have modules attached which contain singular vectors, and in the unitary case with~$c<3$ do not contain subsingular vectors.
We have to subtract singular vectors and their associated submodules by hand, in order to get irreducible unitary representations.
The easiest way to keep track of the complicated embedding structures of these modules is by computing their charged characters,
\begin{equation}
\chi_{(l,m,s)}(q,z)=\Tr_{\Hilb_{(l,m,s)}}q^{L_0-\frac{c}{24}}z^{J_0}\ .
\end{equation}
Explicit expressions can be found e.g. in~\cite{Ravanini:1987yg,Qiu:1987ux,Gepner:1987qi,Kiritsis:1986rv}, and are recalled in appendix~\ref{app:characters}.
We quote here some of the results.

In the NS sector for $|m|\leq l$ the characters read
\begin{align}
\chi^{\text{NS}}_{l,m}(q,z)\equiv &\left( \chi_{(l,m,0)}+ \chi_{(l,m,2)}\right)(q,z)\nonumber\\
= &\ q^{\frac{(l+1)^{2}-m^2}{4(k+2)}-\frac18}\,z^{-\frac{m}{k+2}}\left[\prod_{n=0}^{\infty}\frac{(1+q^{n+\frac12}z)(1+q^{n+\frac12}z^{-1})}{(1-q^{n+1})^2}\right]\times \Gamma^{(k)}_{lm}(\tau,\nu)\ ,
\label{ch3:minmod-NS-character}
\end{align}
and in the R sector (for $|m|\leq l+1$)
\begin{align}
\chi^{R}_{l,m}(q,z)\equiv &\left(\chi_{(l,m,1)}+\chi_{(l,m,-1)}\right)(q,z)\nonumber\\
= \ &q^{\frac{(l+1)^{2}-m^2}{4(k+2)}}\,z^{-\frac{m}{k+2}}(z^{\frac12}+z^{-\frac12})\left[\prod_{n=0}^{\infty}\frac{(1+q^{n+1}z)(1+q^{n+1}z^{-1})}{(1-q^{n+1})^2}\right]\times \Gamma^{(k)}_{lm}(\tau,\nu)\ .
\label{ch3:minmod-R-characters}
\end{align}
We have summarised the structure of the singular vectors in $\Gamma^{(k)}_{lm}$ (defined in~equation~\eqref{def-Gamma}): it is of the form
\begin{equation}
\Gamma_{lm}^{(k)}=1+ (\text{subtractions from singular vectors}) \ .
\end{equation}

The characters can be also written as follows
\begin{equation}\label{ch3:characters-minmod-coset-notation}
\Tr_{\Hilb_{(l,m,s)}} q^{L_0-\frac{c}{24}}z^{J_0}=\chi_{(l,m,s)}(q,z)=\sum_{j=0}^{k-1}C^{(k)}_{l,m-s+4j}(q)\Theta_{2m+(4j-s)(k+2),2k(k+2)}\left(\t,-\tfrac{\n}{2k+4}\right)
\end{equation}
and~$C$ are called branching functions, which can be found for instance in~\cite{Blumenhagen:2009zz}.
The last expression makes modular properties of the characters manifest.

The modular transformation matrix can be inferred by the known modular properties of the~$\Theta$ function (see appendix~\ref{app:characters}) from equation~\eqref{ch3:characters-minmod-coset-notation} (or equivalently by the knowledge of the S-matrices for the algebras entering the definition of the coset, as explained in~\cite{Blumenhagen:2009zz} for example), and reads
\begin{equation}\label{ch3:S-matrix-minmod}
S_{(l',m',s')(l,m,s)}=\frac{1}{k+2} \sin \frac{\pi(l+1)(l'+1)}{k+2}e^{-i\pi\left(\frac{ss'}{2}-\frac{mm'}{k+2}\right)}\ .
\end{equation}

\paragraph{Partition function}
The modular invariant full partition functions of~$N=2$ minimal models have been classified in~\cite{Gannon:1996hp,Gray:2008je}, similarly to the modular invariants for the $\wzw{su}{2}{k}$ WZW models~\cite{Gepner:1986hr,Cappelli:1987hf,Cappelli:1987xt}.
We consider only the so-called A-series in this thesis, which is diagonal, meaning that every holomorphic ground state is coupled only to its anti-holomorphic cousin: 
\begin{equation}
Z_k=\sum_{s=-1}^{2}\sum_{l=0}^{k}\sum_{\substack{m=s-l\\ l+m+s\ \text{even}}}^{s+l}\chi_{(l,m,s)} (q,z)\bar\chi_{(l,m,s)}(\bar q,\bar z)\ .
\end{equation}
It can be seen as a sort of type-0 GSO projection (discussion in section~\ref{app:characters:sec:GSO} for explanations) of the fully supersymmetric partition function.
In formulas: the NS part of the A-series minimal model partition function is given by
\begin{equation}\label{Z-MM-A}
\begin{split}
Z^{\text{NS}}_{k}=&\sum_{l=0}^{k}\sum_{\substack{m=-l\\ l+m\ \text{even}}}^l\chi_{(l,m,0)} (q,z)\bar\chi_{(l,m,0)} (\bar{q},\bar{z}) +\chi_{(l,m,2)} (q,z)\bar\chi_{(l,m,2)} (\bar{q},\bar{z})\\
=& \frac12\sum_{l=0}^{k}\sum_{\substack{m=-l\\ l+m\
\text{even}}}^l\left(\chi^{\text{NS}}_{l,m}
(q,z)\bar\chi^{\text{NS}}_{l,m}(\bar{q},\bar{z})+\chi^{\text{NS}}_{l,m}
(q,-z)\bar\chi^{\text{NS}}_{l,m}(\bar{q},-\bar{z})\right)\ ,
\end{split}
\end{equation}
and can be realised as a projection of the trace over the fully supersymmetric NS Hilbert space
$\cH^{\text{NS}}_{k}=\oplus_{|m|\leq l\leq k}\cH^{\text{NS}}_{l,m}\otimes
\cH^{\text{NS}}_{l,m}$, which reads
\begin{equation}\label{Z-MM-SUSY}
\begin{split} 
\mathcal{P}_{k}^{\text{NS}}(\tau,\nu):&=\sum_{l=0}^{k}\sum_{\substack{m=-l\\ l+m\ \text{even}}}^l\left(\chi_{(l,m,0)} (q,z)+\chi_{(l,m,2)} (q,z)\right)\left(\bar\chi_{(l,m,0)} (\bar{q},\bar{z})+\bar\chi_{(l,m,2)} (\bar{q},\bar{z})\right)\\
&=\left|\frac{\vartheta_3(\tau,\nu)}{\eta^3(\tau)}\right|^2\sum_{l=0}^{k}\sum_{\substack{m=-l\\
l+m\ \text{even}}}^l\left|q^{\frac{(l+1)^{2}-m^2}{4(k+2)}}
\Gamma^{(k)}_{lm}(\tau,\nu)\right|^2\, (z\bar{z})^{-\frac{m}{k+2}}\ ,
\end{split}
\end{equation}
with~$\Gamma^{(k)}_{lm}$ given in~\eqref{def-Gamma}.

We have to be careful here, since the bulk partition function does not change if we flip~$m\to-m,s\to-s$, which is the effect of the mirror automorphism~\eqref{ch3:mirror-automorphism} on the minimal model's modules.
Under the action of the mirror automorphism the diagonal modular partition function becomes anti-diagonal
\begin{equation}\label{ch3:partition-anti-diagonal}
Z_k=\sum_{s=-1}^{2}\sum_{l=0}^{k}\sum_{\substack{m=s-l\\ l+m+s\ \text{even}}}^{s+l}\chi_{(l,m,s)} (q,z)\bar\chi_{(l,-m,-s)}(\bar q,\bar z)\ .
\end{equation}
which is the same as the diagonal one at this level.
Nevertheless, as we explain in section~\ref{ch3:sec:BC}, the mirror automorphism acts non-trivially on boundary conditions.
We have therefore to be specific in the choice of the~A-type partition function.
We use the strictly diagonal one.
In section~\ref{ch3:sec:geometry} we will argue that the minimal model at level~$k$ is T-dual to its own~$\Z_{k+2}$ orbifold.
Since T-duality is realised at the algebraic level by the mirror automorphism, the orbifold has an anti-diagonal partition function. 

\section{Bulk correlators}\label{ch3:sec:correlators}

With the help of equation~\eqref{ch3:primary-decomp-minmod} we can infer the expressions of three-point functions of primary fields using the knowledge of structure constants of the~$\wzw{su}{2}{k}$~WZW model (see~\cite{Mussardo:1988av} for details of this construction and ~\cite{Zamolodchikov:1986bd,Dotsenko:1990zb} for the $\wzw{su}{2}{}$ structure constants).
We restrict the attention here on~NS correlators, whose expressions will be used in the following.
In a model with diagonal partition function the independent computations in the NS-sectors are the correlators of three primary fields (all with $s=0$, here~$\phi_{l,m}\equiv\phi_{l,m,0}$), and correlators involving one superdescendant.
The others can be obtained using the OPEs of the superconformal algebra.

The two point function in standard normalisation reads
\begin{equation}
\langle \phi_{l_{1},m_{1}} (z_{1}) \phi_{l_{2},m_{2}} (z_{2})\rangle=\delta_{l_{1},l_{2}}\delta_{m_{1},-m_{2}}\frac{1}{|z_{1}-z_{2}|^{4h_{l_{1},m_{1}}}} \ .
\end{equation}
The correlator for three~$s=0$ primary fields reads~\cite{Mussardo:1988av}
\begin{multline}\label{ch3:three-point-fct-minmod}
\langle\phi_{l_{1},m_{1}} (z_{1},\bar{z}_{1}) \phi_{l_{2},m_{2}}
(z_{2},\bar{z}_{2})\phi_{l_{3},m_{3}} (z_{3},\bar{z}_{3})\rangle\\
=C(\{l_{i},m_{i}\})\delta_{m_{1}+m_{2}+m_{3},0}|z_{12}|^{2(h_{3}-h_{1}-h_{2})}|z_{13}|^{2(h_{2}-h_{1}-h_{3})}|z_{23}|^{2(h_{1}-h_{2}-h_{3})}
\end{multline}
with 
\begin{equation}
C (\{l_{i},m_{i}\})=\begin{pmatrix}
\frac{l_{1}}{2} & \frac{l_{2}}{2} & \frac{l_{3}}{2}\\
\frac{m_{1}}{2} & \frac{m_{2}}{2}& \frac{m_{3}}{2}
\end{pmatrix}^{\!2}
\sqrt{(l_{1}+1)(l_{2}+1)(l_{3}+1)}\, d_{l_{1},l_{2},l_{3}}\ .
\label{3ptcoefficient}
\end{equation}
Here, $\begin{pmatrix}
j_{1}&j_{2}&j_{3} \\
\mu_{1}&\mu_{2}&\mu_{3}
\end{pmatrix}$ denotes the Wigner 3j-symbols, and
$d_{l_{1},l_{2},l_{3}}$ is a product of Gamma functions,
\begin{equation}\label{defofd}
d_{l_{1},l_{2},l_{3}}^{2}
=\frac{\Gamma(1+\rho)}{\Gamma(1-\rho)}P^{2}(\tfrac{l_{1}+l_{2}+l_{3}+2}{2})
\prod_{k=1}^{3} \frac{\Gamma(1-\rho(l_{k}+1))}{\Gamma(1+\rho(l_{k}+1))}
\frac{P^{2}(\frac{l_{1}+l_{2}+l_{3}-2l_{k}}{2})}{P^{2}(l_{k})}
\end{equation}
with
\begin{equation}\label{def_P}
\rho=\frac{1}{k+2}\quad ,\quad
P(l)=\prod_{j=1}^{l}\frac{\Gamma(1+j\rho)}{\Gamma(1-j\rho)} \ .
\end{equation}
The same methods allow the computation of correlators involving super-descendants~\cite[appendix C]{Fredenhagen:2012rb}. 
The result reads
\begin{align}
& \langle (\bar{G}^{+}_{-\frac{1}{2}}G^{+}_{-\frac{1}{2}}\phi_{l_{1},m_{1}}) (z_{1},\bar{z}_{1})
\phi_{l_{2},m_{2}} (z_{2},\bar{z}_{2})\phi_{l_{3},m_{3}}
(z_{3},\bar{z}_{3}) \rangle \nonumber\\
&\quad  = \frac{2(k+2)}{(l_1-|m_1|+2)(l_1+|m_1|)} 
\begin{pmatrix} \frac{l_{2}+m_{2}}{2} \\ \frac{l_1-|m_1|+2}{2} \end{pmatrix}
\begin{pmatrix} \frac{l_{2}-m_{2}}{2}+\frac{l_1-|m_1|+2}{2} \\ \frac{l_1-|m_1|+2}{2}\end{pmatrix}
\begin{pmatrix} l_{1} \\ \frac{l_1-|m_1|+2}{2} \end{pmatrix}^{\!-1}\nonumber\\
&\qquad \times 
\begin{pmatrix}
\frac{\tilde{l}_{1}}{2} & \frac{l_{2}}{2} & \frac{l_{3}}{2} \\[1mm]
-\frac{\tilde{l}_{1}}{2} & \frac{m_{2}}{2}-\frac{l_1-|m_1|+2}{2} & \frac{m_{3}}{2}
\end{pmatrix}^{\!2}
\sqrt{\big(\tilde{l}_{1}+1\big) \big(l_{2}+1\big) \big(l_{3}+1\big)}\, 
d_{\tilde{l}_{1},l_{2},l_{3}}\nonumber\\
&\qquad  \times 
|z_{12}|^{2
\left(h_{l_{3},m_{3}}-(h_{l_{1},m_{1}}+1/2)-h_{l_{2},m_{2}}\right)} 
|z_{23}|^{2 \left((h_{l_{1},m_{1}}+1/2)-h_{l_{2},m_{2}}-h_{l_{3},m_{3}}\right)}\nonumber\\
&\qquad \times 
|z_{13}|^{2 \left(h_{l_{2},m_{2}}-
(h_{l_{1},m_{1}}+1/2)-h_{l_{3},m_{3}}\right)}\ .
\end{align}

\section{Superconformal boundary conditions}\label{ch3:sec:BC}

Two classes of maximally symmetric boundary conditions can be consistently defined for~$N=2$ SCA~(in this context firstly introduced in~\cite{Maldacena:2001ky} but already analysed in~\cite{Ooguri:1996ck} for supersymmetric non-linear sigma-models), based on the mirror automorphism:
\begin{equation}\label{ch3:mirror-automorphism-def-BC-section}
\O_M\ :\  J_n\longmapsto -J_n\ ,\qquad G_{r}^{\pm}\longmapsto G_r^{\mp}\ .
\end{equation}
They are conventionally named
\begin{itemize}
\item{A-type boundary conditions:}
\begin{align}\label{ch3:A-type-BC}
\begin{array}{rcrcc}
(L_n-\overline{L}_{-n})\,|i\rangle\!\rangle_A & = & (J_n-\overline{J}_{-n})\,|i\rangle\!\rangle_A & =&  0 \\[4pt]
(G^+_r+i\eta\,\overline{G}{}^-_{-r})\,| i\rangle\!\rangle_A  &= &(G^-_r+i\eta\,\overline{G}{}^+_{-r})\,| i\rangle\!\rangle_A & = & 0
\end{array}
\end{align}
\item{B-type boundary conditions:}
\begin{align}\label{ch3:B-type-BC}
\begin{array}{rcrcc}
(L_n-\overline{L}_{-n})\,| i\rangle\!\rangle_B & = & (J_n+\overline{J}_{-n})\,|  i\rangle\!\rangle_B &= & 0 \\[4pt]
(G^+_r+i\eta\,\overline{G}{}^+_{-r})\,|  i\rangle\!\rangle_B & = & (G^-_r+i\eta\,\overline{G}{}^-_{-r})\,|  i\rangle\!\rangle_B &= & 0
\end{array}
\ ,
\end{align}
\end{itemize}
where~$\h=\pm1$ discriminates between the two possible choices of spin structures on the plane.
The mirror automorphism maps boundaries of the A-type to the B-type and vice versa. 

\subsection{A-type boundary conditions}
The boundary states are given by coherent superpositions of Ishibashi states, such that they reproduce the correct coupling with the bulk fields. 
In particular the one-point functions of the bulk fields in presence of boundary conditions must reproduce the amplitude between the boundary state associated to those boundary conditions and the bulk fields.
As shortly explained in appendix~\ref{app:vanilla}, one consistent solution to this problem is given by the Cardy states~\cite{Cardy:1989ir}, constructed directly from the S-matrix to get automatically the expected cylinder modular properties. 
In the case at hand, the knowledge of the S-matrix allows us to write explicitly A-type Cardy states
\be\label{ch3:A-type-bc-minmod-generic}
|L,M,S\ket_A = \sum_{(l,m,s)}\frac{S_{(L,M,S)(l,m,s)}}{\sqrt{S_{(0,0,0)(l,m,s)}}}|l,m,s\iket_A\ ,
\ee
where the S-matrix is the one given in equation~\eqref{ch3:S-matrix-minmod}, the Ishibashi states are implicitly defined in~equation~\eqref{ch3:A-type-BC}, and the range of the sum is given by~\eqref{ch3:allowed-range-minmod}, with the identifications of~\eqref{ch3:identifications-minmod}.
These boundary conditions are labeled by the same labels as the representations of the bosonic subalgebra of~$N=2$ SCA, $0\leq L\leq k$, $M\in\Z \mod 2k+4$ and $S\in\Z\mod 4$ with $L+M+S$ even.
One can insert in equation~\eqref{ch3:A-type-bc-minmod-generic} the explicit expression for the S-matrix given in~\eqref{ch3:S-matrix-minmod}
and read off the one-point functions in presence of an A-type boundary condition labeled by the triple~$(L,M,S)$
\begin{equation}\label{ch3:A-type-one-point-fct}
\bra\phi_{l,m,s}(z,\bar z)\ket_{(L,M,S)}^A =\frac{1}{\sqrt{k+2}}\frac{\sin\frac{\pi(l+1)(L+1)}{k+2}}{\sqrt{\sin\frac{\pi(l+1)}{k+2}}}e^{-i\pi (\frac{sS}{2}-\frac{mM}{k+2})}\frac{1}{|z-\bar z|^{2h_{l,m,s}}}\ .
\end{equation}
From the last formula, using the identification rules and performing a modular transformation, it is possible to write down the open string spectrum between two A-type boundary conditions~\cite{Maldacena:2001ky} as
\begin{equation}
{}_A\bra L,M,S|q^{L_0-\frac{c}{24}}|L',M',S'\ket_A=\sum_j N^{j}_{LL'}\chi_{j,M-M',S-S'}(\tilde q)\ ,
\end{equation}
where~$N$ are the $\kmalg{su}(2)_k$ fusion coefficients (for their expressions and derivations see~\cite{DiFrancesco:1997nk}), and the sum is performed over distinct states.

\subsection{B-type boundary conditions}
As explained in subsection~\ref{ch3:subsection-bulk-geometry}, T-duality maps the~$N=2$ minimal model at level~$k$ to its own~$\Z_{k+2}$ orbifold.
B-type boundary conditions are obtained~\cite{Maldacena:2001ky} by applying a T-duality transformation to the boundary conditions:
in the orbifold construction it is possible to define bulk branes as superposition of branes in the parent theory invariant under the action of the orbifold group.
In our case, we can take superpositions of A-type boundary conditions invariant under~$\Z_{k+2}$; under T-duality they are mapped to B-type boundary conditions.

From the very definition in equation~\eqref{ch3:B-type-BC} we see that B-type boundary conditions cannot couple to charged bulk fields in the case of diagonal models.
Furthermore, it can be shown~\cite{Maldacena:2001ky} that they only couple to~NS sectors in this case\footnote{In anti-diagonal models the situation is opposite: B-type boundary conditions couple only to charged fields, and to R-R fields as well, as commented in subsection~\ref{ch4:sec:branes:subs:contorbi}.} (for some details, see appendix~\ref{app:vanilla}).
They are labeled by two integers~$(L,S)$ with~$0\leq L\leq k$ and~$S=0,1$, and the one-point functions in their presence read
\be 
\bra \phi_{l,m,s}(z,\bar z)\ket ^B_{(L,S)}=\sqrt 2 \frac{\sin\frac{\pi(l+1)(L+1)}{k+2}}{\sqrt{\sin\frac{\pi(l+1)}{k+2}}}\d_{m,0}e^{-i\pi\frac{sS}{2} }\frac{1}{|z-\bar z|^{2h_{l,m,s}}}\ .
\ee

\section{WZW description and geometry}\label{ch3:sec:geometry}

Conformal field theories with affine chiral currents can be realised as sigma-models on (simple and simply connected in this chapter) Lie group manifolds~\cite{Witten:1984ar,Gepner:1986wi,DiFrancesco:1997nk}, and these CFTs are called~Wess-Zumino-Witten (WZW) models. 
The idea is that the fields~$g$, defined on the two-dimensional Riemann surface~$\Sigma$, take value on (a representation of) a Lie group~$G$
\begin{equation}\label{ch3:WZW-map-general}
g:\ \Sigma\longrightarrow G\ .
\end{equation}
The group is equipped with a Killing form~$\bra,\ket=\frac{1}{g_G}\Tr_{\text{Ad}}$ on the tangent space
where~$g_G$ is the dual Coxeter number of the Lie algebra~$\mathfrak{g}$, and the trace is taken on the adjoint representation of the Lie algebra.
\paragraph{Sigma-models on Lie groups: WZW models}
In analogy with the free boson example of section~\ref{ch1:sec:free-boson} we write down a free action for the fields
\be\label{ch3:kinetic-WZW}
S_0=\frac{k}{4\pi i} \int_{\S}\,\bra g^{-1}\de g,g^{-1}\debar g\ket\ dz\wedge d\bar z
\ee
where the choice of the normalisation will become sensible in the following.
This time the non-flatness and non-commutativity of the target space spoils in general the conformal invariance of the theory. 
We must supplement the free theory with the (topological) Wess-Zumino~(WZ) term~\cite{Witten:1984ar}, which compensates the variation of the free action under a conformal transformation. 
The right conformal invariant action reads
\begin{equation}\label{ch3:WZW-action-general}
S_{\text{WZW}}=S_0+\frac{k}{4\pi i}\left(\int_{B_3}\ \underbrace{\frac{1}{6}\bra g^{-1}d g,[g^{-1}d g,g^{-1}d g]\ket}_{:=\omega}\right)\ ,
\end{equation}
where the WZ-form\footnote{From a string-theoretical perspective, the WZ-form is analogous to the $H=dB$~field, where~$B$ is the Kalb-Ramond two-form~(see e.g.~\cite{PolchinskiBookII:1998}).} has been called~$\omega$\, and~$B_3$ is a three-dimensional ball whose boundary is~$\Sigma$.
The freedom of choosing different~$B_3$ such that~$\de B_3=\S$ produces ambiguities in the definition of the quantum theory: the action changes as follows going from~$B_3$ to~$B'_3$
\be\label{ch3:ambig-general-WZW}
\D S_{\text{WZW}}=\frac{k}{4\pi i}\int_{\tilde g(\tilde B_3)}\tilde g^*\omega
\ee
where~$\tilde B_3$ is the manifold without boundary obtained by gluing together~$B_3$ and~$B'_3$ with opposite orientations along the common boundary~$\S$, and~$\tilde g^*\omega$ is the pull-back by~$\tilde{g}$ (the extension to~$B_3-B_3'$ of the map~$g$ defined in~\eqref{ch3:WZW-map-general}) of the three-form~$\omega$.
To avoid anomalous ambiguities the variation of equation~\eqref{ch3:ambig-general-WZW} must be set (\`a la Dirac) to be an integer multiple of~$2\pi i$, in order to give an uniquely defined euclidean path integral.
For every root~$\a$ of the Lie algebra~$\mathfrak{g}$ associated to~$G$ we have
\begin{equation}
\frac{1}{4\pi i}\int_{SU(2)_{\a}}\tilde g ^* \omega =-\underbrace{\left(\frac{2}{|\a|^2}\right)}_{\in \Z} 2\pi i\ ,
\end{equation}
where~$SU(2)_{\a}$ indicates the subgroup obtained by exponentiation of the Cartan triangular subalgebra~$(\frac{2\a^i H_i }{|\a|^2},E^{\a},E^{-\a})$.
We conclude than that the level of the WZW model is quantised:~$k\in\Z$ (all the details for this construction can be found in~\cite{Gawedzki:1999bq}).
Keeping this in mind, $S_{\text{WZW}}$ defines a~CFT.

\paragraph{Cosets as gauged~WZW models}
Our interest in this description is that coset models can be interpreted as sigma-models as well: they are described by WZW models with an appropriate gauge symmetry.
In particular the coset theory $G/H$ is defined as the WZW model on the Lie group~$G$, coupled to the gauge field associated to the local transformation
\be
g(z,\bar z)\mapsto h(z,\bar z) g(z,\bar z) h^{-1}(z,\bar z)\ ,\qquad g\in G\ ,\ h\in H\ ,\ H\subset  G\  .
\ee
Since the gauge symmetry acts by conjugation on the group manifold itself, the orbits have as well a geometric meaning: gauging a symmetry accounts to identify geometrically points of the target manifold lying on the same orbit.
It is worth to mention here that, in contrast with standard gauge theory, conformal invariance forbids the kinetic term for the gauge field.

The analysis that follows from these preliminary considerations is exact for everything concerning the classical theory. However, the conclusions we are going to draw for the quantum field theory are valid only in the large level~$k$ regime, since this is the region of the moduli space in which
stringy effects are negligible and the (super)-gravity approximation holds, hence we can safely talk about geometric quantities~\cite{Felder:1999ka}.

\subsection{Bulk geometry of $N=2$ minimal models: the supersymmetric bell}\label{ch3:subsection-bulk-geometry}
As explained in sections~\ref{ch3:sec:MM-parafermions} and~\ref{ch3:app:sec:KS-models}, $N=2$~minimal models admit a description in terms of the parafermionic coset~$\frac{\wzw{su}{2}{}}{\wzw{u}{1}{}}$ supplemented with the right amount of fermions.
The sigma-model description corresponds to the $U(1)$-gauged~$\wzw{su}{2}{k}$ WZW model, with fermions on the tangent bundle.
One of the effects of introducing fermions is to shift the level~$k$ of the original~$\wzw{su}{2}{k}$ WZW model to~$k+2$.

We start with a standard parameterisation of~$SU(2)$, which is isometric to a three-sphere. Any group element $g\in SU(2)$ is given by
\begin{equation}
g(\theta,\varphi,\tilde \varphi)=\left(
\begin{array}{cc}
\cos \theta e^{i\tilde \varphi} & \sin \theta e^{i \varphi}\\
-\sin \theta e^{-i\varphi} & \cos{\theta} e^{-i\tilde\varphi}
\end{array}
\right)\ ,\qquad 0\leq\theta\leq\frac{\pi}{2}\ ,\ 0\leq \varphi,\tilde\varphi< 2\pi\ .
\end{equation}
In this parameterisation the metric on~$S^3$ with radius~$R=\sqrt {k+2}$ reads
\begin{equation}
ds^2=(k+2)\left(d\theta^2+\sin ^2\theta d\varphi^2+\cos^2\theta d\tilde\varphi^2\right)\ ,
\end{equation}
and the WZ-form defined in~\eqref{ch3:WZW-action-general} is
\begin{equation}
\omega=\cos\theta \sin\theta \ d \theta \wedge d\varphi\wedge d\tilde \varphi\ .
\end{equation}
It is easy to integrate the WZ-form since (as long as~$ \theta \neq \frac{\pi}{2}$)
\begin{equation}
\omega=d\left(\frac{1}{2}\sin^2\theta\,d\varphi\wedge d\tilde{\varphi}\right)\ ,
\end{equation}
and from equation~\eqref{ch3:WZW-action-general} to write down the full action, which reads
\begin{equation}\label{ch3:SU(2)-full-action}
S=\frac{k+2}{8\pi}\int_{\S} d^2z\ \left[\de\theta\de\bar\theta+\tan^2\theta\ \de\varphi\debar \varphi + \cos^2\theta\left(\de\tilde\varphi+\tan^2\theta \de\varphi\right) \left(\debar\tilde\varphi-\tan^2\theta \debar\varphi\right)\right]\ .
\end{equation}

\paragraph{Gauging the model}
The classical action~\eqref{ch3:SU(2)-full-action} has two manifest global symmetries, namely real translations in~$\varphi$ and~$\tilde\varphi$. 
These two angles parameterise two different~$U(1)$ subgroups of~$SU(2)$, conventionally called the vector $U(1)$ described by~$\tilde\varphi$, and the axial one, parameterised by~$\varphi$.
These names are inherited by chiral four-dimensional gauge theories. 
The analogy comes from the fact that gauging the vector symmetry the axial one becomes anomalous and vice versa, as in the four dimensional case. Nevertheless the two models (vectorially and axially gauged) are mutually T-dual. We will be more precise later in this section.
We choose to gauge the vector symmetry: the minimal substitution by the introduction of the appropriate (anti)-holomorphic gauge field~$A(\bar A)$ accounts then to promote
\begin{equation}
\de\tilde\varphi\longrightarrow D\tilde\varphi =\de\tilde\varphi+ A(z)\ ,\qquad \debar\tilde\varphi\longrightarrow \bar D\tilde\varphi=\debar\tilde\varphi+\bar A(\bar z)\ .
\end{equation}
We do not add a kinetic term for the gauge fields since this would spoil conformal invariance.
The action for the gauged model becomes then
\begin{equation}
S=\frac{k+2}{8\pi}\int_{\S} d^2z\ \left[\de\theta\de\bar\theta+\tan^2\theta\ \de\varphi\debar \varphi + \cos^2\theta\left(\de\tilde\varphi+\tan^2\theta \de\varphi+A\right) \left(\debar\tilde\varphi-\tan^2\theta \debar\varphi+\bar A\right)\right]\ .
\end{equation}
The gauge fields are auxiliary, and can be path-integrated away, with the effect of removing completely the~$\tilde\varphi$-dependent term in the action and of producing a dilaton shift~\cite{Gawedzki:1988nj,Schwarz:1992te}.
The result reads
\begin{equation}\label{ch3:action-su(2)/u(1)-sigma-model}
S=\frac{1}{8\pi}\int d^2z\left[(k+2)\left(\de\theta \debar\theta +\tan^2\theta\de\varphi\debar \varphi\right)-\frac 12 R^{(2)}\log \cos^2\theta\right]\ .
\end{equation}

\paragraph{Bell geometry}
We can then read off the geometry (informally called the bell-geometry) of the supersymmetric parafermionic sigma-model described in equation~\eqref{ch3:action-su(2)/u(1)-sigma-model}
\begin{equation}\label{ch3:geometry-su(2)/u(1)}
\text{bell:}\ \ 
\left\{
\begin{array}{rl}
ds^2=&\ \frac{k+2}{1-\rho^2}\left(d\rho^2+\rho^2 d\varphi^2\right)\\
e^{\Phi(\rho,\varphi)}=&\ g_s(0)(1-\rho^2)^{-1/2}
\end{array}
\right.\ ,
\end{equation}
with~$\rho=\sin\theta$, and~$g_s(0)$ the string coupling of the original WZW model.

The space is topologically a disc, with a curvature singularity at the boundary~$\rho=1$; the radial geodesic distance from the centre to the boundary is finite~$d_{01}=\frac{\pi}{2}\sqrt {k+2}$, but the circumference at radius~$\rho=1$ is divergent as 
\begin{equation}
C_{\rho=1}=\lim_{\rho\to1}\ 2\pi\rho \ \sqrt{\frac{k+2}{1-\rho^2}}\ .
\end{equation}
\paragraph{Anomalous $U(1)$}
The~$\varphi$ coordinate in the action~\eqref{ch3:action-su(2)/u(1)-sigma-model} is cyclic, suggesting a residual~$U(1)$ global symmetry. 
This expectation is wrong, since quantum corrections make this symmetry anomalous. 
The divergence of the current associated with the rotations of an angle~$\varphi$ is proportional to the field strength of the gauge field associated to $\tilde{\varphi}$-rotations as $\de_a j^{a}\sim (k+2) F$.
The variation of the action under the global symmetry $\varphi\mapsto\varphi+\a$ is then proportional to $\a (k+2)\int_{\S} F$, which is a topological term. 
The anomaly can be treated following the same lines explained under equation~\eqref{ch3:ambig-general-WZW}: the integral counts (up to $2\pi$ prefactors) the cycles of the second integer homology of the Riemann surface~$\S$.
To avoid ambiguities we are led to restrict~$\a = \frac{n}{k+2}$ with~$n$ integer.
Since~$\a$ is an angle, the resulting discrete symmetry is a cyclic group.
The axial~$U(1)$ global symmetry is thus broken down to~$\Z_{k+2}$.

\paragraph{Wave functions of the light states}
If~$k$ is large enough, the three-sphere has a very big radius, and the longest wave-length excitations of the string are effectively point-like (like gravitational waves).
In particular the target space effective lagrangian for the operators with lowest conformal dimensions reads
\begin{equation}\label{ch3:eff-lagrangian-parafermion}
\mathcal{L}_{\text{eff}}=\frac12 e^{-2\Phi}\sqrt gg^{ab}\de_a \psi\de_b\psi\ ,
\end{equation}
where the indices run in the directions~$\varphi,\rho$ and~$\psi$ is the wave function of the ``lightest'' excitations. 
The equations of motion coming from the lagrangian~\eqref{ch3:eff-lagrangian-parafermion}
\begin{equation}\label{ch3:laplacian-parafermion}
\left( -\frac{1}{2}\nabla^{2} + (\nabla \Phi)\cdot \nabla\right)\psi_{l,m}
(\rho ,\varphi) = 2h_{l,m} \,\psi_{l,m} (\rho ,\varphi) 
\end{equation}
can be solved in terms of the gaussian hypergeometric function~$\,\!_2F_1$. The solution reads (see~\cite{Maldacena:2001ky} for the details of this computation)
\begin{equation}
\psi_{l,m} (\rho ,\varphi) = \rho^{|m|} e^{im\varphi} {}_{2}F_{1}
\left(\tfrac{|m|+l}{2}+1,\tfrac{|m|-l}{2};|m|+1;\rho^{2}\right)\ ,
\end{equation}
with
\begin{equation}
h_{l,m} = \frac{l (l+2)-m^{2}}{4 (k+2)} \ .
\end{equation}

\paragraph{T-dual orbifold description}
As mentioned before, the theory just described does not enjoy at the quantum level the residual global~$U(1)$ axial symmetry, since quantum corrections break it down to~$\Z_{k+2}$.
Nevertheless this discrete abelian symmetry remains, and a natural possible use thereof is to orbifold it~(the orbifold construction is reviewed in chapter~\ref{ch:orbifolds}).
The detailed analysis of the partition function for the parafermions suggests that the~$\Z_l$ orbifold of the theory is the same as its~$\Z_{l'}$ orbifold, provided that~$ll'=k+2$ (again~\cite{Maldacena:2001ky} for details).
In particular in the case~$l'=1$ one is led to the equivalence between the (supersymmetric) parafermion itself and its~$\Z_{k+2}$ orbifold.
Further analyses lend support that the mentioned correspondence reflects a T-duality transformation: the vectorially gauged supersymmetric parafermion is the T-dual of the~$\Z_{k+2}$ orbifold of the axially gauged model.
At the geometric level one can explicitly realise the duality transformation by defining a new radial coordinate
\begin{equation}
\rho'=\sqrt{1-\rho^2}\ ,
\end{equation}
in terms of which the T-dual metric and dilaton read
\begin{equation}
\text{T-bell:}\ \ 
\left\{
\begin{array}{rl}
ds'^2=&\ \frac{k+2}{1-\rho'^2}\left(d\rho'^2+\rho'^2 d\varphi'^2\right)\\
e^{\Phi'(\rho',\varphi')}= &\ \frac{g_s(0)}{\sqrt{k+2}}(1-\rho'^2)^{-1/2}\ ,
\end{array} 
\right.
\end{equation}
provided the identification
\begin{equation}
\varphi'\equiv\varphi'+\frac{2\pi}{k+2}\ .
\end{equation}
The string coupling gets rescaled by~$(k+2)^{-1/2}$; the coordinate singularity of~$\rho\to1$ is mapped in the T-dual description to an orbifold fixed point in~$\rho'=0$.
This fact will be crucial in the analysis of chapter~\ref{ch:our-contorbi}: although the bell geometry 
blows up at the boundary, this becomes a tractable orbifold singularity in the T-dual picture. 
It turns out that all the different regions of the T-fold are explorable.

\subsection{D-branes on $N=2$ minimal models in the geometric picture}\label{ch3:sec:geometry:subsec:branes}

Since we have now a geometric description of the bulk minimal models, we can ask about the geometry of the boundary conditions as well. 
This analysis provides us with very useful pictures in the large~$k$ regime, which is going to be of our interest in the following.
The methods that have been used to find out the geometry of the D-branes in this background are explained in~\cite{Maldacena:2001ky}. Very briefly:
\begin{itemize}
\item 
One can extremise the Dirac-Born-Infeld effective action~\cite{Fradkin:1985qd,Abouelsaood:1986gd} for the embedding geometry of the brane in the target space, usually starting with an ansatz.
\item
Another possibility is to study the scattering amplitude of (superposition of) bulk excitations with the boundary: since we know the wave function for bulk fields from the analysis of the generalised Laplacian of equation~\eqref{ch3:laplacian-parafermion}, we can choose a simple localised profile for them (a~$\d$-functional localised on the target manifold for instance) and write it in terms of eigenfunctions of the Laplacian. 
We can thus identify which bulk labels correspond to the elementary choice of the profile, and we can rewrite the one-point function in presence of a boundary in terms of the eigenfunctions of the target geometry.
We recognise in this way the locus of the brane (further explanations for the analogous WZW case can be found in~\cite{Schomerus:2002dc}).
\end{itemize}
\paragraph{A-type branes}

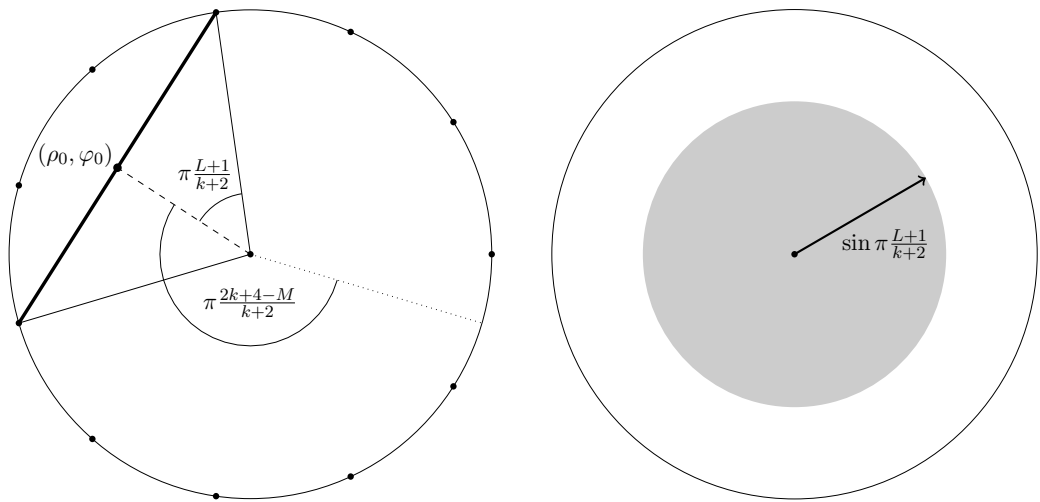
\begin{figure}[h] \centering \subfloat[][Illustration of an A-type brane, black straight line stretched between special points. Here $k=9, L=2, M=10$.]
{
\resizebox{6.5cm}{6.5cm}{%
\begin{tikzpicture}
\draw (0,0) circle (4cm);
\foreach \x/\y in {4. /0.,3.36501 /2.16256,1.66166 /3.63853,-0.569259 /3.95929,-2.61944 /3.023,-3.83797 /1.12693,-3.83797 /-1.12693,-2.61944 /-3.023,-0.569259 /-3.95929,1.66166 /-3.63853,3.36501 /-2.16256} \fill (\x,\y) circle (1.5pt);
\fill (0,0) circle (1.5pt);
\draw [line width=1.6pt] (-0.569259,3.95929) -- (-3.83797,-1.12693); 
\draw (-0.569259,3.95929) -- (0,0);
\draw (0,0) -- (-3.83797,-1.12693);
\fill (-2.20362, 1.41618) circle (2.0pt); 
\draw [dashed, line width=0.4pt] (0,0) -- (-2.20362, 1.41618);
\node [above left] at (-2.14, 1.3) {$(\rho_0,\varphi_0)$};
\draw (-0.142315, 0.989821) arc (98.1818:147.273:1); 
    \node[] at (120:1.5)  {$\pi \frac{L+1}{k+2}$};
\draw [dotted, line width=0.4pt] (0,0) -- (3.83797, -1.12693); 
\node[] at (270:0.8)  {$\pi \frac{2k+4-M}{k+2}$};
\draw [domain=147.273:343.636] plot ({1.5*cos(\x)}, {1.5*sin(\x)});  
\end{tikzpicture}        
}
} 
\quad
\subfloat[][Illustration of a B-type brane, gray disc centred in the origin, with radius growing as $L\leq \frac k2$.]
{
\resizebox{6.5cm}{6.5cm}{%
\begin{tikzpicture}
\draw (0,0) circle (4cm);
\fill [color=gray!40] (0,0) circle (2.5cm);
\fill (0,0) circle (1.5pt);
\draw [->,line width=1pt] (0,0) -- (30:2.5cm);
\node [below] at (20:1.6cm) {$\sin \pi\frac{L+1}{k+2}$};
\end{tikzpicture}   
}
}
\caption{D-branes in $N=2$ minimal models.}  \label{ch3:fig:branes-A-B}
\end{figure}

A-type boundary conditions are labeled by integers $(L,M,S)$, where $0\leq L \leq k$, $M$ is $2k+4$-periodic, $S\in \{-1,0,1,2\}$, and $L+M+S$ is even. 
Setting for simplicity~$S=0$, we find that A-type boundary conditions
correspond to branes that are straight lines stretched between special points on the boundary of the disc, at angles~$\varphi=\frac{M\pi}{k+2}$: they are described by the equation
\begin{equation}
\rho \cos (\varphi -\varphi_{0}) = \rho_{0} \ ,
\end{equation}
where
\begin{equation}\label{ch3:A-branes-geometry-in-terms-of-bound-labels}
\rho_{0} = \cos \frac{\pi (L+1)}{k+2} \quad , \quad 
\varphi_{0} = \frac{\pi M}{k+2} \ . 
\end{equation}
$(\rho_{0},\varphi_{0})$ are the coordinates of the point on the brane that is closest to the origin.
Boundary labels $(L,M,S)$ and $(k-L,M+k+2,S+2)$ describe the same boundary condition, so that we can always choose $L\leq k/2$ such that~$\rho_{0}$ is positive (see figure~\ref{ch3:fig:branes-A-B} (a)).
\paragraph{B-type branes}
B-type boundary conditions are labeled by two integers $(L,S)$ where
$0\leq L \leq k$ and $S=0,1$. 
Geometrically these correspond to two-dimensional discs where the
coordinate of the boundary is $\rho_{1}=\sin \frac{\pi
(L+1)}{k+2}$ (see figure~\ref{ch3:fig:branes-A-B} (b)).
\vspace{0.5cm}

\newpage
\begin{subappendices} 
\section{Kazama-Suzuki Grassmannian cosets}\label{ch3:app:sec:KS-models}
In this section we give a brief review of the Kazama-Suzuki models~\cite{Kazama:1988uz,Kazama:1989qp} since they are mentioned and used in several places of this thesis.

These \CFT possess~$N=2$ supersymmetry, and are described by the bosonic coset
\begin{equation}
\frac{\widehat{g}_k\oplus \widehat{so}(2d)_1}{\widehat{h}_{k+g_G-g_H}}~,
\end{equation}
with $d=\frac12{\rm dim}(G/H)$, $g_G$ and $g_H$ the dual Coxeter numbers of $G$ and $H$, respectively; the $\widehat{so}(2d)_1$ factor arises from the fermions.

The Grassmannian cosets
\begin{equation}
GC(n,k)=\frac{\widehat{su}(n+k)_1\oplus \widehat{so}(2nk)_1}
{\widehat{su}(n)_{k+1}\oplus \widehat{su}(k)_{n+1}\oplus 
\widehat{u}(1)_{kn(k+n)(k+n+1)}}
\end{equation}
 are the simplest among this class, and enjoy level-rank duality~\cite{Gepner:1988wi,Lerche:1989uy,Naculich:1997ic} so that they can be described by the following coset as well:
\begin{equation}\label{ch3:grassmannian-coset-def}
GC(n,k)=\frac{\widehat{su}(n+1)_k\oplus \widehat{so}(2n)_1}
{\widehat{su}(n)_{k+1}\oplus \widehat{u}(1)_{n(n+1)(k+n+1)}}\ .
\end{equation}
The central charge is $c =\frac{3nk}{ k+n+1}$.

Grassmannian coset models enjoy $W_{n+1}$ $N=2$ superconformal symmetry, and are rational.

\subsection{Ideas for coset construction}
To build the bulk spectrum of the theory we have to follow the GKO coset recipe~\cite{Goddard:1985vk,Goddard:1986ee}.
Suppose we are given a group~$G$ and an embedding of a subgroup~$H\hookrightarrow G$: we want to study the theory, denoted by~$\widehat{\mathfrak {g}}_k /\widehat{\mathfrak{h}}_{k'}$, whose spectrum is defined by the branching of the decomposition of the representations of the numerator in terms of representations of the denominator.
For a~$\widehat{\mathfrak {g}}_k /\widehat{\mathfrak{h}}_{k'} $ theory the decomposition reads
\be\label{ch3:coset-hilb-decomposition}
\Hilb_{\mathfrak{g}}^L=\bigoplus_{L'}\Hilb^{(L,L')}_{\mathfrak g / \mathfrak h}\otimes \Hilb_{\mathfrak h}^{L'}\ ,
\ee
where $L$ and $L'$ denote sectors of the representation spaces of the numerator and denominator theory respectively.
The branching~$\Hilb^{(L,L')}_{\mathfrak g / \mathfrak h}$ of this decomposition is the coset theory bulk space state.  
The torus partition function is reconstructed from the torus partition function of the \WZW model of the numerator by decomposing the characters of the numerators in terms of characters of the denominator.
Generators closing the Virasoro algebra are given by the formula
\be 
L^{\mathfrak{g}/\mathfrak{h}}_m=L_m^{\mathfrak{g}}-L_m^{\mathfrak{h}}\ ,
\ee
so that it is easy to write the spectrum of the coset theory once known the spectrum of the numerator and denominator theories:
\be\label{ch3:coset-spectrum-general}
h_{(L,L')}^{\mathfrak{g}/\mathfrak{h}}=h^{\mathfrak{g}}_L-h^{\mathfrak{h}}_{L'}+n\ ,\qquad n\in \Z_{\geq 0}\ .
\ee
The integer shifts~$n$ denote the level~(like in section~\ref{ch3:sec:unitarity-spectrum-minmod}) of the branching of the representations, depend on the levels~$k,k'$ and on the labels~$r,r'$ of numerator and denominator theories, and are in general difficult to compute~\cite{Dunbar:1992gh}. 
Nevertheless, in practical cases, coset models are subjected to several identifications between labels~\cite{Gepner:1989jq}, and this fact makes sometimes possible to identify many representations of the full spectrum to sectors with~$n=0$.
We introduce an \textit{identification current}~$\mathcal{J}$ defined in terms of simple currents\footnote{A simple current~$J$ of a \CFT is a primary field with the simple fusion rules
\be
J \cdot \phi = \phi'\ ,
\ee
which means that for any primary field~$\phi$, its operator product expansion with $J$ only produces fields from a single conformal family~\cite{Schellekens:1989am}.
Fusing~$J$ with itself one gets only the field~$J^2$, and in case of rational models (like the~\WZW models on compact Lie groups), since the number of primaries is limited, there must be an integer~$M$ such that~$J^M=\id$.}~\cite{Schellekens:1989uf} of~$\kmalg{g}_k$ and~$\kmalg{h}_{k'}$.
The identification currents realise on primary fields outer automorphisms of the chiral algebra characterising the \CFT.
The outer automorphism group of an affine Lie algebra~$\widehat{\mathfrak g}_k$ is isomorphic to the centre of the group~$G$ (see~\cite[chapter  14]{DiFrancesco:1997nk}).

\paragraph{Selection rules and identifications}
Let us be a little more specific at this point: 
\smallskip
\noindent
\begin{itemize}
\item
In general in equation~\eqref{ch3:coset-hilb-decomposition} some of the representations in~$L'$ of the denominator do not appear as subsectors of~$L$. 
This simply means that~$\Hilb^{(L,L')}_{\mathfrak g / \mathfrak h}$ may be empty for some~$(L,L')$. 
In other words, in the character decomposition corresponding to equation~\eqref{ch3:coset-hilb-decomposition}, the branching functions are sometimes zero.
This fact induces \textit{selection rules} on the space of states.
The problem can be formulated in terms of identification currents~$\mathcal{J}$ corresponding to the outer automorphisms of the algebras appearing in the coset.
In presence of a common centre for the groups associated to the chiral algebras of the numerator and denominator\footnote{\label{ch3:fnote:comment-on-centers-coset}Sometimes (like in the case of interest of this appendix) the homomorphism~$H\hookrightarrow G$ is not strictly speaking an embedding: the map can be non-injective. 
In this case computing the common centre might be misleading: we have to be careful in identifying which elements of~$H$ are mapped into the identity of~$G$; this set might be different and bigger than the centre of the numerator. 
A better notion to consider here is therefore the pre-image of the centre of the numerator.}, the states in the decomposition of equation~\eqref{ch3:coset-hilb-decomposition} must transform the same~\cite{Moore:1989yh} under the action of the identification currents\footnote{The problem is more involved in presence of fixed points in the action of the identification currents on the fields of the theories~($\mathcal{J}\cdot \phi = \phi$). 
We will avoid this complication assuming that the action is fixed-point free. This is true for Grassmannian cosets.}~\cite{Moore:1989yh,Schellekens:1989uf}.
One can state this condition in terms of the so-called monodromy charge associated to the identification current~$\cJ$, defined as~
$$
Q_{\mathcal{J}}(L)=h_{\mathcal{J}}+h_L-h_{\mathcal{J}\cdot L}\mod 1\ .
$$
The sectors of the coset theory must obey the following rule:
\be\label{ch3:selection-coset-general}
Q_{\mathcal{J}}(L)\= Q_{\mathcal{J}'}(L')\ ,
\ee
where~$\mathcal{J},\mathcal{J'}$ are the identification currents of the numerator and denominator algebras.
\item
Furthermore, in presence of a common centre (with the care explained in footnote~\ref{ch3:fnote:comment-on-centers-coset} of this section), some sectors of the coset theories are identified, and we do not want to count them twice in the decomposition of equation~\eqref{ch3:coset-hilb-decomposition}. 
This fact induces \textit{identifications} on the space of states.
We can state the identification rules in terms of identification currents~$\mathcal{J},\mathcal{J}'$, as
\be
\Hilb^{(L,L')}_{\mathfrak g / \mathfrak h}\cong \Hilb^{(\mathcal{J}\cdot L,\mathcal{J}'\cdot L')}_{\mathfrak g / \mathfrak h}\ .
\ee
\end{itemize}
We become more precise in the example of the Grassmannian model that follows.
 
\subsection{Bulk spectrum Grassmannian coset}
We concentrate ourselves here on the theory $\frac{\widehat{su}(n+1)_k\oplus \widehat{so}(2n)_1}
{\widehat{su}(n)_{k+1}\oplus \widehat{u}(1)_{n(n+1)(k+n+1)}}$ defined via the group embedding
\begin{equation}\label{ch3:KS-embedding-general}
i:U(n)\longrightarrow SU(n+1)\ ,\qquad
i (h,\xi) = \begin{pmatrix}
h\xi & 0\\
0& \xi^{-n}
\end{pmatrix} \in SU (n+1) \ ,
\end{equation}
where $h\in SU (n)$ is a $n\times n$-matrix, and $\xi \in U (1)$ is a phase. This is not strictly speaking an embedding since an element~$(\xi_0^{-1}\id,\xi_0)\in U(n)$ with $\xi_0$ arbitrary phase is mapped to the identity $i (\xi_0^{-1}\id,\xi_0)=\id$ for each $\xi_0^{n}=1$.
This means that the homomorphism is surjective but non-injective (it is a so-called epimorphism).
The issue is solved once we quotient out~$\Z_n$.

\paragraph{Notations for~$su(n)$}
We use the following notations for ~$\widehat{su}(n)_k$: an affine weight~$\hat{\L}$ is expanded in the basis of fundamental weights~$\omega_i$ as follows
\be
\hat{\L}=\L_0\omega_0+\L_1\omega_1+\dots \L_{n-1}\omega_{n-1}\ ,\qquad \hat{\L}=[\L_0,\underbrace{\L_1,\dots,\L_{n-1}}_{:=\L=[\underline{\L}]}]\ ,
\ee
where~$\{\L_i\}$ are the affine Dynkin labels, with~$\L_0=k-\underline{\L}\cdot \underline{\theta}$, and $\underline{\theta}=(1,1,\dots,1)$ is the highest root of $su(n)$.
Since~$\L_0$ can be constructed in terms of the weights of~$su(n)$, integrable highest-weight representations are characterised by non-negative integral Dynkin labels of~$su(n)$ (dominant representations), where only a finite number of dominant weights is allowed at given~$k$:
\begin{equation} 
\sum_{j=1}^{n-1}\L_j\leq k\ .
\end{equation}
It is then possible to decompose the highest weight of the finite dimensional algebra in terms of partitions
\be
\L=\{l_1,l_2,\dots l_{n-1}\}\qquad l_i=\L_i+\dots +\L_{n-1}\ .
\ee
To each partition is associated a Young diagram~$\{l_1,l_2\dots l_{n-1}\}$, where the~$i$-th entry gives the number of boxes in the~$i$-th row.
We denote with~$|\Lambda|=\sum _{j=1}^{n-1} j  {\L_j}$ the number of boxes of the tableau associated with the finite dimensional representation~$\L$.
\paragraph{Representations of the Grassmannian coset}
Specialising the decomposition~\eqref{ch3:coset-hilb-decomposition} to our model
\begin{equation}\label{ch3:KS-hilb-decomposition}
{\cal H}_{\mathfrak{su}(n+1)}^{\Lambda}\otimes {\cal H}_{\mathfrak{so}(2n)}^{\S}=\sum_{\lambda,\mu}
{\cal H}^{\Lambda,\S}_{\l,\m}\otimes \left[{\cal H}^{\l}_{\mathfrak{su}(n)}\otimes {\cal H}^{\m}_{\mathfrak{u}(1)}\right]
\end{equation}
we see that in our theory the sectors are labeled by~$(\L,\S;\l,\m)$, where~$\L$ is a dominant weight of~$\widehat{su}(n+1)_k$,~$\S$ one of the four dominant weights of~$\widehat{so}(2n)_1$ ($\S=0$ singlet~$o$, $\S=2$ vector~$v$ in the NS sector, $\S=1$ spinor~$s$, $\S=-1$ co-spinor~$c$ for the R sector, details can be found in~\cite{Blumenhagen:2009zz}),~$\l$ labels dominant weights of~$\widehat{su}(n)_{k+1}$, and~$\mu$ is an integer labelling the primaries of the free boson compactified at radius~$\sqrt{n(n+1)(k+n+1)}$~(see section~\ref{ch1:sec:free-boson}).

The set~$Z$ that gives rise to the identification currents is given by the elements of the pre-image of the centre of~$SU(n+1)$ (see footnote~\ref{ch3:fnote:comment-on-centers-coset} of this section), namely the set given by elements obtained as $i^{-1}(\Z_{n+1})$. 
It turns out to be generated by $(e^{-2\pi i/n}\id,e^{2\pi i/n}e^{2\pi i/ (n+1)})$, so that it reads $Z=\left \{(\xi^{-1}\id,\xi \eta)\  \middle |\  \xi^{n}=1,\eta^{n+1}=1 \right \}$, which is isomorphic to~$\Z_{n(n+1)}$.

The simple current corresponding to~$Z$ is associated to the quadruple~\cite{Schellekens:1989uf}
$$
\cJ_{0}=(\!\!\underbrace{k\omega_{1}}_{:=\cJ_{n+1}},v; \underbrace{(k+1)\omega_{1}}_{:=\cJ_{n}},k+n+1)
$$
where~$\omega _1$ is the first fundamental weight of both~$su(n+1)$ and $su(n)$. Let us see how it acts on the representations we are interested in.
On highest-weight states of~$su(n)$ the identification simple current~$\cJ_n$ permutes the order of the Dynkin labels of the representation:
\be\label{ch3:action-simple-current-on-irrep}
\cJ_n[\underbrace{k-\underline{\L}\cdot\underline{\theta}}_{\L_0},\L_1,\dots \L_{n-1}]=[\L_{n-1},\L_0,\L_1\dots,\L_{n-2}]\ .
\ee
On~$\hat{u}(1)_k$ representations it simply shifts the~$u(1)$ label~$\mu\mapsto\mu + k+n+1$. 
The current acts on the representations of~$so(4)$ as fusion with the vector; by the known fusion rules of~$\widehat{so}(2n)_1$ the singlet gets mapped to the vector and vice versa, while the spinor goes to the co-spinor and vice versa.

\paragraph{Selection rules} 
The monodromy charges of the numerator and denominator must be equal to give non-empty coset sectors; this means
\begin{equation}
Q_{\cJ_{n+1}} (\Lambda) +Q_{v} (\Sigma)=Q_{\cJ_{n}} (\lambda)+Q_{k+n+1}
(\mu)\ ,
\end{equation} 
which reads
\begin{equation}\label{ch3:KS-selection-rules}
\frac{|\Lambda|}{n+1}-\frac{|\lambda|}{n}+\frac{\mu}{n(n+1)}+\frac{|\S|}{2}\ \in\ \Z ,
\end{equation}
where $|\S|=0$ for $o$ and $v$ representations, $|\S|=1$ for $s$ and $c$.
\paragraph{Identifications}
The simple current gives us the following identifications for the labels:
\begin{align}\label{ch3:KS-identifications}
\begin{array}{ll}
\kmalg{su}(n+1)_k\ :\ & \L_1\equiv k-\sum_{j=1}^{n}\L_j\ ,\quad \L_i\equiv \L_{i-1}\quad\text{for}\quad 2\leq i\leq n\ ,\\
\kmalg{so}(2n)_1\ :\ & \S\equiv \S+2\mod{4}\ ,\\
\kmalg{su}(n)_{k+1}\ :\ &\l_1\equiv k+1-\sum_{j=1}^{n-1}\l_j\ ,\quad \l_i\equiv \l_{i-1}\quad\text{for}\quad 2\leq i\leq n-1\ ,\\
\kmalg{u}(1)_{n(n+1)(n+k+1)}\ :\ & \mu\equiv \mu+k+n+1\mod n(n+1)(n+k+1)\ .
\end{array}
\end{align}
\paragraph{Spectrum}
The spectrum can be determined knowing the expression of the~$L_0$ eigenvalue in terms of weights in the various component theories~\cite{Kazama:1988uz}, for example from the formula~\eqref{ch3:coset-spectrum-general}.
The~$U(1)$-charge can be obtained by a careful definition of the primary fields which generate the~$N=2$~SCA (see for details~\cite{Kazama:1988uz}).
The conformal dimension $h$ and the $U(1)$ charge $Q$ of the primary state $(\Lambda,\S;\lambda,\mu)$ is
\begin{equation}\label{ch3:KS-spectrum-grassmannian}
\begin{split}
& h=\ \frac{1}{2(k+n+1)}\left[\bra \Lambda,\Lambda +2\rho_{su(n+1)}\ket
   -\bra\lambda,\lambda +2\rho_{su(n)}\ket
   -\frac{\mu^2}{n(n+1)}\right]
   +h_{\S} \mod 1\\
& Q=\ -\frac{\mu}{n+k+1}+Q_{\S}\mod 2\ ,
\end{split}
\end{equation}
where
\begin{align}
\begin{array}{lll}
h_{\S}&=(0,\frac 12,\frac n8,\frac n8) \qquad&{\rm for}\quad \S=(0,2,1,-1)\ ,\\
Q_{\S}&=(0,1,\frac n2,\frac n 2-1)\quad&{\rm for}\quad \S=(0,2,1,-1)\ .
\end{array}
\end{align}
We denoted with $\rho_{g}$ the Weyl vector of~$g$; $\bra ,\ket $ denotes the scalar product on the weight space.
\paragraph{Spectral flow}
Since these models enjoy $N=2$ superconformal symmetry, their states are mapped to each other by spectral flow.
In particular it is possible to define a simple current corresponding to the spectral flow automorphism of the algebra
\begin{equation}\label{ch3:spectral-flow-simple-current-grassmannians}
\cJ_{\text{SF}}=(0,s;0,\tfrac{n(n+1)}{2})\ .
\end{equation}
Acting once on the ground states of the Grassmannian coset, it maps the NS sector to the R sector and vice versa, and changes the charge of the state as follows:
\begin{equation}\label{ch3:action-spectral-flow-simple-current-grassmannians}
(\L,\S;\l,\m)\xrightarrow{\cJ_{\text{SF}}}(\L,s\times \S;\l,\m+\tfrac{n(n+1)}{2})\ .
\end{equation}
\paragraph{Anti-chiral primaries}
Since we use the convention that the $U(1)$ charge is negative for positive $\m$, we discuss here and in the following, for the ease of notation, anti-chiral primaries.
Chiral primaries are obtained by conjugating the charge in all the following formulas.
In a Grassmannian coset an anti-chiral primary field can be expressed through the embedding projectors of the denominator finite algebra into the numerator finite algebra (see~\cite[section 13.7]{DiFrancesco:1997nk}), as follows
\begin{equation}\label{ch3:chiral-primaries-KS-general}
\phi_a\sim (\L,0;\cP_1\L,\cP_0\L)\ ,
\end{equation}
where $\cP_1$ and $\cP_0$ are the projection matrix mapping weights of $su(n+1)$ onto weights of $su(n)$ and $u(1)$ respectively.
For the embedding chosen in equation~\eqref{ch3:KS-embedding-general}, the projection matrices act as follows on the highest weights of $su(n+1)$
\begin{equation}\label{ch3:sec:KS:projectors-for-chiral-primaries}
\cP_1[\L_1,\dots,\L_n]=[\L_1,\dots,\L_{n-1}]\ ,\qquad \cP_0[\L_1,\dots,\L_n]=\L_1+2\L_2+\dots+n\L_n\ .
\end{equation}

\subsection{$N=2$ minimal models}\label{ch3:appKS:subsec:minmod}

Minimal models are obtained setting~$n=1$ in the above formulas for the general Grassmannian Kazama-Suzuki coset~\eqref{ch3:grassmannian-coset-def}:
\begin{equation}
GC(1,k)=\frac{\widehat{su}(2)_k\oplus \widehat{so}(2)_1}
{\widehat{su}(1)_{k+1}\oplus\widehat{u}(1)_{2(k+2)}}\ .
\end{equation}
The first consideration is that the integral representations of the~$\widehat{so}(2)_1$ algebra are in one-to-one correspondence with representations of the free boson on a circle of radius 2, namely~$\widehat{u}(1)_{4}$.
Furthermore, in the denominator the algebra~$\widehat{su}(n)$ becomes trivial if~$n=1$. 
The Grassmannian coset becomes then
\begin{equation}
\text{\MM{k}}\equiv GC(1,k)=\frac{\widehat{su}(2)_k\oplus \widehat{u}(1)_4}
{\widehat{u}(1)_{2(k+2)}}\ .
\end{equation}
The Hilbert space decomposition of equation~\eqref{ch3:KS-hilb-decomposition} reads in this case
\begin{equation}
\Hilb^{l}_{\mathfrak{su}(2)_{k}}\otimes\Hilb^{s}_{\mathfrak{u}(1)_{4}}=\sum_{m}\Hilb^{l,s}_m\otimes \Hilb^m_{\mathfrak{u}(1)_{2(k+2)}}\ ,
\end{equation}
so that primary states in the coset are labeled by triples~$(l,m,s)$, where~$l$ is the dominant weight of~$su(2)_k$,~$m$ is the label of the primaries of the free boson at radius~$\sqrt{2(k+2)}$ (see section~\ref{ch1:sec:free-boson}), and~$s=0,2,1,-1$ labels~$\kmalg{u}(1)_4$ representations ($s=0$ singlet and~$s=2$ vector in the NS sector, $s=1$ spinor and~$s=-1$ co-spinor in the R sector).
The embedding homomorphism is given by
\begin{equation}
i:U(1)\longrightarrow SU(2)\ ,\qquad
i (\xi) = \begin{pmatrix}
\xi & 0\\
0& \xi^{-1}
\end{pmatrix} \in SU (2)\ ,
\end{equation}
where~$\xi$ is a phase.
The common centre between~$U(1)$ and $SU(2)$ is given by $Z=\Z_2$
\begin{equation}
Z=\ \begin{pmatrix}
e^{n\pi i} & 0\\
0& e^{n\pi i}
\end{pmatrix}\qquad \text{with}\qquad n=0,1\  ,
\end{equation}
and the associated outer automorphism of the affine Dynkin diagram acts simply by exchanging the two nodes of the diagram.
The simple current is then easily (in the triple~$(l,m,s)$ notation)
\be
\cJ_0=(k,k+2,2)\ ,
\ee
and acts on the affine highest weights as follows
\begin{align}
\begin{array}{ll}
\kmalg{su}(2)_k\ :\  & \cJ_0[\L_0,\L_1]=\cJ_0[k-\L_1,\L_1]=[\L_1,k-\L_1]\ ,\\
\kmalg{u}(1)_{4} \ :\ & \cJ_0[s]=[s+2]\ ,\\
\kmalg{u}(1)_{2(k+2)} \ :\ & \cJ_0[m]=[m+k+2]\ .
\end{array}
\end{align}

\paragraph{Selection and identification rules}

The selection rules~\eqref{ch3:KS-selection-rules} coming from the condition on the monodromy charges read
\begin{equation}
\frac{l}{2}+\frac m2 +\frac s2\in \Z \quad \Lra \quad l+m+s \quad \text{even}\ . 
\end{equation}
The identification rules can be read off from equation~\eqref{ch3:KS-identifications}:
\begin{align}\label{ch3:MM-identifications}
\begin{array}{ll}
\kmalg{su}(2)_{k}\ :\ &  l\equiv k-l\ ,\\
\kmalg{u}(1)_{4}\ :\ & s\equiv s+2\mod 4\ ,\\
\kmalg{u}(1)_{2(k+2)}\ :\ & m\equiv m+k+2\mod 2(k+2)\ .
\end{array}
\end{align}

\paragraph{Spectrum and~$U(1)$ charge}
We just read from equations~\eqref{ch3:KS-spectrum-grassmannian}
\begin{equation}\label{ch3:KS-spectrum-minmod}
\begin{split}
h=&\frac{1}{2(k+2)}\left[\frac{l}{2}(l+2)-\frac{m^2}{2}\right]
   +h_{s} \mod 1\\
Q=&-\frac{m}{k+2}+Q_{s}\mod 2\ ,
\end{split}
\end{equation}
where
\begin{eqnarray}
h_{s}&=&\left (0,\frac 12,\frac 18,\frac 18\right ) \qquad{\rm for}\quad s=(0,2,1,-1)\ ,\\
Q_{s}&=&\left (0,1,\frac 12,-\frac 1 2\right )\quad{\rm for}\quad s=(0,2,1,-1)\ .
\end{eqnarray}
If one furthermore imposes~$|l-m|\leq s$ the representations lie in the so-called standard range, i.e. there is no integer shift in equations~\eqref{ch3:KS-spectrum-minmod}. With the identifications~\eqref{ch3:MM-identifications} one can hope to be able to map any representation in the standard range.
This is in general not possible (as stressed in~\cite{Maldacena:2001ky} and in~\cite{Fredenhagen:2012rb} for instance).
\paragraph{Spectral flow}
The spectral flow simple current of equation~\eqref{ch3:spectral-flow-simple-current-grassmannians} reads simply (in $(l,m,s)$ notations)
\begin{equation}
\cJ_{\text{SF}}=(0,1,1)\ ,
\end{equation}
and acts on irreducible representation as given in equation~\eqref{ch3:action-spectral-flow-simple-current-grassmannians}
\begin{equation}
(l,m,s)\xrightarrow{\cJ_{\text{SF}}}(l,s+1,m+1)\ .
\end{equation}
\paragraph{Anti-chiral primaries}
The projection matrix $\cP_1$ annihilates the highest weight of the $su(2)$ representation, whence $\cP_0$ acts simply as the identity.
The anti-chiral primaries are given in the $(l,m,s)$ notation
\begin{equation}
\phi_a\sim (l,l,0)\ .
\end{equation}

\subsection{$SU(3)/U(2)$ Kazama-Suzuki models}

The second simplest example is the one obtained setting~$n=2$ in the definition~\eqref{ch3:grassmannian-coset-def}
\begin{equation}
GC(2,k)=\frac{\widehat{su}(3)_k\oplus \widehat{so}(4)_1}
{\widehat{su}(2)_{k+1}\oplus \widehat{u}(1)_{6(k+3)}}\ .
\end{equation}
The Hilbert space decomposition~\eqref{ch3:KS-hilb-decomposition} reads in this case
\begin{equation}
{\cal H}_{\mathfrak{su}(3)}^{(\L_1,\L_2)}\otimes {\cal H}_{\mathfrak{so}(4)}^{\S}=\sum_{\lambda,\mu}
{\cal H}^{(\L_1,\L_2),\S}_{\l,\m}\otimes \left[{\cal H}^{\l}_{\mathfrak{su}(2)}\otimes {\cal H}^{\m}_{\mathfrak{u}(1)}\right]
\end{equation}
so that primary states in the coset are labeled by quadruples~$((\L_1,\L_2),\S;\l,\m)$, where~$(\L_1,\L_2)$ are dominant weights of~$\widehat{su}(3)_k$,~$\S$ labels the spinorial factor ($\S=0$ singlet and~$\S=2$ vector in the NS sector, $\S=1$ spinor and~$\S=-1$ co-spinor in the R sector), $\l$ is the dominant weight of~$\widehat{su}(2)_{k+1}$ and finally~$\m$ is the label of the primaries of the free boson at radius~$\sqrt{6(k+3)}$.
The homomorphism mapping the denominator inside the numerator is given by
\begin{equation}
i:U(2)\longrightarrow SU(3)\ ,\qquad
i (h,\xi) = \begin{pmatrix}
h\xi & 0\\
0& \xi^{-2}
\end{pmatrix} \in SU (3)\ ,
\end{equation}
where~$h\in SU(2)$ is a $2\times 2$ matrix.
This is not an embedding since~$(\xi_0^{-1}\id,\xi_0)\in U(2)$ is mapped to the identity for $\xi_0^2=1$, which admits two solutions, namely~$\xi_0=1,-1$.
The pre-image of the centre of~$SU(3)$ is given by $Z=\Z_6$, which reads in our notations
\begin{equation}
Z=\ \begin{pmatrix}
e^{-n\pi i}\id_{2\times 2} & 0\\
0& e^{2\pi i\left(\frac {n'}{3}+\frac n2\right)}
\end{pmatrix}\qquad \text{with}\quad n=0,1\ ,\ n'=0,1,2\  ,
\end{equation}
and the associated outer automorphism of the affine Dynkin diagram acts by permuting the nodes of the affine diagrams of the algebras appearing.
The simple current is (written as a quadruple~$(\L_1,\L_2),\S;\l,\m$)
\be 
\cJ=((k,0),v;k+1,k+3)
\ee
and acts on the affine highest weights as follows
\begin{align}
\begin{array}{ll}
\kmalg{su}(3)_k\ :\  & \cJ_0[\L_0,\L_1,\L_2]=\cJ_0[k-\L_1-\L_2,\L_1,\L_2]=[\L_2,k-\L_1-\L_2,\L_1]\ ,\\
\kmalg{so}(4)_{1} \ :\ & \cJ_0[\S]=[\S+2]\ ,\\
\kmalg{su}(2)_{k+1}\ :\ & \cJ_0[\l_0,\l]=\cJ_0[k+1-\l,\l]=[\l,k+1-\l]\ ,\\
\kmalg{u}(1)_{6(k+3)} \ :\ & \cJ_0[\m]=[\m+k+3]\ .
\end{array}
\end{align}

\paragraph{Selection and identification rules}

The selection rules~\eqref{ch3:KS-selection-rules} coming from the condition on the monodromy charges read then
\begin{equation}
\frac{\L_1+2\L_2}{3}-\frac{\l}{2}+\frac{\m}{6}+\frac{\S}{2}\ \in\  \Z\ .
\end{equation}
The identification rules can be read off from equation~\eqref{ch3:KS-identifications}:
\begin{align}
\begin{array}{ll}
\kmalg{su}(3)_k\ :\  & (\L_1,\L_2)\equiv (k-\L_1-\L_2,\L_1)\ ,\\
\kmalg{so}(4)_{1} \ :\ & \S\equiv \S+2\mod 4\ ,\\
\kmalg{su}(2)_{k+1}\ :\ & \l\equiv k+1-\l \ ,\\
\kmalg{u}(1)_{6(k+3)} \ :\ & \m \equiv \m +k+3\mod{6(k+3)}\ .
\end{array}
\end{align}

\paragraph{Spectrum and~$U(1)$ charge}
We just read from equations~\eqref{ch3:KS-spectrum-grassmannian}
\begin{equation}
\begin{split}
h=&\frac{1}{2(k+3)}\left[\frac 23\left(\L_1^2+\L_2^2+\L_1\L_2\right)+2\left(\L_1+\L_2\right)-\frac{\l}{2}(\l+2)-\frac{\m^2}{6}\right]
   +h_{\S} \mod 1\\
Q=&-\frac{\m}{k+3}+Q_{\S}\mod 2\ ,
\end{split}
\end{equation}
where
\begin{align}
\begin{array}{llll}
h_{\S}&=\left (0,\frac 12,\frac 14,\frac 14\right ) \qquad&{\rm for}&\quad \S=(0,2,1,-1)\ ,\\[4mm]
Q_{\S}&=\left (0,1,1,0\right )\quad&{\rm for}&\quad \S=(0,2,1,-1)\ .
\end{array}
\end{align}
\paragraph{Spectral flow}
The simple current corresponding to the spectral flow automorphism reads in this case (again from equation~\eqref{ch3:spectral-flow-simple-current-grassmannians})
\begin{equation}
\cJ=(0,s;0,3)\ ,
\end{equation}
and its action on the irreducible representation is (from equation~\eqref{ch3:action-spectral-flow-simple-current-grassmannians})
\begin{equation}
((\L_1,\L_2),\S;\l,\m)\xrightarrow{\cJ_{\text{SF}}}((\L_1,\L_2),s\times \S;\l,\m+3)\ .
\end{equation}
\paragraph{Anti-chiral primaries}
It is very easy to read the form of anti-chiral primary states off equations~\eqref{ch3:chiral-primaries-KS-general} and~\eqref{ch3:sec:KS:projectors-for-chiral-primaries}:
\begin{equation}
\phi_a\sim ((\L_1,\L_2),0;\L_1,\L_1+2\L_2)\ .
\end{equation}

\end{subappendices}

\chapter{Limit of~$N=2$ minimal models: geometry}\label{ch:geometry}

In this chapter we start looking at the large level limits of~$N=2$ superconformal minimal models, by analysing the behaviour of the geometry in the limit.
This is the best approach to get a feeling of what happens in the limit, although we will need to study several other~CFT structures to substantiate the suggestions coming from the geometric analysis.
This chapter is intended as a heuristic appetiser to the more technically detailed follow-ups.

\section{Two different limit theories}\label{ch4:sec:intro}

The examples of limit theories that we have sketched in chapter~\ref{ch:limit-theories} all share the characteristic of being uniquely defined (with the case of Virasoro minimal models being an exception at first sight, as noticed in~\ref{ch1:sec:RW}, problem solved with the continuous orbifold interpretation as explained in section~\ref{ch2:sec:RW-as-contorbi}).
The case at hand, the large~$k$ limit of~$N=2$ superconformal unitary minimal models, for which the central charge approaches~$c=3$, is more subtle: the spectrum of~\MM{k} is characterised by two independent quantum numbers, the conformal weight and the~$U(1)$ charge, we have then the freedom of choosing how to rescale the charge while taking the limit.
It turns out that it is possible to define consistently two different limit theories: 
\begin{itemize}
\item[1]
One can keep the charge~$Q$ fixed in the limit and define averaged fields (in the spirit described in chapter~\ref{ch:limit-theories}) 
\begin{equation}\label{ch4:contorbi-averaged-fields}
^{(1)}\Phi^{(k)}_{\!f} = \sum_{i} f (h_{i},Q_{i})\, \phi^{(k)}_{i} \ ,
\end{equation}
corresponding to a certain test function $f(h,Q)$, which makes the two labels~$h,Q$ finite in the limit. 
This leads to the limit theory constructed in~\cite{Fredenhagen:2012rb} with a continuous spectrum of charged primaries: we suggest that it corresponds to an~$N=2$ supersymmetric continuous orbifold~$\C/U(1)$~\cite{Fredenhagen:2012bw}; the details of this correspondence will be explored in chapter~\ref{ch:new-theory} and~\ref{ch:our-contorbi}.
\item[2]
One can differently rescale the charges and define averaged fields
\begin{equation}\label{ch4:free-averaged-fields}
^{(2)}\Phi^{(k)}_{\!f} = \sum_{i} f(h_{i},Q_{i}(k+2)) \,\phi^{(k)}_{i} \ .
\end{equation}
Because of the new scaling, this theory has chargeless primary fields, and we show in chapter~\ref{ch:free-limit}, based on the results of~\cite{Fredenhagen:2012bw}, that this limit is equivalent to a free theory of two uncompactified bosons and two fermions. 
The discrete quantum label that comes out from the rescaled charge of the primaries can be interpreted as the eigenvalue of the rotation operator on the plane spanned by the two bosons. 
In defining the limit theory we must make use of the freedom to redefine the ingredient fields $\phi^{(k)}_{i}$ of $^{(2)}\Phi^{(k)}_{\!f}$ by individual phases, differently from the conventions used for the definition of $^{(1)}\Phi^{(k)}_{\!f}$ in the other limit construction analysed in~\cite{Fredenhagen:2012rb}.  
\end{itemize}

\section{Geometric interpretation of the limits: bulk}\label{ch4:sec:bulk-geometry}

We look now at what happens to the sigma-model description of~$N=2$ minimal models when~$k\to\infty$.
The appearance of the two different limit theories mentioned before can be understood from a geometric point of view.
The sigma-model interpretation of the bulk minimal models has been given in section~\ref{ch3:sec:geometry}; we quote here the results for reference.
The target space geometry reads
 \begin{equation}\label{ch4:bell-geometry}
 \text{bell:}\ \ 
 \left\{
 \begin{array}{rl}
 ds^2=&\ \frac{k+2}{1-\rho^2}\left(d\rho^2+\rho^2 d\varphi^2\right)\\
 e^{\Phi(\rho,\varphi)}=&\ g_s(0)(1-\rho^2)^{-1/2}
 \end{array}
 \right.
 \qquad 0\leq\rho< 1\ ,\ 0\leq\varphi\leq 2\pi\ ,
 \end{equation}
 and the spectrum and wave functions for the bulk NS~primaries are given by
 \begin{equation}\label{ch4:laplacian-parafermion}
 \left( -\frac{1}{2}\nabla^{2} + (\nabla \Phi)\cdot \nabla\right)\psi_{l,m}
 (\rho ,\varphi) = 2\underbrace{\frac{l(l+2)-m^2}{4(k+2)}}_{h_{l,m}} \,\psi_{l,m} (\rho ,\varphi) 
 \end{equation}
 and
 \begin{equation}\label{ch4:minmod-wave-function}
 \psi_{l,m} (\rho ,\varphi) = \rho^{|m|} e^{im\varphi} {}_{2}F_{1}
 \left(\tfrac{|m|+l}{2}+1,\tfrac{|m|-l}{2};|m|+1;\rho^{2}\right)\ .
 \end{equation}
The target space is topologically a disc with infinite circumference, but with finite radius~$\frac{\pi}{2}\sqrt{k+2}$, which goes to infinity in the limit $k\to\infty$.

\smallskip

\subsection{Bulk geometric limit: free theory}\label{ch4:sec:bulk-geometry:sub:free-limit}

We take~$k\to\infty$ now: If we concentrate the attention on the region around the centre the metric approaches the flat metric on the plane. 
To see this, we can rescale the radius as
\begin{equation}
\rho '=\sqrt{k+2}\,\rho \ ,
\end{equation}
such that the metric reads
\begin{equation}
ds^{2} = \frac{1}{1-\rho '^{2}/ (k+2)}\left(d\rho'^{2}+\rho '^{2}d\varphi
^{2}\right) \ .
\end{equation}
Keeping $\rho '$ fixed while taking the limit $k\to \infty$ leads to the flat metric on the plane.

Taking the geometric limit to the flat plane, the wavefunctions $\psi_{l,m}$ should approach the eigenfunctions of the flat Laplacian; these are given in radial coordinates by
\begin{equation}\label{ch4:flat-wave-function}
\psi^{\text{flat}}_{p,m} (\rho ',\varphi) = e^{im\varphi} J_{|m|} (p\rho ')\ , 
\end{equation}
where $J$ is a Bessel function of the first kind,~$p$ is the radial momentum and~$m$ is the eigenvalue of the angular momentum in the direction orthogonal to the plane.
They satisfy
\begin{equation}
-\frac{1}{2}\,\nabla_{\!\text{flat}}^{2} \psi^{\text{flat}}_{p,m} (\rho ',\varphi) =
\frac{p^{2}}{2}\,\psi^{\text{flat}}_{p,m} (\rho ',\varphi) \ .
\end{equation}
Let us see how it works: 
First of all, in order to match the angular dependence of $\psi_{l,m}$ of equation~\eqref{ch4:minmod-wave-function} with the one of $\psi^{\text{flat}}_{p,m}$ of equation~\eqref{ch4:flat-wave-function}, one has to keep the label $m$ fixed in the limit. 
Then, for the eigenvalue $h_{l,m}$ to approach $h_{p}=\frac{p^{2}}{4}$, the label $l$ has to grow with the square root of $k$, namely $l\approx p \sqrt{k+2}$. 
Consequently, the wavefunctions $\psi_{l,m}$ behave in the limit
\begin{align}
(k+2)^{|m|/2}\psi_{l,m} &= \rho '^{|m|}e^{im\varphi}\, {}_{2}F_{1}
(\tfrac{|m|+l}{2}+1,\tfrac{|m|-l}{2};|m|+1;\tfrac{\rho'^{2}}{k+2})\\
& = e^{im\varphi} \sum_{n=0}^{(l -|m|)/2} \frac{\left(\frac{l+|m|}{2}
\right)_{n} \left(\frac{-l+|m|}{2} \right)_{n}}{n!(|m|+1)_{n}}
(\rho')^{2n+|m|} (k+2)^{-n} \\
& = e^{im\varphi} \sum_{n=0}^{(l-|m|)/2} \frac{(-1)^{n}}{n!
(|m|+1)_{n}} \left(\frac{l-|m|-2n+2}{2\sqrt{k+2}} \right) \dotsb
\left(\frac{l-|m|}{2\sqrt{k+2}} \right)\nonumber\\
& \qquad \times 
\left(\frac{l+|m|}{2\sqrt{k+2}} \right) \dotsb
\left(\frac{l+|m|+2n-2}{2\sqrt{k+2}} \right) (\rho')^{2n+|m|} \\
& \sim  e^{im\varphi} \sum_{n=0}^{\infty} 
\frac{(-1)^{n}p^{2n}2^{-2n}}{n!(|m|+1)_{n}} (\rho')^{2n+|m|}\\
& \sim  e^{im\varphi} \,J_{|m|} (p\rho') \ .
\end{align}
We have shown that up to an overall normalisation the wavefunctions $\psi_{l,m}$ approach the wavefunctions of the free theory.

In conclusion, this preliminary analysis suggests that the theory at hand is a superconformal free theory of two bosons and two real fermions: the label~$m$ of the minimal model has to be kept finite in the limit, and it can be interpreted as the eigenvalue of~$SO(2)$ rotations; the label~$l$ scales with~$\sqrt{k+2}$.
Detailed~CFT computations confirm this expectation, as we show in chapter~\ref{ch:free-limit}.

\subsection{Bulk geometric limit: continuous orbifold}\label{ch4:sec:bulk-lim:subs:contorbi}

In contrast with what we have just explained, we can rescale the charge in a different (and more natural) way: we can choose to hold fixed an integer label~$n=\frac12(l-|m|)$ while scaling~$m\sim k$ to keep the~$U(1)$ charge finite in taking the~$k\to\infty$ limit~\cite{Fredenhagen:2012rb}. 
With this scaling the~$U(1)$ charge becomes real and the conformal dimension of the NS~primaries reads
\begin{equation}
h_{Q,n}=|Q|\left(n+\frac12\right)\ ,\qquad -1< Q< 1\ ,\ \ \ n\in\Z\ .
\end{equation}
\smallskip

Apart from the angular part, the wavefunction $\psi_{l,m}$ in equation~\eqref{ch4:minmod-wave-function} is a polynomial in $\rho$ containing $n+1$ terms with powers ranging from $\rho^{|m|}$ to $\rho^{|m|+2n}$. 
If $|m|$ is large, the wavefunctions get localised around $\rho =1$, where the metric and the dilaton diverge and the sigma-model description becomes singular.
However, as already observed in section~\ref{ch3:sec:geometry}, under T-duality the minimal
model is mapped to its own $\mathbb{Z}_{k+2}$ orbifold
\begin{equation}\label{ch4:orbifold-identification}
\begin{split}
d\tilde{s}^{2} & =
\frac{k+2}{1-\tilde{\rho}^{2}}\left(d\tilde{\rho}^{2}+\tilde{\rho}^{2}d\tilde{\varphi}^{2}\right)\\
e^{\tilde{\Phi} -\Phi_{0}} & = \frac{1}{\sqrt{k+2}}
\frac{1}{\sqrt{1-\tilde{\rho}^{2}}} \\
\tilde{\varphi} & \equiv  \tilde{\varphi} +\frac{2\pi}{k+2} \ .
\end{split}
\end{equation}
The singular region around $\rho =1$ is mapped to the region close to the conical singularity of the~$\Z_{k+2}$ orbifold at $\tilde{\rho}=0$. 
This suggests that taking the limit of minimal models in this way (which corresponds to the one described in~\cite{Fredenhagen:2012rb}) corresponds to taking the limit in the orbifolded model by focussing on the region around $\tilde{\rho}=0$.
In the T-dual picture we can again rescale the radius as $\tilde{\rho}'=\sqrt{k+2}\tilde{\rho}$ and keep $\tilde{\rho}'$ fixed in the limit. 
The metric $d\tilde{s}^{2}$ in these new coordinates approaches again the flat metric on the plane if we keep~$\tilde{\rho}'$ finite, but according to~\eqref{ch4:orbifold-identification} all angles have to be identified. 
This suggests that the resulting limit theory is the theory on a flat plane $\mathbb{R}^{2}$ orbifolded by the rotation group $SO (2)$.

In conclusion, this preliminary analysis suggests that the theory obtained in this way is an~$SO(2)$ continuous orbifold of the flat plane: the label~$m$ of the minimal model scales with~$k$ in the limit, while the difference~$l-|m|$ is kept finite.
Detailed CFT computations confirm this expectation, as will be explained in chapter~\ref{ch:our-contorbi}.

\section{Geometric interpretation: boundary}\label{ch4:sec:limit-geometry-BC}

Let us sketch what happens to the geometry of D-branes when~$k\to\infty$.
\subsection{Branes in the free theory limit}\label{ch4:sec:branes:subs:free-limit}
\paragraph{A-type}
D-branes of the A-type correspond to straight lines (as explained in \ref{ch3:sec:BC}): If they are distant from the origin, then in the free theory limit, since we concentrate around the centre, we are not able to see them.
However we can rescale the boundary labels in such a way that the brane stays close enough to the centre; in this way they are not ``swept away'' when~$k\to\infty$, and we can reproduce the expected one-dimensional branes of a free theory of two bosons and two fermions.
In formulas, we recall from subsection~\ref{ch3:sec:geometry:subsec:branes} that the distance from the origin of an A-brane is given in terms of the boundary data as~$\rho_0=\cos \frac{\pi(L+1)}{k+2}$. 
The angle between the normal to the brane and the $x$-axis is given by~$\varphi_0=\frac{\pi M}{k+2}$.
Now, rescaling the radial coordinate as in the free theory limit, we have
\be
\rho_0'=\sqrt{k+2}\cos \frac{\pi(L+1)}{k+2}\ ,
\ee
and this tends to a constant~$R$ if
\be 
L=\frac12(k+2)-\frac{R}{\pi}\sqrt{k+2}+\order{1}\ .
\ee
Analogously, we can scale the boundary label~$M$ in such a way that the angle~$\varphi_0$ stays constant in the limit
\be
M=\frac{k+2}{\pi}\varphi_0+\order{1}\ .
\ee
In conclusion one-dimensional branes survive in the free theory limit, and are labeled by~$R\in \R$ and an angle~$\varphi_0$ giving the position on the plane of flat one-dimensional D-branes.
\paragraph{B-type}
B-type boundary conditions correspond in the geometric picture to two-dimensional discs of radius~$\rho_1$ depending on the boundary label~$L$ as~$\rho_1=\sin \frac{\pi (L+1)}{k+2}$.
In the free theory limit we expect that we can define two limits: one for which the disc shrinks to a point, and the resulting brane will be zero-dimensional, and one for which the disc covers the whole plane, corresponding to a space-filling brane in the limit.
This expectation is confirmed, and we postpone the details to chapter~\ref{ch:free-limit}.

\subsection{Branes in the continuous orbifold limit}\label{ch4:sec:branes:subs:contorbi}

The continuous orbifold limit is based on a T-duality transformation, that allows us to explore the region close to the curvature singularity on the boundary of the disc.
T-duality is realised at the level of the algebra as the mirror automorphism~$\O_M$ defined in equation~\eqref{ch3:mirror-automorphism}.
As already mentioned in the preamble of section~\ref{ch3:sec:BC}, the mirror automorphism exchanges the type of D-branes, since it flips the sign of the~$U(1)$ current's generators and exchange the two holomorphic supercurrents.
Moreover it exchanges the sign of the anti-holomorphic sectors' charge and spin alignment, in our conventions chosen in equation~\eqref{ch3:partition-anti-diagonal}.

We have to be especially careful then: B-type boundary conditions couple only to chargeless fields in a diagonal model, but also to charged fields in an anti-diagonal model.
Reversely, A-type boundary conditions couple in the T-dual picture to chargeless fields.

\paragraph{A-type}
B-type branes are mapped by T-duality to one-dimensional straight lines. 
They are not present in the limit, since the limit theory possesses only one chargeless field compared to a continuum of charged ones, so that in all amplitudes the contribution coming from the coupling to these branes is negligible.
In the continuous orbifold picture they correspond to bulk branes, which are outnumbered by fractional branes since there is a continuum of twisted sectors.

\paragraph{B-type}
A-type one-dimensional D-branes are mapped by T-duality to B-type two-dimensional discs, and they couple to charged bulk fields. 
If the labels~$L,M$ in equation~\eqref{ch3:partition-anti-diagonal} do not grow fast enough with~$k$, then the branes are very short and close to the boundary, and get mapped to very small discs close to the centre of the T-bell.
They can be recognised to be the point-like fractional branes of the continuous orbifold, as confirmed by the comparison of subsection~\ref{ch7:sec:branes:subs:point-like}.
They come in discrete families.

If the label~$L$ scales with~$k$, then the one-dimensional branes are close to a diameter of the bell.
They get mapped by T-duality to discs covering the whole T-bell.

\chapter{The free theory limit}
\label{ch:free-limit}

\smallskip
\noindent
The discussion in chapter~\ref{ch:geometry} suggests us to compare the limit of
averaged fields~$^{(2)}\Phi^{(k)}_{\!f}$ of~$N=2$ minimal models with the free theory of two uncompactified real bosons and two free real fermions.
This limit accounts to scale the label~$l$ with~$\sqrt k$, and to keep~$m$ finite, so that the primary fields of the new theory become chargeless.
In particular we find the following expression for the limit of the spectrum and of the~$U(1)$ charge (we restrict to the standard range, and we set~$s=0$):
\begin{align}
\begin{array}{c}
\lim\limits_{k\to\infty}\\[0.5cm]
\text{FREE THEORY}
\end{array}
\quad
\boxed{
\begin{array}{lll}
h_{l,m,0}=\frac{l(l+2)-m^2}{4(k+2)}\quad &\xrightarrow{l=p\sqrt{k+2}}\qquad & h_{p}=\frac{p^2}{4}\\
Q_{m,0}=-\frac{m}{k+2}\quad&\xrightarrow{m\ \text{finite}}\qquad & Q=0\\
-l\leq m\leq l\ ,\ 0\leq l\leq k\quad&\xrightarrow{\phantom{m\ \text{finite}}}\qquad & m\in\Z\ ,\ p\in\R_{\geq 0}
\end{array}
}
\end{align}

\section{Partition function}\label{ch5:sec:partition-fct}
In this section we want to show how to reproduce the torus partition function of the $N=2$ supersymmetric free theory of two uncompactified free bosons and two free fermions~\cite{Gaberdiel:2004nv} as the limit of the partition function of~$N=2$ minimal models.
We recall here the definition of the fully supersymmetric partition function in the NS sector, before GSO projection, given in section~\ref{app:characters:sec:GSO}:
\begin{equation}
\mathcal{P}^{\text{NS}}_{k}= \Tr_{\cH_{k}^{\text{NS}}}
\left( q^{L_{0}-\frac{c}{24}}z^{J_{0}}\,
\bar{q}^{\bar{L}_{0}-\frac{c}{24}}\bar{z}^{\bar{J}_{0}} \right)\ ,
\qquad
\mathcal{H}^{\text{NS}}_{k} = \bigoplus_{0\leq l\leq k}\bigoplus_{\substack{|m|\leq l\\ l+m\ \text{even}}}
\mathcal{H}_{l,m}\otimes \mathcal{H}_{l,m} \ .
\end{equation}
In the $k\to\infty$ limit the partition function diverges: there are infinitely many states approaching the same conformal weight and charge. 
This can be seen by focussing on the Neveu-Schwarz ground states (the leading contribution in equation~\eqref{minmod-NS-character})
\begin{equation}
\mathcal{P}^{\text{NS,g.s.}}_{k} =  
\sum_{\substack{|m|\leq l \leq k \\ l+m\ \text{even}}}
(q\bar{q})^{\frac{(l+1)^{2}-m^{2}}{4 (k+2)}}
(z\bar{z})^{-\frac{m}{k+2}}\ .
\end{equation}
We introduce the positive integer variable $n=\frac{1}{2} (l-|m|)$, and explicitely perform the summation over $m$:
\begin{align}
\mathcal{P}^{\text{NS,g.s.}}_{k}  & = \sum_{n=0}^{\lfloor
\frac{k}{2}\rfloor}  (q\bar{q})^{\frac{(2n+1)^{2}}{4 (k+2)}}
\left(\sum_{m=0}^{k-2n} + \sum_{m=-k+2n}^{-1} \right) 
(q\bar{q})^{\frac{(2n+1)|m|}{2(k+2)}}(z\bar{z})^{-\frac{m}{k+2}}\\
&=  \sum_{n=0}^{\lfloor
\frac{k}{2}\rfloor}(q\bar{q})^{\frac{(2n+1)^{2}}{4 (k+2)}}
\bigg[\frac{1-(q\bar{q})^{\frac{2n+1}{2 (k+2)}
(k-2n+1)}(z\bar{z})^{-\frac{k-2n+1}{k+2}}}{1
-(q\bar{q})^{\frac{2n+1}{2 (k+2)}}(z\bar{z})^{-\frac{1}{k+2}}}\nonumber\\
&\qquad \qquad \qquad \qquad  
+\frac{1-(q\bar{q})^{\frac{2n+1}{2 (k+2)}
(k-2n+1)}(z\bar{z})^{\frac{k-2n+1}{k+2}}}{1
-(q\bar{q})^{\frac{2n+1}{2 (k+2)}}(z\bar{z})^{\frac{1}{k+2}}} 
\bigg] \ ,
\label{partfuncdiverges}
\end{align}
where we have written $\lfloor x \rfloor$ for the greatest integer smaller or equal $x$. 
Now we cut off the summation over $n$ by $n\leq \Lambda\sqrt{k+2}$ with $0<\Lambda <1$. 
In this range, the denominators in~\eqref{partfuncdiverges} go to zero, and the summands are divergent as $\frac{k+2}{n+\dotsb}$. 
The sum over $n$ gives a logarithmic divergence, such that the partition function scales as $(k+2)\log(k+2)$. 
This infinity signals an infinite degeneracy of states. 
Part of it can be eliminated by introducing additional quantum numbers that lift the degeneracy. 
This is nevertheless in general not enough: there can be a divergence due to the emergence of a continuous spectrum, in which case we should regularise the partition function by appropriately rescaling the density of states.
This is the same kind of phenomenon we have come across while describing the~$R\to\infty$ limit of the free boson on a circle of section~\ref{ch1:sec:free-boson-limit}: when $R\to\infty$ the partition function diverges as the volume~$R$. 
The considerations about the rescaling of the partition function in the continuum limit hold here as well.

The divergence associated to the appearance of infinitely many chargeless fields can be cured by keeping track of the quantum label~$m$ in the limit.
As mentioned in chapter~\ref{ch:geometry}, in the theory of a free uncompactified boson expressed in radial coordinates, $m$ corresponds to the eigenvalue of the angular momentum operator $M$; it is sensible then to insert a sort of ``chemical potential'' operator $e^{i\varphi M}$ in the sum over states: in this way the partition function becomes a formal power series in $e^{i\varphi}$ and $e^{-i\varphi}$, and the coefficient of $e^{im\varphi}$ gives the contribution of states with a definite angular momentum eigenvalue~$m$. 
As explained in details in section~\ref{ch3:sec:geometry}, in the bell geometry there is an analogous classical rotational symmetry (real shifts of the angle~$\varphi$).
Quantum corrections break this~$U(1)$ symmetry down to~$\Z_{k+2}$.
The discrete rotation by an angle $\delta \varphi=2\pi i\frac{r}{k+2}$ ($r$ integer) is realised in this case by the operator $g^{r}$, where $g$ acts on $\Hilb_{l,m}\otimes\Hilb_{l,m}$ by multiplication with the phase $e^{2\pi i\frac{m}{k+2}}$. 
To mimic the insertion of $e^{i\varphi M}$ in the free field theory, we introduce the operator $g^{\lfloor\frac{\varphi}{2\pi} (k+2)\rfloor}$ in the partition function, such that states with a given $m$ will get the phase $e^{im\varphi}$ for large~$k$.

The regularised partition function therefore becomes
\begin{equation}
\cP^{\text{NS}}_{k,(\varphi)}= \left|\frac{\vartheta_3(\tau,\nu)}{\eta^3(\tau)}\right|^2 \sum_{m=-k}^{k} e^{2\pi i\frac{m}{k+2}\lfloor \frac{\varphi}{2\pi} (k+2)\rfloor} (z\bar{z})^{-\frac{m}{k+2}} \sum_{\substack{l=|m|\\ l+m\ \text{even}}}^{k}(q\bar{q})^{\frac{(l+1)^{2}-m^{2}}{4 (k+2)}} \big|\Gamma_{lm}^{(k)} (\tau,\nu )\big|^{2} \ ,
\end{equation}
where we used the minimal model characters given in~\eqref{minmod-NS-character}. 
$\Gamma_{lm}^{(k)}$ is defined in~equation~\eqref{def-Gamma}, it is of the form
\begin{equation}
\Gamma_{lm}^{(k)}=1+ (\text{subtractions from singular vectors}) \ ,
\end{equation}
and its large $k$ behaviour is analysed in~\eqref{limit-Gamma}. 
The contribution of a fixed $m$ is then
\begin{equation}
\cP^{\text{NS}}_{k,(\varphi,m)}\approx \left|\frac{\vartheta_3(\tau
,\nu)}{\eta^3(\tau)}\right|^2 e^{im\varphi} 
\frac{\sqrt{k+2}}{2} \int dp\, (q\bar{q})^{p^{2}/4} \ ,
\end{equation}
where we have made use of the Euler-MacLaurin sum formula\footnote{We use here the $B_1=+\frac12$ convention on Bernoulli numbers, and $f^{(0)}(x)\equiv f(x)$} (see e.g.~\cite[appendix D]{Andrews:book})
\begin{equation}\label{Euler-Maclaurin-formula}
\sum_{a<k\leq b}f(k)=\int_{a}^bdx\ f(x)+\sum_{j=1}^{M}\frac{B_j}{j!}\left(f^{(j-1)}(b)-f^{(j-1)}(a)\right)+\text{Rem}_M[f]
\end{equation}
 to convert the sum over $l$ into an integral over $p=l/\sqrt{k+2}$. 
As explained in section~\ref{app:characters:sec:char-limits}, for fixed $m$ and large $l$ all singular vectors disappear and $\Gamma_{lm}^{(k)}\to 1$. 
To get the physical partition function, the trace over the GSO projected Hilbert space, we have to combine $\cP$ evaluated at $\nu$ and at 
$\nu+\frac{\pi}{2}$ (as explained in section~\ref{app:characters:sec:GSO}), and we find
\begin{multline}
\frac{1}{\sqrt{k+2}} \left[\cP^{\text{NS}}_{k,(\varphi)} (\tau ,\nu) +
\cP^{\text{NS}}_{k,(\varphi)} \left(\tau ,\nu +\tfrac{\pi}{2}\right) \right] \\
\to 
\frac{1}{2}\left(\left|\frac{\vartheta_3(\tau,\nu)}{\eta^3(\tau)}\right|^2
+\left|\frac{\vartheta_4(\tau,\nu)}{\eta^3(\tau)}\right|^2
\right) \sum_{m\in\mathbb{Z}} e^{im\varphi} \int_{0}^{\infty}dp\,
(q\bar{q})^{p^{2}/4} \ .
\end{multline}
This is exactly the NS part  of the partition function of two free uncompactified bosons and two fermions (see e.g.~\cite{Blumenhagen:2009zz}[chapter 12.2]), weighted by the rotation operator $e^{iM\varphi}$. 
The~$(k+2)^{-\frac 12}$ rescaling is due to the infinite degeneracy of states at a fixed~$m$, and it can be understood by looking at the density of the smeared fields, as explained in the following section, and as explained in the free boson example of section~\ref{ch1:sec:free-boson-limit}: in the interval $[p,p+\Delta p]$ there are $\frac{\sqrt{k+2}}{2}\Delta p$ ground states contributing to the partition function (see~\eqref{densofstates}). 
The rescaling therefore corresponds to adjusting the density of states to $1$ per unit interval $\Delta p$. 

\section{Fields and two-point function}\label{ch5:sec:fields-and-two-point-fct}

\paragraph{Free theory}

The primary fields $\Phi^{\text{free}}_{\mathbf{p}}$ in the NS sector are labelled by a complex momentum~$\mathbf{p}$, they have conformal weight $h=\frac{|\mathbf{p}|^{2}}{4}$ and $U (1)$ charge $Q=0$. 
We can define a new ``radial'' basis,
\begin{equation}\label{ch5:def-radial-basis-free-fields}
\Phi^{\text{free}}_{p,m} = \sqrt{\frac{p}{2\pi}}\int d\varphi\ \Phi^{\text{free}}_{pe^{i\varphi}}\, e^{im\varphi} \ ,
\end{equation}
where $p=|\mathbf p|$, and the factor in front of the integral reproduces the canonical normalisation of the two-point function:
\begin{equation}\label{ch5:two-pt-fct-free-fields}
\langle \Phi^{\text{free}}_{p_{1},m_{1}}(z_{1},\bar{z}_{1})\Phi^{\text{free}}_{p_{2},m_{2}} (z_{2},\bar{z}_{2})\rangle= (-1)^{m_{1}}\delta (p_{1}-p_{2}) \delta_{m_{1}+m_{2},0}\frac{1}{|z_{1}-z_{2}|^{p_{1}^{2}}} \ .
\end{equation}
We expect that the two-point function of the fields in the limit theory approaches equation~\eqref{ch5:two-pt-fct-free-fields}.
\paragraph{Limit theory}

We define the primary fields $\Phi_{p,m}$ of the limit theory, which arise from averaged fields $^{(2)}\Phi^{(k)}_{\!f}$ by choosing appropriate averaging functions $f$ (following the strategy explained in chapter~\ref{ch:limit-theories}). 
In the NS sector we introduce the averaged fields for the $k^{\text{th}}$ minimal model as
\begin{equation}
\Phi^{\epsilon ,k}_{p,m} = \frac{1}{\left|N (p,\epsilon,k,m)\right|} \sum_{l\in N (p,\epsilon ,k,m)} \phi_{l,m} \ ,
\end{equation}
where we denote NS primaries with~$\phi_{l,m}\equiv \phi_{l,m,s=0}$.
The set~$N(p,\e,k,m)$ is defined as follows
\begin{equation}
N (p,\epsilon ,k,m) = \left\{ l: l+m\ \text{even}\, ,\ p-\frac{\epsilon}{2} <\frac{l}{\sqrt{k+2}} < p + \frac{\epsilon}{2}  \right\} \ .
\end{equation}
and contains all allowed labels $l$ that are close to $p\sqrt{k+2}$.
For large $k$ this corresponds to using the (discontinuous) averaging function
$$
f_{p,\epsilon} (h)=\left\{\!\begin{array}{ll}
\!1/\epsilon & \text{for}\ |p-2\sqrt{h}|<\epsilon /2\\
\!0 & \text{else}
\end{array} \right.
$$
and a rescaling of the fields by $2/\sqrt{k+2}$.
For large $k$ the number of elements in $N (p,\epsilon ,k,m)$ (assuming $p-\frac{\epsilon}{2}>0$) scales as
\begin{equation}\label{densofstates}
\left|N (p,\epsilon ,k,m)\right| = \epsilon \frac{\sqrt{k+2}}{2} + \mathcal{O} (1) \ .
\end{equation}
The quantum labels~$(h,Q)$ of these smeared fields approach in the limit~$(\frac{p^2}{4},0)$.
Their correlators are defined (suppressing the obvious $z,\bar z$ dependence in the left hand side) as
\begin{equation}
\langle \Phi_{p_{1},m_{1}}\dots \Phi_{p_{r},m_{r}} \rangle = \lim_{\epsilon \to 0} \lim_{k\to \infty} \beta (k)^{2} \alpha (k)^{r}\langle \Phi_{p_{1},m_{1}}^{\epsilon ,k}(z_{1},\bar{z}_{1})\dotsb \Phi_{p_{r},m_{r}}^{\epsilon ,k} (z_{r},\bar{z}_{r})\rangle \ ,
\end{equation}
with normalisation factors $\alpha (k)$ for each field, and an overall normalisation $\beta^{2} (k)$ for the vacuum on the sphere.
In this case we have to use also the freedom of redefining the fields~$\phi_{l,m}$ by individual phases, and we do it directly at the level of the minimal model's fields:
\begin{equation}\label{conventionchange}
\phi_{l,m} \to (-1)^{\frac{l-m}{2}} \phi_{l,m} \ .
\end{equation}
The necessity of this complication will become clear when we discuss three-point correlators in section~\ref{ch5:sec:three-point-fct}.
With this change of sign, we write the two-point function of the minimal model as
\begin{align}
\langle \phi_{l_{1},m_{1}} (z_{1},\bar{z}_{1})\phi_{l_{2},m_{2}}
(z_{2},\bar{z}_{2})\rangle &=
(-1)^{\frac{l_{1}-m_{1}+l_{2}-m_{2}}{2}}\,
\delta_{l_{1},l_{2}}\, \delta_{m_{1}+m_{2},0} \,\frac{1}{|z_{12}|^{4h_{1}}} \nonumber\\
&= (-1)^{m_{1}}\, \delta_{l_{1},l_{2}}\, \delta_{m_{1}+m_{2},0}\,
\frac{1}{|z_{12}|^{4h_{1}}} \ .
\end{align}
As we have explained in details in chapter~\ref{ch:limit-theories}, we can now take the $k\to\infty$ and $\e\to 0$ limits, and fix the coefficients~$\a$ and $\b$ in order to get finite two-point functions:
\begin{equation}\label{ch5:alpha-beta}
\alpha(k)\beta (k) = \frac{(k+2)^{1/4}}{\sqrt{2}} \ .
\end{equation}
We obtain
\begin{equation}
\left\langle \Phi_{p_{1},m_{1}}
(z_{1},\bar{z}_{1})\Phi_{p_{2},m_{2}} (z_{2},\bar{z}_{2})\right\rangle
= (-1)^{m_{1}}\delta (p_{1}-p_{2}) \delta_{m_{1}+m_{2},0}\frac{1}{|z_{1}-z_{2}|^{p_{1}^{2}}}  \ ,
\end{equation}
in perfect agreement with~\eqref{ch5:two-pt-fct-free-fields}.

\section{Three-point function}\label{ch5:sec:three-point-fct}

\paragraph{Free theory}
We restrict the attention to NS primary fields $\Phi^{\text{free}}_{\mathbf{p}}$ of the theory of two real uncompactified bosons and two real free fermions, for which the three-point function takes the very simple form
\begin{multline}
\langle \Phi^{\text{free}}_{\mathbf{p}_{1}} (z_{1},\bar{z}_{1})
\Phi^{\text{free}}_{\mathbf{p}_{2}} (z_{2},\bar{z}_{2})
\Phi^{\text{free}}_{\mathbf{p}_{3}} (z_{3},\bar{z}_{3})\rangle = 
\delta^{(2)} (\mathbf{p}_{1}+\mathbf{p}_{2}+\mathbf{p}_{3})\\
\times  
|z_{12}|^{2 (h_{3}-h_{1}-h_{2})} |z_{23}|^{2 (h_{1}-h_{2}-h_{3})} 
 |z_{13}|^{2 (h_{2}-h_{1}-h_{3})} \ . 
\label{freefield}
\end{multline}
In order to perform the comparison with the limit of minimal models, we have to express the fields in the radial basis defined in equation~\eqref{ch5:def-radial-basis-free-fields}.
The three-point functions can therefore be written as follows:
\begin{align}
&\langle \Phi^{\text{free}}_{p_{1},m_{1}} (z_{1},\bar{z}_{1})
\Phi^{\text{free}}_{p_{2},m_{2}} (z_{2},\bar{z}_{2})
\Phi^{\text{free}}_{p_{3},m_{3}} (z_{3},\bar{z}_{3})\rangle\nonumber\\
&\  = \sqrt{\frac{p_{1}p_{2}p_{3}}{(2\pi)^{3}}} \int d\varphi_{1}
d\varphi_{2} d\varphi_{3}\
e^{im_{1}\varphi_{1}+im_{2}\varphi_{2}+im_{3}\varphi_{3}}
 \langle \Phi^{\text{free}}_{p_{1}e^{i\varphi_{1}}} (z_{1},\bar{z}_{1})
\Phi^{\text{free}}_{p_{2}e^{i\varphi_{2}}} (z_{2},\bar{z}_{2})
\Phi^{\text{free}}_{p_{3}e^{i\varphi_{3}}} (z_{3},\bar{z}_{3})\rangle\nonumber\\
&\  =  \sqrt{\frac{p_{1}p_{2}p_{3}}{(2\pi)^{3}}} |z_{12}|^{2 (h_{3}-h_{1}-h_{2})} |z_{23}|^{2 (h_{1}-h_{2}-h_{3})} 
 |z_{13}|^{2 (h_{2}-h_{1}-h_{3})}\nonumber\\
&\qquad \times \int d\varphi_{1}
d\varphi_{2} d\varphi_{3}\
e^{im_{1}\varphi_{1}+im_{2}\varphi_{2}+im_{3}\varphi_{3}}\, 
\delta^{(2)}
(p_{1}e^{i\varphi_{1}}+p_{2}e^{i\varphi_{2}}+p_{3}e^{i\varphi_{3}}) \\
&\ = \sqrt{\frac{p_{1}p_{2}p_{3}}{2\pi}} |z_{12}|^{2 (h_{3}-h_{1}-h_{2})} |z_{23}|^{2 (h_{1}-h_{2}-h_{3})} 
 |z_{13}|^{2 (h_{2}-h_{1}-h_{3})}\nonumber\\
&\qquad \times \delta_{m_{1}+m_{2}+m_{3}}\int
d\varphi_{2} d\varphi_{3}\
e^{im_{2}\varphi_{2}+im_{3}\varphi_{3}}\,
\delta^{(2)}(p_{1}+p_{2}e^{i\varphi_{2}}+p_{3}e^{i\varphi_{3}}) \ .
\end{align}
\begin{figure}
\begin{center}
\includegraphics{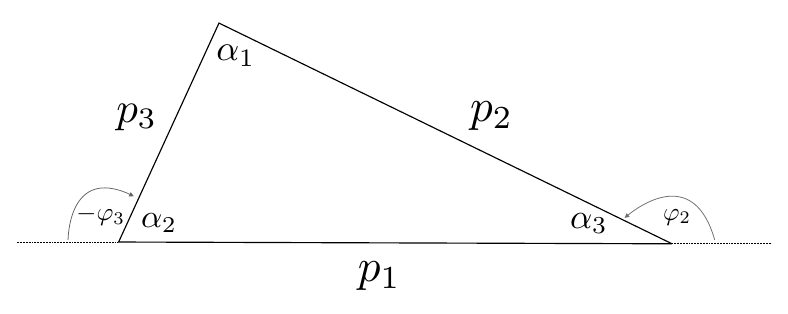}
\end{center}
\caption{\label{fig:free-angles}The triangle spanned by $p_{1}$,
$p_{2}e^{i\varphi_{2}}$ and $p_{3}e^{i\varphi_{3}}$.}
\end{figure}
The integral over the angles $\varphi_{2}$ and $\varphi_{3}$ contains the selection rule
\begin{equation}\label{ch5:closure-triang-free-theory-three-point}
p_{1}+p_{2}e^{i\varphi_{2}}+p_{3}e^{i\varphi_{3}}=0\ ,\qquad
\begin{array}{lll}
q_{1} &:= p_{1} + p_{2}\cos \varphi_{2} + p_{3} \cos \varphi_{3} & = 0\\
q_{2} &:= p_{2}\sin \varphi_{2} + p_{3}\sin \varphi_{3} & = 0  
\end{array}\ ,
\end{equation}
which can be geometrically visualised as the closure of the triangle~$\mathbf{p}_1\mathbf{p}_2\mathbf{p}_3$ in the~$\mathbf{p}$-plane (see figure~\ref{fig:free-angles}).
Moreover, an obvious necessary condition is that
\begin{equation}
|p_{2}-p_{3}| \leq p_{1} \leq p_{2}+p_{3}\ .
\end{equation}
The angles $\varphi_{i}$ take values in the interval $[-\pi,\pi]$. 
For any solution $(\varphi_{2},\varphi_{3})$ there is another solution $(-\varphi_{2},-\varphi_{3})$ that corresponds to the triangle reflected at the side $p_{1}$. 
If $\varphi_{2}>0$, then $\varphi_{3}<0$, and the relation to the angles of the triangle is given like in figure~\ref{fig:free-angles}:
\begin{align}
\varphi_{2} &= \alpha_{1}+\alpha_{2}  &
\varphi_{3} &= \alpha_{2} -\pi \ .
\end{align}
The integral can be therefore explicitly evaluated by plugging in the values of $\varphi_{2}$ and $\varphi_{3}$ for the two solutions, and dividing this by the Jacobian determinant
\begin{equation}
\left| \det \left( \frac{\partial q_{i}}{\partial \varphi_{j}} \right)_{i,j}\right| = 
p_{2} p_{3} |\sin (\varphi_{2}-\varphi_{3})| = 2 A (p_{1},p_{2},p_{3})
\ ,
\end{equation}
where $A (p_{1},p_{2},p_{3})$ is the area of the triangle (see equation~\eqref{area}).
We find at the end
\begin{align}\label{ch5:three-point-free}
&\langle \Phi^{\text{free}}_{p_{1},m_{1}} (z_{1},\bar{z}_{1})
\Phi^{\text{free}}_{p_{2},m_{2}} (z_{2},\bar{z}_{2})
\Phi^{\text{free}}_{p_{3},m_{3}} (z_{3},\bar{z}_{3})\rangle\nonumber\\
&\quad = \sqrt{\frac{p_{1}p_{2}p_{3}}{2\pi}} |z_{12}|^{2 (h_{3}-h_{1}-h_{2})} |z_{23}|^{2 (h_{1}-h_{2}-h_{3})} 
 |z_{13}|^{2 (h_{2}-h_{1}-h_{3})}\nonumber\\
&\qquad \times \delta_{m_{1}+m_{2}+m_{3}}
\frac{\cos (m_{2}\alpha_{1}-m_{1}\alpha_{2} +\pi (m_{1}+m_{2}))}{A
(p_{1},p_{2},p_{3})}\ .
\end{align}
Obviously, if the triangle with sides~$\mathbf{p}_1\mathbf{p}_2\mathbf{p}_3$ does not exist, the correlator is simply zero.

\paragraph{Limit theory}

The three-point functions in the limit theory are obtained from the three-point functions in the minimal models, given in equation~\eqref{ch3:three-point-fct-minmod}.
In the limit~$k\to\infty$ and $l_{i}\approx p_{i}\sqrt{k+2}$, the product of gamma functions summarised in the function~$d_{l_1,l_2,l_3}$ approaches~1.
In this regime the correlators read
\begin{multline}\label{ch5:three-point-general-free-limit}
\langle\phi_{l_{1},m_{1}} (z_{1},\bar{z}_{1}) \phi_{l_{2},m_{2}}
(z_{2},\bar{z}_{2})\phi_{l_{3},m_{3}} (z_{3},\bar{z}_{3})\rangle
= (-1)^{\frac{l_{1}+l_{2}+l_{3}}{2}} \begin{pmatrix}
\frac{l_{1}}{2} & \frac{l_{2}}{2} & \frac{l_{3}}{2}\\
\frac{m_{1}}{2} & \frac{m_{2}}{2}& \frac{m_{3}}{2}
\end{pmatrix}^{2}\\
\times \sqrt{(l_{1}+1)(l_{2}+1)(l_{3}+1)} \, 
\delta_{m_{1}+m_{2}+m_{3},0}|z_{12}|^{2(h_{3}-h_{1}-h_{2})}
|z_{13}|^{2(h_{2}-h_{1}-h_{3})}|z_{23}|^{2(h_{1}-h_{2}-h_{3})}
\ .
\end{multline}
$\begin{pmatrix}
j_{1}&j_{2}&j_{3} \\
\mu_{1}&\mu_{2}&\mu_{3}
\end{pmatrix}$ are the Wigner 3j-symbols, whose asymptotic behaviour in various regimes is analysed in details in appendix~\ref{app:wigner}. 
Here we need specifically equation~\eqref{3j-asympt-app}, which with our entries reads
\begin{multline}
\begin{pmatrix}
\frac{l_{1}}{2} & \frac{l_{2}}{2} & \frac{l_{3}}{2} \\
\frac{m_{1}}{2} & \frac{m_{2}}{2} & \frac{m_{3}}{2}
\end{pmatrix} 
= (k+2)^{-1/2}\frac{(-1)^{\frac{l_{1}-l_{2}-m_{3}}{2}}}{\sqrt{\frac{\pi}{2}A
(p_{1},p_{2},p_{3})}}\\
\times   \cos \left(\frac{l_{1}+l_{2}-l_{3}}{4}\pi
+\frac{m_{2}\alpha_{1}-m_{1}\alpha_{2}}{2} \right) + \mathcal{O} (k^{-1}) \ .
\end{multline}
In the~$k\to \infty$ regime equation~\eqref{ch5:three-point-general-free-limit} becomes then
\begin{multline}
\langle\phi_{l_{1},m_{1}} (z_{1},\bar{z}_{1}) \phi_{l_{2},m_{2}}
(z_{2},\bar{z}_{2})\phi_{l_{3},m_{3}} (z_{3},\bar{z}_{3})\rangle
= (k+2)^{-1/4} \frac{2\sqrt{p_{1}p_{2}p_{3}}}{\pi A
(p_{1},p_{2},p_{3})} \delta_{m_{1}+m_{2}+m_{3},0}\\
\times (-1)^{\frac{l_{1}+l_{2}+l_{3}}{2}}
\cos^{2} \left(\tfrac{l_{1}+l_{2}-l_{3}}{4}\pi
+\tfrac{m_{2}\alpha_{1}-m_{1}\alpha_{2}}{2} \right) 
|z_{12}|^{2(h_{3}-h_{1}-h_{2})}
|z_{13}|^{2(h_{2}-h_{1}-h_{3})}|z_{23}|^{2(h_{1}-h_{2}-h_{3})}
\ .
\end{multline}
To find the right correlators in the limit theory we have to average over the labels~$l_{i}$.
We observe: 
\begin{multline}
(-1)^{\frac{l_{1}+l_{2}+l_{3}}{2}}
\cos^{2} \left(\tfrac{l_{1}+l_{2}-l_{3}}{4}\pi
+\tfrac{m_{2}\alpha_{1}-m_{1}\alpha_{2}}{2} \right) \\
= (-1)^{m_{3}}
\times \left\{\begin{array}{ll}
\cos^{2} \left( \frac{m_{2}\alpha_{1}-m_{1}\alpha_{2}}{2} \right) &
\ \text{for}\ \frac{l_{1}+l_{2}-l_{3}}{2} = 0 \ \text{mod}\  2\\[5pt]
-\sin^{2} \left( \frac{m_{2}\alpha_{1}-m_{1}\alpha_{2}}{2} \right) &
\ \text{for}\ \frac{l_{1}+l_{2}-l_{3}}{2} = 1 \ \text{mod}\  2 \ . 
\end{array} \right. 
\end{multline}
These contributions combine in average to give
\begin{equation}\label{average}
\frac{1}{2} (-1)^{m_{3}} \left(\cos^2
\tfrac{m_{2}\alpha_{1}-m_{1}\alpha_{2}}{2} -
\sin^2 \tfrac{m_{2}\alpha_{1}-m_{1}\alpha_{2}}{2}\right) = 
\frac{1}{2} (-1)^{m_{3}} \cos \left(
m_{2}\alpha_{1}-m_{1}\alpha_{2}\right) .
\end{equation} 
At the end we arrive to the expression
\begin{multline}
\langle\Phi_{p_{1},m_{1}} (z_{1},\bar{z}_{1}) \Phi_{p_{2},m_{2}}
(z_{2},\bar{z}_{2})\Phi_{p_{3},m_{3}} (z_{3},\bar{z}_{3})\rangle
 = \beta^{2}(k)\alpha^{3}(k) (k+2)^{-1/4} \,\delta_{m_{1}+m_{2}+m_{3},0} \,(-1)^{m_{3}} \\
\times \frac{\sqrt{p_{1}p_{2}p_{3}}}{\pi}\,\frac{\cos (m_{2}\alpha_{1}-m_{1}\alpha_{2})}{A
(p_{1},p_{2},p_{3})}\,
|z_{12}|^{2 (h_{3}-h_{1}-h_{2})} |z_{23}|^{2 (h_{1}-h_{2}-h_{3})} 
 |z_{13}|^{2 (h_{2}-h_{1}-h_{3})} \ ,
\end{multline}
which reproduces exactly the correlator of the free theory~\eqref{ch5:three-point-free} if we set
\begin{align}
\alpha (k) &= \sqrt{2\pi} (k+2)^{-1/4} &
\beta (k) &= \frac{1}{2\sqrt{\pi}} (k+2)^{1/2} \ .
\end{align}
The redefinition of the individual minimal model primaries $\phi_{l,m}$ by the sign $(-1)^{\frac{l-m}{2}}$ is crucial in matching the expressions. 
Without these phases, the averaging in~\eqref{average} gives simply $\frac{1}{2} (-1)^{m_{3}}$, so that the three-point functions would have a rather trivial dependence on the labels $m_{i}$.

\section{A-type boundary conditions}\label{ch5:sec:A-type-bc}

\paragraph{Free theory}
\begin{figure}
\begin{center}
\includegraphics{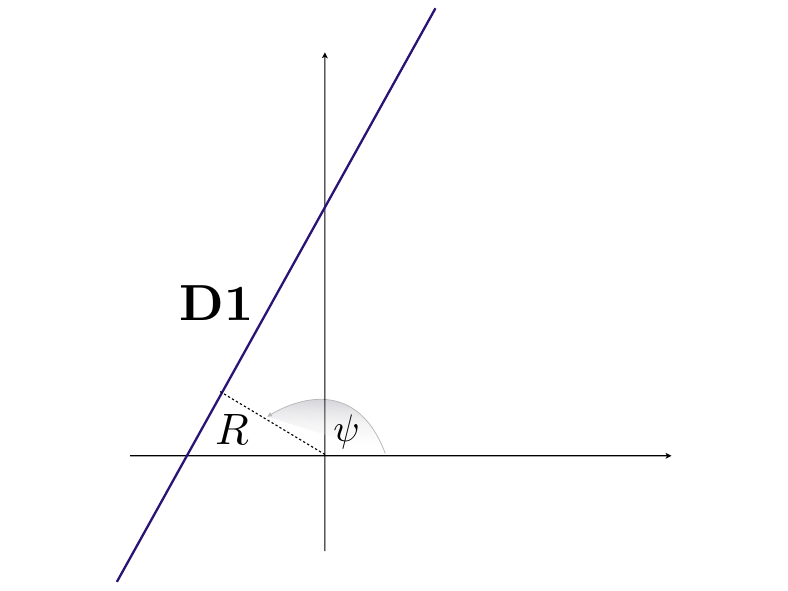}
\end{center}
\caption{\label{fig:coord-free}Illustration of the boundary condition
that corresponds to a one-dimensional brane, and the distance $R$ and
the angle $\psi$ that determine its position.}
\end{figure}
In a free theory a boundary condition corresponding to a $d$-dimensional brane in a $D$-dimensional target space that only couples to the NS-NS sector has the one-point function (see~\cite{Blumenhagen:2013fgp} for instance)
\[
\big\langle e^{i\vec{p}\cdot\vec{X}}\big\rangle = 2^{-\frac{D}{4}} (\alpha
')^{\frac{D-2d}{4}} \delta^{(d)} (\vec{p}_{\parallel})e^{i\vec{R}\cdot
\vec{p}_{\perp}} |z-\bar{z}|^{-2h_{p}}\ ,
\]
where the conformal weight is $h_{p}=\frac{\alpha'p^{2}}{4}$.
The GSO projection of the full Hilbert space allows either the even- or the odd-dimensional branes to couple to the R-R sector; in this case there is an additional factor of $2^{-1/2}$. 
In the free theory under analysis the position of a one-dimensional brane is given by a vector $Re^{i\psi}$, which gives its shortest distance from the origin~$R$, and by the oriented angle~$\psi$ between the horizontal axis and the normal to the brane, as sketched in figure~\ref{fig:coord-free}.
In the NS sector the one-point functions are then, in our conventions, given by
\begin{equation}
\big\langle \Phi^{\text{free}}_{pe^{i\varphi}}
(z,\bar{z})\big\rangle^{\!A}_{\!R,\psi} = \frac{1}{2}\,
\delta (p \cos (\psi -\varphi)) \, e^{iR p \sin (\psi -\varphi)}
\frac{1}{|z-\bar{z}|^{2h_{p}}} \ ,
\end{equation}
where the prefactor $1/2$ is a sign of the implicit choice of GSO projection: also the R-R fields couple to the one-dimensional brane.
In the radial basis~\eqref{ch5:def-radial-basis-free-fields}, the one-point function is therefore
\begin{align}
\big\langle \Phi^{\text{free}}_{p,m} (z,\bar{z})\big\rangle^{\!A}_{\!R,\psi} &=
\sqrt{\frac{p}{2\pi}} \int d\varphi\ e^{im\varphi} \langle
\Phi^{\text{free}}_{pe^{i\varphi}} (z,\bar{z})\rangle^{A}_{R,\psi} \\
&= \frac{1}{\sqrt{2\pi p}} e^{im\psi} \cdot \left\{\begin{array}{ll}
\cos Rp & \text{for $m$ even}\\[4pt]
i \sin Rp & \text{for $m$ odd}
\end{array} \right.\ .
\label{ffonepointA}
\end{align}

\paragraph{Limit theory}
A-type boundary conditions for minimal models have been described in section~\ref{ch3:sec:BC}, and we have already discussed the geometry of the limit in subsection~\ref{ch4:sec:branes:subs:free-limit}.
We want to become more precise here: we recall from that discussion, that it is possible to keep the one-dimensional branes at fixed distance~$R$ and at a given angle~$\varphi_0$ if we rescale the boundary labels as
\begin{equation}\label{ch5:choice-bound-labels-free-limit}
L = \frac{1}{2} (k+2) - \frac{R}{\pi}\sqrt{k+2} + \mathcal{O} (1) \ , \qquad M = \frac{k+2}{\pi} \varphi_{0} + \mathcal{O} (1) \ .
\end{equation}
We expect $\varphi_{0}$ to coincide with the angle $\psi$ of figure~\ref{fig:coord-free} up to a possible additive shift. 

A-type one-point functions in the NS sector have been given in equation~\eqref{ch3:A-type-one-point-fct}; setting there $s=0$ and multiplying by the sign~$(-1)^{\frac{l-m}{2}}$ coming from the field redefinition of equation~\eqref{conventionchange}, we get
\begin{equation}
\langle \phi_{l,m} (z,\bar{z})\rangle^{A}_{(L,M,S)} 
= \frac{(-1)^{\frac{l-m}{2}}}{\sqrt{k+2}} \frac{\sin \frac{\pi (l+1)
(L+1)}{k+2}}{\sqrt{\sin \frac{\pi (l+1)}{k+2}}} e^{\pi i
\frac{Mm}{k+2}} \frac{1}{|z-\bar{z}|^{2h_{l,m}}} \ .
\end{equation}
If we plug in the boundary labels chosen in~\eqref{ch5:choice-bound-labels-free-limit}, we find that the one-point functions behave as follows
\begin{equation}\label{Atypeonepoint}
\langle \phi_{l,m} (z,\bar{z})\rangle^{A}_{(L,M,S)} 
= \frac{(k+2)^{-1/4}}{2\sqrt{\pi p}}
\left(e^{iR\frac{l+1}{\sqrt{k+2}}} - e^{i\pi (l+1) -iR\frac{l+1}{\sqrt{k+2}}}
  \right)
e^{i (\varphi_{0}-\frac{\pi}{2})m}
 \frac{1}{|z-\bar{z}|^{2h_{l,m}}} \ .
\end{equation}
To obtain the expression for the one-point functions of the limit field we multiply this last expression by $\alpha(k)\beta(k)$ given in~\eqref{ch5:alpha-beta}, and take the limit $k\to \infty$ while keeping $m$ constant and scaling $l\approx p\sqrt{k+2}$. 
We get
\begin{equation}
\big\langle \Phi_{p,m}\big\rangle^{\!A}_{\!R,\varphi_{0}} = \frac{1}{\sqrt{2\pi p}}
e^{i (\varphi_{0}-\frac{\pi}{2})m} \cdot \left\{\begin{array}{ll}
\cos Rp & \text{for $m$ even}\\
i\sin Rp & \text{for $m$ odd}
\end{array} \right.\ ,
\end{equation}
which precisely matches the free field theory one-point function~\eqref{ffonepointA} once we identify $\psi =\varphi_{0}-\frac{\pi}{2}$.

\section{B-type boundary conditions}
\label{ch5:sec:B-type}

\paragraph{Free theory}

A zero-dimensional brane sitting at the origin of the plane is characterised by the following simple one-point function for NS primaries
\begin{equation}
\big\langle \Phi_{\mathbf{p}}^{\text{free}}
(z,\bar{z})\big\rangle^{\!B}_{\!(0)} =
\frac{1}{\sqrt{2}} |z-\bar{z}|^{-2h_{p}} \ ,
\end{equation}
which reads in radial basis
\begin{equation}\label{ch5:one-point-fct-free-zero-dim-branes}
\big\langle \Phi_{p,m}^{\text{free}} (z,\bar{z})\big\rangle^{\!B}_{\!(0)} = 
\sqrt{\pi p} \,\delta_{m,0} \frac{1}{|z-\bar{z}|^{2h_{p}}} \ .
\end{equation}
Note that the zero-dimensional brane does not couple to the R-R fields, because we have chosen the GSO projection such that the one-dimensional brane described in section~\ref{ch5:sec:A-type-bc} couples to them. 
The prefactor is thus simply $2^{-D/4}=2^{-1/2}$.

\smallskip
\smallskip

We find here two-dimensional space filling branes as well; they come in a one-parameter family whose parameter is the intensity of a constant background electric field (see for example~\cite{Abouelsaood:1986gd,DiVecchia:1999fx} and the discussion
in~\cite{Gaberdiel:2004nv}), with field strength~$F_{\mu\nu}=\begin{pmatrix}0 & f\\ -f & 0 \end{pmatrix}$.
The label we choose is an angle~$\phi\in(-\pi,\pi)$, in terms of which $\sin \phi =\frac{2f}{1+f^{2}}$ and $\cos\phi =\frac{1-f^{2}}{1+f^{2}}$.
These boundary conditions are characterised by the one-point amplitude
\begin{equation}
\big\langle \Phi_{\mathbf{p}}^{\text{free}}(z,\bar{z})\big\rangle^{\!B}_{\!\phi} = \frac{1}{\sqrt{2}\cos \frac{\phi}{2}}\delta^{(2)} (\mathbf{p}) \ .
\end{equation}
It is convenient at this point to work with the $\d$-distribution evaluated on a test function $\zeta (\mathbf{p})$, and to analyse a smeared one-point function
\begin{equation}
\Big\langle \int d^{2}p\ \zeta (\mathbf{p})\Phi_{\mathbf{p}}^{\text{free}}(z,\bar{z}) \Big\rangle^{\!\!B}_{\!\!\phi} = \frac{1}{\sqrt{2}\cos \frac{\phi}{2}}\zeta (0) \ ,
\end{equation}
which in the radial basis reads
\begin{align}
\Big\langle \int dp\ \sum_{m} \zeta_{p,-m} \Phi_{p,m}^{\text{free}}
(z,\bar{z})\Big\rangle^{\!\!B}_{\!\!\phi}
=\Big\langle \int d^{2}p\ \zeta (\mathbf{p})\Phi_{\mathbf{p}}^{\text{free}}
(z,\bar{z}) \Big\rangle^{\!\!B}_{\!\!\phi} &= \frac{1}{\sqrt{2}\cos \frac{\phi}{2}}\zeta (0) \\
&=
\frac{1}{\sqrt{2}\cos \frac{\phi}{2}}\frac{\zeta_{p,0}}{\sqrt{2\pi p}}\bigg|_{p=0}\ ,
\end{align}
where 
\begin{equation}
\zeta_{p,m} = \sqrt{\frac{p}{2\pi}} \int d\varphi\ e^{im\varphi} \, \zeta
(pe^{i\varphi}) \ .
\end{equation}
We can reformulate this as 
\begin{subequations}
\begin{align}
\big\langle \Phi_{p,m}^{\text{free}} (z,\bar{z})\big\rangle^{\!B}_{\!\phi} &= 0 \qquad
\text{for}\ m\not= 0\\
\Big\langle \sqrt{2\pi}\int_{0}^{\infty}dp\ \sqrt{p}\,\chi
(p)\Phi_{p,0}^{\text{free}} (z,\bar{z})\Big\rangle^{\!\!B}_{\!\!\phi} &= \frac{1}{\sqrt{2}\cos \frac{\phi}{2}}\chi (0) \ ,
\label{D2free}
\end{align}
\end{subequations}
for suitable test functions $\chi$ on the positive real line.
\paragraph{Limit theory}

We have discussed the geometry of this limit in subsection~\ref{ch4:sec:branes:subs:free-limit}: we expect the B-type discs to correspond to zero-dimensional free theory branes, if the boundary labels scale in such a way that the discs shrink to a point at the origin.
The one-point functions of NS primaries have been given in section~\ref{ch3:sec:BC}\footnote{Note that the sign $(-1)^{\frac{l-m}{2}}$ that one expects from the field
redefinition~\eqref{conventionchange} is absorbed by a sign hidden inside the definition of the B-type Ishibashi states in~\cite{Maldacena:2001ky}.} and read
\begin{equation}
\langle \phi_{l,m} (z,\bar{z})\rangle^{B}_{(L,S)} =
\sqrt{2}\,\frac{\sin \frac{\pi (l+1) (L+1)}{k+2}}{\sqrt{\sin \frac{\pi
(l+1)}{k+2}}} \delta_{m,0} |z-\bar{z}|^{-2h_{l,m}} \ .
\end{equation}

We can keep the label $L$ fixed, such that the radius of the disc, $\rho_{1}'=\sqrt{k+2}\sin\frac{\pi (L+1)}{k+2}$, goes to zero. 
The corresponding one-point function reads
\begin{equation}\label{ch5:one-point-fct-zero-dim-BC}
\big\langle \Phi_{p,m} (z,\bar{z})\big\rangle^{\!B}_{\!(L,S)} = \sqrt{\pi p} (L+1)\delta_{m,0} |z-\bar{z}|^{-2h_{p}} \ ,
\end{equation}
and correspond to a superposition of the zero-dimensional free theory branes of equation~\eqref{ch5:one-point-fct-free-zero-dim-branes}.
Remarkably this is an integer multiple of the one-point function for $L=0$: it describes a stack of coincident $L+1$ elementary branes. 
This confirms the fact that in $N=2$ minimal models B-type boundary conditions with $L>0$ can be obtained from a superposition of boundary conditions with $L=0$ by boundary RG flows, which are short when $k\to\infty$~\cite{Fredenhagen:2003xf}.

\smallskip
\smallskip

\noindent
We expect the space filling electric branes of the free theory to come from discs covering the whole two-dimensional space in the limit of minimal model's geometry.
These are labelled by $(L,S)$ where $L$ is scaled linearly with $k$, $L=\lfloor \Lambda (k+2)\rfloor$. 
The minimal model one-point functions behave with this choice of parameters as
\begin{equation}
\langle \phi_{l,m} (z,\bar{z})\rangle^{B}_{(\lfloor \Lambda
(k+2)\rfloor,S)} 
\approx  \sqrt{\frac{2 (k+2)}{\pi (l+1)}}\, \sin \left(\pi \Lambda (l+1) \right) \delta_{m,0} |z-\bar{z}|^{-2h_{l,m}} \ .
\end{equation}
For large~$l$ the sine oscillates rapidly, so that the only non zero contribution is given by~$l=0$, namely~$p=0$ in the limit theory parameter: the one-point function of
$\Phi_{p,0}$ is suppressed for non-zero $p$.
This is expected from the analysis of the zero-dimensional branes in the free theory.
To evaluate the contribution at $p=0$, we consider also in this case the one-point function for fields smeared by a test function $\chi$,
\begin{align}
&\Big\langle \sqrt{2\pi}\int_{0}^{\infty} dp\ \sqrt{p}\,\chi (p)\,
\Phi_{p,0} (z,\bar{z})\Big\rangle^{\!\!B}_{\!\!(\lfloor \Lambda
(k+2)\rfloor,S)}\nonumber\\
&\qquad = \lim_{k\to\infty} \sqrt{2\pi}\,\sqrt{2}\,(k+2)^{-1/4} \sum_{l\
\text{even}} \left(\frac{l+1}{\sqrt{k+2}}\right)^{\frac{1}{2}}\, \chi
\left(\frac{l+1}{\sqrt{k+2}} \right)
\langle \phi_{l,0} (z,\bar{z})\rangle^{B}_{(\lfloor \Lambda
(k+2)\rfloor,S)} \nonumber\\
&\qquad = \lim_{k\to\infty}  2 \sqrt{2}\sum_{l\ \text{even}} \sin
\left(\pi\Lambda (l+1) \right) \,\chi
\left(\frac{l+1}{\sqrt{k+2}} \right) |z-\bar{z}|^{-2h_{l,0}} \nonumber\\
&\qquad = \frac{\sqrt{2}}{\sin \pi \Lambda}\chi (0) \ ,
\end{align}
which equals twice the result in eq.\ \eqref{D2free} if we set 
\begin{equation}\label{identificationofphi}
\phi=\pm 2\pi\big(\Lambda-\tfrac{1}{2}\big)\ .
\end{equation}
The limiting boundary condition is therefore not elementary. 
It results in a superposition of two two-dimensional branes of the free theory. 
By looking at the relative spectrum of the zero-dimensional brane one finds that the two branes have opposite electric field (corresponding to the two signs of~$\phi$ in~\eqref{identificationofphi}). 
This is consistent with the identification of B-type boundary states in minimal models under $L\leftrightarrow k-L$, which becomes now the identification $\Lambda\leftrightarrow 1-\Lambda$, corresponding to a switch of the sign in~\eqref{identificationofphi}.
 
\chapter{New~$c=3$ theory}\label{ch:new-theory}

Following the considerations in subsection~\ref{ch4:sec:bulk-lim:subs:contorbi}, we can study a limiting procedure different from the one presented in chapter~\ref{ch:free-limit}, namely the one dictated by the averaging fields~${}^{(1)}\Phi^{(k)}_1$ of~$N=2$ minimal models of section~\ref{ch4:sec:intro}.
In this limit we scale the label~$m$ of minimal models with~$k$ as~$k$ becomes large, keeping the difference~$l-|m|$ finite.
The central charges $c=3\frac{k}{k+2}$ approaches~$c=3$ in the limit.
In this chapter we analyse the behaviour of the defining structures of minimal models in this limit.
We get a new non-rational theory with~$c=3$, with some similarity to the theory of Runkel and Watts described in section~\ref{ch1:sec:RW}.
We follow closely reference~\cite{Fredenhagen:2012rb}.
The next chapter will be devoted to the comparison of this new theory to a supersymmetric continuous orbifold~\cite{Fredenhagen:2012bw}.

\section{The spectrum}\label{ch6:sec:spectrum}
We recall here for clarity the expressions for the spectrum of ground states of the~$N=2$ unitary minimal models (see e.g. subsection~\ref{ch3:appKS:subsec:minmod}).
\paragraph{The spectrum of the~$k^{\text{th}}$ model} 
The unitary representations of the bosonic subalgebra of the $N=2$ superconformal algebra are labelled by
three integers $(l,m,s)$, where
\begin{equation}
0\leq l \leq k\quad ,\quad m\equiv m+2k+4 \quad ,\quad s\equiv s+4 \ .
\end{equation}
Only those triples $(l,m,s)$ are allowed for which $l+m+s$ is even, and triples are identified according to the relation
\begin{equation}\label{cosetident}
(l,m,s) \equiv (k-l,m+k+2,s+2) \ .
\end{equation}
The conformal weight and the $U (1)$ charge of the vectors in a representation $\mathcal{H}_{(l,m,s)}$ are given by 
\begin{align}
h & \in h_{l,m,s} + \mathbb{N}  &  h_{l,m,s} &= \frac{l(l+2)-m^{2}}{4 (k+2)}+\frac{s^{2}}{8}\\
Q & \in Q_{m,s} + 2\mathbb{Z} & Q_{m,s} & =-\frac{m}{k+2}+\frac{s}{2} \ .
\end{align}
We consider models with a diagonal spectrum, i.e.\ with equal left-and right-moving weights, $\bar{h}=h$, and charges, $\bar{Q}=Q$, of the ground states. 
The conformal weight and the $U (1)$ charge of the ground states of $\mathcal{H}_{l,m,s}$ are exactly given by $h_{l,m,s}$ and $Q_{m,s}$ (without integer shifts) if the labels satisfy
\begin{equation}
|m-s| \leq l \ ,
\end{equation}
which is the standard range. 
$\Hilb_{l,m,0}\oplus\Hilb_{l,m,2}$ constitues the NS representation of the full superconformal algebra.
The corresponding primaries are denoted here by~$\phi_{l,m}$.
$\Hilb_{l,m,1}\oplus\Hilb_{l,m,-1}$ as well is a representation of the full superconformal algebra, the R sector.
The corresponding primaries are denoted here by~$\psi^{\pm}_{l,m}$.
The ground states in the Ramond sector have labels $(l,l+1,1)$ and the corresponding field is denoted~$\psi_l^0$.
\paragraph{The limit}
We want to take the limit $k\to \infty$, keeping the $U (1)$ charge and the conformal weight fixed in the limit.
Let us first consider the primary states in the NS sector, namely states with~$s=0$.
For a fixed charge~$Q$ we have to scale $m$ with $k$ such that
\begin{equation}
m \approx -Q (k+2)\ .
\end{equation}
As already noticed in subsection~\ref{ch4:sec:bulk-lim:subs:contorbi}, when the level $k$ becomes large, the spectrum of $U (1)$-charges becomes continuous in the range $-1< Q< 1$. 
On the other hand, the label $l$ is determined by $h_{l,m,0}$ and $Q_{m,0}$ by 
\begin{equation}\label{l_intermsof_hq}
l = \sqrt{(k+2)^{2}Q_{m,0}^{2} + 4 (k+2)h_{l,m,0}+1}-1 \ .
\end{equation}
Keeping $Q_{m,0} \approx Q \not= 0$ and $h_{l,m,0}\approx h$ fixed, the label $l$ scales as 
\begin{equation}\label{l_in_the_limit}
l = |m| + 2 \frac{h}{|Q|} -1 + \mathcal{O} (1/k) \ .
\end{equation}
The label $l$ thus differs from the linearly growing $|m|$ only by a fixed finite number, which has to be an even integer (due to the selection rules of the coset, again refer to the subsection~\ref{ch3:appKS:subsec:minmod}),
\begin{equation}\label{relation_l_and_m}
l = |m| + 2n \quad ,\quad n=0,1,2,\dotsc \ .
\end{equation}
Whereas $|Q|$ can take any value between $0$ and $1$, we see by comparing~\eqref{l_in_the_limit} and~\eqref{relation_l_and_m} that the ratio $h/|Q|$ can only take discrete values,
\begin{equation}
h_{Q,n} = (2n+1) |Q|/2 \ ,
\end{equation}
and $n=0$ corresponds to chiral primary and anti-chiral primary fields.

\begin{figure}
\begin{center}
\includegraphics[width=7.5cm]{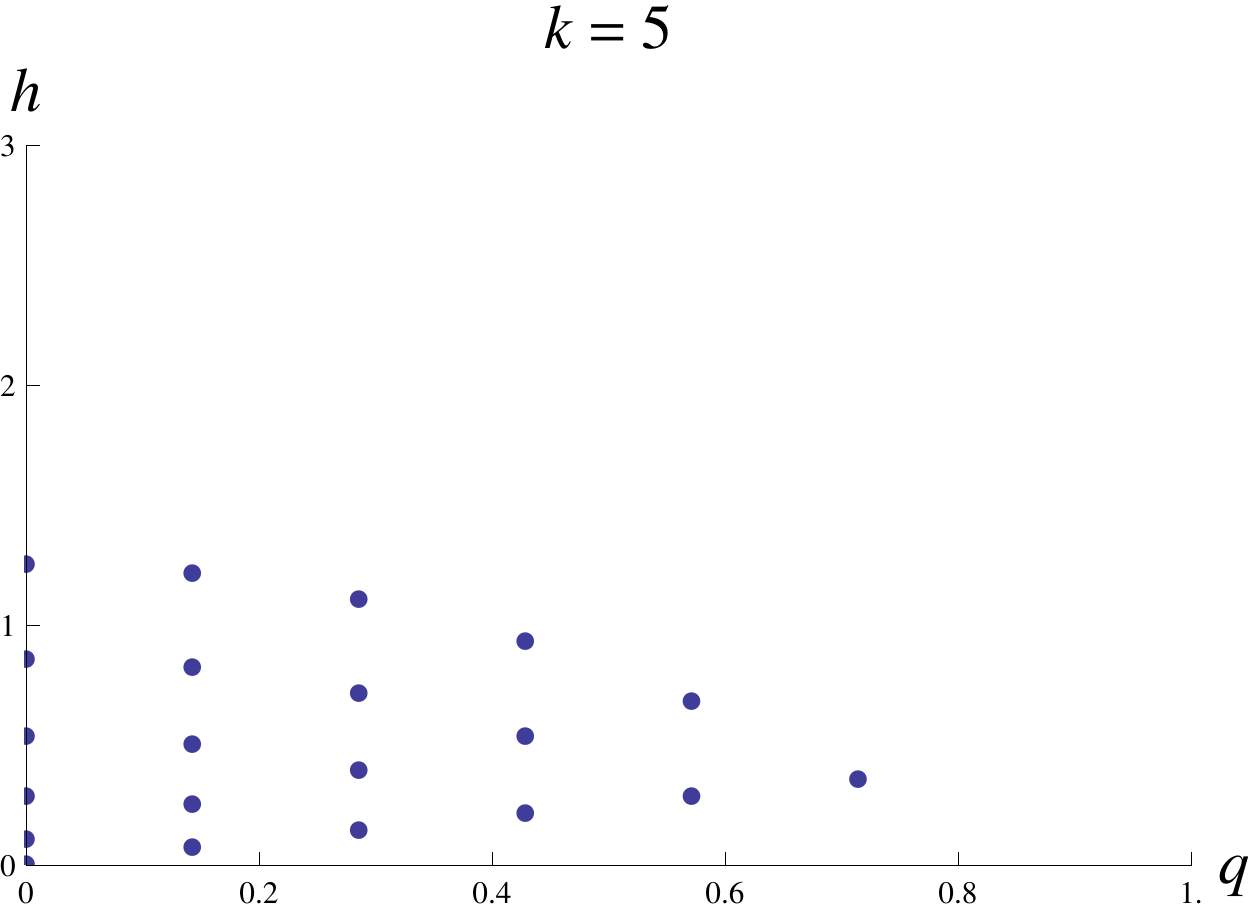}\ \  
\includegraphics[width=7.5cm]{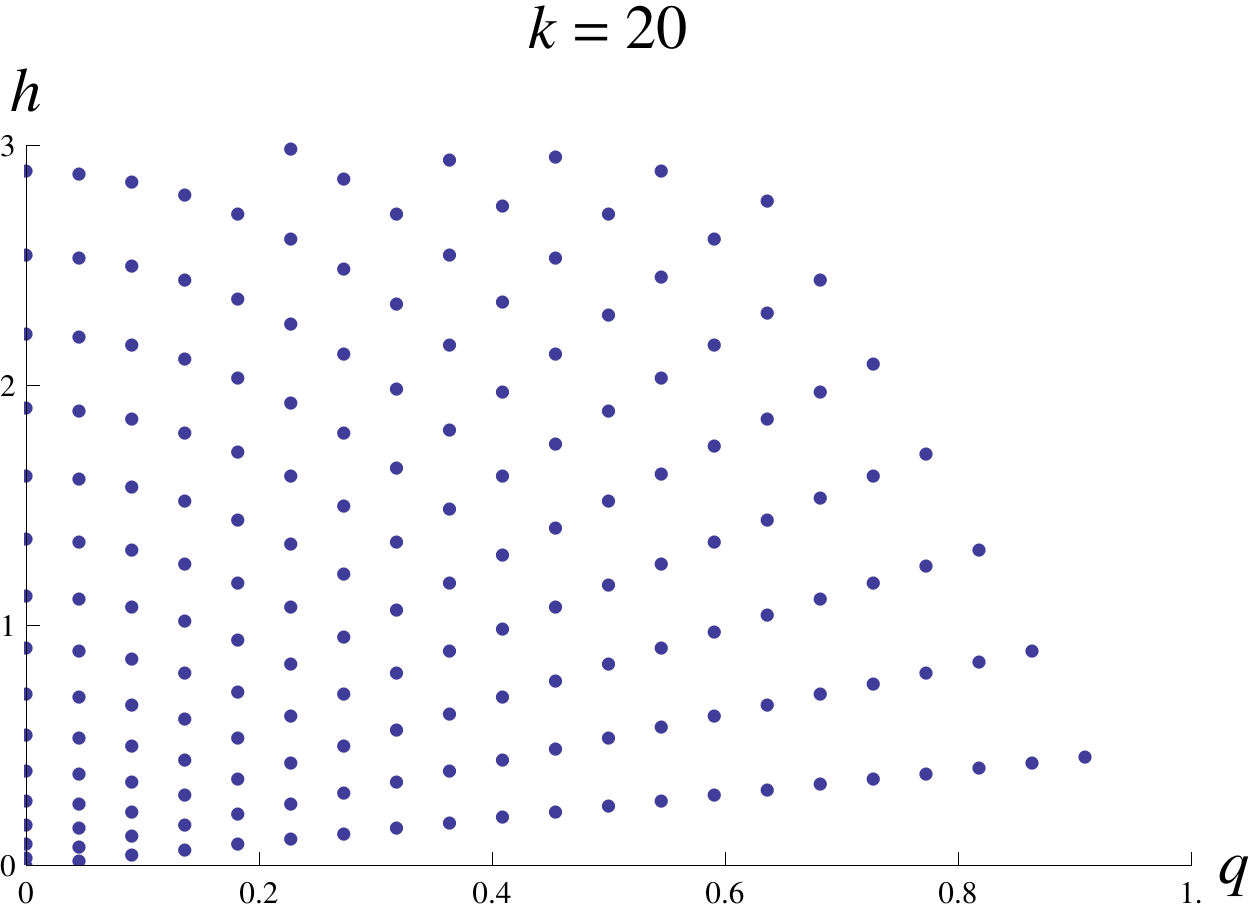}\\[8mm]
\includegraphics[width=7.5cm]{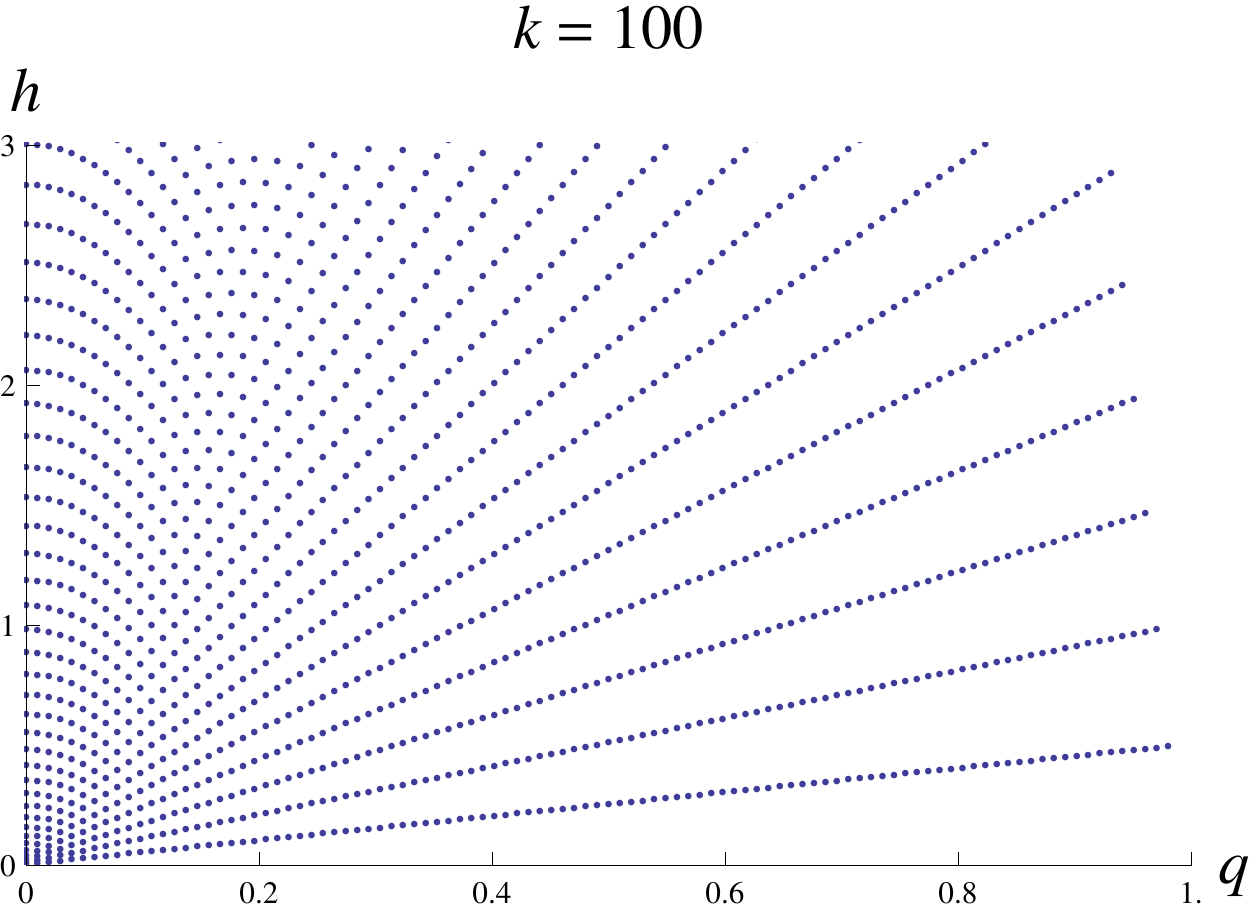}\ \ 
\includegraphics[width=7.5cm]{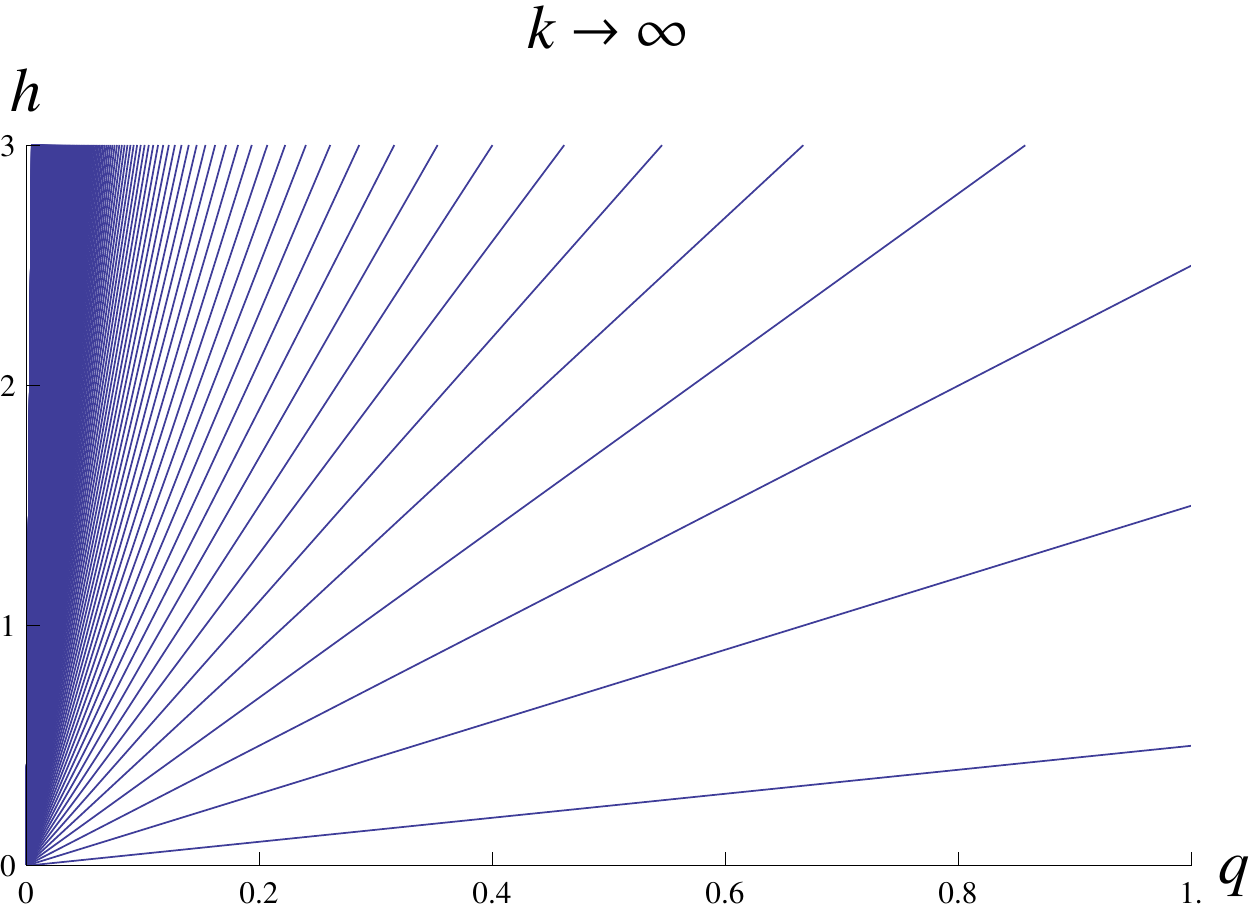}
\end{center}
\caption{\label{fig:spectrum}Behaviour of the spectrum of primary fields in the NS sector for large levels $k$: when one plots the values of the conformal weight $h$ and of the $U (1)$ charge $Q$ as dots in the $h$-$Q$-plane, one observes that the points assemble along straight lines starting from the origin. 
Notice that we only plotted the points corresponding to positive charge $Q$ (the negative charged part is just the mirror picture) and we truncated the conformal weights by $h\leq 3$.}
\end{figure}

We summarise the spectrum of this new non-rational theory in the following table:
\begin{align}
\begin{array}{c}
\lim\limits_{k\to\infty}\\[0.5cm]
\text{``NEW'' THEORY}
\end{array}
\quad
\boxed{
\begin{array}{cll}
h_{l,m,0}=\frac{l(l+2)-m^2}{4(k+2)}\quad &\xrightarrow[m=-Q(k+2)]{l=|m|+2n}\quad & h_{Q,n}=\frac{|Q|}{2}\left(n+\frac 12\right)\\
\substack{|m|\leq l\ ,\ l\leq k \\ l+m\ \text{even}}&\xrightarrow{\phantom{m=-Q(k+2)}}\quad & n\in\Z_{\geq 0}\ ,\ Q\in[-1,1]
\end{array}
}
\end{align}
In the $h$-$Q$-plane, the NS spectrum is thus concentrated on lines going through the origin (see figure~\ref{fig:spectrum}), and the fields $\Phi_{Q,n}$ are labelled by their continuous $U
(1)$-charge $Q$ and a discrete label $n$.
\smallskip

By a similar analysis we find in the Ramond sector on the one hand the Ramond ground states leading to fields $\Psi^{0}_{Q}$ with $h=\frac{1}{8}$ and $-\frac{1}{2}<Q<\frac{1}{2}$ built from fields $\psi^{0}_{l}$ with $l\approx (k+2)(\frac{1}{2}-Q)$. 
In addition there are the fields $\Psi^{\pm}_{Q,n}$ with $-\frac{1}{2}<\pm Q<\frac{3}{2}$ and 
\begin{equation}
h^{\pm}_{n} (Q) =\frac{1}{8} + n \left| Q\mp \frac{1}{2}\right| \ .
\end{equation}
They are obtained from fields $\psi^{\pm}_{l,m}$ with $l=|m|+2n-1$ and $m\approx - (k+2) \left( Q\mp \frac{1}{2}\right)$.

\section{Partition function}\label{ch6:sec:partition-fct}

We now want to study the behaviour of the partition function in the regime we have discussed in the previous section: $n=\frac{l-|m|}{2}$ is a fixed non-negative integer, and $|m|$ scales with $k$.  
The NS contribution to the partition function for the $A_{k+2}$ minimal model reads (see section~\ref{ch3:sec:paraf:subs:characters-partition-fct} and appendix~\ref{app:characters} for notations and details)
\begin{equation}\label{type-0-MM-Z}
Z^{\text{NS}}_{k} =
\frac12\sum_{l=0}^k\sum_{\substack{m=-l\\ m+l\ \text{even}}}^l  \left[
\chi^{\text{NS}}_{l,m}\bar
\chi^{\text{NS}}_{l,m}(q,z)+\chi^{\text{NS}}_{l,m}\bar \chi^{\text{NS}}_{l,m}(q,-z)\right]\ ,
\end{equation}
where $z=e^{2\pi i\nu}$.
We first analyse the partition function before taking the (GSO-like) projection, i.e.\ the corresponding trace is taken over the full supersymmetric Hilbert space~(comments about it in section~\ref{app:characters:sec:GSO}), 
\begin{equation}\label{redundant-Z-MM}
\begin{split} 
\mathcal{P}^{\text{NS}}_{k}  &= \sum_{l=0}^{k}\sum_{\substack{m=-l\\ m+l\ \text{even}}}^l \chi^{\text{NS}}_{l,m}\bar \chi^{\text{NS}}_{l,m}(q,z)\\
&=\left|\frac{\vartheta_3(\tau,\nu)}{\eta^3(\tau)}\right|^2\sum_{l=0}^{k}\sum_{m=-l}^l\left|q^{\frac{(l+1)^{2}-m^2}{4(k+2)}}
\Gamma^{(k)}_{lm}(\tau,\nu)\right|^2\ .
\end{split}
\end{equation}
For large level $k$ we expect a divergence due to the large number of almost chargeless fields~\cite{Fredenhagen:2012bw}, since for large~$n$ the fields tend to become chargeless (they approach the free theory limit of chapter~\ref{ch:free-limit}), and their contribution to the partition function hides the one due to the charged fields, which is our interest here. 
For this reason we insert the factor $(1-e^{2\pi irJ_{0}})$ in the trace, and arrive at (we set $\nu =0$ in
the following)
\begin{equation}\label{PintermsofI}
\mathcal{P}^{\text{NS},(r)}_{k} =\left|\frac{\vartheta_3(\tau,0)}{\eta^3(\tau)}\right|^2
\left[
\sum_{n=0}^{\lfloor\frac k2\rfloor}\mathcal I^{(r)}_{k,n}
\right]
\end{equation}
with
\begin{equation}\label{reg-MM}
\mathcal{I}^{(r)}_{k,n}:=2\sum_{m=1}^{k-2n}(q\bar
q)^{\frac{1}{k+2}\left(n+\frac{1}{2}\right)^{2}+\frac{m}{k+2}\left(n+\frac12\right)}
\left| \Gamma^{(k)}_{m+2n,m} (\tau ,0)\right|^2
\left(1-\cos \left( 2\pi r \tfrac{m}{k+2}\right)\right)\ .
\end{equation}
For large level $k$, the main contribution comes from small $n$ and large $m$: the regularisation factor $(1-\cos (\cdot))$ is small unless $m$ is of order $k$, while the exponent containing the conformal weight tells us that for large $m$ only small values of $n$ contribute significantly. 
In this limit, only one singular vector survives in $\Gamma^{(k)}_{lm}$ (the one present in the $c=3$ representations of type $I$ in appendix~\ref{app:characters}).  
Using the Euler-MacLaurin formula of equation~\eqref{Euler-Maclaurin-formula} to convert the sum over~$m$ into an integral over~$Q$, we obtain
\begin{align}
\mathcal{I}^{(r)}_{k,n} &\approx 2\sum_{m=1}^{k-2n}(q\bar
q)^{\frac{m}{k+2}\left(n+\frac12\right)}
\left| \frac{1-q^{m+2n+1}}{(1+q^{n+\frac12})(1+q^{m+n+\frac12})}\right|^2
\left(1-\cos \left( 2\pi r \frac{m}{k+2}\right)\right) \\
&\approx 2 (k+2) \int_{0}^{1} dQ\, (q\bar q)^{Q\left(n+\frac12\right)} 
\left| \frac{1}{(1+q^{n+\frac12})}\right|^2 
\left(1-\cos \left( 2\pi r Q \right)\right)\ .
\end{align}
Inserting this into~\eqref{PintermsofI} we find
\begin{equation}
\lim_{k\to\infty} \frac{1}{k+2} \mathcal{P}^{\text{NS},(r)}_{k} 
=\left|\frac{\vartheta_3(\tau,0)}{\eta^3(\tau)}\right|^2\sum_{n=0}^{\infty}\left|\frac{1}{1+q^{n+\frac12}}\right|^2\int ^1_{-1}dQ \ (q\bar q)^{|Q|(n+\frac12)}\left(1-e^{2\pi ir Q}\right)\ .
\end{equation}
The rescaling by a factor $1/(k+2)$ can be understood as follows: for a fixed $n$ and a given small interval $[Q,Q+\Delta Q]$ there are roughly $(k+2)\Delta Q$ ground states in the $k^{\text{th}}$ minimal model that contribute with approximately the same weight $(q\bar{q})^{|Q|(n+\frac{1}{2})}$. 
The rescaling thus corresponds to a  rescaling of the density of states to $1$ per unit interval $\Delta Q$.
It is now easy to write down the projected version of the regularised partition function (see appendix~\ref{app:characters:sec:GSO}), which reads in the NS sector
\begin{equation}\label{ch6:projected-part-fct-NS}
Z ^{\text{NS},(r)}=\frac12\left(
\left|\frac{\vartheta_3(\tau,0)}{\eta^3(\tau)}\right|^2+\left|\frac{\vartheta_4(\tau,0)}{\eta^3(\tau)}\right|^2\right)
\sum_{n=0}^{\infty}\left|\frac{1}{1+q^{n+\frac12}}\right|^2\int ^1_{-1}dQ \ (q\bar q)^{|Q|(n+\frac12)}\left(1-e^{2\pi ir Q}\right)\ .
\end{equation}

\section{Fields and correlators}\label{ch6:sec:fields-and-correlators}

We now want to become more precise about how the limit of the fields is taken.
We proceed in the spirit of chapter~\ref{ch:limit-theories}, defining averaged fields, and taking the limit at the end to get the correlators.
We focus here on the NS sector, the construction in the R sector is analogous. 
\paragraph{Averaged fields}
For the fields $\Phi_{Q,n}$ with $0<|Q|<1$ we proceed as follows. 
We first define averaged fields,
\begin{equation}
\Phi_{Q,n}^{\epsilon ,k} = \frac{1}{|N (Q,\epsilon ,k)|} \sum_{\substack{m \in N (Q,\epsilon ,k)\\ l = |m|+2n}}\phi_{l,m} \ ,
\end{equation} 
where the set $N (Q,\epsilon ,k)$ contains all labels $m$ such that the
corresponding charge $Q_{m}$ is close to $Q$, more precisely
\begin{equation}
N (Q,\epsilon ,k) = \left\{m \left| Q-\frac{\epsilon}{2}< -\frac{m}{k+2} <Q+\frac{\epsilon}{2}\right. \right\} \ .
\end{equation}
The cardinality of the set is
\begin{equation}
|N (Q,\epsilon ,k)| = \epsilon (k+2) + \mathcal{O} (1) \ .
\end{equation}
We assume that $\epsilon$ is small enough such that $|Q|\pm\frac{\epsilon}{2}$ is still between $0$ and $1$.

The correlator of fields in the limit theory is then defined (recall the various examples of chapter~\ref{ch:limit-theories}) as
\begin{equation}\label{defofcorrelators}
\langle \Phi_{Q_{1},n_{1}} (z_{1},\bar{z}_{1}) \dotsb \Phi_{Q_{r},n_{r}} (z_{r},\bar{z}_{r})\rangle= \lim_{\epsilon \to 0} \lim_{k\to \infty} \beta (k)^{2} \alpha (k)^{r}\langle \Phi_{Q_{1},n_{1}}^{\epsilon ,k}(z_{1},\bar{z}_{1})\dotsb \Phi_{Q_{r},n_{r}}^{\epsilon ,k} (z_{r},\bar{z}_{r})\rangle \ .
\end{equation}
We stress that also in this case, as in the computation of chapter~\ref{ch:free-limit}, we could allow $\alpha$ to depend also on the field labels $Q,n$. 
It turns out that this is not necessary here. 
The $k$-dependence of $\alpha$ and $\beta$ are determined such that we obtain finite correlators in the limit. 
Obviously we need at least two correlators with a different number of fields to determine the $k$-dependence of both factors $\alpha$ and $\beta$.

\section{Two-point function}\label{ch6:sec:two-point-fct}

Let us analyse now the two-point function. 
Two-point functions for the minimal models in the~$s=0$ sector are 
\begin{equation}
\langle \phi_{l_{1},m_{1}} (z_{1}) \phi_{l_{2},m_{2}} (z_{2})\rangle=\delta_{l_{1},l_{2}}\delta_{m_{1},-m_{2}}\frac{1}{|z_{1}-z_{2}|^{4h_{l_{1},m_{1}}}} \ .
\end{equation}
This becomes in the limit theory:
\begin{align}
\langle \Phi_{Q_{1},n_{1}} (z_{1},\bar{z}_{1}) \Phi_{Q_{2},n_{2}} (z_{2},\bar{z}_{2})\rangle
&= \lim_{\epsilon \to 0}\lim_{k\to \infty} \alpha (k)^{2}\beta (k)^{2}
\langle \Phi_{Q_{1},n_{1}}^{\epsilon ,k} (z_{1},\bar{z}_{1})
\Phi_{Q_{2},n_{2}}^{\epsilon ,k} (z_{2},\bar{z}_{2})\rangle  \\
&= \lim_{\epsilon \to 0}\lim_{k\to \infty} \frac{\alpha (k)^{2}\beta
(k)^{2}}{\epsilon^{2} (k+2)^{2}}
\sum_{m \in N (Q_{1},\epsilon ,k)\cap N (-Q_{2},\epsilon ,k)}
 \frac{\delta_{n_{1},n_{2}}}{|z_{1}-z_{2}|^{4h_{|m|+2n_{1},m}}}\ .
\end{align}
As explained in details in the analogous example of section~\ref{ch1:sec:free-boson-limit}, the conformal weight $h_{|m|+2n_{1},m}$ approaches $h_{n_{1}} (Q_{1}) =(2n_{1}+1) |Q_{1}|/2$ in the limit, and the sum over $m$ can be replaced by the cardinality of the overlap
\begin{equation}
|N (Q_{1},\epsilon ,k)\cap N (-Q_{2},\epsilon ,k)| = (k+2) (\epsilon-|Q_{1}+Q_{2}|) \theta (\epsilon-|Q_{1}+Q_{2}|) + \mathcal{O} (1) \ .
\end{equation}
In the limit $\epsilon \to 0$ we obtain a $\delta$-distribution,
\begin{equation}\label{delta}
\frac{\epsilon -|x|}{\epsilon^{2}}\, \theta (\epsilon -|x|)\, \to\, \delta(x)
\ . 
\end{equation}
With the choice 
\begin{equation}\label{ch6:alpha-beta}
\alpha (k)\beta (k)= \sqrt{k+2}
\end{equation}
to absorb the $k$-dependent pre-factor, we find the two-point function in a standard normalisation,
\begin{equation}
\langle \Phi_{Q_{1},n_{1}} (z_{1},\bar{z}_{1}) \Phi_{Q_{2},n_{2}} (z_{2},\bar{z}_{2})\rangle= \delta_{n_{1},n_{2}} \delta (Q_{1}+Q_{2})\frac{1}{|z_{1}-z_{2}|^{4h_{n_{1}} (Q_{1})}} \ .
\end{equation} 

\section{Three-point functions}\label{ch6:sec:three-point-fct}

\subsection{Correlators of primary fields}

As discussed in section~\ref{ch3:sec:correlators}, the correlators of three primary fields in minimal models  are closely related to the
three-point functions of the $\widehat{su} (2)_k$ Wess-Zumino-Witten model derived
in~\cite{Zamolodchikov:1986bd,Dotsenko:1990zb}. Similar methods allow
the computation of correlators involving superdescendants (as explicitly shown in~\cite[Appendix C]{Fredenhagen:2012rb}).
We rewrite here equation~\eqref{ch3:three-point-fct-minmod}: the correlator of three primary fields in the Neveu-Schwarz
sector in a model with diagonal spectrum reads
\begin{multline}
\langle\phi_{l_{1},m_{1}} (z_{1},\bar{z}_{1}) \phi_{l_{2},m_{2}}
(z_{2},\bar{z}_{2})\phi_{l_{3},m_{3}} (z_{3},\bar{z}_{3})\rangle\\
=C(\{l_{i},m_{i}\})\delta_{m_{1}+m_{2}+m_{3},0}|z_{12}|^{2(h_{3}-h_{1}-h_{2})}|z_{13}|^{2(h_{2}-h_{1}-h_{3})}|z_{23}|^{2(h_{1}-h_{2}-h_{3})}
\end{multline}
with 
\begin{equation}
C (\{l_{i},m_{i}\})=\begin{pmatrix}
\frac{l_{1}}{2} & \frac{l_{2}}{2} & \frac{l_{3}}{2}\\
\frac{m_{1}}{2} & \frac{m_{2}}{2}& \frac{m_{3}}{2}
\end{pmatrix}^{\!2}
\sqrt{(l_{1}+1)(l_{2}+1)(l_{3}+1)}\, d_{l_{1},l_{2},l_{3}}\ .
\end{equation}
Here, $\begin{pmatrix}
j_{1}&j_{2}&j_{3} \\
\mu_{1}&\mu_{2}&\mu_{3}
\end{pmatrix}$ denotes the Wigner 3j-symbols, and
$d_{l_{1},l_{2},l_{3}}$ is a product of Gamma functions,
\begin{equation}\label{ch6:d-function-gamma}
d_{l_{1},l_{2},l_{3}}^{2}
=\frac{\Gamma(1+\rho)}{\Gamma(1-\rho)}P^{2}(\tfrac{l_{1}+l_{2}+l_{3}+2}{2})
\prod_{k=1}^{3} \frac{\Gamma(1-\rho(l_{k}+1))}{\Gamma(1+\rho(l_{k}+1))}
\frac{P^{2}(\frac{l_{1}+l_{2}+l_{3}-2l_{k}}{2})}{P^{2}(l_{k})}
\end{equation}
with
\begin{equation}
\rho=\frac{1}{k+2}\quad ,\quad
P(l)=\prod_{j=1}^{l}\frac{\Gamma(1+j\rho)}{\Gamma(1-j\rho)} \ .
\end{equation}

We want to understand the limit\footnote{In~\cite{D'Appollonio:2003dr} a related limit of WZW models $\widehat{su} (2)_{k}$ has been considered.} of this expression when $k\to \infty$ while the labels $l_{i}$ and $m_{i}$ grow such that the conformal weight $h$ and the $U (1)$-charge $Q$ stay constant. 
In particular we have
\begin{equation}
l_{i} = |m_{i}| +2n_{i}\quad \text{and} \quad m_{i} = -Q_{(m_{i})}
(k+2)\ ,
\end{equation}
where $n_{i}$ is a fixed integer, and $Q_{(m_{i})}$ lies in an $\epsilon$-interval around $Q_{i}$, hence it stays approximately constant in the limit.

The Wigner 3j-symbols enforce the condition $m_{1}+m_{2}+m_{3}=0$ as well as $l_{i_{1}}\leq l_{i_{2}}+l_{i_{3}}$ for any permutation $\{i_{1},i_{2},i_{3} \}$ of $\{1,2,3 \}$. 
For definiteness we assume now that
\begin{equation}
m_{1},m_{2}\geq 0 \quad ,\quad m_{3}=-m_{1}-m_{2} \leq 0\ .
\end{equation}
For large $|m_{i}|$ the conditions on the $l_{i}$ translate into a single condition on the $n_{i}$,
\begin{equation}
l_{3}\leq l_{1}+l_{2} \Rightarrow n_{3}\leq n_{1}+n_{2} \ .
\end{equation}
When we consider the asymptotic behaviour of the three-point coefficient~\eqref{3ptcoefficient} for large $k$, there are two parts which have to be treated carefully. 
One is the Wigner 3j-symbol whose limits are discussed in appendix~\ref{app:wigner}. 
The other is the limit of the products of Gamma functions, where $P (l)$ becomes an infinite product when $l$ goes to infinity.
 
The infinite products of Gamma functions in the numerator and denominator cancel and leave a finite product in the limit. 
We skip the detailed computation here, suggesting the interested reader the discussion in~\cite[Section 3 and Appendix B]{Fredenhagen:2012rb}.
The outcome is that the coefficient $d_{l_{1},l_{2},l_{3}}$ behaves in the limit as
\begin{equation}\label{ch6:d-coefficient-limit}
d_{l_{1},l_{2},l_{3}} = \left(\prod_{j=1}^{3}\frac{\Gamma
(1+Q_{(m_{j})})}{\Gamma (1-Q_{(m_{j})})}
\right)^{\!-\frac{1}{2}\sum_{i=1}^{3}\sigma_{i} (2n_{i}+1)}\left(1+\mathcal{O} (\tfrac{1}{k}) \right) \ ,
\end{equation}
where $\sigma_{i}=\text{sgn} (Q_{i})$ denotes the sign of the corresponding charge.
In this form the expression is valid without any assumptions on which of the charges are positive or negative.

The asymptotic behaviour of the 3j-symbols in this regime is derived in section~\ref{app:wigner:sec:transition-region}. 
For $m_{i}$ linearly growing with $k$ and $m_{1},m_{2}\geq 0$, $m_{3}\leq 0$, it is given by (see~\eqref{app:3jasymptotic})
\begin{equation}
\begin{pmatrix}
\frac{|m_{1}|}{2}+n_{1} & \frac{|m_{2}|}{2}+n_{2} & \frac{|m_{3}|}{2}+n_{3}\\
\frac{m_{1}}{2} & \frac{m_{2}}{2} & \frac{m_{3}}{2}
\end{pmatrix} = (-1)^{m_{1}+n_{3}+n_{2}} \left(|m_{3}| \right)^{-1/2}
d^{J}_{M',M}
(\beta) \cdot (1+\mathcal{O} (\tfrac{1}{k}))\ ,
\end{equation}
where $d^{J}_{M',M} (\beta)$ is the Wigner $d$-matrix and 
\begin{equation}\label{betaJMM}
\cos \beta
=\frac{|m_{1}|-|m_{2}|}{|m_{1}|+|m_{2}|}\ , \
J=\frac{n_{1}+n_{2}}{2}\ ,\ M' =
-\frac{n_{1}+n_{2}}{2}+n_{3}\ ,\ M=\frac{n_{1}+n_{2}}{2}-n_{2}
\ .
\end{equation}

Putting everything together, the three-point coefficient $C(\{l_{i},m_{i}\})$ given in~\eqref{3ptcoefficient} has the limiting behaviour
\begin{equation}
C (\{l_{i},m_{i}\}) \sim (k+2)^{1/2} \mathcal{C} (\{Q_{(m_{i})},n_{i}\})\ ,
\end{equation}
where $\mathcal{C}$ is a smooth function of the charges $Q_{i}$. 
For $Q_{1},Q_{2}<0$ and $Q_{3}>0$ it is given by
\begin{equation}\label{Ccal}
\mathcal{C} (\{Q_{i},n_{i} \}) = \left( \frac{|Q_{1}Q_{2}|}{|Q_{3}|}\right)^{1/2}(d^{J}_{M',M}(\beta))^{2} \left(\prod_{j=1}^{3}\frac{\Gamma(1+Q_{j})}{\Gamma (1-Q_{j})}\right)^{\!n_{1}+n_{2}-n_{3}+\frac{1}{2}} \ ,
\end{equation}
with $\cos \beta = \frac{|Q_{1}|-|Q_{2}|}{|Q_{1}|+|Q_{2}|}$ and $J,M,M'$ given in~\eqref{betaJMM}. 
Notice that $\mathcal{C}$ in this case is non-zero only for $n_{1}+n_{2}\geq n_{3}$.

Now we are ready to work out the limit of the 3-point function. 
It is given by 
\begin{align}
& \langle \Phi_{Q_{1},n_{1}} (z_{1},\bar{z}_{1}) \Phi_{Q_{2},n_{2}} (z_{2},\bar{z}_{2}) \Phi_{Q_{3},n_{3}} (z_{3},\bar{z}_{3})\rangle\nonumber\\
&\qquad 
= \lim_{\epsilon \to 0} \lim_{k\to \infty} 
\beta (k)^{2} \alpha (k)^{3} \langle \Phi_{Q_{1},n_{1}}^{\epsilon ,k}
(z_{1},\bar{z}_{1})\Phi_{Q_{2},n_{2}}^{\epsilon ,k} (z_{2},\bar{z}_{2})
\Phi_{Q_{3},n_{3}}^{\epsilon ,k} (z_{3},\bar{z}_{3})\rangle \\
& \qquad = \lim_{\epsilon \to 0} \lim_{k\to \infty} \frac{\beta (k)^{2}
\alpha (k)^{3}}{\epsilon^{3} (k+2)^{3}} \sum_{\{ m_{i}\in N
(Q_{i},\epsilon ,k)\}} C (\{|m_{i}|+n_{i},m_{i} \})
\delta_{m_{1}+m_{2}+m_{3},0}\nonumber\\
&\qquad \qquad \times 
|z_{12}|^{2(h_{3}-h_{1}-h_{2})}|z_{13}|^{2(h_{2}-h_{1}-h_{3})}|z_{23}|^{2(h_{1}-h_{2}-h_{3})} 
\ .
\end{align}
As we have already discussed the limit of the coefficient~$C$, it only remains to determine the factor coming from the sum over the labels~$m_i$, namely given by the cardinality of the set
\begin{equation}
N_{123}= \{(m_{1},m_{2},m_{3}) \in N (Q_{1},\epsilon ,k) \times N (Q_{2},\epsilon ,k) \times N (Q_{3},\epsilon ,k):m_{1}+m_{2}+m_{3}=0
\} \ .
\end{equation}
This is almost identical to the computation discussed in section~\ref{ch1:sec:free-boson-limit}, and the result reads
\begin{equation}
|N_{123}| = (k+2)^{2}\epsilon^{3} \d_{\e}\left(\sum_{i}Q_{i}\right) +\mathcal{O} (k+2) \ ,
\end{equation}
where~$\d_{\e}$ is a~$\d$-family which approaches a $\d$-distribution as~$\e\to 0$.
By using the condition~\eqref{ch6:alpha-beta} we can absorb the remaining $k$-dependence by setting
\begin{equation}
\alpha (k) = (k+2)^{-1/2} \quad ,\quad \beta (k) = (k+2)  \ .
\end{equation}
The total result is then
\begin{multline}
\langle \Phi_{Q_{1},n_{1}} (z_{1},\bar{z}_{1}) \Phi_{Q_{2},n_{2}} (z_{2},\bar{z}_{2}) \Phi_{Q_{3},n_{3}} (z_{3},\bar{z}_{3})\rangle \\
= \mathcal{C} (\{Q_{i},n_{i} \}) \delta ({\textstyle \sum_{i}Q_{i}})
|z_{12}|^{2(h_{3}-h_{1}-h_{2})}|z_{13}|^{2(h_{2}-h_{1}-h_{3})}|z_{23}|^{2(h_{1}-h_{2}-h_{3})} 
\end{multline}
with $\mathcal{C}$ given in~\eqref{Ccal}. 

\subsection{Correlators involving superdescendants}

Now we want to show that also the three-point function of two
primaries and one superdescendant (which corresponds to the odd fusion
channel~\cite{Mussardo:1988av}) has a well-defined limit. We will
limit ourselves to the case of a superdescendant obtained by acting
with $G^{+}$, the discussion for $G^{-}$-descendants is analogous. As derived in detail in~\cite[Appendix C]{Fredenhagen:2012rb}, such a correlator is given by
\begin{align}
& \langle (\bar{G}^{+}_{-\frac{1}{2}}G^{+}_{-\frac{1}{2}}\phi_{l_{1},m_{1}}) (z_{1},\bar{z}_{1})
\phi_{l_{2},m_{2}} (z_{2},\bar{z}_{2})\phi_{l_{3},m_{3}}
(z_{3},\bar{z}_{3}) \rangle \nonumber\\
&\quad = \frac{k+2}{2 (n_{1}+1) (l_{1}-n_{1})}
\begin{pmatrix} \frac{l_{2}+m_{2}}{2} \\ \frac{l_{1}-m_{1}}{2}+1 \end{pmatrix}
\begin{pmatrix} \frac{l_{1}+l_{2}-m_{1}-m_{2}}{2}+1 \\ \frac{l_{1}-m_{1}}{2}+1\end{pmatrix}
\begin{pmatrix} l_{1} \\ \frac{l_{1}-m_{1}}{2}+1 \end{pmatrix}^{\!-1}\nonumber\\
&\qquad  \times \begin{pmatrix}
\frac{k-l_{1}}{2} & \frac{l_{2}}{2} & \frac{l_{3}}{2} \\
-\frac{k-l_{1}}{2} & \frac{m_{1}+m_{2}-l_{1}}{2}-1 & \frac{m_{3}}{2}
\end{pmatrix}^{\!2}
\sqrt{(k-l_{1}+1) (l_{2}+1) (l_{3}+1)}\, 
d_{k-l_{1},l_{2},l_{3}}\nonumber\\
&\qquad  \times 
|z_{12}|^{2 (h_{l_{3},m_{3}}-(h_{l_{1},m_{1}}+1/2)-h_{l_{2},m_{2}})} 
|z_{23}|^{2 ((h_{l_{1},m_{1}}+1/2)-h_{l_{2},m_{2}}-h_{l_{3},m_{3}})}\nonumber\\
&\qquad \times 
|z_{13}|^{2 (h_{l_{2},m_{2}}- (h_{l_{1},m_{1}}+1/2)-h_{l_{3},m_{3}})}\ ,
\label{oddcorrelator}
\end{align}
where $l_{i}\geq |m_{i}|$ and we assume that $m_{1},m_{2}>0$ and
$m_{3}<0$.  

To determine the limit we first simplify the prefactor (that we call $A$)
in~\eqref{oddcorrelator} by expressing the 3j-symbol with the help
of~\eqref{app:extremal3j},
\begin{align}
A &=  \frac{k+2}{2 (n_{1}+1) (l_{1}-n_{1})}
\begin{pmatrix} \frac{l_{2}+m_{2}}{2} \\ \frac{l_{1}-m_{1}}{2}+1 \end{pmatrix}
\begin{pmatrix} \frac{l_{1}+l_{2}-m_{1}-m_{2}}{2}+1 \\ \frac{l_{1}-m_{1}}{2}+1\end{pmatrix}
\begin{pmatrix} l_{1} \\ \frac{l_{1}-m_{1}}{2}+1 \end{pmatrix}^{\!-1}\nonumber\\
&\qquad  \times \begin{pmatrix}
\frac{k-l_{1}}{2} & \frac{l_{2}}{2} & \frac{l_{3}}{2} \\
-\frac{k-l_{1}}{2} & \frac{m_{1}+m_{2}-l_{1}}{2}-1 & \frac{m_{3}}{2}
\end{pmatrix}^{\!2} \nonumber\\
& = \frac{k+2}{2 (n_{1}+1) (l_{1}-n_{1})}\, \frac{(\frac{l_{1}+m_{1}}{2}-1)!}{l_{1}!(\frac{l_{1}-m_{1}}{2}+1)!} 
\nonumber\\
&\qquad \times \frac{(\frac{-k+l_{1}+l_{2}+l_{3}}{2})!(\frac{l_{3}+m_{3}}{2})!(\frac{l_{2}+m_{2}}{2})!(k-l_{1})!}{(\frac{k-l_{1}+l_{2}+l_{3}}{2}+1)!(\frac{k-l_{1}-l_{2}+l_{3}}{2})!(\frac{k-l_{1}+l_{2}-l_{3}}{2})!(\frac{l_{3}-m_{3}}{2})!(\frac{l_{2}-m_{2}}{2})!}
\ .
\end{align}
In the limit we set $l_{i}=|m_{i}|+2n_{i}$ where the $n_{i}$ are kept constant, and the $m_{i}$ are sent to infinity growing linearly in $k$. By using~\eqref{limitoffactorials} we get
\begin{align}
A & = \frac{k+2}{2 (n_{1}+1) (m_{1}+n_{1})} \,\frac{n_{3}!}{(n_{1}+1)!(n_{2})!(-n_{1}-n_{2}+n_{3}-1)!}\nonumber\\
&\quad \times \frac{(m_{1}+n_{1}-1)!(k-m_{1}-2n_{1})!(m_{2}+n_{2})!(-m_{3}+n_{1}+n_{2}+n_{3}+1)!}{(m_{1}+2n_{1})!(k-m_{1}-n_{1}+n_{2}-n_{3}+1)!(m_{2}-n_{1}+n_{2}+n_{3})!(-m_{3}+n_{3})!}\nonumber\\
& = \frac{1}{2 (k+2) (n_{1}+1)}\, \frac{n_{3}!}{ (n_{1}+1)!n_{2}!(-n_{1}-n_{2}+n_{3}-1)!}\nonumber\\
& \qquad \times
|Q_{1}|^{-n_{1}-2}|1+Q_{1}|^{-n_{1}-n_{2}+n_{3}-1}|Q_{2}|^{n_{1}-n_{3}}|Q_{3}|^{n_{1}+n_{2}+1}\left(
1+\mathcal{O} (\tfrac{1}{k})\right)\ ,
\end{align}
where $Q_{i}=-\frac{m_{i}}{k+2}$ is kept fixed in the limit. 
The asymptotic form of $d_{k-l_{1},l_{2},l_{3}}$ can be evaluated by the same methods explained in ~\cite[Appendix B]{Fredenhagen:2012rb} to obtain equation~\eqref{ch6:d-coefficient-limit}.
The result reads
\begin{multline}
d_{k-l_{1},l_{2},l_{3}} =\left(\frac{\Gamma (1+|1+Q_{1}|)\Gamma
(1-|Q_{2}|)\Gamma (1+|Q_{3}|)}{\Gamma (1-|1+Q_{1}|)\Gamma
(1+|Q_{2}|)\Gamma (1-|Q_{3}|)}
\right)^{\!n_{1}+n_{2}-n_{3}+\frac{1}{2}} \cdot \left(1+\mathcal{O}
(\tfrac{1}{k+2}) \right)\ .
\end{multline}
The final result for the three-point correlator of two primaries and one superdescendant in the limit theory is then given by
\begin{align}
&\langle
(G^{+}_{-\frac{1}{2}}\bar{G}^{+}_{-\frac{1}{2}}\Phi_{Q_{1},n_{1}})
(z_{1},\bar{z}_{1}) \Phi_{Q_{2},n_{2}}
(z_{2},\bar{z}_{2})\Phi_{Q_{3},n_{3}} (z_{3},\bar{z}_{3})\rangle = \nonumber\\
&\quad = \frac{1}{2 (n_{1}+1)}\,
\frac{n_{3}!}{(n_{1}+1)!n_{2}!(n_{3}-n_{1}-n_{2}-1)!}
|1+Q_{1}|^{-n_{1}+n_{2}+n_{3}-\frac{1}{2}}
|Q_{2}|^{n_{1}-n_{3}+\frac{1}{2}}|Q_{3}|^{n_{1}+n_{2}+\frac{3}{2}}\nonumber\\
& \qquad \times |Q_{1}|^{-n_{1}-2}\left(\frac{\Gamma (1+|1+Q_{1}|)\Gamma
(1-|Q_{2}|)\Gamma (1+|Q_{3}|)}{\Gamma (1-|1+Q_{1}|)\Gamma
(1+|Q_{2}|)\Gamma (1-|Q_{3}|)} \right)^{\!n_{1}+n_{2}-n_{3}+\frac{1}{2}}\nonumber\\
& \qquad \times \delta (1+Q_{1}+Q_{2}+Q_{3}) |z_{12}|^{2
(h_{3}-h_{1}-\frac{1}{2}-h_{2})}|z_{13}|^{2
(h_{2}-h_{1}-\frac{1}{2}-h_{3})}|z_{23}|^{2
(h_{1}+\frac{1}{2}-h_{2}-h_{3})} \ ,
\end{align}
where we assumed that $Q_{1},Q_{2}<0$ and $Q_{3}>0$. The
generalisation to other cases is straightforward. As in a
superconformal theory all three-point functions are determined if
the three-point correlators of three primaries and the correlators of
two primaries and one superdescendant are given, this result shows that all
three-point functions of the limit theory are well-defined.

\section{Boundary conditions and one-point functions}\label{ch6:sec:BC}

In this section we investigate the limit of bulk one-point functions on the upper half plane. 
As stressed in section~\ref{ch3:sec:BC} and in subsection~\ref{ch3:sec:geometry:subsec:branes}, in a diagonal model only chargeless fields can couple to B-type boundary conditions.
In the limit theory analysed in this chapter, the uncharged ground state is outnumbered by the (continuous) spectrum of the charged ones, so that we do not find any B-type boundary conditions in this limit theory. 
On the other hand we find two families of A-type boundary conditions in the limit.

Recall (subsection~\ref{ch3:sec:BC}) that A-type boundary conditions are characterised by the following one-point functions
\begin{equation}
\langle \phi_{l,m} (z,\bar{z}) \rangle_{(L,M,S)} = 
(k+2)^{-1/2} \frac{\sin \frac{\pi (l+1) (L+1)}{k+2}}{\sqrt{\sin
\frac{\pi (l+1)}{k+2}}} e^{\pi i \frac{mM}{k+2}} 
|z-\overline{z}|^{-2 h_{l,m}}\ . 
\end{equation}
When we take the limit $k\to \infty$ we have some freedom of what to do with the boundary labels. 
There are two natural choices: either we keep the boundary labels constant in the limit, or we scale them in the same way as we scale the field labels. 
Both lead to sensible expressions.

We have already discussed in subsection~\ref{ch4:sec:branes:subs:contorbi} about the geometry described by these two different ways of scaling the boundary labels.
In the T-bell geometry these boundary conditions correspond to discs centred in the origin; if the boundary labels are held constant in the limit, the discs become very small and approach the orbifold singularity.
These~\textit{discrete} boundary conditions are associated to fractional branes in the continuous orbifold picture (as confirmed by detailled computations in chapter~\ref{ch:our-contorbi}), naturally coming in discrete families parameterised by the irreducible finite representations of the orbifold group (see the introduction to these matters in chapter~\ref{ch:orbifolds}).
In contrast to that, if we appropriately scale the boundary labels, the discs preserve finite radius, and since the region of the T-bell of our concern in the limit is the one close to the centre, they become space-filling branes in the limit.
The latter are characterised by a continuous electric field, thus come naturally in a~\textit{continuous} family.

\subsection{Discrete boundary conditions}\label{ch6:sec:BC:subs:discrete}

First we will take the limit such that the boundary labels are kept
fixed. The one-point function in the limit is then
\begin{align}
\langle \Phi_{Q,n} (z,\bar{z})\rangle_{(L,M,S)} & = \lim_{\epsilon
\to 0} \lim_{k\to\infty} \alpha (k)\beta (k) \langle
\Phi_{Q,n}^{\epsilon ,k} (z,\bar{z}) \rangle_{(L,M,S)} \nonumber\\
& =  \lim_{\epsilon \to 0} \lim_{k\to\infty} \frac{\alpha (k)\beta
(k)}{\epsilon (k+2)^{\frac{3}{2}}} \sum_{m\in N (Q,\epsilon ,k)} 
\frac{\sin \frac{\pi (|m|+2n+1) (L+1)}{k+2}}{\sqrt{\sin
\frac{\pi (|m|+2n+1)}{k+2}}} e^{\pi i \frac{mM}{k+2}} 
|z-\overline{z}|^{-2 h_{|m|+2n,m}}\nonumber\\
& = \frac{\sin (\pi |Q| (L+1))}{\sqrt{\sin (\pi |Q|)}} e^{-\pi i QM} 
|z-\overline{z}|^{-2 h_{n} (Q)} \ . 
\end{align}
For the Ramond fields one finds
\begin{align}
\langle \Psi^{0}_{Q} (z,\bar{z}) \rangle_{(L,M,S)} &= 
 \frac{\sin (\pi |\frac{1}{2}-Q| (L+1))}{\sqrt{\sin (\pi
 |\frac{1}{2}-Q|)}} e^{\pi i (\frac{1}{2}-Q) M}e^{-\pi i\frac{S}{2}}  
|z-\overline{z}|^{-1/4}\\
\langle \Psi^{\pm}_{Q,n} (z,\bar{z}) \rangle_{(L,M,S)} &= 
 \frac{\sin (\pi |\frac{1}{2}\mp Q| (L+1))}{\sqrt{\sin (\pi
 |\frac{1}{2}\mp Q|)}} e^{\pi i (\pm\frac{1}{2}-Q) M}e^{\mp \pi i\frac{S}{2}}  
|z-\overline{z}|^{-2 h^{\pm}_{n} (Q)} \ .
\end{align}
These boundary conditions are not independent. Using the trigonometric
identity
\begin{equation}
\sin \left(\pi |Q| (L+1) \right) = \sin \left(\pi |Q| \right)
\sum_{j=0}^{L} e^{i\pi Q (L-2j)} \ ,
\end{equation}
we see that
\begin{equation}
\langle \ \cdot\ \rangle_{(L,M,S)} = \sum_{j=0}^{L} \langle
\ \cdot\ \rangle_{(0,M+L-2j,S)} \ .
\end{equation}
All boundary conditions are therefore superpositions of boundary conditions with $L=0$, and the elementary boundary conditions are $(0,M,S)$. 
As already noticed under equation~\eqref{ch5:one-point-fct-zero-dim-BC}, this can be compared to the situation in minimal models before taking the limit, where all boundary conditions can be obtained by boundary renormalisation group flows from superpositions of those with $L=0$~\cite{Fredenhagen:2001nc,Maldacena:2001ky}. 
These flows become shorter when the level $k$ grows, and in the limit the boundary conditions can be identified.

\subsection{Continuous boundary conditions}\label{ch6:sec:BC:subs:continuous}

Now we will scale the boundary labels in the same way as we did for
the field labels. We introduce a continuous parameter $\cQ$, $0<|\cQ|<1$,
and a discrete parameter $N\in\mathbb{N}_{0}$, and instead of
considering fixed boundary labels in the limit, we consider a sequence
of boundary conditions $B_{k} (\cQ,N)$ of the form
\begin{equation}
B_{k}(\cQ,N) = (|\lfloor -\cQ (k+2) \rfloor|+2N,\lfloor -\cQ
(k+2)\rfloor,0) \ ,
\end{equation}
where $\lfloor x \rfloor$ denotes the greatest integer smaller or
equal to $x$.
The one-point function in the limit is then
\begin{align}
\langle \Phi_{Q,n} (z,\bar{z})\rangle_{(\cQ,N)} & = \lim_{\epsilon
\to 0} \lim_{k\to\infty} \alpha (k)\beta (k) \langle
\Phi_{Q,n}^{\epsilon ,k} (z,\bar{z}) \rangle_{B_{k} (\cQ,N)} \\
& =  \lim_{\epsilon \to 0} \lim_{k\to\infty} \frac{\alpha (k)\beta
(k)}{\epsilon (k+2)^{\frac{3}{2}}} \sum_{m\in N (Q,\epsilon ,k)} 
\frac{\sin \frac{\pi (|m|+2n+1) (|\lfloor -\cQ (k+2)\rfloor|+2N+1)}{k+2}}{\sqrt{\sin
\frac{\pi (|m|+2n+1)}{k+2}}} \nonumber \\
&\qquad \times e^{\pi i \frac{m \lfloor -\cQ (k+2)\rfloor}{k+2}} 
|z-\overline{z}|^{-2 h_{|m|+2n,m}} \ .
\end{align}
We observe that the arguments of the sine function in the numerator
and of the exponential diverge when $k$ is sent to infinity, so that
we get strongly oscillating expressions. Their combination behaves as
\begin{align}
& 2i\sin \frac{\pi (|m|+2n+1) (|\lfloor -\cQ (k+2)\rfloor|+2N+1)}{k+2} 
e^{\pi i \frac{m \lfloor -\cQ (k+2)\rfloor}{k+2}}\nonumber\\
& \sim\left(e^{i\frac{\pi |m||\lfloor -\cQ
(k+2)\rfloor|}{k+2}}e^{i\frac{\pi [|m| (2N+1)+ (2n+1)|\lfloor -\cQ
(k+2)\rfloor|]}{k+2}} - e^{-i\frac{\pi |m||\lfloor -\cQ
(k+2)\rfloor|}{k+2}}e^{-i\frac{\pi [|m| (2N+1)+ (2n+1)|\lfloor -\cQ
(k+2)\rfloor|]}{k+2}}\right) \nonumber \\
& \quad \times e^{i \frac{\pi m \lfloor -\cQ (k+2)\rfloor}{k+2}}\\
& \sim \left\{ \begin{array}{ll}
\left(e^{2i\frac{\pi |m||\lfloor -\cQ
(k+2)\rfloor|}{k+2}}e^{i\pi [|Q| (2N+1)+ (2n+1)|\cQ|]} - e^{-i\pi [|Q| (2N+1)+ (2n+1)|\cQ|]}\right) & \text{for} \ Q\cQ>0\\[3mm]
\left(e^{i\pi [|Q| (2N+1)+ (2n+1)|\cQ|]} - e^{-2i\frac{\pi |m||\lfloor -\cQ
(k+2)\rfloor|}{k+2}}e^{-i\pi [|Q| (2N+1)+ (2n+1)|\cQ|]}\right) & \text{for} \ Q\cQ<0 \ .
\end{array} \right. 
\end{align}
Upon taking the average over $m$ the strongly oscillating term is
suppressed, and in the limit only the other term survives. The final
result is therefore
\begin{equation}
\langle \Phi_{Q,n} (z,\bar{z})\rangle_{(\cQ,N)} = 
\frac{1}{2i\sqrt{\sin (\pi |Q|)}}|z-\bar{z}|^{-2 h_{n} (Q)}\times \left\{\begin{array}{ll}
- e^{-i\pi [|Q| (2N+1)+ (2n+1)|\cQ|]} & \text{for} \ Q\cQ>0\\[2mm]
e^{i\pi [|Q| (2N+1)+ (2n+1)|\cQ|]} & \text{for} \ Q\cQ<0\ .
\end{array} \right.
\end{equation}
Similarly, in the Ramond sector we find
\begin{align}
\langle \Psi^{0}_{Q} (z,\bar{z})\rangle_{(\cQ,N)} &= 
\frac{e^{-\pi i\frac{S}{2}}}{2i\sqrt{\sin (\pi |\frac{1}{2}-Q|)}}|z-\bar{z}|^{-1/4}\times \left\{\begin{array}{ll}
- e^{-i\pi|\frac{1}{2}-Q| (2N+1)} & \text{for} \ \cQ>0\\[2mm]
e^{i\pi |\frac{1}{2}-Q| (2N+1)} & \text{for} \ \cQ<0 
\end{array} \right. \\
\langle \Psi^{\pm}_{Q,n} (z,\bar{z})\rangle_{(\cQ,N)} &= 
\frac{e^{\mp \pi i\frac{S}{2}}|z-\bar{z}|^{-2 h^{\pm}_{n} (Q)}}{2i\sqrt{\sin (\pi |Q\mp\frac{1}{2}|)}}\times \left\{\begin{array}{ll}
- e^{-i\pi [|Q\mp \frac{1}{2}| (2N+1)+ 2n|\cQ|]} & \text{for} \ (Q\mp
\frac{1}{2})\cQ>0\\[2mm]
e^{i\pi [|Q\mp \frac{1}{2}| (2N+1)+ 2n|\cQ|]} & \text{for} \ (Q\mp
\frac{1}{2}) \cQ<0\ .
\end{array} \right.
\end{align}


\chapter{Continuous orbifold interpretation}\label{ch:our-contorbi}

In chapter~\ref{ch:new-theory}, along the lines of~\cite{Fredenhagen:2012rb}, we have constructed a limit of minimal models where the field labels $l$ and $m$ are sent to infinity such that both the conformal weight and the $U(1)$ charge are kept fixed. 
The resulting theory contains a spectrum of primary fields that is continuous in the $U(1)$ charge. 
As discussed in chapter~\ref{ch:geometry} the geometry of the limit suggests that this new $c=3$~theory can be interpreted as a continuous orbifold (structure presented in section~\ref{ch2:sec:contorbi}) of a free~CFT.
In particular we have conjectured that the right formulation should be given in terms of an~$N=2$ supersymmetric theory of two uncompactified bosons and two 
fermions orbifolded by the rotation group~$SO (2)\simeq U (1)$. 
The~$U(1)$ charge of the parent theory serves as a twist parameter.

\section{The orbifold}\label{ch7:sec:orbifold}

Notations and conventions follow closely the ones in~\cite{Gaberdiel:2004nv}.  
We start by defining the real bosonic coordinates $X^1(z,\bar z),X^2(z,\bar z)$ and their fermionic counterparts $\psi^1(z,\bar z),\psi^2(z,\bar z)$. 
We rearrange the fields to work on the complex plane with one free complex fermion, namely defining
\begin{subequations}
\begin{align}
\phi &=\tfrac{1}{\sqrt 2}(X^1+iX^2) & \phi^* &=\tfrac{1}{\sqrt 2}(X^1-iX^2)\\[4pt]
\psi &=\tfrac{1}{\sqrt 2}(\psi^1+i\psi^2) & \psi^* &=\tfrac{1}{\sqrt 2}(\psi^1-i\psi^2)
\ ,
\end{align}
\end{subequations}
such that the mode expansion of the (holomorphic) fields reads
\begin{subequations}
\begin{align}
\de\phi &=-i\sum_{\substack{m\in\Z}}\alpha_m z^{-m-1} & \de\phi^* &=-i\sum_{\substack{m\in\Z}}\alpha^*_m z^{-m-1}\\
\psi &=\sum_{\substack{r\in\Z+\eta}}\psi_r z^{-r-\frac{1}{2}} & \psi^*&=\sum_{\substack{r\in\Z+\eta}}\psi^*_r z^{-r-\frac{1}{2}}\ ,
\end{align}
\end{subequations}
where $\eta=0,\frac12$ in the Ramond and Neveu-Schwarz sector respectively. 
The antiholomorphic case is analogous. 
For simplicity we will restrict the following discussion to the Neveu-Schwarz sector. 
The modes respect the algebra of one free complex boson and one free Neveu-Schwarz complex fermion:
\begin{subequations}
\begin{align}  
  [\alpha_m, \alpha^*_n] &= m\,\delta_{m,-n}  &  \{\psi_r,\psi^*_s\} &= \delta_{r,-s}\\[4pt] 
  [\alpha_m,\alpha_n] &= [\alpha^*_m, \alpha^*_n] = 0 & \{\psi_r,\psi_s\}&=\{\psi^*_r,\psi^*_s\}=0 \ .
\end{align}
\end{subequations}
We can explicitly realise the $N =2$ superconformal algebra by defining the generators through our holomorphic fields as 
\begin{subequations}\label{SUSY-generators}
\begin{align}
T&=-\de\phi\de\phi^*-\frac12(\psi^*\de\psi+\psi\de\psi^*) & J&=-\psi^*\psi\\
G^+&=i\sqrt 2\,\psi\de\phi^* & G^-&=i\sqrt 2\,\psi^*\de\phi \ ,
\end{align}
\end{subequations}
and similarly for their antiholomorphic counterparts.
We write here for reference the explicit realisation of the zero mode of the energy momentum tensor
\begin{equation}\label{ch7:L0-realisation}
L_0 = \sum_{m} :\alpha_{-m}\alpha^*_{m}: +
    \sum_{s}s\ :\psi^{*}_{-s}\psi_s:\ .
\end{equation}

We want to end up with an $N=2$ theory; we therefore choose the action of the orbifold group in such a way that the currents in~\eqref{SUSY-generators} are invariant under the transformation and supersymmetry is not broken. 
In particular we choose the $U(1)$ action on the fields as follows
\begin{subequations}
\begin{align}
U(\theta)\cdot\phi&=e^{i\theta}\phi & U(\theta)\cdot\phi^*&=e^{-i\theta}\phi^* \\[4pt]
U(\theta)\cdot\psi&=e^{i\theta}\psi & U(\theta)\cdot\psi^*&=e^{-i\theta}\psi^*\ ,
\end{align}
\end{subequations}
so that in terms of the coordinates $X^1,X^2$ on the plane it is realised by the rotation matrix
\begin{align}
U(\theta)\cdot\vec X\equiv\mathcal{R}_{\theta}\cdot \vec X=\left(
\begin{array}{cc}
\cos{\theta}&-\sin{\theta}\\
\sin{\theta}& \cos{\theta}
\end{array}
\right)\cdot\left(
\begin{array}{c}
X^1\\
X^2
\end{array}
\right)\ .
\end{align}
The action of the group on the field modes is thus
\begin{subequations}
\begin{align}
\alpha_n&\mapsto e^{i\theta}\alpha_n & \alpha^*_n&\mapsto e^{-i\theta}\alpha^*_n\\[4pt]
\psi_r&\mapsto e^{i\theta}\psi_r & \psi^*_r&\mapsto e^{-i\theta}\psi^*_r\ .
\end{align}
\end{subequations}

\section{Partition function}\label{ch7:sec:partition-fct}

We now want to determine the partition function of the orbifold. 
We first look at the Neveu-Schwarz part, and work with the full supersymmetric Hilbert space. 
To compare with the minimal models we will later perform a (GSO-like) projection by $\frac{1}{2} (1+(-1)^{F+\tilde{F}})$ onto states of even fermion number (see section~\ref{app:characters:sec:GSO}).

By inserting a twist operator we obtain the $\theta$-twined characters
\begin{equation}\label{th0-comp}
\torus{U(\theta)}{\id}\,=
\Tr_{\cH_{0}^{\text{NS}}}\left(
U(\theta)q^{L_0-\frac18}\bar{q}^{\bar{L}_0-\frac18}\right) \ ,
\end{equation}
where we denoted by $\mathcal{H}_{0}^{\text{NS}}$ the (unprojected) Neveu-Schwarz part of the Hilbert space of the parent free theory (in harmony with the conventions summarised in section~\ref{ch2:sec-app:notations}).

The orbifold group acts non-trivially on the vacua labelled by the momentum on the plane,
\begin{equation}
|\vec{p}\ \rangle\ \longmapsto\ |\mathcal{R}_{\theta}\cdot\vec{p}\ \rangle\ ,
\end{equation}
so that the momentum dependent part of equation~\eqref{th0-comp} becomes
\begin{equation}\label{int-zero-modes}
\int d^2p\  \delta^2(\mathcal{R}_{\theta}\cdot \vec p-\vec p)\ (q\bar{q})^{\frac{|\vec{p}|^2}{4}}=\int d^2p\ \frac{1}{\det({\mathcal{R}_{\theta}-1)}} \delta^2(\vec p)\ (q\bar{q})^{\frac{|\vec{p}|^2}{4}} \ .
\end{equation}
The momentum independent part can be computed as follows: looking only at the holomorphic sector, the sum over $m$ in the realisation of $L_0$ of equation~\eqref{ch7:L0-realisation}, being at the exponent, can be brought down as an infinite product; starting from $m=1$, the trace is taken over the space of states generated by a single mode creation operator, in formulas
\begin{equation}
\prod_{m=1}^{\infty}(1+e^{i\theta}q^m+e^{2i\theta}q^{2m}+\dots)(1+e^{-i\theta}q^m+e^{-2i\theta}q^{2m}+\dots)=\prod_{m=1}^{\infty}\frac{1}{1-q^{m}e^{i\theta}}\frac{1}{1-q^{m}e^{-i\theta}}\ .
\end{equation}
The fermionic NS part, taking into account the right periodicity, gives us
\begin{equation}
\prod_{n=1}^{\infty}(1+q^{n-\frac12}e^{i\theta})(1+q^{n-\frac12}e^{-i\theta})\ .
\end{equation}
Collecting all the pieces together, the $\theta$-twined character is then
\begin{align}\label{th0-c}
\torus{U(\theta)}{\id}\,&=
\Tr_{\mathcal{H}_{0}^{\text{NS}}}\left( U(\theta)q^{L_0-\frac18}\bar{q}^{\bar{L}_0-\frac18}\right)\nonumber\\
&=\int d^2p\  \frac{\delta^2(\vec p)}{\det({\mathcal{R}_{\theta}-1)}}\ (q\bar q)^{\frac{|\vec{p}|^2}{4}}\left|q^{-\frac18}\prod_{n=0}^{\infty}\frac{(1+q^{n+\frac12}e^{i\theta})(1+q^{n+\frac12}e^{-i\theta})}{(1-q^{n+1}e^{i\theta})(1-q^{n+1}e^{-i\theta})}\right|^2\nonumber\\
&=\left|\frac{\vartheta_3(\tau,\frac{\theta}{2\pi})}{\vartheta_1(\tau,\frac{\theta}{2\pi})}\right|^2\ .
\end{align}
We then act with a modular S-transformation on the complex modulus of the torus ($\tau\mapsto -\frac{1}{\tau}$) to get from the $\theta$-twined untwisted character to the character of the $\theta$-twisted sector,
\begin{align}
\torus {U(\theta)}{\id}\quad\overset{S}{\longmapsto}\quad\torus{\id}{U(\theta)}\ .
\end{align}
We can benefit from known transformation properties of the $\vartheta$-functions, summarised in appendix~\ref{app:characters}, in particular
\begin{equation}\label{modular-thetas}
\frac{\vartheta_3(-\frac{1}{\tau},\nu)}{\vartheta_1(-\frac{1}{\tau},\nu)}=i\frac{\vartheta_3(\tau,\nu\tau)}{\vartheta_1(\tau,\nu\tau)}\ ,
\end{equation}
so that the $\theta$-twisted sector reads
\begin{align}
\torus{\id}{U(\theta)}\,=&\,\Tr_{\mathcal{H}_{\theta}^{\text{NS}}}\left( q^{L_0-\frac18}\bar{q}^{\bar{L}_0-\frac18}\right)
=\left|\frac{\vartheta_3(\tau,\frac{\tau\theta}{2\pi})}{\vartheta_1(\tau,\frac{\tau\theta}{2\pi})}\right|^2\\
=&\left|q^{-\frac18+\frac{\theta}{4\pi}}
\prod_{n=0}^{\infty}\frac{(1+q^{n+\frac12+\frac{\theta}{2\pi}})(1+q^{n+\frac12-\frac{\theta}{2\pi}})}{(1-q^{n+\frac{\theta}{2\pi}})(1-q^{n+1-\frac{\theta}{2\pi}})}\right|^2
\ .
\end{align}
We can now get the $\theta'$-twined character over the $\theta$-twisted sector by acting once more with the orbifold group on the modes. 
We get the following:
\begin{align}\label{thpth}
\torus{U(\theta')}{U(\theta)}\,&=\Tr_{\mathcal{H}_{\theta}^{\text{NS}}}\left( U(\theta')q^{L_0-\frac18}\bar{q}^{\bar{L}_0-\frac18}\right)\nonumber\\
&=\left|q^{-\frac18+\frac{\theta}{4\pi}}\prod_{n=0}^{\infty}\frac{(1+q^{n+\frac12+\frac{\theta}{2\pi}}e^{i\theta'})(1+q^{n+\frac12-\frac{\theta}{2\pi}}e^{-i\theta'})}{(1-q^{n+\frac{\theta}{2\pi}}e^{i\theta'})(1-q^{n+1-\frac{\theta}{2\pi}}e^{-i\theta'})}\right|^2\nonumber\\
&=\left|\frac{\vartheta_3(\tau,\frac{\tau\theta+\theta'}{2\pi})}{\vartheta_1(\tau,\frac{\tau\theta+\theta'}{2\pi})}\right|^2\ ,
\end{align}
which is the expression we are interested in.

The contribution of a $\theta$-twisted sector to the unprojected partition function is therefore obtained by integrating equation~\eqref{thpth} over the twisting parameter~$\theta'$,
\begin{align}
\mathcal{P}_{\theta-\text{tw}}^{\text{NS}}&=\frac{1}{2\pi}\int_{0}^{2\pi}d\theta'\,\torus{U(\theta')}{U(\theta)}= \int_{0}^{2\pi}\frac{d\theta'}{2\pi}\ \Tr_{\mathcal{H}_{\theta}^{\text{NS}}}\left(U(\theta')q^{L_0-\frac18}\bar{q}^{\bar{L}_0-\frac18} \right)\\
&=\int_{0}^{2\pi}\frac{d\theta'}{2\pi}\left|\frac{\vartheta_3(\tau,\frac{\tau\theta+\theta'}{2\pi})}{\vartheta_1(\tau,\frac{\tau\theta+\theta'}{2\pi})}\right|^2\ .
\label{Z-twisted-implicit}
\end{align}
Using some identities in~\cite[Appendix C]{Gaberdiel:2004nv} the modular functions can be recast in the form
\begin{equation}\label{th-identity}
\frac{\vartheta_3(\tau,\nu)}{\vartheta_1(\tau,\nu)}=-2i\frac{\vartheta_3(\tau,0)}{\eta^3(\tau)}\sum_{n=0}^{\infty}\cos{\left[2\pi(n+1/2)(\nu-\tau/2)\right]}\ \frac{q^{\frac n2+\frac14}}{1+q^{n+\frac12}}\ ,
\end{equation}
so that the integral~\eqref{Z-twisted-implicit} becomes
\begin{equation}\label{Z-twisted-explicit}
\mathcal{P}_{\theta-\text{tw}}^{\text{NS}}=4\left|\frac{\vartheta_3(\tau,0)}{\eta^3(\tau)}\right|^2\sum_{n,\bar{n}=0}^{\infty}\frac{q^{\frac n2+\frac14}\bar{q}^{\frac {\bar n}{2}+\frac14}}{(1+q^{n+\frac12})(1+\bar{q}^{\bar n+\frac12})}I^{\theta}_{n,\bar{n}}
\end{equation}
with
\begin{align}
I^{\theta}_{n,\bar{n}}&=\int_0^{2\pi} \frac{d\theta'}{2\pi}\cos\left[(n+\tfrac{1}{2}) (\tau(\theta-\pi)+\theta') \right]\, \cos\left[ (\bar n+\tfrac{1}{2})(\bar{\tau}(\theta-\pi)+\theta')\right]\\
&=\frac{\delta_{n,\bar n}}{2}\ \cos\left[ (n+\tfrac{1}{2})(\pi-\theta)(\tau-\bar{\tau})\right]\ .
\label{thp-integral}
\end{align}
Inserting~\eqref{thp-integral} into~\eqref{Z-twisted-explicit}, evaluating the sum over $\bar{n}$, and combining the cosine with the~$q,\bar q$~dependent part of the numerator, we arrive at
\begin{equation}
\mathcal{P}_{\theta-\text{tw}}^{\text{NS}}=\left|\frac{\vartheta_3(\tau,0)}{\eta^3(\tau)}\right|^2
\sum_{n=0}^{\infty}\frac{q^{\frac{\theta}{2\pi}(n+\frac12)}\bar{q}^{\frac{\theta}{2\pi}(n+\frac12)}+q^{(1-\frac{\theta}{2\pi})(n+\frac12)}\bar{q}^{(1-\frac{\theta}{2\pi})(n+\frac12)}}
{(1+q^{n+\frac12})(1+\bar{q}^{n+\frac12})}\ .
\end{equation}
The unprojected supersymmetric partition function is then obtained by integrating over all twisted sectors 
\begin{align}\label{cont-orb-Z-infty}
\mathcal{P}^{\text{NS}}_{\mathbb{C}/U(1)}&=\int_0^{2\pi}\frac{d\theta}{2\pi}
\mathcal{P}_{\theta-\text{tw}}^{\text{NS}}=
\sum_{n=0}^{\infty}\int_{-2\pi}^{2\pi}\frac{d\theta}{2\pi}\frac{q^{\frac{|\theta|}{2\pi}(n+\frac12)}\bar{q}^{\frac{|\theta|}{2\pi}(n+\frac12)}}{(1+q^{n+\frac12})(1+\bar{q}^{n+\frac12})}\nonumber\\
&=\left|\frac{\vartheta_3(\tau,0)}{\eta^3(\tau)}\right|^2\sum_{n=0}^{\infty}\left|\frac{1}{1+q^{n+\frac12}}\right|^2\int ^1_{-1}dQ \ (q\bar q)^{|Q|(n+\frac12)}\nonumber\\
&=\sum_{n=0}^{\infty}\int_{-1}^{1}dQ\left|\chi^{I}_{|Q|(n+\frac12),Q}\right|^2\ ,
\end{align} 
where we used the definitions of appendix~\ref{app:characters} for the $c=3$ character $\chi^I$.

The last expression is ill-defined: the integration over $Q$ gives
\begin{equation}
 \mathcal{P}^{\text{NS}}_{\mathbb{C}/U(1)}=\frac{1}{2\pi\tau_2}\left|\frac{\vartheta_3(\tau,0)}{\eta^3(\tau)}\right|^2\sum_{n=0}^{\infty}\frac{1-(q\bar q)^{n+\frac12}}{\big|1+q^{n+\frac12}\big|^2}\ \frac{1}{n+\frac12}\ ,
\end{equation} 
which exhibits a logarithmic divergence when we sum over $n$. 
The fields that contribute to this divergence are the chargeless ones, as one can see by looking at the large $n$ asymptotic behaviour of the function~\eqref{cont-orb-Z-infty}: the fraction in front of the integral tends to one and the integrand localises around~$Q\sim0$. 
Therefore a sensible regulator would screen away the untwisted fields. 
We define
\begin{equation}\label{cont-orb-Z-reg}
\mathcal{P}^{\text{NS},(r)}_{\mathbb{C}/U(1)}:=\left|\frac{\vartheta_3(\tau,0)}{\eta^3(\tau)}\right|^2\sum_{n=0}^{\infty}\left|\frac{1}{1+q^{n+\frac12}}\right|^2\int ^1_{-1}dQ \ (q\bar q)^{|Q|(n+\frac12)}\left(1-e^{2\pi ir Q}\right)\ ,
\end{equation}
which corresponds to inserting $1-e^{2\pi irJ_{0}}$ in the trace, where $J_{0}$ is the zero mode of the $U(1)$ current $J(z)$. 
We see explicitly that this cures the logarithmic divergences of the sum in equation~\eqref{cont-orb-Z-infty} by performing the integral over the twist $Q$,
\begin{multline}
 \mathcal{P}^{\text{NS},(r)}_{\mathbb{C}/U(1)}=\left|\frac{\vartheta_3(\tau,0)}{\eta^3(\tau)}\right|^2\\
\times \sum_{n=0}^{\infty}\left|\frac{1}{1+q^{n+\frac12}}\right|^2\left[\frac{1-(q\bar q)^{n+\frac12}}{2\pi\tau_2(n+\frac12)}-\frac{1-e^{2\pi ir}(q\bar q)^{n+\frac12}}{4\pi\tau_2(n+\frac12)-2\pi ir} -\frac{1-e^{-2\pi ir}(q\bar q)^{n+\frac12}}{4\pi\tau_2(n+\frac12)+2\pi ir} \right]\ .
\end{multline}
The summand is suppressed by $n^{-2}$ for large $n$, and the series converges.

From equation~\eqref{cont-orb-Z-reg} it is easy to write down the (GSO-like) projected version of the regularised partition function, which reads in the Neveu-Schwarz sector
\begin{equation}\label{cont-orb-Z-reg-proj}
Z _{\mathbb{C}/U(1)}^{\text{NS},(r)}=\frac12\left(
\left|\frac{\vartheta_3(\tau,0)}{\eta^3(\tau)}\right|^2+\left|\frac{\vartheta_4(\tau,0)}{\eta^3(\tau)}\right|^2\right)
\sum_{n=0}^{\infty}\left|\frac{1}{1+q^{n+\frac12}}\right|^2\int ^1_{-1}dQ \ (q\bar q)^{|Q|(n+\frac12)}\left(1-e^{2\pi ir Q}\right)\ .
\end{equation}

We find perfect agreement with~\eqref{ch6:projected-part-fct-NS}, the regulated projected partition function of minimal models in the limit studied in chapter~\ref{ch:new-theory}.
This lends strong support to our interpretation of the~$c=3$ new limit theory as a~$\C /U(1)$ supersymmetric continuous orbifold.

\section{Boundary conditions}\label{ch7:sec:BC}

As reviewed in section~\ref{ch2:sec:discrete-orbifold-generalities}, the technologies to study boundary conditions on discrete orbifold models is well developed (see e.g.\ \cite{Billo:2000yb} and references
therein), and essentially they are also applicable for the continuous orbifold we are considering (see also~\cite{Gaberdiel:2011aa}). 

For continuous orbifolds one meets the phenomenon that the untwisted
fields are in a sense outnumbered by the twisted fields -- in the
partition function~\eqref{cont-orb-Z-reg-proj} the untwisted,
chargeless fields give a contribution of measure zero. 
Therefore the
only interesting boundary conditions are those that couple to the
twisted sectors, i.e.\ fractional boundary states. To obtain those we
have to start from boundary conditions in the plane 
that are invariant under the action of the orbifold group. In our
case, these are the boundary conditions corresponding to a point-like
brane at the origin of the plane, and the boundary conditions
corresponding to space-filling branes.
\subsection{Zero-dimensional branes}\label{ch7:sec:branes:subs:point-like}
Let us focus on the point-like brane. 
The fractional boundary conditions are labelled by irreducible representations of the orbifold group $U(1)$, i.e.\ by an integer $m$. 
The relative spectrum for two such boundary conditions labelled by $m$ and $m'$ follows from the usual orbifold rules (see equation~\eqref{ch2:self-energy-open-string-continuous}),
\begin{equation}\label{relopenstring}
\mathcal P_{m,m'}(\tilde \t)=\int_{0}^{2\pi}\frac{d\theta}{2\pi}\ \chi_m(\theta)\chi^{*}_{m'}(\theta)\ \Tr_{\mathcal{H}_0^{\text{open}}}\left[U(\theta)\tilde q^{L_0-\frac18}\right]\ ,
\end{equation}  
where $\tilde q=e^{2\pi i\tilde \tau}$ and $\chi_m (\theta)=e^{im\theta}$ is a $U(1)$ group character. 
$\mathcal{H}_0^{\text{open}}$ denotes the Hilbert space of boundary fields for the point-like brane in the parent free theory, which is just given by the free Neveu-Schwarz vacuum representation. Note that depending on the projection of the bulk spectrum, the point-like boundary condition could couple to the Ramond-Ramond sector, in which case the boundary spectrum would be projected by $\frac{1}{2} (1\pm (-1)^{F})$. 
The unprojected spectrum will be denoted by $\mathcal{P}_{m,m'}$ as discussed in section~\ref{app:characters:sec:GSO}. 
Evaluating~\eqref{relopenstring} we find
\begin{equation}\label{open-string-ampl}
\begin{split}
\mathcal P_{m,m'}(\tilde \t)=&\int_{0}^{2\pi}\frac{d\theta}{2\pi}e^{i(m-m')\theta}\ 2\sin{\frac{\theta}{2}}\ \frac{\vartheta_3(\tilde\tau,\frac{\theta}{2\pi})}{\vartheta_1(\tilde\tau,\frac{\theta}{2\pi})}\\
=&
-4i\frac{\vartheta_3(\tilde\tau,0)}{\eta^3(\tilde\tau)}\sum_{n=0}^{\infty}\frac{q^{\frac n2+\frac14}}{1+q^{n+\frac12}}
\ \int_{0}^{2\pi}\frac{d\theta}{2\pi}\ \sin{\frac{\theta}{2}}\ e^{i(m-m') \theta}\cos{(n+\tfrac{1}{2})(\theta-\pi\tilde\tau)}\ ,
\end{split}
\end{equation}
where we have made again use of equation~\eqref{th-identity}. We can explicitly evaluate the integral,
\begin{multline}
\int_{0}^{2\pi}\frac{d\theta}{2\pi}\ \sin{\frac{\theta}{2}}\ e^{i\Delta m \theta}\cos{(n+\tfrac{1}{2})(\theta-\pi\tilde\tau)}\\
=\frac{1}{4i}\left[\tilde q^{\frac12(n+\frac12)}\left(\delta_{\Delta m,n}-\delta_{\Delta m-1,n}\right)+\tilde q^{-\frac12(n+\frac12)}\left(\delta_{-\Delta m+1,n}-\delta_{-\Delta m,n}\right)\right]\ ,
\end{multline}
where $\Delta m=m-m'$. Inserting this into~\eqref{open-string-ampl} we
find that the spectrum is given by single $N=2$ characters: in the
notations of appendix~\ref{app:characters} we obtain
\begin{subequations}
\begin{align}
\mathcal P_{m,m}(\tilde \t)&=
\frac{\vartheta_3(\tilde \tau,0)}{\eta^3(\tilde \tau)}\left(\frac{1-\tilde q^{\frac12}}{1+\tilde q ^{\frac12}}\right)=\chi^{\text{vac}}_{0,0}(\tilde q)\label{ch7:contorbi-D0-open-string-self-energy-vacuum}\\
\intertext{and (for $m\not=m'$)}
\mathcal P_{m,m'}(\tilde \t) &=
\frac{\vartheta_3(\tilde \tau,0)}{\eta^3(\tilde \tau)}\ \tilde q^{|\Delta m|-\frac12}\left[\frac{1-\tilde q}{(1+\tilde q^{|\Delta m|-\frac12})(1+\tilde q^{|\Delta m| +\frac12})}\right]
=\chi^{III^{\pm}}_{|\Delta m|-\frac12,\,\pm 1}(\tilde q)\ ,
\end{align}
\end{subequations}
where the upper sign applies for $\Delta m>0$ and vice versa.
This result can now be compared to the limit of minimal
models. In section~\ref{ch6:sec:BC} two types of boundary conditions have been identified. 
The first one is obtained by keeping the boundary labels fixed while taking the limit. 
Only for $L=0$ one obtains elementary boundary conditions. 
The label $S$ can be fixed to even values for a fixed gluing condition for the supercurrents, and the two remaining choices $S=0,2$ determine the overall sign of the Ramond-Ramond couplings (thus distinguishing brane and anti-brane).
The relative spectrum for two such boundary conditions reads~\cite{Maldacena:2001ky}
\begin{equation}
Z^{(k)}_{(0,M,S),(0,M',S')}(\tilde{\t})=\chi_{(0,M-M',S-S'+2)}(\tilde q)\ .
\end{equation} 
This is a projected part of the full supersymmetric character
$\chi^{\text{NS}}_{0,M-M'}$. For $M=M'$ this is the minimal model
vacuum character, which for $k\to\infty$ goes to the $c=3$ vacuum
character. For $M\not= M'$, using the field identifications of equation~\eqref{ch3:MM-identifications} the labels can be brought to the
standard range, $(0,M-M')\sim (k,M-M'\mp (k+2))$, where the sign
depends on $M-M'$ being positive or negative. In the limit
$k\to\infty$ the corresponding character approaches a type $III$
character (see~\eqref{limit-III-minus} and~\eqref{limit-III-plus}),
\begin{equation}
\lim_{k\to\infty} \chi^{\text{NS}}_{0,M-M'\mp (k+2)} = 
\chi^{III^{\pm}}_{\frac{M-M'}{2}-\frac12,\pm 1}\ .
\end{equation}
The unprojected part of the boundary spectrum thus coincides with the
spectrum for the fractional boundary conditions in the continuous
orbifold upon identifying $M=2m$. On the other hand, the spectrum in
the limit of minimal models is projected. To get agreement we
therefore need that the point-like boundary conditions in the
continuous orbifold model couple to the Ramond-Ramond sector, which
specifies the necessary (GSO-like) projection in the Ramond-Ramond
sector. Note that this is precisely opposite from the projection that
we need in the free field theory limit, which is in accordance with
the T-duality that we use in the geometric interpretation of the
equivalence of a minimal model and its $\mathbb{Z}_{k+2}$ orbifold
(see the discussion at the end of subsection~\ref{ch4:sec:branes:subs:contorbi}).
\smallskip

In subsection~\ref{ch6:sec:BC:subs:discrete}, instead of the boundary spectrum, the one-point
functions have been determined. To make contact to these results, we
perform a modular transformation to get the boundary state overlap: we rewrite the boundary
partition function~\eqref{open-string-ampl} in terms of the modulus
$\tau=-\frac{1}{\tilde{\tau}}$ using the known
transformation properties~\eqref{modular-thetas},
\begin{equation}
\mathcal P_{m,m'}(\tilde{\t})=
2i\int_{0}^{2\pi}\frac{d\theta}{2\pi}e^{i(m-m')\theta}\sin{\frac{\theta}{2}}\
\frac{\vartheta_3(\tau,\frac{\tau\theta}{2\pi})}{\vartheta_1(\tau,\frac{\tau\theta}{2\pi})}\
.
\end{equation}
The ratio of $\vartheta$-functions can be rewritten using \mbox{eq.\ \eqref{th-identity}},
\begin{equation}
\begin{split}
\frac{\vartheta_3(\tau,\frac{\tau\theta}{2\pi})}{\vartheta_1(\tau,\frac{\tau\theta}{2\pi})}=&-2i\frac{\vartheta_3(\tau,0)}{\eta^3(\tau)}\sum_{n=0}^{\infty}\cos{\left[2\pi(n+1/2)\left(\frac{\tau\theta}{2\pi}-\tau/2\right)\right]}\ \frac{q^{\frac n2+\frac14}}{1+q^{n+\frac12}}\\
=&-i\frac{\vartheta_3(\tau,0)}{\eta^3(\tau)}\sum_{n=0}^{\infty}
\frac{q^{(n+\frac12)\frac{\theta}{2\pi}} + q^{(n+\frac12)(1-\frac{\theta}{2\pi})}     }{1+q^{n+\frac12}}\ ,
\end{split}
\end{equation}
so that we obtain
\begin{equation}\label{overlap-contorbi}
\begin{split}
\mathcal P_{m,m'}(\tilde \t)=&
\frac{\vartheta_3(\tau,0)}{\eta^3(\tau)}\int_{0}^{2\pi}\frac{d\theta}{2\pi}e^{i(m-m')\theta}\,
2\sin{\frac{\theta}{2}}\,\sum_{n=0}^{\infty}
\frac{q^{(n+\frac12)\frac{\theta}{2\pi}} + q^{(n+\frac12)(1-\frac{\theta}{2\pi})}}{1+q^{n+\frac12}}\\
=& \sum_{n=0}^{\infty}\int_{-1}^{+1}dQ\ 2\sin \left(\pi |Q| \right)
e^{2\pi i (m-m')Q}\,\chi^{I}_{|Q|(n+\frac12),Q}(q)\ .\end{split}
\end{equation}
If we do the same analysis for the projected spectrum, we find
\begin{multline}
Z_{m,m'}(\tilde{\t})=\sum_{n=0}^{\infty}\int_{-1}^{+1}dQ\ \sin\left(\pi |Q|\right)
e^{2\pi
i(m-m')Q}\left(\chi^{\text{NS}}_{|Q|(n+\frac12),Q}(q)+\chi^{\text{R}}_{\frac{1}{8}+|Q|(n+1),Q}(q)\right)\\
+ \int_{-\frac{1}{2}}^{\frac{1}{2}}dQ\  \sin \left(\pi
\big|Q-\tfrac{1}{2}\big| \right) e^{2\pi i (m-m')(Q-\frac{1}{2})}\,
\chi^{\text{R}^{0}}_{\frac{1}{8},Q} (q)
\ .
\end{multline}
Comparing with the formulas presented in subsection~\ref{ch6:sec:BC:subs:discrete}, we find perfect agreement with the one-point functions given there for the discrete A-type boundary states of the limit theory for $L=0$ and with the identification $M=2m$. 

\subsection{Space-filling branes}\label{ch7:sec:branes:subs:space-filling}

Along similar lines let us discuss boundary conditions that correspond to two-dimensional branes. As we discussed in section~\ref{ch5:sec:B-type} in the parent free theory there is a one-parameter family of space-filling branes that differ in the strength of a constant electric background field, which can be labelled by an angle~$\phi$. 
In the orbifold the boundary conditions obtain an additional integer label $m$ that determines the corresponding representation of $U(1)$. 
The unprojected part of the annulus partition function with such a two-dimensional boundary condition labelled by $\phi$ and $m$, and a zero-dimensional boundary condition labelled by $m'$ is then (using again~\eqref{th-identity})
\begin{align}
\mathcal{P}_{(\phi,m),m'} (\tilde{\t}) &= \int_{0}^{2\pi} \frac{d\theta}{2\pi}
e^{i (m-m'+\frac{1}{2})\theta} \, i \frac{\vartheta_{3}
(\tilde{\tau},\frac{\theta+ (\phi
+\pi)\tilde{\tau}}{2\pi})}{\vartheta_{1} (\tilde{\tau},\frac{\theta+ (\phi
+\pi)\tilde{\tau}}{2\pi})} \\
& = \frac{\vartheta_{3}(\tilde{\tau},0)}{\eta^{3}(\tilde{\tau})}
\frac{\tilde{q}^{\frac{\pi \mp \phi}{2\pi}|\Delta m
+\frac{1}{2}|}}{1+\tilde{q}^{|\Delta m+\frac{1}{2}|}} \ ,
\label{relspectrum}
\end{align}
where the upper sign corresponds to $\Delta m=m-m'\geq 0$, and the lower one to $\Delta m<0$. 
These are the type $I$ characters $\chi^{I^{\pm}}_{(n+\frac{1}{2})|Q|,Q}$ for charge $|Q|=\frac{\pi \mp \phi}{2\pi}$ and $n=|\Delta m+\frac{1}{2}|-\frac{1}{2}$. 
Note that this is precisely the result we expect from the limit of minimal
models: in subsection~\ref{ch6:sec:BC:subs:continuous} we constructed a continuous family of A-type boundary states labelled by $\cQ,N$ as a limit of minimal model boundary states with labels
\begin{equation}
(L,M,S) = (|\lfloor -\cQ (k+2)\rfloor| +2N, \lfloor-\cQ (k+2)\rfloor ,0) \ ,
\end{equation}
where $\lfloor x\rfloor$ denotes the greatest integer smaller or equal $x$.
Their relative spectrum (without projection) to a boundary condition $(0,M',0)$ with fixed $M'$ is simply given by $\chi^{\text{NS}}_{L,M-M'}$, and in the limit we find (see appendix~\ref{app:characters:sec:limit}) 
\begin{equation}
\chi^{\text{NS}}_{|\lfloor -\cQ (k+2)\rfloor|+2N,\lfloor -\cQ (k+2)\rfloor -M'}
\to \left\{ \begin{array}{ll}
\chi^{I^{+}}_{|\cQ|\,|N-\frac{M'}{2}+\frac{1}{2}|,\cQ} & \cQ>0,\ N\geq \frac{M'}{2} \\[6pt]
\chi^{I^{+}}_{|\cQ-1|\,|N-\frac{M'}{2}+\frac{1}{2}|,\cQ-1} & \cQ>0,\ N<\frac{M'}{2}\\[6pt]
\chi^{I^{-}}_{|\cQ|\,|N+\frac{M'}{2}+\frac{1}{2}|,\cQ} & \cQ<0,\ N\geq -\frac{M'}{2}\\[6pt]
\chi^{I^{-}}_{|\cQ+1|\,|N+\frac{M'}{2}+\frac{1}{2}|,\cQ+1} & \cQ<0,\ N<-\frac{M'}{2}
\end{array}\right. \ . 
\end{equation}
These are the type $I$ characters that we found above in~\eqref{relspectrum} if we identify 
\begin{align}
\phi &=2\pi \big(-\cQ\pm \tfrac{1}{2}\big) & m &=-\tfrac{1}{2} \pm
\big(N+\tfrac{1}{2}\big) \ ,
\end{align}
where the upper sign applies for $\cQ>0$, and the lower for $\cQ<0$.

\chapter{Limits of Kazama-Suzuki models}\label{ch:limit-KS}

Defining limit theories is an appealing framework for many reasons, as extensively argued in various places in this thesis.
This last chapter is devoted to give elements for a possible generalisation of the results in this work. 
We present additional (unpublished) material to substantiate the idea that continuous orbifolds of free theories may be a unifying language for supersymmetric limit theories.

\section{Kazama-Suzuki models: limit and continuous orbifold}\label{ch8:sec:correspondence}
The obvious extension of the work presented in chapters~\ref{ch:geometry}-\ref{ch:our-contorbi} and in the research papers~\cite{Fredenhagen:2012rb,Fredenhagen:2012bw}, is to generalise this construction to more complicated Kazama-Suzuki models, briefly presented in section~\ref{ch3:app:sec:KS-models}: we would like to understand the theories emerging as the large level limit of supersymmetric Grassmannian cosets.
The additional appeal of the limits of these models is related to the recent interest in specific classes of two-dimensional CFTs, proposed to be holographically dual to higher-spin three-dimensional gauge theories~\cite{Campoleoni:2010zq,Gaberdiel:2010pz,Gaberdiel:2012uj}.
The $N=2$ supersymmetric version of these ideas, proposed in~\cite{Creutzig:2011fe}, and further elaborated in~\cite{Candu:2012jq,Candu:2012tr}, relates the large level and large rank limits of the supersymmetric Grassmannian cosets of section~\ref{ch3:app:sec:KS-models}, to the $N=2$ supersymmetric higher-spin Vasiliev theory on AdS${}_3$. 
The large level limits analysed here would correspond to a small 't-Hooft coupling limit of a finite rank boundary theory.
Since these models enjoy level-rank duality, it is conceivable that some constructions involving large level limits appear also in large rank limit theories, which are the most analysed in this context as large $N$ boundary theory in the AdS${}_3$/CFT${}_2$ gauge-gravity duality.

We conjecture in this section that the large level limit of Grassmannian cosets corresponds (with a suitable choice of averaged fields) to an $N=2$ continuous orbifold, in complete analogy with the results presented for~$N=2$ minimal models.
Precisely we propose here the formal limit (once taken care of appropriately scaling the labels)
\begin{equation}\label{ch8:formal-limit-KS-contorbi}
\left.\lim\limits_{k\to\infty}\frac{\widehat{su}(n+1)_k\oplus \widehat{so}(2n)_1}
{\widehat{su}(n)_{k+1}\oplus \widehat{u}(1)_{n(n+1)(k+n+1)}}\right|_{\text{coset}}\sim \left.\frac{\left(\C^n+\text{fermions}\right)}{U(n)}\right|_{\text{orbifold}}\ .
\end{equation}
The limit of the central charge is $c=\lim\limits_{k\to\infty}\frac{3nk}{ k+n+1}=3n$.

\section{Boundary conditions match}\label{ch8:sec:BC-match}

The $su(n+1)/u(n)$ Grassmannian coset models are rational in the bulk sector with respect to $N=2$ supersymmetric $W_{n+1}$-algebras.
The boundary conditions that we consider here are the ones preserving one copy of the full chiral algebra of the theory, specifically A-type boundary conditions (following the terminology outlined in section~\ref{ch3:sec:BC}), using the strictly diagonal modular invariant in the closed string sector; these are realised by Cardy boundary states.
With these conventions, in the minimal model's case, these boundary conditions would correspond to the straight lines of subsection~\ref{ch3:sec:geometry:subsec:branes}, hence to discs in the T-dual orbifold picture.
We want to compare then A-type limiting boundary conditions to point-like fractional branes in the continuous orbifold.
The Cardy states are labeled in the same way as the primary fields in the bulk theory (as reviewed in appendix~\ref{app:vanilla}), with the same identifications and selection rules.
Recalling section~\ref{ch3:app:sec:KS-models}, they are labeled by the set $(\L,\S;\l,\m)$, with $\L,\S,\l$ dominant weights of~$\widehat{su}(n+1),\widehat{so}(2n),\widehat{su}(n)$ respectively, and~$\m$ indicating the~$U(1)$ charge of the free boson on the circle of radius $\sqrt{n(n+1)(n+k+1)}$.
In general Grassmannian cosets, boundary states with $su(n+1)$ labels $(\L_1,\dots \L_n)$ can be obtained as superposition of $(0,\dots,0)$ boundary states by boundary renormalisation flow processes~\cite{Fredenhagen:2001kw}.
Since the boundary flows become shorter and shorter as $k$ grows, in the $k\to\infty$ regime we expect to find elementary Neveu-Schwarz boundary conditions labeled by the set~$((0,0),\cS;\cL,\cM)$, in the singlet ($\cS=0$, branes) and vector ($\cS=2$ anti-branes) sector respectively.
This is in complete analogy with what we observe in the simpler case of the limits of $N=2$ minimal models, see sections~\ref{ch5:sec:B-type} and~\ref{ch6:sec:BC}.

From these considerations one can draw the conclusion that, in the $k\to\infty$ limit, boundary conditions obtained by keeping finite the labels of Cardy states, are described by the triples $(\cS,\cL,\cM)$, where $\cS$ is an~$\widehat{so}(2n)$ label, and $(\cL,\cM)$ are $\widehat{u}(n)$ labels.

We observe a remarkable agreement with what is expected in the continuous orbifold description: fractional branes in an orbifold are labeled by irreducible representations of the orbifold group~(see sections~\ref{ch2:sec:discrete-orbifold-generalities} and~\ref{ch2:sec:contorbi} for details).
The discrete irreducible representations of $U(n)$ on the right hand side of equation~\eqref{ch8:formal-limit-KS-contorbi} should correspond to the discrete boundary conditions of the limit of the coset, also labeled by $U(n)$ irreducible representations.

This fact is true for $n=1$, as extensively explained in chapter~\ref{ch:our-contorbi}: the $U(1)$ representations' label $m$ characterises the fractional branes of $\C/U(1)$, whence the $U(1)$ charge $M$ of the $U(1)$ in the denominator of the $su(2)/u(1)$ coset labels discrete representations of the limit of minimal models.
The precise identification is in the $n=1$ case $M=2m$.

The detailed analysis in the general Kazama-Suzuki case is still out of range, since it is still not clear how to scale the irreducible representations in the limit.
Nevertheless, explicit computations of the open string spectrum on the orbifold side are still doable (although cumbersome), in the second simplest example, namely for the continuous orbifold $\C^2/U(2)$.
In this case we are able to write down explicit expressions for the open string spectrum of point-like fractional branes.
Furthermore we manage to give an explicit proposal for the identifications of the labels of the boundary condition in the limit theory, with labels for the boundary conditions in the continuous orbifold.

\section{Limit of boundary conditions on the $SU(3)/U(2)$ model}\label{ch8:sec:BC-su(3)}
The limits we are interested in are based on the large $k$ behaviour of the rational sequence of Grassmannian cosets
\begin{equation}
\lim\limits_{k\to\infty}\frac{\widehat{su}(3)_k\oplus \widehat{so}(4)_1}
{\widehat{su}(2)_{k+1}\oplus \widehat{u}(1)_{6(k+3)}}\ .
\end{equation}
The central charge approaches the value $c=6$ in the $k\to\infty$ limit, and the spectrum of the primaries reads (recall section~\ref{ch3:app:sec:KS-models})
\begin{equation}\label{ch8:spectrum-repr-generic}
h=\frac{1}{2(k+3)}\left[\frac 23\left(\L_1^2+\L_2^2+\L_1\L_2\right)+2\left(\L_1+\L_2\right)-\frac{\l}{2}(\l+2)-\frac{\m^2}{6}\right]
   +h_{\S}+n\ .
\end{equation}
The analysis of the limit of open string spectra corresponds to study the large~$k$ behaviour of representations in the space~$\Hilb^{(0,0),\cS}_{\cL,\cM}$ in the decomposition
\begin{equation}\label{ch8:decomp-Hilbert-space}
{\cal H}_{\mathfrak{su}(3)}^{(0,0)}\otimes {\cal H}_{\mathfrak{so}(4)}^{\cS}=\sum_{\cL,\cM}
{\cal H}^{(0,0),\cS}_{\cL,\cM}\otimes \left[{\cal H}^{\cL}_{\mathfrak{su}(2)}\otimes {\cal H}^{\cM}_{\mathfrak{u}(1)}\right]\ .
\end{equation}
We restrict here the attention for simplicity to the NS-sector, $\cS=0,2$.
At finite $k$ the conformal weight associated to an arbitrary representation of this kind reads
\begin{equation}\label{ch8:spectrum-repr-00}
h=-\frac{1}{2(k+3)}\left[\frac{\cL}{2}(\cL+2)+\frac{\cM^2}{6}\right]
   +h_{\cS} + n\ ,
\end{equation}
with selection rules summarised in
\begin{equation}\label{ch8:ident-selec-repr-00}
 -\frac{\cL}{2}+\frac{\cM}{6}+\frac{\cS}{2}\in\Z\ ,
\end{equation}
and $h_{\cS}=(0,\12)$ in the NS-sector for $\cS=0$ and $\cS=2$ respectively.
Moreover, the label $\cM$ gets identified as $\cM\sim \cM+6(k+3)$, according to the discussion of $\kmalg{u}(1)$ representations at the end of section~\ref{ch1:sec:free-boson}.
It is apparent how crucial the integer shifts $n$ in equation~\eqref{ch8:spectrum-repr-00} are, if we want to guess something about the limit of the spectrum. 
Unfortunately, as already mentioned in section~\ref{ch3:app:sec:KS-models}, these integers are not known explicitly.
Nevertheless, $n=0$ for some representations: the vacuum (by definition), and the ones associated to ground states of $\widehat{su}(3)$ explicitly contained in the decomposition of ground states of $\widehat{su}(2)\oplus \widehat{u}(1)$.
\subsection{The $c=6$ vacuum}
The (unprojected) overlap between two elementary boundary states 
indicated here by $B_{r}=((0,0),\cS;\cL,\cM)$ ($r$ is a multi-index), is given by the character of the corresponding representation in the open string sector, in formulas
\begin{equation}
Z_{B_{r}B_{r'}}(\tilde \t)=\chi^{c=6}_{f(r,r')}(\tilde q)\ ,
\end{equation}
with $f(r,r')$ some (still unknown) combination of the multi-indices characterising the $c=6$ sector under analysis.
The (unprojected) self energy of an open string stretched between the two simplest branes labeled by~$B_0=((0,0),0;0,0)$ is given by the vacuum character of the chiral algebra preserved by the boundary, namely
\begin{equation}
Z_{B_0B_0}(\tilde \t)=\chi_{\text{vac}}(\tilde q)\ .
\end{equation}
In our case we have to compute the character of the vacuum of $N=2$, $W_3$-algebra. 
In the limit $k\to\infty$ the character can be explicitly computed as follows: the generators of $W_3$ $N=2$ \SCA are organised in two $N=2$ super multiplets, whose components read
\begin{align}\label{ch8:generators-W3-N=2}
\begin{array}{lllll}
 & G^+(z) &       &      &\\
J(z)&         & T(z)   &      &\\
 & G^-(z)  &      &      &\\
 && & W^{(5/2)+}(z) &\\
 && W^{(2)}(z)& & W^{(3)}(z)\\
 && & W^{(5/2)-}(z)
\end{array}
\end{align}
The vacuum is chosen to be invariant under the full algebra, hence it is defined by the following rules
\begin{equation}\label{ch8:singular-vectors-c=6-vacuum}
\begin{split}
 & J_{0}|\text{vac}\ket=G^{\pm}_{-\frac 12}|\text{vac}\ket=L_{-1}|\text{vac}\ket=0\\
& W^{(2)}_{-1} |\text{vac}\ket=W^{(5/2)\pm}_{-\frac{3}{2}}|\text{vac}\ket=W^{(3)}_{-2}|\text{vac}\ket=0\ .
\end{split}
\end{equation}
The modules are the ones created by the negative modes of the generators in equation~\eqref{ch8:generators-W3-N=2}, leaving out the modes listed explicitely in equations~\eqref{ch8:singular-vectors-c=6-vacuum}

It is then easy then to write explicitly the character:
\begin{equation}\label{ch8:vacuum-character-c=6}
\chi^{c=6}_{\text{vac}}(q)=\prod_{n=0}^{\infty}\frac{\left(1+q^{n+\frac 32}\right)\left(1+q^{n+\frac 32}\right)\left(1+q^{n+\frac 52}\right)^2}{\left(1-q^{n+1}\right)\left(1-q^{n+2}\right)^2\left(1-q^{n+3}\right)}=\left[\frac{\vartheta_3(\t,0)}{\eta^3(\t)}\right]^2\frac{\left(1-q^{\12}\right)^4\left(1+q\right)}{\left(1+q^{\frac 32}\right)^2}\ .
\end{equation}
In the first equality in equation~\eqref{ch8:vacuum-character-c=6} one can recognise in the numerator (denominator) the modules associated with the fermionic (bosonic) generators of equation~\eqref{ch8:generators-W3-N=2}, starting from the first mode that does not annihilate the state, following the definition of the vacuum in equation~\eqref{ch8:singular-vectors-c=6-vacuum}.

Characters associated to ground states with higher conformal weights are difficult to analyse with these methods, since a complete classification of singular vectors for general representations of $N=2$ $W_3$ algebra is (to our knowledge) still missing.

\subsection{Other representations}

Another class of representations with $n=0$ is given by the ground states of the numerator which appear explicitly as ground states in the decomposition of equation~\eqref{ch8:decomp-Hilbert-space}.
This means that we restrict the attention to the decomposition of irreducible representations of $su(3)$ in terms of irreducible representations of $su(2)\oplus u(1)$.
Finite highest-weight representations $(\L_1,\L_2)$ of $su(3)$ get decomposed as follows
\begin{equation}\label{ch8:decomposition-su(3)-su(2)-u(1)}
(\L_1,\L_2)\ra \bigoplus_{\g_1=0}^{\L_1} \bigoplus_{\g_2=0}^{\L_2}\big([\g_1+\g_2];2(\L_2-\L_1)+3(\g_1-\g_2)\big)\ ,
\end{equation}
where on the right hand side we indicated $su(2)$ ($u(1)$) representations in squared (round) brackets.
By applying four times the identification rules (and using the cyclicity of the $u(1)$ label), the labels of the representations that correspond to boundary states can be mapped to a more tractable set 
\begin{equation}\label{ch8:boundary-state-in-tractable-labels}
((0,0),0;\cL,\cM)\sim((k,0),0;\cL, \cM-2(k+3))\ .
\end{equation}
Reading from equation~\eqref{ch8:decomposition-su(3)-su(2)-u(1)}, a representation of $su(3)$ with labels $(k,0)$ gets decomposed as follows
\begin{equation}\label{ch8:decomposition-(0,k)}
(k,0)\ra\bigoplus_{\g_1=0}^{k}\big([\g_1];-2k+3\g_1\big)\ .
\end{equation}
A sufficient condition for a state of the form in equation~\eqref{ch8:boundary-state-in-tractable-labels} to have zero shift $n$ in the spectrum, is to belong to the set given in equation~\eqref{ch8:decomposition-(0,k)}, a condition that in terms of boundary labels reads
\begin{equation}
\cM-2(k+3)\= -2k+3\cL\quad\Lra\quad \cM\=3(\cL+2)\ .
\end{equation}
Therefore, the conformal weight of the representation corresponding to this boundary condition,
\be\label{ch8:special-reps-KS}
((0,0),0;\cL,3(\cL+2))\sim ((k,0),0,\cL,3(\cL+2)-2(k+3))
\ee 
is given by the formula~\eqref{ch8:spectrum-repr-generic}, with $h_{\cS=0}=0$, and $n=0$.
It reads:
\begin{equation}
h=\frac{k}{k+3}(\cL+1)-\frac{\cL(1+2\cL)}{2(k+3)}=(\cL+1)+\order{\frac{1}{k}}\ ,
\end{equation}
if the label $\cL$ is kept finite, in analogy with the discrete representations of section~\ref{ch6:sec:BC} in the example of minimal models.
These representations can be compared to the analogues of type-B tiny discs, obtained as fractional branes in the continuous orbifold description of section~\ref{ch7:sec:branes:subs:point-like}.

\section{Continuous orbifold $\C^2/U(2)$}\label{ch8:sec:BC-contorbi}

In this section we present the continuous orbifold that we conjecture to be dual to the limit of $SU(3)/U(2)$ Kazama-Suzuki model.
We concentrate again on boundary conditions.

The parent theory is the theory of four free real bosons, denoted $X_1(z,\bar z),\dots X_4(z,\bar z)$, accompanied by four free real fermions, $\Psi^1(z,\bar z)\dots \Psi^4(z,\bar z)$.
We rearrange the coordinates and the fermions in order to work on the complex plane $\C^2$:
\begin{subequations}
\begin{align*}
\phi_1 &=\tfrac{1}{\sqrt 2}(X^1+iX^2) & \phi_1^* &=\tfrac{1}{\sqrt 2}(X^1-iX^2)\quad & \phi_2 &=\tfrac{1}{\sqrt 2}(X^3+iX^4) & \phi_2^* &=\tfrac{1}{\sqrt 2}(X^3-iX^4)\\[4pt]
\psi_1 &=\tfrac{1}{\sqrt 2}(\Psi^1+i\Psi^2) & \psi_1^* &=\tfrac{1}{\sqrt 2}(\Psi^1-i\Psi^2)\quad & \psi_2 &=\tfrac{1}{\sqrt 2}(\Psi^3+i\Psi^4) & \psi_2^* &=\tfrac{1}{\sqrt 2}(\Psi^3-i\Psi^4)
\ .
\end{align*}
\end{subequations}
The mode expansion of the (holomorphic) fields reads ($j=1,2$)
\begin{subequations}
\begin{align}
\de\phi_j &=-i\sum_{\substack{m\in\Z}}\alpha^j_m z^{-m-1} & \de\phi_j^{*} &=-i\sum_{\substack{m\in\Z}}\alpha^{j*}_m z^{-m-1}\\
\psi_j &=\sum_{\substack{r\in\Z+\eta}}\psi^j_r z^{-r-\frac{1}{2}} & \psi_j^*&=\sum_{\substack{r\in\Z+\eta}}\psi^{j*}_r z^{-r-\frac{1}{2}}\ ,
\end{align}
\end{subequations}
where $\eta=0,\frac12$ in the Ramond and Neveu-Schwarz sector respectively. 
The antiholomorphic case is analogous. 
For simplicity we will restrict the following discussion to the Neveu-Schwarz sector. 
The modes respect the algebra of two free complex boson and two free Neveu-Schwarz complex fermion:
\begin{subequations}
\begin{align}  
  [\alpha^i_m, \alpha^{j*}_n] &= m\,\delta^{ij}\delta_{m,-n}  &  \{\psi^i_r,\psi^{j*}_s\} &=\delta^{ij} \delta_{r,-s}\\[4pt] 
  [\alpha^i_m,\alpha^j_n] &= [\alpha^{i*}_m, \alpha^{j*}_n] = 0 & \{\psi^i_r,\psi^j_s\}&=\{\psi^{i*}_r,\psi^{j*}_s\}=0 \ .
\end{align}
\end{subequations}
We can explicitly realise the $N =2$ superconformal algebra by defining the generators as 
\begin{subequations}\label{ch8:SUSY-generators-hyperplane}
\begin{align}
T&=-\sum_{\substack{i=1}}^2\left[\de\phi_i\de\phi_i^*+\frac12(\psi_i^*\de\psi_i+\psi_i\de\psi_i^*)\right] & J&=-\sum_{\substack{i=1}}^2\ \psi_i^*\psi_i\\
G^+&=\phantom{-}\sum_{\substack{i=1}}^2 i\sqrt 2\,\psi_i\de\phi_i^* & G^-&=\phantom{-}\sum_{\substack{i=1}}^2i\sqrt 2\,\psi_i^*\de\phi_i \ ,
\end{align}
\end{subequations}
and similarly for their antiholomorphic counterparts.
A group element $g\in U(2)$, parameterised by real angles $\vec{x}^{T}=(x_1,\dots, x_4)$ acts on $\C^2$, parameterised by the complex fields $\phi_1(z,\bar z),\phi_2(z,\bar z)$, as follows:
\begin{align}
U(g(\vec{x}))\cdot
\left(\begin{array}{l}
\phi_1\\
\phi_2
\end{array}\right)
=\left(
\begin{array}{ll}
e^{ix_4}(\cos x_1+i\sin x_1\cos x_2 )& \sin x_1\sin x_2 e^{i(x_3+x_4)}\\
-\sin x_1\sin x_2 e^{i(-x_3+x_4)} & e^{ix_4}(\cos x_1-i\sin x_1\cos x_2 )
\end{array}
\right)\cdot
\left(\begin{array}{l}
\phi_1\\
\phi_2
\end{array}\right)\ ,
\end{align}
where the parameters on the group run in the range $x_1\in[0,2\pi],\ x_2\in[0,\frac{\pi}{2}],\ x_3\in[0,2\pi],\ x_4\in [0,\pi]$.
We want to end up with an $N=2$ theory; we therefore choose the action $U(g)$ of the orbifold group on the fields in such a way that the currents in~\eqref{ch8:SUSY-generators-hyperplane} stay invariant.

\subsection{Point-like fractional branes in the continuous orbifold}

The boundary spectrum of the supersymmetric continuous orbifold $\C^2/U(2)$ can be computed using the methods outlined in section~\ref{ch2:sec:contorbi}.
We concentrate on the open string spectrum between fractional branes, labeled by the finite irreducible representation multi-indices $r,r'$ of the orbifold Lie group $G$:
\begin{equation}\label{ch8:open-string-spectrum-contorbi-general}
Z_{rr'}(\tilde \t)=\frac{1}{|G|}\int_{G}d\m (g)\;\chi_r(g)\chi^*_{r'}(g)\Tr_{\Hilb^{\text{open}}_0 }\ U(g) \ \tilde q^{L_0-\frac 14}\ .
\end{equation} 
Here $\Hilb^{\text{open}}_0$ is the Hilbert space for point-like branes in the parent theory, namely the sum of two free NS vacuum representations.
The volume of the group is $|U(2)|=4\pi ^3$, the Haar measure reads in our coordinates $d\mu(g(\vec{x})) =dx_1dx_2dx_3dx_4\sin^2 x_1\sin x_2$.
The $U(2)$ group characters are labeled by $U(2)$ representation labels $l,m$, the $SU(2)$ and $U(1)$ quantum numbers respectively, and in our coordinates read
\begin{equation}
\chi_{l,m}=\frac{\sin(1+l)x_1}{\sin x_1}e^{imx_4}\ ,
\end{equation}
with $l+m$ even.
They enjoy the following composition rule
\begin{equation}
\chi_{l,m}\ \chi^{*}_{l',m'}=\sum_{L=|l-l'|}^{l+l'}\chi_{L,m-m'}\ .
\end{equation}
It is possible to diagonalise the representative $U(g)$ inside the trace of equation~\eqref{ch8:open-string-spectrum-contorbi-general} acting by conjugation with appropriate group elements:
since the Virasoro generator $L_0$ is by definition invariant under the action of orbifold group elements, we can write
\begin{equation}
\Tr_{\Hilb^{\text{open}}_0 }\ U(g(\vec{x})) \ \tilde q^{L_0-\frac 14}=\Tr_{\Hilb^{\text{open}}_0 }
\left(\begin{array}{ll}
e^{i(x_4-x_1)} & 0\\
0 & e^{i(x_4+x_1)}
\end{array}\right)
\tilde q^{L_0-\frac 14}
\end{equation}
Inside the trace, the action of the group on $\phi_1,\phi_2$ is equivalent to the action of two independent $U(1)$ acting on $\phi_1$ and $\phi_2$ respectively, parameterised by the angles $\theta_1\equiv x_4-x_1$ and $\theta_2\equiv x_4+x_1$.
In complete analogy with equation~\eqref{th0-c}, we see that the trace becomes the product of ratios of theta functions, and we can write
\begin{equation}
Z_{lm,l'm'}(\tilde \t) = \sum_{L=|l-l'|}^{l+l'}Z_{L(m-m'),0}(\tilde \t)
\end{equation}
\begin{equation*}
Z_{L(m-m'),0}(\tilde{\t})=\frac{2}{\pi^2}\int_{\T^2}d^2\theta\;\sin(1+L)\tfrac{\theta_1-\theta_2}{2}\sin\tfrac{\theta_1-\theta_2}{2}e^{i\frac{\theta_1+\theta_2}{2}(m-m')}4\sin\tfrac{\theta_1}{2} \sin\tfrac{\theta_2}{2}\frac{\vartheta_3(\tilde{\t},\frac{\theta_1}{2\pi})}{\vartheta_1(\tilde{\t},\frac{\theta_1}{2\pi})}\frac{\vartheta_3(\tilde{\t},\frac{\theta_2}{2\pi})}{\vartheta_1(\tilde{\t},\frac{\theta_2}{2\pi})}\ .
\end{equation*}
The integral can be solved explicitly by means of the following formula~\cite[appendix A]{Sugawara:2012ag}
\begin{equation}
\frac{\vartheta_3(\t,\nu)}{\vartheta_1(\t,\nu)}=-i\frac{\vartheta_3(\t,0)}{\eta^3(\t)}\sum_{n\in\Z}\frac{e^{2\pi i \nu\left(n+\12\right)}}{1+q^{n+\12}}\ .
\end{equation}
The result reads
\begin{equation}\label{ch8:open-string-spectrum-contorbi}
Z_{lm,l'm'}(\tilde \t) = \sum_{L=|l-l'|}^{l+l'}\chi^{c=6}_{L,m-m'}(\tilde q)\ ,
\end{equation}
where
\begin{equation}\label{ch8:type3-characters-c=6}
\begin{split}
&\chi^{c=6}_{L,M}(q) := q^{-\frac52+\frac32M-\frac12L}\left[\frac{\vartheta_3(\t,0)}{\eta^3(\t)}\right]^2\\
&\frac{(q-1)^3(q+1)(q^{1+L}-1)}
{(1+q^{-\frac32+\frac{M-L}{2}})(1+q^{-\frac12+\frac{M-L}{2}})(1+q^{\frac12+\frac{M-L}{2}})(1+q^{-\frac12+\frac{M+L}{2}})(1+q^{\frac12+\frac{M+L}{2}})(1+q^{\frac32+\frac{M+L}{2}})}\ .
\end{split}
\end{equation}
The set of characters presented in equation~\eqref{ch8:type3-characters-c=6} are the characters of irreducible representations of the unprojected NS open string spectrum of point-like fractional branes in the $N=2$ supersymmetric continuous orbifold $\C^2/U(2)$.

The (unprojected) open string spectrum can be analysed by studying the leading behaviour of the characters of equation~\eqref{ch8:type3-characters-c=6}, for the allowed range of the parameters $L,M$. 
The leading exponent of $q$ gives the value of the conformal weight of the ground state corresponding to the chosen $L,M$ labels.
$L\in\Z_{>0}$, $M\in\Z$, and we can recognise the following records:
\begin{align}\label{ch8:table-leading-behav-contorbi-characters}
\begin{array}{c|c}
\text{range} & \text{leading term}\\[3pt]
\hline\hline\\[-6pt]
|M|\leq L-2 & q^{L-1}\\[4pt]
|M|=L & q^{L-\frac12}\\[4pt]
|M|=L+2 & q^{L+1}\\[4pt]
|M|>L+2 & q^{-\frac 52 -\frac 12(-3|M|+L)}
\end{array}
\ .
\end{align}
\subsection{Comparison with the limit of KS models}
The first check one can perform is the identification in equation~\eqref{ch8:open-string-spectrum-contorbi} of the vacuum character at $c=6$ of~\eqref{ch8:vacuum-character-c=6}.
It is easy to find 
\begin{equation}
\chi^{c=6}_{0,0}(q)=\chi^{c=6}_{\text{vac}}(q)\ ,
\end{equation}
so that $l=l'=0,m=m'$ in equation~\eqref{ch8:open-string-spectrum-contorbi} corresponds to the open string fully supersymmetric representation at $\cL=0,\cM=0$ when $k\to\infty$.
This means that the character with $L=M=0$ must correspond to the character with $\cL=\cM=0$.

Another possible check is to compare the labels of the states of the series of equation~\eqref{ch8:special-reps-KS}, whose conformal weight is~$\cL+1$, with some representations in table~\eqref{ch8:table-leading-behav-contorbi-characters}.
The ground states of the series $M=L+2$ have the right conformal weight, given the identification $L=\cL$.
Moreover, since $\cM=3(\cL+2)$ for these states, we find plausible the identification $3M=\cM$, which reproduces $M=L+2$.
From the identification rules~\eqref{ch8:ident-selec-repr-00} of the coset, we confirm that $M+L$ must be even, as expected for a $U(2)$ representation.
We propose then, that the (unprojected) continuous orbifold's open string spectrum between point-like fractional branes reproduces the (unprojected) open string amplitude between Cardy A-type branes for the Grassmannian coset at $n=2$ for large level $k$, given the following identifications:
\begin{equation}
\cL \equiv L\ ,\quad \cM\equiv 3 M\ ,\quad \text{with}\quad L+M\ \ \text{even}\ .
\end{equation}

\chapter*{Summary and outlook}\markboth{SUMMARY AND OUTLOOK}{SUMMARY AND OUTLOOK}\label{ch:outlook}
\addcontentsline{toc}{chapter}{Summary and outlook}

In this thesis we have reported results concerning two-dimensional conformal field theories (CFTs) emerging as the limit of inverse sequences of renormalisation group flows in the space of two-dimensional theories.
In this context we have substantiated the idea that continuous orbifolds constitute a class of CFTs suitable to the description of non-rational and non-free CFTs emerging as limit theories.
They are by definition non-rational theories, but still manageable.
In particular we have analysed in detail limits of $N=2$ supersymmetric CFTs, minimal models and Kazama-Suzuki models.
In order to introduce the reader to the following technical material, which constitutes the core of our work, we have collected in the first three chapters some general ideas and known technology.

\section*{Summary}

The first two chapters present the main elements of the following discourse: the idea of taking limits of sequences of CFTs is introduced starting from general considerations about the procedure itself, followed by the very simple example of the free boson; the limit theory of Runkel and Watts (the example that has inspired our analysis) is then briefly reviewed.
Continuous orbifold has been introduced step-by-step in the second chapter, by starting with simple examples of discrete orbifolds, and generalising to the more involved constructions that we have used in the following.
We have tried in this part to present known (or very conceivable) results in an original way.

The third chapter is an extensive introduction or a small review of $N=2$ minimal models, which is to date the best understood class of interacting $N=2$ CFTs.
Although the topic is well-known to practitioners, we have encountered some difficulties in finding the material that we have actually used, since the evolution of ideas and techniques in this field has spread over almost thirty years by now.
Hence we have decided to collect this set of data here.
\smallskip

The main contributions of this work start with the fourth chapter, where we observe from a closer distance the large level limit of $N=2$ minimal models.
Already in this chapter we introduce interesting features of the limit: in general we expect in the limit from any sequence of CFTs based on an extended conformal algebra, that more CFTs can emerge (from the different ways of rescaling the labels of the representations in the large level regime).
We present a geometric interpretation of this phenomenon, which leads to the proposed continuous orbifold interpretation of one limit CFT, and to a free theory of two bosons and two fermions as another limit CFT: scaling the labels differently amounts to concentrating on different regions of the sigma-model geometry.
In chapter~\ref{ch:free-limit} we report detailed CFT data in the free theory limit, and directly compare them with the expected results from the free theory.
In chapter~\ref{ch:new-theory} we construct explicitly the new limit theory.
A non-trivial and promising result is, for example, that the structure constants converge to a simple (yet non trivial) hypergeometric function if we scale the labels appropriately: as explained in appendix~\ref{app:wigner}, in general limits in this region of the parameter space do not give such clean expressions.
in chapter~\ref{ch:our-contorbi} we construct the spectrum of open and closed string excitations of the supersymmetric continuous orbifold~$\C/U(1)$.
The comparison between the two sets of data agrees completely, and gives strength to our proposal.

Large part of these results have been published in~\cite{Fredenhagen:2012rb,Fredenhagen:2012bw}.
\smallskip

The last chapter contains the analysis of the open string spectrum (the technology used is reviewed in appendix~\ref{app:vanilla}) of more complicated $N=2$ supersymmetric CFTs, with emphasis on $su(3)$ Kazama-Suzuki models.
This is unpublished material, which lends support to the general conjecture that $N=2$ continuous orbifolds of the form $\C^{n}/U(n)$ are the right CFTs to describe limits of Grassmannian coset $N=2$ supersymmetric CFTs.

\section*{Outlook}
More work remains to be done in this context, which could lead in our opinion to very interesting developments.

We would like to extend the analysis of chapter~\ref{ch:limit-KS} of the limit of $su(3)$ Kazama-Suzuki models, and of the corresponding continuous orbifold.
In our opinion, a better understanding of this second simplest example would be already enough to give elements for a concrete proposal for the behaviour of the large level limit of $su(n)$ Kazama-Suzuki models.
This might be of interest in the context of holographic higher-spin dualities~\cite{Gaberdiel:2012uj}, since the $N=2$ supersymmetric version~\cite{Candu:2012tr} of the triality of Gaberdiel and Gopakumar~\cite{Gaberdiel:2012ku} (a sort of level-rank duality for CFTs with $W_{\infty}$ chiral algebra) should connect large $k$ $W_n$ Kazama-Suzuki CFTs with large $n$ $W_k$ ones.

We also think that the technology to construct continuous orbifolds of free theories can be considerably improved and extended.
In particular, we have recently observed a striking similarity between the integrals that one has to compute to determine the open and closed string spectra in these models, and some elliptic hypergeometric integrals intensively studied by Spiridonov~\cite{Spiridonov2008}; these integrals find in turn application in the analysis of dualities between four- and six-dimensional supersymmetric conformal field theories through their superconformal indices~\cite{Spiridonov:2009za}.
Although this is nothing more than a curious coincidence at this stage, in our opinion and as the history of our field has taught us, one should carefully consider this kind of similarities.


Another interesting perspective that we would like to mention here is the connection of our limits to $N=2$ Liouville theory (see~\cite{Nakayama:2004vk,Hosomichi:2004ph} for reviews) in the $c\to3$ limit.
This is work in progress: we have been able to match the bulk spectrum of the supersymmetric orbifold $\C/U(1)$ to the limit of the discrete representations of $N=2$ Liouville theory, and the spectrum of the free theory to the limit of the continuous representations.
Moreover, three-point functions for the discrete series of representations appearing in $N=2$ Liouville theory tend in the limit to a hypergeometric function with the same dependence on the labels of the representations of the limit theory analysed in chapter~\ref{ch:our-contorbi}.
Unfortunately the part of the correlator summarised in the $d$ function of equation~\eqref{ch6:d-function-gamma} cannot be reproduced yet.

If the connection with Liouville theory could be made precise, this would open the possibility of embedding continuous orbifolds into type II string theory.
We just want to mention a couple of inspiring facts that could guide the development: The context is the backreaction of massless bulk fields to the presence of a ring of $k$ NS5-branes spread evenly on a circle~\cite{Seiberg:1997zk}.
This background is described by the coset $\frac{\hat{su}(2)_k\oplus \hat{sl}(2,\R)_k}{\hat u(1)\oplus \hat u(1)}$~\cite{Israel:2004ir}, which is the product of an $N=2$ minimal model with the T-dual of $N=2$ Liouville theory (the cigar background described e.g. in the review~\cite{Schomerus:2005aq}).
The $k\to\infty$ limit would correspond to the holographic description of the weak coupling regime of little string theory in the double scaling limit~\cite{Giveon:1999px}.
It could be of great interest to understand whether a continuous orbifold could find its place in this picture.

\appendix
\chapter{Characterology}\label{app:characters}
In this appendix we collect conventions and various formulas used in the main text to compute~CFT quantities on the torus.
We use the conventions
\be
q=e^{2\pi i \tau}, \quad z=e^{2\pi i \nu}\ .
\ee

In writing characters we prefer to use $q,z$ as arguments.
In writing modular functions we use instead $\t,\nu$.
Partition functions (for the ease of reading) are usually left with implicit arguments; in case of non-obvious dependence, we normally prefer $\t,\nu$.

\section{Theta functions, and their modular properties}\label{app:characters:sec:thetas}

Throughout the text we use $\vartheta$ and $\eta$ functions with the following conventions:
\begin{align*}
\begin{array}{rrll}
 \vartheta_1(\tau,\nu)\ =&\!\!\!\! -iz^{\frac12}q^{\frac18} \!\!\!\!&\prod\limits_{n=0}^{\infty}(1-q^{n+1}z)(1-q^{n}z^{-1})(1-q^{n+1})&= -i\sum\limits_{n\in\Z}(-1)^nq^{\frac 12(n+\frac 12)^2}z^{n+\frac 12}\\
 \vartheta_2(\tau,\nu)\ = & z^{\frac12}q^{\frac18}&\prod\limits_{n=0}^{\infty}(1+q^{n+1}z)(1+q^{n}z^{-1})(1-q^{n+1})&= \sum\limits_{n\in\Z}q^{\frac 12(n+\frac 12)^2}z^{n+\frac 12}\\
\vartheta_3(\tau,\nu)\ =& &\prod\limits_{n=0}^{\infty}(1+q^{n+\frac12}z)(1+q^{n+\frac12}z^{-1})(1-q^{n+1})&= \sum\limits_{n\in\Z}q^{\frac 12n^2}z^{n}\\
\vartheta_4(\tau,\nu)\ =& &\prod\limits_{n=0}^{\infty}(1-q^{n+\frac12}z)(1-q^{n+\frac12}z^{-1})(1-q^{n+1})&=\sum\limits_{n\in\Z}(-1)^nq^{\frac 12n^2}z^{n}\\
\eta(\tau)\ =& q^{\frac{1}{24}}&\prod\limits_{n=0}^{\infty}(1-q^{n+1})\ .&
\end{array}
\end{align*}

In the main text we use also the following functions:
\begin{flalign*}
\vartheta_p(\tau)=&\frac{q^{\frac{p^2}{4}}}{\eta(\t)}\\
\Theta_{l,k}(\t,\n)=&\sum_{n\in\Z}q^{k(n+\frac{l}{2k})^2}z^{k(n+\frac{l}{2k})}
\end{flalign*}
\paragraph{Modular transformations}
In this paragraph we collect some modular properties of the functions just defined.
In particular we want to write down the behaviour under modular T-transformations ($\t\mapsto\t+1$) and modular S-transformations ($\t\mapsto-\frac{1}{\t}$).
We denote here and in the main text
\begin{equation*}
\tilde{\tau}=- \frac{1}{\tau}\ ,\quad \tilde{q}=e^{2\pi i\tilde\tau}=e^{-\frac{2\pi i}{\tau}}\ ;\qquad \tilde{\nu}=\frac{\nu}{\tau}\ ,\quad \tilde{z}=e^{2\pi i\tilde\nu}=e^{\frac{2\pi i\nu}{\tau}}\ .
\end{equation*}
The Dedekind~$\h$-function behaves as follows:
\begin{equation*}
\h(\t+1)=e^{\frac{i\pi}{12}}\h(\t)\ ,\qquad \h\left(-\frac{1}{\t}\right)=\sqrt{-i\t}\h(\t)\ .
\end{equation*}
The Jacobi~$\vartheta$-functions behave as follows:
\begin{align*}
\begin{array}{ll}
 \vartheta_1(\tau+1,\nu)=e^{\frac{i\pi}{4}}\vartheta_1(\t,\n)\ , &\qquad \vartheta_1\left(-\frac{1}{\t},\frac{\nu}{\t}\right)=-i\sqrt{-i\t}e^{\frac{i\pi\nu^2}{\tau}}\vartheta_1(\t,\nu) \\
 \vartheta_2(\tau+1,\nu)=e^{\frac{i\pi}{4}}\vartheta_2(\t,\n)\ , &\qquad \vartheta_2\left(-\frac{1}{\t},\frac{\nu}{\t}\right)=\sqrt{-i\t}e^{\frac{i\pi\nu^2}{\tau}}\vartheta_4(\t,\nu) \\
\vartheta_3(\tau+1,\nu)=\vartheta_4(\t,\n) \ ,&\qquad \vartheta_3\left(-\frac{1}{\t},\frac{\nu}{\t}\right)=\sqrt{-i\t}e^{\frac{i\pi\nu^2}{\tau}}\vartheta_3(\t,\nu) \\
\vartheta_4(\tau+1,\nu)=\vartheta_3(\t,\n) \ ,& \qquad\vartheta_4\left(-\frac{1}{\t},\frac{\nu}{\t}\right)= \sqrt{-i\t}e^{\frac{i\pi\nu^2}{\tau}}\vartheta_2(\t,\nu)\ .
\end{array}
\end{align*}
The function $\vartheta_p$ behaves as follows:
\begin{equation*}
\vartheta_p(\t+1)=e^{i\frac{\pi}{2}\left(p^2-\frac{1}{6}\right)}\vartheta_p(\t)\ ,\qquad \vartheta_p\left(-\frac{1}{\t}\right)=\frac{1}{\sqrt 2}\int_{-\infty}^{+\infty}ds\ e^{i\pi p s}\vartheta_s(\t)\ .
\end{equation*}
The  Ka\v c-Peterson $\Theta$-function behaves as follows:
\begin{equation*}
\Theta_{l,k}(\t+1,\nu)=e^{i\pi \frac{l^2}{2k}}\Theta_{l,k}(\t,\nu)\ ,\qquad \Theta_{l,k}\left(-\frac{1}{\t},\frac{\nu}{\tau}\right)=e^{i\pi \frac{\nu^2}{2\tau}}\sqrt{-i\t}\sum_{l'=-k+1}^k\frac{e^{i\pi \frac{ll'}{k}}}{\sqrt{2k}}\Theta_{l',k}(\t,\nu)\ .
\end{equation*}

\section{Characters}\label{app:characters:sec:char-limits}
In this section we collect explicit formulas for the characters of the representations that are used in the core of the thesis.
We put the accent on limits of characters.

\subsection{Limit of Virasoro characters}
For reference and as a warm-up example we write formulas for the characters of Virasoro algebra at~$c=1$, the characters for Virasoro unitary minimal models, and we compare the limit of the latter with the former.

The character over a representation $\cV_h$ of the Virasoro algebra at central charge $c$ is defined:
\begin{equation}
\chi^{(c)}_h(q)=\Tr_{\mathcal{V}_h}q^{L_0-\frac{c}{24}}=\frac{q^{h+\frac{1-c}{24}}}{\eta(\t)}\ .
\end{equation}
At central charge~$c=1$, if we restrict to unitary modules, irreducible representations are labeled by a positive integer~$r$. 
Furthermore, to impose unitarity, we have to subtract at each level $r$ a submodule associated to a singular vector~\cite{DiFrancesco:1997nk}; the resulting irreducible character reads
\begin{equation}
\chi_{h_r}^{c=1}(q)\equiv \hat{\chi}_r(q)=\frac{q^{\hat{h}_r}}{\eta(\t)}\left(1-q^r\right)=\vartheta_{r-1}(q)-\vartheta_{r+1}(q)\ ,\qquad \hat{h}_r\equiv\frac{(r-1)^2}{4}\ .
\end{equation}
Irreducible representations for unitary Virasoro minimal models at central charge $c_k=\frac{k^2+5k}{k^2+5k+6}$ are labeled by two integers $r=1\dots k+1,\ s=1\dots k+2$, as summarised in equations~\eqref{ch1:spectrum-MM}.
Their conformal weights read
\begin{equation}\label{app:characters:spectrum-MM}
 h_{r,s}=\frac{\left[r(k+3)-s(k+2)\right]^2-1}{4(k+2)(k+3)}\ ,\qquad h_{r,s}\equiv h_{k+2-r,k+3-s}\ ,
 \end{equation}
and their characters are~\cite{Feigin:1982tg,RochaCaridi:1986gw}
\begin{equation}\label{app:characters:vir-minmod-char}
\chi^{(k)}_{r,s}(q)=\frac{q^{\frac{1-c}{24}}}{\eta(\tau)}\sum_{m\in \Z}\left(q^{h_{r+2m(k+2),s}}-q^{h_{r+2m(k+2),-s}}\right)\ .
\end{equation}

The $c\to 1$ limit can be performed in two ways.

In the Roggenkamp-Wendland~\cite{Roggenkamp:2003qp} scaling (which appears in the boundary spectrum analysis of Runkel and Watts as well~\cite{Runkel:2001ng,Graham:2001tg}) we keep fixed and finite the labels $r,s$, while sending $k\to\infty$.
We easily find the following relations
\begin{equation}
\lim\limits_{c\to 1}h_{r,s}=\hat{h}_{|r-s|+1}\ ,\qquad \lim\limits_{c\to 1}h_{r,1}=\lim\limits_{c\to 1}h_{1,r}=\hat{h}_{r}\ .
\end{equation}
The $c\to 1$ limit of the Virasoro minimal model's characters is thus, as it is easy to verify,
\begin{equation}
\lim\limits_{k\to \infty}\chi^{(k)}_{r,s}(q)=\sum_{n=1}^{\min{(r,s)}}\hat{\chi}_{|r-s|+2n-1}(q)\ ,\qquad \lim\limits_{k\to \infty}\chi^{(k)}_{r,1}=\lim\limits_{k\to\infty}\chi^{(k)}_{1,r}(q)=\hat{\chi}_{r}(q)\ .
\end{equation}

The Runkel-Watts bulk case is different: the difference of the labels $r-s$ is kept finite, while $r+s$ is scaled with $k+3$.
Then
\begin{equation}
\lim\limits_{c\to 1}h_{r,s}=\frac{1}{4}\left[\frac{r+s}{2(k+3)}+(r-s)\right]^2=\frac{x^2}{4}\ .
\end{equation}
The characters approach a $\vartheta$ function as defined in section~\ref{app:characters:sec:thetas}:
\begin{equation}
\lim\limits_{k\to \infty}\chi^{(k)}_{r,s}(q)=\vartheta_x(q)\ .
\end{equation}

\subsection{$c=3$ characters of $N=2$ superconformal algebra}

A general character over a sector $\mathcal{H}_{h,Q}$ of the $\mathcal{N}=2$ superconformal algebra is defined:
\begin{equation}
\chi_{h,Q}^{\phantom {NS}}(q,z)=\Tr_{\mathcal{H}_{h,Q}}q^{L_0-\frac{c}{24}}z^{J_0}\ .
\end{equation}
We discuss here the characters of the unitary fully supersymmetric irreducible representations at $c=3$.
The Verma modules of the $N=2$ superconformal algebra contain several singular submodules,\footnote{In general there are also subsingular submodules, but they do not show up for unitary representations~\cite{Klemm:2003vn}.} which have to be taken into account. 
The structure of the singular submodules can be read off from the embedding diagrams of the representations (for further details we refer to~\cite{Kiritsis:1986rv,Eholzer:1996zi}); we will follow the classification of~\cite{Klemm:2003vn}. 
Let us explain the procedure of explicitely constructing the characters in the example of the representations of type
$I^{\pm}$ in the notations of the aforementioned paper; the labels
satisfy $\frac{h}{Q}\in \Z+\frac12$, with positive $h$ and
$Q\not\in\mathbb{Z}$. 
In this case we have only one charged singular vector. 
The singular vectors at level $\frac{h}{|Q|}=n+\frac12$ can be recognised to be\footnote{One can,
for instance, follow the spectral flow of Neveu-Schwarz null vectors
starting from the (anti)chiral primaries.}
\begin{align}
\begin{array}{ll}
G^-_{\frac12}G^-_{\frac32}\dots G^{-}_{\frac{h}{|Q|}-1}G^{+}_{-\frac{h}{|Q|}}G^{+}_{-\frac{h}{|Q|}+1}\dots G^{+}_{-\frac32} G^{+}_{-\frac12}|n,Q\ket\quad \text{for}\ Q>0\\
G^+_{\frac12}G^+_{\frac32}\dots
G^{+}_{\frac{h}{|Q|}-1}G^{-}_{-\frac{h}{|Q|}}G^{-}_{-\frac{h}{|Q|}+1}\dots
G^{-}_{-\frac32} G^{-}_{-\frac12}|n,Q\ket\quad \text{for}\ Q<0
\end{array}\ ,
\end{align}
and they have relative charge $+1$ and $-1$, respectively. 
In the character of the irreducible representation we have to subtract the contribution of the submodule associated to them. 
The result is
\begin{equation}
\chi^{I^{\pm}}_{n,Q}(q,z)=q^{(n+\frac12)|Q|-\frac18}z^Q\left[\prod_{m=0}^{\infty}\frac{(1+q^{m+\frac12}z)(1+q^{m+\frac12}z^{-1})}{(1-q^{m+1})^2}\right]\left(1-\frac{q^{n+\frac12}z^{\text{sgn}Q}}{1+q^{n+\frac12}z^{\text{sgn}Q}}\right)\ .
\end{equation}
The other cases are analogous, and we can write:
\begin{subequations}\label{c=3-characters}
\begin{itemize}
\item \textbf{Vacuum:} ($Q=h=0$) 
\begin{flalign}
&\chi^{\text{vac}}_{0,0}(q,z)= \frac{\vartheta_{3}(\tau,\nu)}{\eta^{3}(\tau)} 
\left(1-\frac{q^{\frac12}z}{1+q^{\frac12}z}-\frac{q^{\frac12}z^{-1}}{1+q^{\frac12}z^{-1}}\right)&
\end{flalign}
\item \textbf{Type ${\boldsymbol{0}}$:} ($Q=0\, ,\ h\in\R\setminus\{0\}$)
\begin{flalign}
&\chi^{0}_{h,0}(q,z)=q^{h}
\frac{\vartheta_{3}(\tau,\nu)}{\eta^{3}(\tau)}  &
\end{flalign}
\item \textbf{Type ${\boldsymbol{I^{\pm}}}$:} ($0<|Q|<1\, ,\
h=|Q|(n+\frac12)\, ,\  n\in \Z_{\geq 0}$)
\begin{flalign}
&\chi^{I^{\pm}}_{|Q|(n+\frac12),Q}(q,z)=q^{(n+\frac12)|Q|}z^Q
\frac{\vartheta_{3}(\tau,\nu)}{\eta^{3}(\tau)}  
\left(1-\frac{q^{n+\frac12}z^{\text{sgn}Q}}{1+q^{n+\frac12}z^{\text{sgn}Q}}\right)&
\end{flalign}
\item \textbf{Type ${\boldsymbol{II}}^{\pm}$:} ($Q=\pm1\, ,\
h\in\R_{\geq0}$)
\begin{flalign}
&\chi^{II^{\pm}}_{h,Q}(q,z)=q^{h}z^Q
\frac{\vartheta_{3}(\tau,\nu)}{\eta^{3}(\tau)} \left(1-q^{|Q|}\right)&
\end{flalign}
\item \textbf{Type ${\boldsymbol{III^{\pm}}}$:} ($Q=\pm1\, ,\
h\in\Z +\frac12$)
\begin{flalign}
&\chi^{III^{\pm}}_{h,Q}(q,z)=q^{h}z^Q
\frac{\vartheta_{3}(\tau,\nu)}{\eta^{3}(\tau)}
\left(1-q-
\frac{q^{h}z^{\text{sgn}(Q)}}{1+q^{h}z^{\text{sgn}(Q)}}+
\frac{q^{h+2}z^{\text{sgn}(Q)}}{1+q^{h+1}z^{\text{sgn}(Q)}}
\right)&
\end{flalign}
\end{itemize}
\end{subequations}

Ramond characters can be obtained from the Neveu-Schwarz characters by
spectral flow (see e.g.\ \cite{Lerche:1989uy}). We
give an example: let us denote spectral flowed operators and sectors
by an upper label $\eta$, which indicates the amount of spectral flow
units to use. Under a flow of $\eta=\pm 1/2$, primary vectors of the
Neveu-Schwarz sector become Ramond
primaries, and the same happens for Neveu-Schwarz
singular vectors, which flow to Ramond singular vectors. The Ramond
characters can then be computed using the formula
\begin{equation}
\chi_{h^{\eta},Q^{\eta}}^{\phantom {\text{NS}}}(q,z)=\Tr_{\mathcal{H}_{h^{\eta},Q^{\eta}}}q^{L_0-\frac{c}{24}}z^{J_0}=\Tr_{\mathcal{H}_{h,Q}}q^{L^{\eta}_0-\frac{c}{24}}z^{J^{\eta}_0}\ ,
\end{equation}
with the spectral flowed operators
\begin{align}
 L_n^{\eta}&=L_n-\eta J_n+\frac{c}{6}\eta^2\delta_{n,0} &
 J_n^{\eta}&=J_n-\frac{c}{3}\eta\delta_{n,0}\ .
\end{align}
For $c=3$ and $\eta=\frac12$, $L^{1/2}_{0}=L_0-\frac12J_0+\frac18$ and $J^{1/2}_0=J_0-\frac12$, we have
\begin{equation}
\chi_{h^{1/2},Q^{1/2}}^{\phantom {\text{NS}}}(q,z)=q^{\frac18}z^{-\frac12}\chi_{h,Q}(q,q^{-\frac12}z)\ .
\end{equation}
Starting e.g.\ from the type $I$ characters in the Neveu-Schwarz
sector we find the characters
\begin{align}\label{R0-character}
\chi^{\text{R}^0}_{\frac18,Q}(q,z)&=\frac{z^{Q}}{z^{1/2}-z^{-1/2}}
\frac{\vartheta_{2} (\tau,\nu)}{\eta^{3} (\tau)} \ ,\
-\frac{1}{2}<Q<\frac{1}{2} \\
\chi^{\text{R}}_{\frac18+n|Q|,Q}(q,z)&=
\frac{q^{n|Q|}z^{Q}}{1+q^{n}z^{\text{sgn}(Q)}}
\frac{\vartheta_{2} (\tau,\nu)}{\eta^{3} (\tau)}  \ ,\ 0<|Q|<1 \ ,\
n\geq 1\ ,
\end{align}
where in the first character the lowest lying state is a Ramond ground
state, whereas in the second character there are two lowest lying
states of charges $Q\pm \frac{1}{2}$.

\subsection{$N=2$ minimal model characters}

Unitary irreducible representations for the bosonic subalgebra of the
$N=2$ superconformal algebra at central charge $c=3\frac{k}{k+2}$ are
labelled by three integers $(l,m,s)$ with $0\leq l\leq k$, $m\equiv
m+2k+4$, $s\equiv s+4$, and $l+m+s$ even. Not all triples label
independent representations, and they are identified according to
\begin{equation}\label{fieldidentification}
(l,m,s) \sim (k-l,m+k+2,s+2) \ .
\end{equation}
The character over a sector labeled by $(l,m,s)$ is defined
\be
\chi_{(l,m,s)}=\Tr_{\Hilb_{(l,m,s)}} q^{L_0-\frac{c}{24}}z^{J_0}\ .
\ee

A useful way of writing the characters for $N=2$ minimal models, based on the results of~\cite{Qiu:1987ux,Gepner:1987qi}, is the following
\begin{equation}\label{app:characters:minmod-kac-peterson-form}
\chi_{(l,m,s)}(q,z)=\sum_{j=0}^{k-1}C^{(k)}_{l,m-s+4j}(q)\Theta_{2m+(4j-s)(k+2),2k(k+2)}\left(\t,-\frac{\n}{2k+4}\right)\ .
\end{equation}
Modular properties are explicit in this formulation (thanks to the known behaviour of $\Theta$-functions, recalled in section~\ref{app:characters:sec:thetas}).

Representations of the full superconformal algebra are obtained by combining representations labelled by $(l,m,s)$ and $(l,m,s+2)$.
In the Neveu-Schwarz sector for $|m|\leq l$ the characters can be written~\cite{Ravanini:1987yg}
\begin{align}
\chi^{\text{NS}}_{l,m}(q,z):=&\left( \chi_{(l,m,0)}+ \chi_{(l,m,2)}\right)(q,z)\nonumber\\
= &\ q^{\frac{(l+1)^{2}-m^2}{4(k+2)}-\frac18}\,z^{-\frac{m}{k+2}}\left[\prod_{n=0}^{\infty}\frac{(1+q^{n+\frac12}z)(1+q^{n+\frac12}z^{-1})}{(1-q^{n+1})^2}\right]\times \Gamma^{(k)}_{lm}(\tau,\nu)\ ,
\label{minmod-NS-character}
\end{align}
and in the Ramond sector (for $|m|\leq l+1$)
\begin{align}
\chi^{R}_{l,m}(q,z):=&\left(\chi_{(l,m,1)}+\chi_{(l,m,-1)}\right)(q,z)\nonumber\\
= \ &q^{\frac{(l+1)^{2}-m^2}{4(k+2)}}\,z^{-\frac{m}{k+2}}(z^{\frac12}+z^{-\frac12})\left[\prod_{n=0}^{\infty}\frac{(1+q^{n+1}z)(1+q^{n+1}z^{-1})}{(1-q^{n+1})^2}\right]\times \Gamma^{(k)}_{lm}(\tau,\nu)\ ,
\label{minmod-R-characters}
\end{align}
where the structure of the singular vectors is summarised in $\Gamma^{(k)}_{lm}$,
\begin{align}
 \Gamma^{(k)}_{lm}(\tau,\nu) = &\sum_{p=0}^{\infty}q^{(k+2)p^2+(l+1)p}\left(1-\frac{q^{(k+2)p+\frac{l+m+1}{2}}z}{1+q^{(k+2)p+\frac{l+m+1}{2}}z}-\frac{q^{(k+2)p+\frac{l-m+1}{2}}z^{-1}}{1+q^{(k+2)p+\frac{l-m+1}{2}}z^{-1}}\right)\nonumber\\
  - & \sum_{p=1}^{\infty}q^{(k+2)p^2-(l+1)p}\left(1-\frac{q^{(k+2)p-\frac{l+m+1}{2}}z^{-1}}{1+q^{(k+2)p-\frac{l+m+1}{2}}z^{-1}}-\frac{q^{(k+2)p-\frac{l-m+1}{2}}z}{1+q^{(k+2)p-\frac{l-m+1}{2}}z}\right)\ .
\label{def-Gamma}
\end{align}

\subsection{Limits of $N=2$ minimal model characters}\label{app:characters:sec:limit}

In the limit $k\to\infty$ in the expression~\eqref{def-Gamma} for
$\Gamma^{(k)}_{lm}$ in each sum only the first summand can contribute,
\begin{align}\label{limit-Gamma}
 \Gamma^{(k)}_{lm}(\tau,\nu) \approx &\left( 1-\frac{q^{\frac{l+m+1}{2}}z}{1+q^{\frac{l+m+1}{2}}z}-\frac{q^{\frac{l-m+1}{2}}z^{-1}}{1+q^{\frac{l-m+1}{2}}z^{-1}}\right)\nonumber\\
 & - q^{k-l+1}\left(1-\frac{q^{\frac{2k-l-m+3}{2}}z^{-1}}{1+q^{\frac{2k-l-m+3}{2}}z^{-1}}-\frac{q^{\frac{2k-l+m+3}{2}}z}{1+q^{\frac{2k-l+m+3}{2}}z}\right)
 \ , 
\end{align}
and the precise behaviour of the character depends on the details
of how $l$ and $m$ behave in the limit.

For our analysis we need to consider the following cases in the Neveu-Schwarz sector:
\begin{enumerate}
\item $l=m=0$: The limit character is simply the $N=2$ vacuum
character,
\begin{equation}
\lim_{k\to\infty} \chi^{\text{NS}}_{0,0} = \chi^{\text{vac}}_{0,0}\ .
\end{equation}
\item $l+m=2n$ finite, $m/ (k+2)\to -Q$, $0<Q<1$: Only one singular
vector survives and we find
\begin{equation}
\lim_{k\to \infty} \chi^{\text{NS}}_{|m|+2n,m}
= \chi_{Q (n+\frac{1}{2}),Q}^{I^{+}} \ .
\end{equation}
\item $l-m=2n$ finite, $m/ (k+2)\to -Q$, $-1<Q<0$: Only one singular
vector survives and we find
\begin{equation}
\lim_{k\to \infty} \chi^{\text{NS}}_{m+2n,m}
= \chi_{|Q| (n+\frac{1}{2}),Q}^{I^{-}} \ .
\end{equation}
\item $l+m=2n$ finite, $l=k$: The first summand in~\eqref{limit-Gamma}
gives one positively charged singular vector, the second produces one uncharged one and adds one positively charged singular submodule. We find
\begin{equation}\label{limit-III-minus}
\lim_{k\to\infty} \chi^{\text{NS}}_{k,-k+2n} 
= \chi_{n+\frac{1}{2},1}^{III^{+}} \ .
\end{equation}
\item $l-m=2n$ finite, $l=k$: Analogously to the previous case we obtain
\begin{equation}\label{limit-III-plus}
\lim_{k\to\infty} \chi^{\text{NS}}_{k,k-2n} 
= \chi_{n+\frac{1}{2},-1}^{III^{-}} \ .
\end{equation}
\end{enumerate}
There are several other cases, depending on the behaviour of $l\pm m$
for large $k$; in these other situations the limiting character
decomposes into a sum of $N=2$ characters. We illustrate this in the
example of fixed labels $l,m$: in this instance the
conformal weights and $U(1)$ charge of all the primary fields approach
zero, the second line of equation~\eqref{limit-Gamma} gets suppressed,
but the first line stays finite. The character then takes the
form
\begin{equation}
\lim_{k\to\infty}\chi^{\text{NS}}_{l,m} (q,z) =\frac{\vartheta_3(\tau,\nu)}{\eta^3(\tau)}\left( 1-\frac{q^{\frac{l+m+1}{2}}z}{1+q^{\frac{l+m+1}{2}}z}-\frac{q^{\frac{l-m+1}{2}}z^{-1}}{1+q^{\frac{l-m+1}{2}}z^{-1}}\right)\ .
\end{equation}
Noticing the relation
\begin{equation}
\frac{\vartheta_3(\tau,\nu)}{\eta^3(\tau)}
\left(\frac{q^{n+\frac{1}{2}}z^{\pm 1}}{1+q^{n+\frac{1}{2}}z^{\pm 1}}
- \frac{q^{n+\frac{3}{2}}z^{\pm 1}}{1+q^{n+\frac{3}{2}}z^{\pm 1}}
\right) = \chi^{III^{\pm}}_{n+\frac{1}{2},\pm 1} (q,z) \ ,
\end{equation}
it is easy to show that
\begin{equation}
\lim_{k\to\infty}\chi^{\text{NS}}_{l,m} =
\chi^{\text{vac}}_{0,0}+\sum_{j=0}^{\frac{l+m}{2}-1}\chi^{III^+}_{\frac{l+m}{2}-(\frac12+j),1}+\sum_{j=0}^{\frac{l-m}{2}-1}\chi^{III^-}_{\frac{l-m}{2}-(\frac12+j),-1}\ .
\end{equation}
Following similar lines it is possible to show that this kind of
decomposition is common to all the cases we have not listed
explicitly.

\section{Supersymmetric partition functions and GSO projections}\label{app:characters:sec:GSO}

The toroidal partition function of a CFT contains the information about which right movers couple to which left movers.
In other words, it gives us the collection of non-chiral fields present in the spectrum.
Modular transformations map certain combinations of characters among each other, and the requirement of modular invariance for the partition function drastically constraints its form.

A fully supersymmetric partition function keeps all the possible combinations between holomorphic and anti-holomorphic fields, regardless of their spin alignment. 
For $N=2$ minimal models this means to couple a holomorphic representation of the full superconformal algebra with its anti-holomorphic counterpart.
We indicate the fully supersymmetric partition function with $\cP _k=\cP^{\text{NS}}_k+\cP^{\text{R}}_k$. 
The NS part reads:
\begin{equation}
\begin{split} 
\mathcal{P}_{k}^{\text{NS}}(\tau,\nu)&=\sum_{l=0}^{k}\sum_{\substack{m=-l\\ l+m\ \text{even}}}^l\left(\chi_{(l,m,0)} (q,z)+\chi_{(l,m,2)} (q,z)\right)\left(\bar\chi_{(l,m,0)} (\bar{q},\bar{z})+\bar\chi_{(l,m,2)} (\bar{q},\bar{z})\right)\\
&=\left|\frac{\vartheta_3(\tau,\nu)}{\eta^3(\tau)}\right|^2\sum_{l=0}^{k}\sum_{\substack{m=-l\\
l+m\ \text{even}}}^l\left|q^{\frac{(l+1)^{2}-m^2}{4(k+2)}}
\Gamma^{(k)}_{lm}(\tau,\nu)\right|^2\, (z\bar{z})^{-\frac{m}{k+2}}\ ,
\end{split}
\end{equation}
and the R part:
\begin{equation}
\begin{split} 
\mathcal{P}_{k}^{\text{R}}(\tau,\nu)&=\sum_{l=0}^{k}\sum_{\substack{m=-l\\ l+m\ \text{even}}}^l\left(\chi_{(l,m,1)} (q,z)+\chi_{(l,m,-1)} (q,z)\right)\left(\bar\chi_{(l,m,1)} (\bar{q},\bar{z})+\bar\chi_{(l,m,-1)} (\bar{q},\bar{z})\right)\\
&=\left|\frac{\vartheta_2(\tau,\nu)}{\eta^3(\tau)}\right|^2\sum_{l=0}^{k}\sum_{\substack{m=-(l+1)\\
l+m\ \text{odd}}}^{l+1}\left|q^{\frac{(l+1)^{2}-m^2}{4(k+2)}}
\Gamma^{(k)}_{lm}(\tau,\nu)\right|^2\, (z\bar{z})^{-\frac{m}{k+2}}\ .
\end{split}
\end{equation}
It is immediate to see that~$\cP$ is not modular invariant since, for example, under the modular T-transformation, $\vartheta_3\overset{T}{\mapsto} \vartheta _4$.
There is a countable amount of choices of modular partition functions (whose description and classification~\cite{Gannon:1996hp,Gray:2008je} is inherited from the analysis of $SU(2)_k$ modular invariants pioneered in~\cite{Cappelli:1987hf}).
In this thesis we are interested in the so-called A-series of minimal models, whose representatives have a diagonal modular invariant partition function:
\begin{equation}
Z_{k}(\tau,\nu)=\sum_{s=-1}^{2}\sum_{l=0}^{k}\sum_{\substack{|m-s|\leq l\\ l+m+s\
\text{even}}}\chi_{(l,m,s)}
(q,z)\bar\chi_{(l,m,s)}(\bar{q},\bar{z})\ .
\end{equation}
In the string theory language a constraint on the allowed closed string excitations corresponds to a GSO-projection, and a diagonal partition function corresponds to a projection of type-0 (since no NS-sector is combined to any R-sector, we do not realise target-space fermions in this way, thus we break supersymmetry in the target-space; for details see~\cite{Gaberdiel:2000jr}).

Sometimes it is easier to firstly compute the fully supersymmetric partition function, and then to perform the GSO-projection at the end.
This is the approach we adopt in this thesis: we define ``target-space'' fermion number operators $(-1)^F,(-1)^{\tilde{F}}$, acting on the holomorphic and anti-holomorphic sectors respectively, and we retain only the subsectors of the theory that are left invariant by the type-0 GSO-projections
\begin{align}\label{app:characters:type-0-GSO}
\begin{array}{rcl}
\text{NS-sector:}& \qquad & P_{\text{GSO}}=\12 \left(1+(-1)^{F+\bar F}\right)\ ,\\
\text{R-sector:}& \qquad & P_{\text{GSO}}=\12 \left(1\pm(-1)^{F+\bar F}\right)\ .
\end{array}
\end{align} 
The operators $(-1)^F,(-1)^{\bar{F}}$ are defined such that their eigenvalue on a generic descendant is either 1 or -1 depending on the number of times one has acted with fermionic generators on the corresponding ground state to produce the descendant.
In our context the operators $F,\bar F$ should map states with label $s$, to states with label $s+2$. 

We can realise operators that work similarly in terms of the~$U(1)$ generator $J_0$, in particular for the holomorphic NS sector we have
\begin{equation}\label{app:characters:def-similar-fermion-numbers}
\begin{split}
\Tr_{\mathcal H_{l,m,0}\oplus \mathcal H_{l,m,2}}\left[\12\left(1+(-1)^{J_0}\right)q^{L_0-\frac{c}{24}}z^{J_0}\right]=& \frac12(\chi^{NS}_{l,m}(q,z)+e^{i\pi \frac{m}{ k+2}}\chi^{NS}_{l,m}(q,-z))\ ,\\
\Tr_{\mathcal H_{l,m,0}\oplus \mathcal H_{l,m,2}}\left[\12\left(1-(-1)^{J_0}\right)q^{L_0-\frac{c}{24}}z^{J_0}\right]=& \frac12(\chi^{NS}_{l,m}(q,z)-e^{i\pi \frac{m}{ k+2}}\chi^{NS}_{l,m}(q,-z))\ .
\end{split}
\end{equation}
Up to a charge correction on the second addendum on the right hand side of equations~\eqref{app:characters:def-similar-fermion-numbers}, these operators do the right job.
In terms of fully supersymmetric characters the GSO-projected partition function thus reads (in the NS sector for example)
\begin{equation}
\begin{split}
Z^{\text{NS}}_{k}(\tau,\nu)=&\sum_{l=0}^{k}\sum_{\substack{m=-l\\ l+m\ \text{even}}}^l\chi_{(l,m,0)} (q,z)\bar\chi_{(l,m,0)} (\bar{q},\bar{z}) +\chi_{(l,m,2)} (q,z)\bar\chi_{(l,m,2)} (\bar{q},\bar{z})\\
=&\ \frac12\sum_{l=0}^{k}\sum_{\substack{m=-l\\ l+m\
\text{even}}}^l\left(\chi^{\text{NS}}_{l,m}
(q,z)\bar\chi^{\text{NS}}_{l,m}(\bar{q},\bar{z})+\chi^{\text{NS}}_{l,m}
(q,-z)\bar\chi^{\text{NS}}_{l,m}(\bar{q},-\bar{z})\right)\\
=&\ \12\left(\cP_{k}^{\text{NS}}(\t,\n)+ \cP_{k}^{\text{NS}}(\t,\n+\tfrac 12)\right)\ .
\end{split}
\end{equation}

The boundary theory is also constrained by the choice of the bulk partition function, since the boundary states only couple to physical sectors of the closed string theory. 
In particular open strings stretched between different D-branes can couple or not to the R-R fields, depending on the choice of the sign in equation~\eqref{app:characters:type-0-GSO}, and on the dimensionality of the D-brane~\cite{Gaberdiel:2002jr}.
In this thesis we fix the convention that odd-dimensional branes (of the A-type in strictly diagonal models, using the terminology explained in section~\ref{ch3:sec:BC}) couple to R-R closed strings.
In the string theory language, this accounts to choose a type 0B  projection, namely the minus sign in equations~\eqref{app:characters:type-0-GSO}.

\section{A non-trivial modular transformation}
In this section we want to show an alternative proof of a central identity of this thesis, namely (compare with equations~\eqref{ch7:contorbi-D0-open-string-self-energy-vacuum} and~\eqref{overlap-contorbi}) the modular transformation of the $c=3$ vacuum character~$\chi^{\text{vac}}_{0,0}(q)$ defined in the list~\eqref{c=3-characters}.
The modular transformed vacuum character is given in terms of $(\tilde{\t},\tilde{z})$ by
\begin{equation}\label{app:char:modular-transf-vacuum-LHS}
\chi^{\text{vac}}_{0,0}(\tilde q)=\frac{\vartheta_3(\tilde \t,\tilde \n)}{\eta^3(\tilde \t)}\left(1-\frac{\tilde q^{\12}\tilde z}{1+\tilde q^{\12}\tilde z}-\frac{\tilde q^{\12}\tilde z^{-1}}{1+\tilde q^{\12}\tilde z^{-1}}\right)\ .
\end{equation}
The claim is:
\begin{equation}\label{app:char:modular-transf-vacuum-RHS}
\chi^{\text{vac}}_{0,0}(\tilde q)=\frac{\vartheta_3( \t, \n)}{\eta^3( \t)}\sum_{n\in\Z_{\geq 0}}\int_{-1}^{1}dQ\;2\sin{\pi |Q|}q^{\left(n+\12\right)|Q|}z^Q\left(1-\frac{q^{n+\12}z^{\sgn Q}}{1+q^{n+\12}z^{\sgn Q}}\right)\ .
\end{equation}
\paragraph{Left hand side}
Let us start from equation~\eqref{app:char:modular-transf-vacuum-LHS}: noticing that
\begin{equation}
\frac{1-\tilde q^{\12}\tilde z}{1+\tilde q^{\12}\tilde z}=-i\tan \frac{\pi}{2}(\tilde\t+2\tilde\n)\ ,
\end{equation}
we can recast it as
\begin{equation}\label{app:char:modular-transf-last-LHS}
\begin{split}
\eqref{app:char:modular-transf-vacuum-LHS}=&\ \frac{1}{2\t}\frac{\vartheta_3( \t, \n)}{\eta^3( \t)}\left(\tan \frac{\pi}{2}(\tilde \t+2\tilde \n)+\tan \frac{\pi}{2}(\tilde \t-2\tilde \n)\right)\\
=&\ \frac{1}{\pi \t}\frac{\vartheta_3( \t, \n)}{\eta^3( \t)}\left[\sum_{n\geq 0}\ 
\frac{\frac{\tilde \t}{2}+\tilde{\n}}{\left(n+\12\right)^2-\left(\frac{\tilde \t}{2}+\tilde \n\right)^2}+
\frac{\frac{\tilde \t}{2}-\tilde{\n}}{\left(n+\12\right)^2-\left(\frac{\tilde \t}{2}-\tilde \n\right)^2}\right]
\end{split}
\end{equation}
where we have used the relation~\cite[first formula of 1.421]{Gradshteyn:book}
\begin{equation}
\tan \frac{\pi}{2}x=\frac{x}{\pi}\sum_{n\geq 0}\frac{1}{\left(n+\12\right)^2-\frac{x^2}{4}}
\end{equation}
to go from the first to the second equality.
\paragraph{Right hand side}
The integral in equation~\eqref{app:char:modular-transf-vacuum-RHS} can be explicitely evaluated by dividing the integration range into positive and negative~$Q$:
\begin{equation}
I_1=\int_{0}^1dQ\;q^{\left(n+\12\right)Q}z^Q\sin{\pi Q}\ ,\qquad I_2=\int_{-1}^0dQ\;q^{-\left(n+\12\right)Q}z^Q(-\sin{\pi Q})\ .
\end{equation}
The integrals are easy to evaluate once written the integrands in terms of exponentials, and the results read:
\begin{align}
\begin{array}{lll}
I_1=&\frac{1}{4\pi}\left(1+q^{\left(n+\12\right)}z\right)&\left[\frac{1}{\left(n+\12\right)\t +\nu +\12}-\frac{1}{\left(n+\12\right)\t +\nu -\12}\right]\ ,\\[4mm]
 I_2=&\frac{1}{4\pi}\left(1+q^{\left(n+\12\right)}z^{-1}\right)&\left[\frac{1}{\left(n+\12\right)\t -\nu +\12}-\frac{1}{\left(n+\12\right)\t -\nu -\12}\right]\ .
 \end{array}
\end{align}
Collecting and rearranging the pieces, equation~\eqref{app:char:modular-transf-vacuum-RHS} becomes
\begin{equation}
\begin{split}
\eqref{app:char:modular-transf-vacuum-RHS}=\ 
&-\frac{1}{2\pi \t}\frac{\vartheta_3( \t, \n)}{\eta^3( \t)}\sum_{n\geq 0}\ \left[
\frac{2\frac{\n}{\t}+\frac{1}{\t}}{\left(n+\12\right)^2-\left(\frac{\nu}{\tau}+\frac{1}{2\t}\right)^2}
+\frac{-2\frac{\n}{\t}+\frac{1}{\t}}{\left(n+\12\right)^2-\left(-\frac{\nu}{\tau}+\frac{1}{2\t}\right)^2}
\right]\\
=\ &-\frac{1}{\pi \t}\frac{\vartheta_3( \t, \n)}{\eta^3( \t)}\sum_{n\geq 0}\ \left[
\frac{\tilde{\nu}-\frac{\tilde \t}{2}}{\left(n+\12\right)^2-\left(\tilde{\nu}-\frac{\tilde \t}{2}\right)^2}
+\frac{-\tilde{\nu}-\frac{\tilde \t}{2}}{\left(n+\12\right)^2-\left(-\tilde{\nu}-\frac{\tilde \t}{2}\right)^2}
\right]\ ,
\end{split}
\end{equation}
which corresponds to the second line of~\eqref{app:char:modular-transf-last-LHS}.
We have then proven the anticipated claim, namely $\eqref{app:char:modular-transf-vacuum-LHS}=\eqref{app:char:modular-transf-vacuum-RHS}$.

\chapter{Asymptotics of Wigner 3j-symbols}
\label{app:wigner}
We want to approximate the Wigner 3j-symbols in the limit of large
quantum numbers, in a specific range of parameters defined by the
limiting procedure which is described in the core of this paper.

\section{Notations and preliminaries}
To set up our notations, let us briefly state the definition of the
Clebsch-Gordan coefficients. A spin $j$ representation $V_{j}$ of $su (2)$
with standard generators $J_{i}$ satisfying
$[J_{i},J_{j}]=i\epsilon_{ijk}J_{k}$ has a natural basis consisting of
the eigenvectors $|j,\mu\rangle$ of the generator $J_{3}$ with
eigenvalue $\mu$. The tensor product of two irreducible
representations can be decomposed into irreducible representations of
the diagonal subalgebra,
\begin{equation}
V_{j_{1}}\otimes V_{j_{2}} = \bigoplus_{j} V_{j} \ ,
\end{equation}
where $|j_{1}-j_{2}|\leq j\leq j_{1}+j_{2}$ and $j+j_{1}+j_{2}$ is an integer.
The Clebsch-Gordan coefficients 
\begin{equation}
\langle j_1,\mu_1;j_2,\mu_2|j_1,j_2,j,\mu\rangle
\end{equation} 
are then given by the overlap of the two natural sets of basis vectors.

Closely related are the Wigner 3j-symbols that are defined as
\begin{equation}
\begin{pmatrix} j_{1}& j_{2} & j_{3} \\
\mu_{1} & \mu_{2} & \mu_{3}
\end{pmatrix}
:=\frac{(-1)^{j_1-j_2-\mu_3}}{\sqrt{2j_3+1}}\langle
j_1,\mu_1;j_2,\mu_2|j_1,j_2,j_3,-\mu_3\rangle \ ,
\end{equation}
with the choice of conventions: $\mu_3=-\mu =-\mu_1-\mu_2$.\\
An explicit expression was obtained by Racah in~\cite{Racah:1942II}
(see e.g.\ \cite[section 8.2, eq.3]{Varsalovic:book}),
\begin{align}
&\begin{pmatrix} j_{1}& j_{2} & j_{3} \\
\mu_{1} & \mu_{2} & \mu_{3}
\end{pmatrix} =(-1)^{j_{1}-j_{2}-\mu_{3}}
\left(\frac{(j_1\!+\!j_2\!-\!j_{3})!(j_1\!-\!j_2\!+\!j_{3})!(-j_1\!+\!j_2\!+\!j_{3})!}{(j_1\!+\!j_2\!+\!j_{3}\!+\!1)!}\right)^{\!1/2}\nonumber\\
&\qquad \times
[(j_1\!+\!\mu_1)!(j_1\!-\!\mu_1)!(j_2\!+\!\mu_2)!(j_2\!-\!\mu_2)!(j_{3}\!+\!\mu_{3})!
(j_{3}\!-\!\mu_{3})!]^{1/2}  \nonumber \\
&\qquad \times
\sum_z  
\frac{(-1)^z}{z!(j_1\!+\!j_2\!-\!j_{3}\!-\!z)!(j_1\!-\!\mu_1\!-\!z)!
(j_2\!+\!\mu_2\!-\!z)!(j_{3}\!-\!j_2\!+\!\mu_1\!+\!z)!(j_{3}\!-\!j_1\!-\!\mu_2\!+\!z)!
}\;,
\label{Racah}
\end{align}
where the sum over $z$ runs over all the values for which the
arguments of the factorials in the denominator are non-negative. In
particular, this formula provides a simple expression if one of the
labels $\mu_{i}$ is extremal, e.g.\ 
\begin{align}
&\begin{pmatrix}
j_{1} & j_{2} & j_{3}\\
-j_{1} & \mu_{2} & \mu_{3}
\end{pmatrix} = \begin{pmatrix}
j_{3} & j_{1} & j_{2} \\
\mu_{3} & -j_{1} & \mu_{2}
\end{pmatrix}\nonumber\\
& \quad = (-1)^{j_{3}-j_{1}-\mu_{2}}
\left(\frac{(-j_{1}\!+\!j_{2}\!+\!j_{3})!(j_{3}\!+\!\mu_{3})!(j_{2}\!+\!\mu_{2})!(2j_{1})!}{(j_{1}\!-\!j_{2}\!+\!j_{3})!(j_{1}\!+\!j_{2}\!-\!j_{3})!(j_{3}\!-\!\mu
_{3})!(j_{2}\!-\!\mu_{2})!(j_{1}\!+\!j_{2}\!+\!j_{3}\!+\!1)!}
\right)^{\!\frac{1}{2}} \ .
\label{app:extremal3j}
\end{align}

\section{Wigner's estimate} 

\begin{figure}[t] \centering \subfloat[][The shaded region is the
projection of the triangle formed by the classical vectors on the
$x$-$y$ plane.]
{\includegraphics[width=.45\columnwidth]{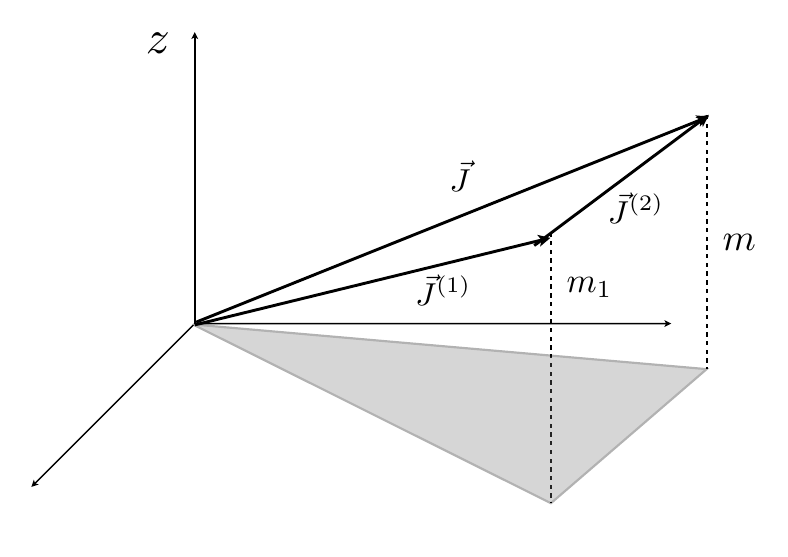}} \quad
\subfloat[][The continuous line is a plot coming from the Wigner
estimate~\eqref{Wigner's-estimate}. The points connected by dashed
lines are the exact values of the 3j-symbols. $n$ ranges from 0 to
200.]
{\includegraphics[width=.45\columnwidth]{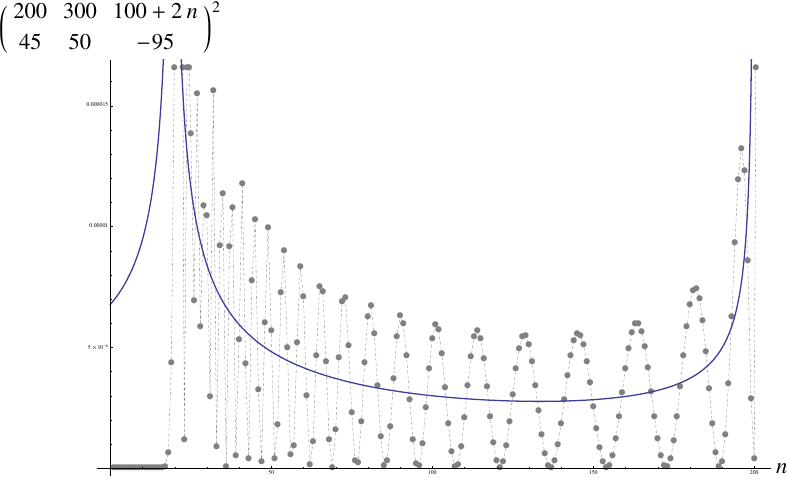}}
\caption{Wigner approximation.}  \label{Wigner-fig}
\end{figure}

For large quantum numbers one expects the Clebsch-Gordan coefficients
to be related to the classical problem of adding angular momenta. This
issue has first been discussed by Wigner in~\cite{Wigner:book}. To
each quantum angular momentum specified by $j_{i},\mu_{i}$ we
therefore associate a vector $\vec{J}^{{(i)}}$ of length squared
$|\vec{J}^{(i)}|^{2}=j (j+1)$ and with specified $z$-component
$J_{z}^{(i)}=\mu_{i}$. The $x$- and $y$- component are not
specified. Classically such angular momenta can be coupled to zero if
they satisfy the condition
$\vec{J}^{(1)}+\vec{J}^{(2)}+\vec{J}^{(3)}=0$. If this is the case,
the triangle their projections form in the $x$-$y$-plane (see
figure~\ref{Wigner-fig} (a)) has an area
\begin{equation}\label{app:defA}
A=\frac{1}{4}\sqrt{(\lambda_1+\lambda_2+\lambda_3) (-\lambda_1+\lambda_2+\lambda_3) (\lambda_1-\lambda_2+\lambda_3) (\lambda_1+\lambda_2-\lambda_3)} \ ,
\end{equation}
where
$\lambda_{i}=\sqrt{|\vec{J}^{(i)}|^{2}-|J_{z}^{(i)}|^{2}}=\sqrt{j_{i}
(j_{i}+1)-\mu_{i}^{2}}$ are the lengths of the projections of
$\vec{J}^{(i)}$ in the $x$-$y$-plane.
The quantum numbers are then said to lie in a \emph{classically
allowed} region. If there are no associated vectors that can be added
to zero, they belong to a \emph{classically forbidden} region; in that
case the ``area'' $A$ in~\eqref{app:defA} is imaginary. If the
projected triangle degenerates ($A=0$), they are said to be in the
\emph{transition region}.

Wigner gave an estimate of the averaged semiclassical behaviour of the
Clebsch-Gordan coefficient in the allowed region~\cite{Wigner:book},
\begin{equation}\label{Wigner's-estimate} |\langle
j_1,\mu_1;j_2,\mu_2|j_1,j_2,j,\mu\rangle|^2_{\text{averaged}}\approx\frac{2j+1}{4\pi|A|}\ .
\end{equation} 
One naturally expects (and it is shown numerically e.g.\ in~\cite{Schulten:1971yv}) that the accuracy of the approximation goes down when the area $A$ is small compared to the typical length squared of the vectors $\vec{J}^{(i)}$.
In other words, the Wigner's estimate works remarkably well as far as we are deep inside the allowed region.
\subsection{Improvement of the Wigner estimate, and issue in the transition region}
The Wigner approximation can be improved using different representations of Clebsch-Gordan coefficients (for details see e.g.\ \cite{Schulten:1971yv,Reinsch:1999}):
An integral formula exists,
\begin{displaymath}
\langle j_1,m_1;j_2,m_2 |j_1,j_2,j,m \rangle =
(-1)^{j+m} \, (2 i)^{j+j_1+j_2} \, \pi^{-2} \,
N_{j_1 \, m_1 \, j_2 \, m_2 \, j \, m} \, 
\end{displaymath}
\begin{equation}\label{app:wigner-integral-rep}
\times \int_{-\pi/2}^{\pi/2} \int_{-\pi/2}^{\pi/2}d\theta  d \phi\,e^{2 i m_1 \phi + 2 i m_2 \theta} \, \sin^{j+j_2-j_1} \theta \,\sin^{j_1+j_2-j} (\theta - \phi) \, \sin^{j+j_1-j_2} \phi
\ ,
\end{equation}
where $N$ is some function of the quantum numbers of order unit in the large quantum numbers regime. 
From~\eqref{app:wigner-integral-rep} one can perform an asymptotic expansion in the large scale of the angular momentum quantum numbers.
This approach works fine, and considerably improves the Wigner estimate, only if the region of interest lies inside the allowed or the forbidden region: in both cases it is possible to use saddle points methods to estimate the large order behaviour of the integral~\eqref{app:wigner-integral-rep}. 
The area of the Wigner triangle of eqn.~\eqref{Wigner's-estimate} plays the role of a ``control parameter'' for the approximation, dividing two very different behaviours (both under control) for the saddles.

In the classically allowed region (positive area of the shaded triangle of figure~\ref{Wigner-fig}(a)) there are two separate simple saddles, which go to complex infinity in the complexified $\theta$-$\phi$ plane, when we reduce the positive area towards zero. 
If we are far enough from the transition region, we can easily choose the paths of steepest descent and perform the asymptotic expansion (all the details can be found in~\cite{Reinsch:1999}).
This is of interest for us, and we quote the result of this saddle point analysis in equation~\eqref{3j-deep}.
 
By smoothly varying the area of the Wigner triangle across zero into the classically forbidden region we recognise a single simple saddle point coming down from complex infinity, at a finite distance from the horizontal axis once the area of the Wigner triangle becomes imaginary.
From this moment on, it is possible to estimate the integral by the steepest descent method as well (for the result see~\cite{Reinsch:1999}).

Difficulties arise if we study the behaviour of the integral in the neighbourhood of the critical zero area: in this region the two saddles of the allowed region are very close to each other, and tend in the limit of zero area to ``scatter'' into the single saddle of the forbidden region. 
This problem is known to mathematicians, and goes under the name of ``expansion in presence of coalescent saddles'' (for a detailed analysis of the matter look at~\cite[Chapter 9]{Bleistein1975}), and the technology is known to overcome it. 
However, in the specific case of 3j-symbols, the two saddles of the classically allowed region coalesce at complex infinity, and moreover they flow there at an``essentially infinite'' velocity varying continuously the area around the critical zero value, making the saddle point analysis impossible (since the saddles disappear at every order in the expansion parameter).

For our application in chapter~\ref{ch:new-theory}, namely to determine the limit of the three-point function for the fields $\Phi_{Q,n}$, we will see that we are precisely in this transition region, and we have to follow a different route to deal with the limit. 
We will be precise in section~\ref{app:wigner:sec:transition-region} of this appendix.

For the correlators of the fields $\Phi_{p,m}$ of chapter~\ref{ch:free-limit}, we are in the classically allowed region, and the (improved) Wigner estimate applies. 
We give the due details in the next section.

\section{Deeply in the allowed region}\label{app:wigner:sec:allowed-region}
\begin{figure}[h!]
  \centering
    \includegraphics{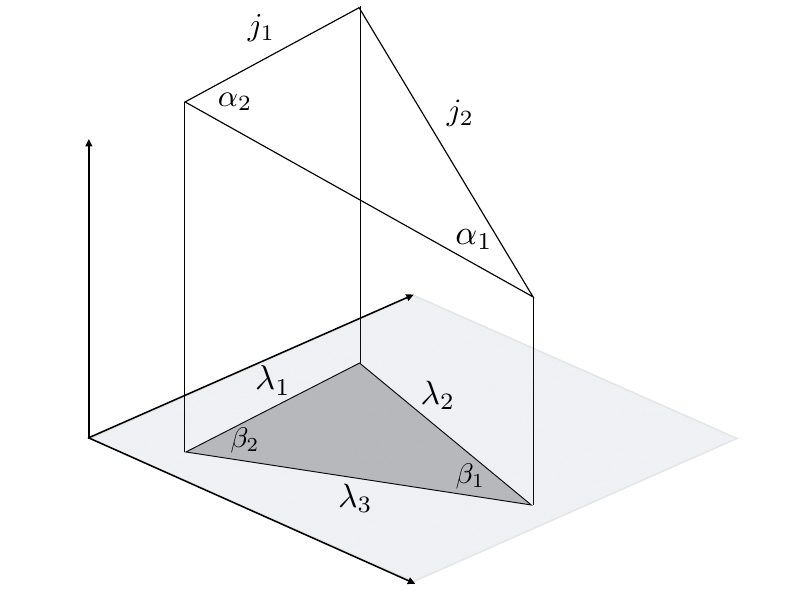}
\caption{\label{fig:PR-angles}Ponzano-Regge angles defined in
equations~(\ref{PR-angles}) and~(\ref{def-alphas}): the $\alpha_i$ are
the internal angles of the triangle formed by the $j_i$ labels; the
$\beta_i$ are the internal angles of the triangle projected on the
xy-plane (where the $\mu_{i}$ measure the z-components of the angular momenta); $\gamma_i$ (not present here) is the angle between the outer normals to the faces adjacent to the edge $j_i$.}
\end{figure}

We are interested in the region of the parameter space of 3j-symbols
in which the angular momentum labels $j_{i}$ scale like $j_i \propto \sqrt{k}$ and the
magnetic labels $\mu_i$ stay finite in the limit of large $k$. In this
range we are deeply inside the classically allowed region,
and we can use the approximation methods derived
in~\cite{Reinsch:1999}. In particular we find there \cite[eq.\ (3.23)]{Reinsch:1999}
\begin{equation}\label{3j-deep}
\begin{pmatrix}
j_1 & j_2 & j_3 \\
\mu_1 & \mu_2 & \mu_3
\end{pmatrix}
\simeq
2 I_{j_1 \, \mu_1 \, j_2 \, \mu_2 \, j_3 \, \mu_3} \,
(-1)^{j_1-j_2-\mu_3}\sqrt{\frac{j_3}{2j_3+1}}\frac{\cos\left[\chi + \frac{\pi}{4} - \pi(j_3+1) \right]
}{(4\pi A (\lambda_{1},\lambda_{2},\lambda_{2}))^{1/2}}\ ,
\end{equation}
where $\chi$ is defined as
\begin{equation}\label{chi-def}
 \chi =(j_1+\tfrac{1}{2})\gamma_1+(j_2+\tfrac{1}{2})
 \gamma_2+(j_3+\tfrac{1}{2})
 \gamma_3+\mu_2 \beta_1-\mu_1 \beta_2\ .
\end{equation}
We use the Ponzano-Regge angles $\gamma_{1,2,3},\beta_{1,2}$
(see~\cite{Ponzano:1968} and figure~\ref{fig:PR-angles}) which through
their cosines read
\begin{subequations}\label{PR-angles}
\begin{align}
\cos\gamma_1 &= {\frac{\mu_3(j_1^2 + j_2^2 - j_3^2) - \mu_2(j_1^2 + j_3^2 -j_2^2)}
{4A (j_{1},j_{2},j_{3}) \lambda_1}} & 
\cos\beta_1 &= {\frac{\lambda_3^2 + \lambda_2^2 - \lambda_1^2}
{2 \lambda_2 \lambda_3}}\\[5pt]
\cos\gamma_2 &= {\frac{\mu_1(j_3^2 + j_2^2 - j_1^2) - \mu_3(j_2^2 + j_1^2 -j_3^2)}
{4A (j_{1},j_{2},j_{3}) \lambda_2}} & 
\cos\beta_2 &= {\frac{\lambda_1^2 + \lambda_3^2 - \lambda_2^2}
{2 \lambda_1 \lambda_3}}\\[5pt]
\cos\gamma_3 &= {\frac{\mu_2(j_1^2 + j_3^2 - j_2^2) - \mu_1(j_3^2 + j_2^2 -j_1^2)}
{4A (j_{1},j_{2},j_{3}) \lambda_3}} \ .& &  
\end{align}
\end{subequations}
Here,
\begin{align}
\lambda_i &= \sqrt{j_i^2-\mu_i^2} \qquad i = 1,2,3 \ ,\\
\intertext{and}
A (x_{1},x_{2},x_{3}) &=\frac{1}{4}
\sqrt{(x_3 + x_1 + x_2) (-x_3 + x_1 + x_2) (x_3 - x_1 + x_2) (x_3 + x_1 - x_2)}
\label{area}
\end{align}
is the area of the triangle with side lengths $x_{i}$.

The quantity $I_{j_1 \, \mu_1 \, j_2 \, \mu_2 \, j_3 \, \mu_{3}}$ appearing in
equation~\eqref{3j-deep} is defined as
\begin{multline}
I_{j_1 \, \mu_1 \, j_2 \, \mu_2 \, j_3 \, \mu_3} =
\sqrt{ \frac{(j_3+1/2)(j_3+j_1+j_2)}{j_3(j_3+j_1+j_2+1)} } \\
\times\ 
\frac{f(j_1+\mu_1)\, f(j_1-\mu_1)\, f(j_2+\mu_2)\, f(j_2-\mu_2)\,
f(j_3+\mu_3)\,f(j_3-\mu_3)}
{f(j_1+j_2+j_3)\, f(j_1+j_2-j_3)\, f(j_1-j_2+j_3)\, f(-j_1+j_2+j_3)} \  ,
\end{multline}
where $f(n)$ is the square root of the ratio of $n!$ to the Stirling
approximation of $n!$, and has the following large $n$ behaviour
\begin{equation}
f(n) = \sqrt{\frac{n!}{\sqrt{2 \pi n}\,n^n e^{-n}}} =1+\frac{1}{24 n}+\mathcal{O} \left(\frac{1}{n^2}\right)\ .
\end{equation}

We now consider the situation where the labels $j_{i}$ are
proportional to $\sqrt{k}$ for large $k$ while keeping $\mu_i$
finite. In this regime we have
\begin{equation}
I = 1+\mathcal O\left(k^{-1/2}\right)\ ,
\end{equation}
and the angles behave as follows:
\begin{subequations}
\begin{align}
\cos\gamma_{1,2,3} & = f_{1,2,3}+\mathcal{O}(k^{-3/2})\\
\cos\beta_1 & =\frac{-j_1^2+j_2^2+j_3^2}{2j_2j_3}+\mathcal{ O}(k^{-1})
& \cos\beta_2&=\frac{j_1^2-j_2^2+j_3^2}{2j_1j_3}+\mathcal {O}
(k^{-1})\ ,
\end{align}
\end{subequations}
where we used the definitions
\begin{subequations}\label{def-fs}
\begin{alignat}{3}
 f_1&=\frac{\mu_3(j_1^2+j_2^2-j_3^2)-\mu_2(j_1^2-j_2^2+j_3^2)}{4A (j_{1},j_{2},j_{3}) j_1} &&\propto k^{-\frac{1}{2}}\\
 f_2&=\frac{\mu_1(-j_1^2+j_2^2+j_3^2)-\mu_3(j_1^2+j_2^2-j_3^2)}{4A (j_{1},j_{2},j_{3}) j_2}&&\propto k^{-\frac{1}{2}}\\
 f_3&=\frac{\mu_2(j_1^2-j_2^2+j_3^2)-\mu_1(-j_1^2+j_2^2+j_3^2)}{4A (j_{1},j_{2},j_{3}) j_3}&&\propto k^{-\frac{1}{2}}\ .
\end{alignat}
\end{subequations}
Inverting~\eqref{PR-angles} and expanding in $k$ we get
$\gamma_{1,2,3}=\frac{\pi}{2}-\frac{f_{1,2,3}}{k^{1/2}}+\mathcal{O}(k^{-3/2})$,
so that $\chi$ of \mbox{eq.\ \eqref{chi-def}} becomes
\begin{align}
 \chi = \frac{\pi}{2}(j_1+j_2+j_3) 
+ \bigg[\frac34\pi&-(j_1f_1+j_2f_2+j_3f_3)-\mu_1\cos^{-1}\frac{j_1^2-j_2^2+j_3^2}{2j_1j_3}\nonumber\\
&+ \mu_2\cos^{-1}\frac{-j_1^2+j_2^2+j_3^2}{2j_2j_3}\bigg]+\mathcal{O}(k^{-1/2})\ .
\end{align}
Since $\sum j_i f_i=0$, we have
\begin{equation}
\begin{split}
&\cos\left[\chi + \frac{\pi}{4} - \pi(j_3+1)
\right]=\cos\left[(j_1+j_2-j_3)\frac{\pi}{2}+\mu_2\alpha_1-\mu_1\alpha_2\right]
\end{split}
\end{equation}
with (see figure~\ref{fig:PR-angles})
\begin{align}\label{def-alphas}
 \alpha_1&:=\arccos \frac{-j_1^2+j_2^2+j_3^2}{2j_2 j_3} &
 \alpha_2&:=\arccos \frac{j_1^2-j_2^2+j_3^2}{2j_1 j_3}\ .
\end{align}
The remaining factor behaves as $\sqrt{\frac{j_3}{2j_3+1}}=\frac{1}{\sqrt2}\left(1+\mathcal O (k^{-1/2})\right)$.\\
Collecting all the pieces we get
\begin{equation}\label{3j-asympt-app}
\begin{pmatrix}
j_1 & j_2 & j_3 \\
\mu_1 & \mu_2 & \mu_3
\end{pmatrix} 
=\frac{(-1)^{j_1-j_2-\mu_3}}{\sqrt{2\pi A
(j_{1},j_{2},j_{3})}}\cos\left[(j_1+j_2-j_3)\frac{\pi}{2}+\mu_2\alpha_1-\mu_1\alpha_2
\right]\left(1+\mathcal{O}(k^{-1/2}) \right)\ .
\end{equation}

\section{On the transition region}\label{app:wigner:sec:transition-region}
When we discuss the limit of the three-point functions for the charged
fields $\Phi_{Q_{i},n_{i}}$ of chapter~\ref{ch:new-theory} we are led to consider the asymptotics of
the 3j-symbols for quantum numbers\footnote{In the main text we use
$l_{i}=2j_{i}$ and $m_{i}=2\mu_{i}$.} $j_{i}=|\mu_{i}|+n_{i}$, where
$n_{i}$ is kept fixed, and the $|\mu_{i}|$ grow linearly in a
parameter $k$.

The 3j-symbol vanishes unless the usual conditions on the addition of
angular momenta are satisfied, namely 
\begin{equation}
\mu_1+\mu_2+\mu_3=0 \quad \text{and}\quad j_{i_{1}}+j_{i_{2}}\geq j_{i_{3}}
\end{equation}
for any permutation $i_{1},i_{2},i_{3}$ of $1,2,3$. If we assume
$\mu_{1},\mu_{2}>0$ and $\mu_{3}<0$, then for large $|\mu_{i}|$ the
conditions on the $j_{i}$ reduce to one condition 
$n_1+n_2\geq n_{3}$.

Because the $z$-components of the angular momenta $\vec{J}^{(i)}$ are
close to maximal in our case, their projections to the $x$-$y$-plane are
short and have lengths 
\begin{equation}
\lambda_{i}=\sqrt{|\mu_{i}| (2n_{i}+1)+n
(n+1)} \ ,
\end{equation}
which only grow with the square root of $k$. This means that
the quantity $A$ given in~\eqref{app:defA}, which describes the area
of the triangle in the $x$-$y$-plane provided it exists, is relatively
small. Thus we are in the transition region between the
classically allowed and the classically forbidden region, and cannot
use the classical Wigner estimate.

Instead we can get the asymptotic behaviour directly from the Racah
formula~\eqref{Racah}. Firstly, we have to understand the
range of $z$ in the sum in~\eqref{Racah}. In the limit of large
$\mu_{i}$ we see that the arguments $j_2+\mu_2-z$ and $j-j_2+\mu_1+z$
do not constrain the sum since they are both surely positive. Bounds
to the summation range are given by the other factorials in the denominator
of equation~\eqref{Racah}, and the summation range is
\begin{equation}
\mathcal{I}:= \{z\in\mathbb{Z}\,|\, z\geq 0,\  z\geq n_1-n_3 ,\ z\leq n_1+n_2-n_3,\
z\leq n_1\}\ .
\end{equation}
Even in the limit of large $|\mu_{i}|$ the summmation range stays
finite, and its lower bound is either zero or $n_1-n_3$ depending on
its sign.

The 3j-symbols can be rewritten as
\begin{equation*}
\begin{split}
& \begin{pmatrix} j_{1}& j_{2} & j_{3} \\
\mu_{1} & \mu_{2} & \mu_{3}
\end{pmatrix} = (-1)^{j_{1}-j_{2}-\mu_{3}}\\
&\times
\underbrace{
\left(\frac{[n_1+n_2-n_3]!}{[2 (|\mu_{1}|+|\mu_{2}|)
+n_1+n_2+n_3+1]!}\right)^{\!1/2}
\times\big((n_{1})!(n_2)!(n_3)![2 (|\mu_{1}|+|\mu_{2}|) +n_3]!\big)^{1/2}}_{\mathbf{N}}\\
&\times
\sum_{z\in\mathcal{I}}\underbrace{
\frac{\left([2|\mu_{1}|+n_1-n_2+n_3]![2|\mu_{1}|+n_1]!\right)^{1/2}}{\left([2|\mu
_{1}|+n_3-n_2+z]!\right)^{1/2}\left([2|\mu_{1}|+n_3-n_2+z]!\right)^{1/2}}
}_{\mathbf A}\\
&\qquad \times \underbrace{
\frac{\left([2|\mu_{2}|+n_2-n_1+n_3]![2|\mu_{2}|+n_2]!\right)^{1/2}}{\left([2|\mu
_{2}|+n_2-z]!\right)^{1/2}\left([2|\mu_{2}|+n_2-z]!\right)^{1/2}}
}_{\mathbf B}
\\
& \qquad \times \underbrace{\frac{(-1)^z}{z!}\frac{1}{[n_1+n_2-n_3-z]![n_1-z]![n_3-n_1+z]!}}_{\mathbf C}
\ .\end{split}
\end{equation*}
Using the fact that $k$ is large we are able to recast parts $A$ and $B$ using that
\begin{equation}\label{limitoffactorials}
\frac{(K+a)!}{K!}=\frac{K!}{K!}\times
(K+1)\dots(K+a)=K^a\left(1+\mathcal{O}\left(\frac{1}{K}\right)
\right)\qquad \text{for large}\ K
\end{equation}
so that the leading contributions read
\begin{equation}
A\approx (2|\mu_{1}|)^{n_1+\frac{n_{2}-n_{3}}{2}-z}\ ,\qquad B\approx
(2|\mu_{2}|)^{z-\frac{n_{1}-n_{3}}{2}}\ .
\end{equation}
Similarly, the factor $N$ can be approximated by
\begin{equation}
N\approx (2|\mu_{1}|+2|\mu_{2}|)^{-\frac{n_{1}+n_{2}+1}{2}}\times\sqrt{n_{1}!n_{2}!n_{3}![n_{1}+n_{2}-n_{3}]!}\ .
\end{equation}
The 3j-symbol then reads
\begin{align}
& \begin{pmatrix} j_{1}& j_{2} & j_{3} \\
\mu_{1} & \mu_{2} & \mu_{3}
\end{pmatrix} = (-1)^{2|\mu_{1}|+n_{1}-n_{2}}\sqrt{n_{1}!n_{2}!n_{3}![n_{1}+n_{2}-n_{3}]!}
(2|\mu_{1}|+2|\mu_{2}|)^{-\frac{n_{1}+n_{2}+1}{2}}\nonumber \\
&\quad \times \sum_{z\in\mathcal{I}} (-1)^{z} \frac{1}{z![n_1+n_2-n_3-z]![n_1-z]![n_3-n_1+z]!} 
(2|\mu_{1}|)^{n_1+\frac{n_{2}-n_{3}}{2}-z}(2|\mu_{2}|)^{z-\frac{n_{1}-n_{3}}{2}}\ .
\end{align}
Introducing the notation
\begin{equation}
J= \frac{n_{1}+n_{2}}{2}\quad ,\quad M=\frac{n_{1}-n_{2}}{2}\quad ,\quad 
M' = -\frac{n_{1}+n_{2}}{2}+n_{3}\ ,
\end{equation}
and
\begin{equation}
\cos \beta = \frac{|\mu_{1}|-|\mu_{2}|}{|\mu_{1}|+|\mu_{2}|}\ ,
\end{equation}
we can express the asymptotic form of the 3j-symbol as
\begin{equation}\label{app:3jasymptotic}
\begin{pmatrix} j_{1}& j_{2} & j_{3} \\
\mu_{1} & \mu_{2} & \mu_{3}
\end{pmatrix} \approx (-1)^{2|\mu_{1}|+n_{3}-n_{2}}
(2|\mu_{1}|+2|\mu_{2}|)^{-\frac{1}{2}} d^{J}_{M',M} (\beta) \ .
\end{equation}
Here, $d^{J}_{M',M} (\beta)$ denotes the Wigner d-matrix~\cite{Wigner:book,Varsalovic:book},
\begin{align}
&d^{J}_{M',M} (\beta) = \sqrt{(J\!+\!M')!(J\!-\!M')!(J\!+\!M)!(J\!-\!M)!} \nonumber \\
&\quad \times \sum_{z}
\frac{(-1)^{M'-M+z}}{(J\!+\!M\!-\!z)!z!(M'\!-\!M\!+\!z)!(J\!-\!M'\!-\!z)!}
\left(\cos \tfrac{\beta}{2} \right)^{2J+M-M'-2z}
\left(\sin \tfrac{\beta}{2} \right)^{M'-M+2z} \ .
\end{align}
The Wigner d-matrix is expressible in terms of standard $_{2}F_{1}$
hypergeometric functions. More precisely, for $n_{1}\leq n_{3}$, we find
\begin{align}
&\begin{pmatrix}j_{1}&j_{2}&j_{3}\\
\mu_{1}&\mu_{2}&\mu_{3}\end{pmatrix} = 
(-1)^{2\mu_1+n_{1}-n_{2}} \frac{1}{(n_{3}-n_{1})!} \sqrt{\frac{n_{2}!n_{3}!}{n_{1}!(n_{1}+n_{2}-n_{3})!}}\,(2|\mu_{1}|+2|\mu_{2}|)^{-\frac{1}{2}}\nonumber\\  
& \qquad \times  \frac{|\mu_{1}|^{n_1+\frac{n_2-n_3}{2}}|\mu_{2}|^{\frac{n_{3}-n_{1}}{2}}}{(|\mu_{1}|+|\mu_{2}|)^{\frac{n_{1}+n_{2}}{2}}} \, {}_2F_1\left(n_3-n_2-n_1,-n_1;n_3-n_1+1;-\frac{|\mu_{2}|}{|\mu_{1}|}\right)\left(1+\mathcal{O}(1/k)\right)\ ,
\end{align}
whereas for $n_{1}\geq n_{3}$ we have
\begin{align}
&\begin{pmatrix}j_{1}&j_{2}&j_{3}\\
\mu_{1}&\mu_{2}&\mu_{3}\end{pmatrix} = 
(-1)^{2\mu_1+n_{1}-n_{2}}\frac{1}{(n_{1}-n_{3})!} \sqrt{\frac{n_{1}!}{n_{2}!n_{3}!}}
\,(2|\mu_{1}|+2|\mu_{2}|)^{-\frac{1}{2}}\nonumber\\  
&\qquad \times \frac{|\mu_{1}|^{\frac{n_{2}+n_{3}}{2}}|\mu_{2}|^{\frac{n_{1}-n_{3}}{2}}}{(|\mu_{1}|+|\mu_{2}|)^{\frac{n_{1}+n_{2}}{2}}}\, {}_2F_1\left(-n_3,-n_2;n_{1}-n_{3}+1;-\frac{|\mu_{2}|}{|\mu_{1}|}\right)\left(1+\mathcal{O}(1/k)\right)\ .
\end{align}

\chapter{Elements of boundary CFT}\label{app:vanilla}

In this appendix we are interested in giving a glimpse of the vast field of boundary conformal field theory (BCFT), since in the main text we make several times use of these technologies and concepts.
We present a very brief and incomplete review of the so-called boundary state formalism.

The introduction of boundary conditions on the world sheet breaks in general the symmetry algebra of the bulk CFT.
We are interested in conformal boundary conditions, hence boundaries on the real line ($z=\bar z$) of a Riemann surface $\S$ parameterised by holomorphic coordinates $z,\bar z$  (in this discussion we restrict our attention to the sphere), which leave invariant at least one copy of the conformal algebra at the locus $z=\bar z$.

The (anti)-holomorphic fields of the bulk CFT correspond to representations of the general chiral symmetry algebra $\cW_L$ ($\cW_R$).
We restrict the analysis to the cases $\cW_L=\cW_R=\cW$, so that the Hilbert space of the bulk CFT decomposes as
\begin{equation}
\Hilb^{\text{bulk}}=\bigoplus_{i,\tilde i\in \cS}\Hilb_i\otimes \Hilb_{\tilde i}\ ,
\end{equation}
where $i,\tilde i$ are labels of holomorphic and anti-holomorphic representations of~$\cW$, respectively, belonging to the bulk spectrum~$\cS$ of the CFT.
Furthermore, we look only at ``maximally symmetric'' boundary conditions, which are boundaries preserving (up to automorphisms) one copy of the algebra $\cW$.
This is formalised as
\begin{equation}\label{vanilla:gluing-generic}
W(z)=\Omega(\overline W(\bar z))\ ,\qquad \text{for}\ z=\bar z\ ,
\end{equation}
where~$W\ (\overline{W})$ is an arbitrary generator of the (anti)-holomorphic copy of (the vertex operator algebra associated to) $\cW$, and~$\Omega$ is an outer automorphism of $\cW$, which must act trivially on the Virasoro field, to exactly preserve conformal invariance (in the example of $N=2$ minimal models analysed in section~\ref{ch3:sec:BC}, B-type boundary conditions are associated to the trivial automorphism~$\O=\id$, whence A-type boundary conditions are given by the mirror automorphism $\O=\O_M$ defined in equation~\eqref{ch3:mirror-automorphism-def-BC-section}).
Equation~\eqref{vanilla:gluing-generic} tells us how the chiral generators of holomorphic and anti-holomorphic sectors are ``glued" together along the boundary at $z=\bar z$.


The boundary state~$|\a\ket_{\O}$ (associated to a boundary condition~$\a$ and to the outer automorphism~$\O$) is a sort of coherent state representation of $\cW$, which summarises the information about how bulk fields couple to the boundary.
The gluing of equation~\eqref{vanilla:gluing-generic} is translated into the following identity for the boundary state 
\begin{equation}\label{vanilla:generic-condition-boundary-states}
\left (W_n- (-1)^{h_{W}}\Omega(\overline {W}_{-n})\right) |\a \ket_{\O}=0\ ,
\end{equation}
where~$W_n$ are the modes of the field~$W(z)$ and~$h_W$ is its conformal dimension.

If the action of the gluing automorphism on the representation labels reads $\omega (\tilde i)=i^+$  ($i^+$ is the label conjugate to $i$, i.e. whose fusion with $i$ gives once the identity), equation~\eqref{vanilla:generic-condition-boundary-states} admits for each sector $i$ of the representation space of the algebra~$\cW$ a (unique up to a constant, and of course depending on $\O$) solution $|i\iket_{\O}$, called Ishibashi state~\cite{Ishibashi:1988kg}.

Ishibashi states couple to bulk fields belonging to the set $\Hilb_i\otimes\Hilb_{\omega^{-1}(i^+)} $ (in the case of $N=2$ minimal models, the mirror automorphism flips the sign of the labels $m,s$, namely $(m,s)\mapsto (-m,-s)$. The conjugated label to $(l,m,s)$ is $(l,-m,-s)$.
Consequently B-type Ishibashi states couple to $\Hilb_{l,m,s}\otimes\Hilb_{l,-m,-s}$ bulk ground states, whence A-type ones couple to ground states belonging to the set $\Hilb_{l,m,s}\otimes\Hilb_{l,m,s}$.
As a consequence, in a strictly diagonal model, B-type boundary conditions couple only to NS-sectors ($s\=0\mod 2$), and to chargeless excitations ($m\=0$), whence A-type boundary conditions couple to all bulk ground states).

The set of Ishibashi states is commonly normalised as
\begin{equation}
{}_{\O}\ibra i| q^{L_0-\frac{c}{24}} |j\iket_{\O}=\d_{i,j}\chi_i(q)\ ,
\end{equation}
and a generic boundary state encoding boundary conditions preserving the algebra~$\cW$ reads
\begin{equation}\label{vanilla:boundary-states}
|\a \ket_{\Omega}=\sum _{i}B^i_{\a}|i\iket_{\Omega}\ ,
\end{equation}
with $B_{\a}^i$ unknown coefficients to be determined.

\begin{figure}[t]
\begin{tikzpicture}   
\fill[bottom color=gray!40, top color=white] (0,0) rectangle (6,4); 
\foreach \x/\y in {0.7/2,1.5/3,2.3/1,4/2.8,5.2/1.2} \fill (\x,\y) circle (1.5pt);
\foreach \x/\y/\z in {0.7/2/1,1.5/3/2,2.3/1/3,4/2.8/4,5.2/1.2/5} \node[below] at (\x,\y) {$\phi_{\z}$};
\draw [line width=1.6pt] (0,0) -- (6,0); 
\node[below] at (3,0) {Boundary};
\fill[bottom color=gray!40, top color=white] (9,0) rectangle (15,4);
\draw [line width=1.6pt] (9,0) -- (15,0); 
\node[below] at (12,0) {Boundary};
\draw [->,line width=1.6pt] (6.5,1.7) -- (8.5,1.7); 
\node[above] at (7.5,1.8) {Bulk OPE};
\fill (12,2) circle (1.5pt);
\node[above right] at (12,2) {$\Phi$};
\end{tikzpicture}
\caption{The bulk field insertions on the left are brought pairwise close to each other, and fused by bulk OPE. If we repeat this procedure for any field insertion we end up with some coefficient times a one-point function: the new piece of data that we need to encode the effect of the boundary, is therefore the one-point function for each bulk primary field.}
\label{vanilla:fig:n-point-to-one-point}    
\end{figure}
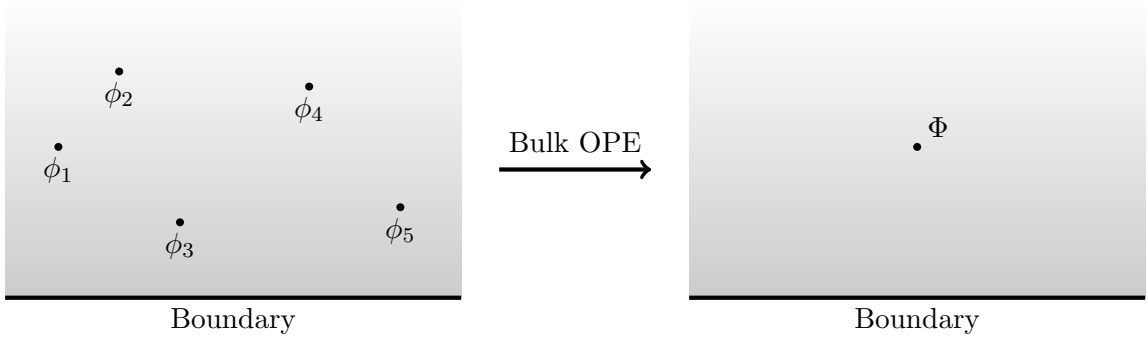

We assume that, locally in the interior, the bulk fields are not affected by the presence of the boundary.
As a consequence, the bulk OPE on the upper half plane are left invariant if $z\neq \bar z$.
The effect on correlators of introducing a boundary on the sphere is hence encoded in the one-point functions of the bulk fields corresponding to states in $\Hilb_i\otimes \Hilb_{\tilde i}$ (as pictorially sketched in figure~\ref{vanilla:fig:n-point-to-one-point}).
The exact proof of this statement can be found in\cite{Recknagel:1998sb,Cardy:1991tv}, and we quote the result here:
\begin{equation}\label{vanilla:coeff-one-pt}
\bra \phi_{i,\tilde i}(z,\bar z)\ket_{\a}=\frac{A^{\a}_{i,\tilde i}\ \delta_{\tilde i,\omega^{-1} (i^+)}}{|z-\bar z|^{2h_i}}\quad \quad  
B^{i+}_{\a}=A_{i\omega^{-1}(i^+)}\equiv A_i^{\a}\ .
\end{equation}
The coefficients of the boundary states in equation~\eqref{vanilla:boundary-states} are subject to two sets of constraints, the first coming from modular invariance~\cite{Cardy:1989ir}, the second from the associativity of the OPE in presence of a boundary~\cite{Cardy:1991tv}.

We discuss the first one here, the so-called Cardy constraint.

Once defined appropriate boundary states through the one-point functions, one computes the tree-level transition amplitude in the closed string sector, namely a cylinder amplitude between two boundary states.
Using world-sheet duality (the fact that a cylinder diagram between two D-branes labeled by~$\a,\b$ is equivalent under a modular S-transformation to an annulus diagram of an open string with Chan-Paton labels~$\a,\b$) one can rephrase the transition between two given boundary states as a one-loop diagram for an open string stretched between the boundaries labeled by $\a$ and $\b$, which we indicate as $Z_{\a\b}$.

We have assumed that the boundary does not break the symmetry described by $\cW$.
The space of states of the boundary CFT must be therefore decomposable into irreducible representations of $\cW$, and the annulus amplitude reads then:
\begin{equation}\label{vanilla:one-loop-open}
Z_{\a \b}(\tilde{q}):=\Tr_{\Hilb^{\text{open}}}\tilde{q}^{L_0-\frac{c}{24}}=\sum_iN^{\b}_{i\a}\chi_i(\tilde q)\ ,
\end{equation}
where $N^{\b}_{i\a}$ is a set of positive integers, encoding the multiplicity of the states circulating the loop, and $L_0$ acts on the open string Hilbert space.
This decomposition gives the (generating function of the) perturbative spectrum of the open string stretched between two D-branes, but it is very implicit.
The definition of boundary states, and world-sheet duality help us in making this amplitude computable.

From the definition of boundary states and the normalisation of Ishibashi states, the tree-level overlap between boundary states reads
\begin{equation}\label{vanilla:tree-level-closed}
{}_{\O}\bra \Theta \b|q^{\12(L_0+\overline{L}_0-\frac{c}{12})}|\a\ket_{\O} =\sum_jA_{j^+}^{\a}A_j^{\b}\chi_j(q)\ ,
\end{equation}
where we have used the identification of the coefficients in equation~\eqref{vanilla:boundary-states} with the one-point functions, as given in equation~\eqref{vanilla:coeff-one-pt}; $L_0$ acts on the bulk Hilbert space now.
The operator $\Theta$ is a CPT conjugation operator.

We can then write the Cardy constraints equating the right hand sides of equations~\eqref{vanilla:one-loop-open} and~\eqref{vanilla:tree-level-closed} by world-sheet duality
\begin{equation}\label{vanilla:cardy-constraints}
\sum_jA_{j^+}^{\a}A_j^{\b}\chi_j(q)=\sum_jA_{j^+}^{\a}A_j^{\b}S_{ji}\chi_i(\tilde q)\=\sum_iN_{i\a}^{\b}\ \chi_i(\tilde q)\ .
\end{equation}

One solution to these constraints is given by the Cardy states~\cite{Cardy:1989ir}
\begin{equation}\label{vanilla:cardy-states}
|\a\ket_{\O}=\sum_i\frac{S_{\a i}}{\sqrt{S_{0i}}}|i\iket_{\O}\ ,
\end{equation}
where $S_{\a\a'}$ are the components of the modular S-matrix of the representations of the algebra preserved by the boundary.
They are labeled like irreducible representations of the algebra~$\cW$. 
Cardy conditions are automatically satisfied in virtue of the Verlinde formula~\cite{Verlinde:1988sn}, relating fusion coefficients (integers in rational theories) with the S-matrix elements:
\begin{equation}
n_{\a\a'}^{\b}=\sum_{\gamma}\frac{S_{\a\g}S_{\a'\g} S^*_{\b\g}}{S_{0\g}}\ .
\end{equation}
Cardy states satisfy Cardy constraints with $N_{i\a}^{\b}=n_{i\a}^{\b}$.


\newpage

\bibliographystyle{nb}

\addcontentsline{toc}{chapter}{Bibliography}

\chapter*{Acknowledgements}\markboth{ACKNOWLEDGEMENTS}{ACKNOWLEDGEMENTS}
\addcontentsline{toc}{chapter}{Aknowledgements}

The last three years have been an extraordinary intense period, and many people have played an important role in this experience.

First and foremost I want to express my sincerest gratitude to my supervisor and friend, Stefan Fredenhagen.
It has been a great fortune to have been exposed for a long time to his deep vision and insights, and to have benefited from his support, his encouragement and advices in so many occasions.
Thanks Stefan!

I also want to thank my official adviser, Hermann Nicolai, who gave me the opportunity to work in the really inspiring environment of the AEI, and my second adviser, Stefan Theisen. 
I am grateful to Christine Gottschalkson, to the IT support team and all the administration crew here at the Institute, for their very unbuerocratic help with every problem with papers and circuits that I had to face on a daily basis.

I am very happy to thank my good friends Stefan Pfenninger, Rouven Frassek and Parikshit Dutta, for sharing our daily struggles and enthusiasms, for extremely valuable conversations and (not less important), for bringing fun into the long days in the office and in the libraries in Berlin. 
I thank all the present and past collegues PhD students here at the AEI, with whom I spent long coffee breaks, funny excursions, and terrible food. In particular: Francesco Caravelli, Philip Fleig, Filippo Guarnieri, Carlos Guedes, Despoina Katsimpouri, Pan Kessel, Michael Koehn, Carlo Meneghelli, Burkhard Schwab, Rui Sun, Johannes Th\"uringen.

During this time I have benefited of uncountable illuminating conversations and discussions with many brilliant physicists.
Among them I would like to aknowledge: Yuri Aisaka, Emil Akhmedov, Costas Bachas, Nicholas Behr, Niklas Beisert, Dario Benedetti, Andrea Campoleoni, Nils Carqueville, Matthias Gaberdiel, Ingo Kirsch, Sebastian Krug, Marcos Mari\~no, Ilarion Melnikov, Cesare Nardini, Rachele Nerattini, Teake Nutma, Daniele Oriti, Sara Pasquetti, Elli Pomoni, Amir Kashani-Poor, Sylvain Ribault, Volker Schomerus, Oliver Schlotterer, Cornelius Schmidt-Colinet, Domenico Seminara, Lorenzo Sindoni, Evgeny Skvortsov, Paulina Suchanek, Massimo Taronna, Jan Troost, Grigory Vartanov, Roberto Volpato.

Last but not least, I thank my family, and all my friends living in various corners of the world.
Without their constant support and trust this work would not have been possible.
\medskip

Thanks!

\markboth{LEBENSLAUF UND PUBLIKATIONSLISTE}{LEBENSLAUF UND PUBLIKATIONSLISTE}
\begin{center}
\phantom{physik macht spass}
\vspace{2.8cm}
{\LARGE \bf Lebenslauf}\\
\vspace*{0.4cm}
{\large Cosimo Restuccia (cosimo.restuccia@aei.mpg.de)}
\end{center}

\vspace*{0.5cm}

\begin{tabular}{cccl}
10/2003 & - & 3/2007 & BSc in Physical Sciences, Universit\`a di Firenze\\
\phantom{1}9/2007 & - & 12/2009 & MSc in Physics and Astrophysics, Universit\`a di Firenze\\
10/2007 & - & \phantom{1}9/2008 & Erasmusjahr, Universit\"at Potsdam und Humboldt Universit\"at\\
11/2008 & - & 12/2009 & Masterarbeit, Max-Planck-Institut f\"ur Gravitationsphysik\\
\phantom{1}1/2010 & - & \phantom{1}4/2013 & Doktorarbeit, Max-Planck-Institut f\"ur Gravitationsphysik
\end{tabular}

\vspace*{3.0cm}

\begin{center}
{\LARGE \bf Publikationsliste}\\
\end{center}

\vspace*{0.5cm}

\begin{itemize}
 \item[-]  S.~Fredenhagen and C.~Restuccia,
  ``The geometry of the limit of N=2 minimal models''
  J.~Phys.\ A {\bf 46} (2013) 045402
  [arXiv:1208.6136 [hep-th]].
  \vspace*{0.2cm}
 \item[-]  S.~Fredenhagen, C.~Restuccia and R.~Sun,
  ``The limit of N=(2,2) superconformal minimal models''
  JHEP {\bf 1210} (2012) 141
  [arXiv:1204.0446 [hep-th]].
    \vspace*{0.2cm}
  \item[-]  S.~Fredenhagen and C.~Restuccia,
  ``DBI analysis of generalised permutation branes''
  JHEP~{\bf 1001} (2010) 065
  [arXiv:0908.1049 [hep-th]].
\end{itemize}

\vspace{3cm}

Cosimo Restuccia\\
Berlin, 30. April 2013

\thispagestyle{empty}

\section*{Selbstst\"andigkeitserkl\"arung}

\vspace{1cm}

Hiermit erkl\"are ich meine Dissertation selbstst\"andig ohne fremde
Hilfe verfasst zu haben und nur die angegebene Literatur und
Hilfsmittel verwendet zu haben.
%
%
%
%
\\[5cm]
Cosimo Restuccia\\
Berlin, 30. April 2013

\end{document}